\documentclass[11pt,hidelinks,a4paper]{article}

\usepackage[utf8]{inputenc}
\usepackage[english]{babel}
\usepackage[T1]{fontenc}

\usepackage{mathtools}
\usepackage{amsmath, amsthm, amssymb,amsfonts}
\usepackage{leftindex}
\usepackage{thmtools,thm-restate}
\usepackage{proof-at-the-end}
\PassOptionsToPackage{hidelinks}{hyperref}
\usepackage[hidelinks]{hyperref}
\usepackage[nameinlink,noabbrev]{cleveref}
\usepackage{placeins}
\usepackage{graphicx}
\usepackage{float}
\usepackage{subcaption}
\usepackage[skip=2pt]{caption}

\usepackage{tikz}
\usetikzlibrary{arrows.meta, positioning}

\usepackage[frozencache]{minted}

\usepackage{dsfont}
\usepackage{physics}
\usepackage{mleftright} %

\usepackage{booktabs}

\usepackage{natbib} 
\usepackage[a4paper, total={16cm, 22cm}]{geometry}

\newtheorem{model}{Model}
\newtheorem{theorem}{Theorem}

\newtheorem{definition}[theorem]{Definition}

\newcommand{\BE}{{\mathbb{E}}}

\newcommand{\BP}{{\mathbb{P}}}

\newcommand{\BR}{{\mathbb{R}}}

\newcommand{\BZ}{{\mathbb{Z}}}

\newcommand{\CD}{{\cal D}}

\newcommand{\CN}{{\cal N}}
\newcommand{\CO}{{\cal O}}

\newcommand{\mX}{{{m_x}}}
\newcommand{\mW}{{{m_w}}}
\newcommand{\mC}{{{m_c}}}
\newcommand{\mD}{{{m_d}}}

\DeclareMathOperator{\Var}{Var} 
\DeclareMathOperator{\Cov}{Cov}

\DeclareMathOperator{\CI}{CI}

\DeclareMathOperator{\AR}{AR}
\DeclareMathOperator{\LM}{LM}
\DeclareMathOperator{\LR}{LR}

\DeclareMathOperator{\Wald}{Wald}
\DeclareMathOperator{\CLR}{CLR}

\DeclareMathOperator*{\argmax}{argmax}
\DeclareMathOperator*{\argmin}{argmin}

\DeclareMathSymbol{\shortminus}{\mathbin}{AMSa}{"39}

\DeclareMathOperator{\diag}{diag}

\DeclareMathOperator*{\plim}{plim}
\DeclareMathOperator{\vecop}{vec}

\newcommand{\lambdamin}[1]{{\lambda_\mathrm{min}\mleft(#1\mright)}}

\def\tX{{\tilde{X}}}

\def\liml{\mathrm{LIML}}
\def\tsls{\mathrm{TSLS}}
\def\ols{\mathrm{OLS}}
\DeclarePairedDelimiterX{\infdivx}[2]{(}{)}{%
  #1\;\delimsize\|\;#2%
}

\newcommand{\Id}{\mathrm{Id}}
\newcommand{\tod}{\overset{d}{\to}}
\newcommand{\toP}{\overset{\BP}{\to}}

\let\phi\varphi
\let\theta\vartheta
\let\epsilon\varepsilon

\let\emptyset\varnothing

\let\leq\leqslant
\let\geq\geqslant

\def\kclass{\mathrm{k}}
\newcommand\numberthis{\addtocounter{equation}{1}\tag{\theequation}}

\pgfkeys{/prAtEnd/malte/.style={end, restate, text proof={Proof}, one big link translated={\hspace{-0.71cm} See proof on page}}}  %
\pgfkeys{/prAtEnd/malte_all_end/.style={normal}}
\pgfkeys{/prAtEnd/malte_intro/.style={end, restate, one big link, text proof={Proof}, text link section}}

\input{./aux/jupyter_preamble}

\author{Malte Londschien
\\
\vspace{0.1cm}\\
{\small Seminar for Statistics, ETH Z\"urich, Switzerland}\\
{\small AI Center, ETH Z\"urich, Switzerland}\\
}

\title{A statistician's guide to weak-instrument-robust inference in instrumental variables regression with illustrations in Python}
\begin{document}

\date{August 2025}
\maketitle

\begin{abstract}
    \noindent We provide an overview of results relating to estimation and weak-instrument-robust inference in instrumental variables regression.
    Methods are implemented in the \texttt{ivmodels} software package for Python, which we use to illustrate results.
\end{abstract}

\section{Introduction}

Instrumental variables regression is essential for empirical economics.
It allows for causal effect estimation in the presence of unobserved confounding, also called endogeneity.
Instruments are unconfounded variables that affect the outcome only through the treatment variable.
They encode variation in the treatment variable, akin to the placebo or treatment assignment in randomized controlled trials.
Instrumental variables regression uses this variation to estimate the causal effect of the treatment on the outcome.

One popular estimator in instrumental variables regression is the two-stage least-squares (TSLS) estimator.
It is computed in two steps.
First, the endogenous treatment variables are regressed on the instruments (the first stage).
Second, the outcome is regressed on the first regression's fitted values (the second stage).
It has desirable properties: If the instruments are strong enough, that is, they encode sufficient variation in the confounded treatment variables, it is consistent and asymptotically Gaussian.
The limiting Gaussian distribution's covariance can be estimated from the data, and the resulting Wald test is often used to construct confidence sets and $p$-values for the causal effect.

Such uncertainty quantification is essential for causal effect estimation to influence policy.
However, as noted by \citet{staiger1997instrumental}, in many applications, the signal-to-noise ratio in the first stage is low.
This leads to the two-stage least-squares estimator being biased towards the biased ordinary least-squares estimator and having a distribution that is not well-approximated by a Gaussian.
\citet{staiger1997instrumental} show that, for a single instrument and endogenous variable, if the expected first-stage $F$-statistic exceeds 10, asymptotically, the two-stage least squares estimator has at most 10\% bias relative to the ordinary least squares estimator and the Wald-based 95\% confidence sets have at least 85\% coverage.
This resulted in the following widely adopted heuristic: If the first-stage $F$-statistic is greater than 10, the instrument is considered strong enough, and Wald-based confidence sets are considered valid.

An alternative to such pre-testing is to use weak-instrument-robust inference.
This relies on tests that have the correct asymptotic distribution, even under weak instrument asymptotics \citep{staiger1997instrumental} where the first stage's signal-to-noise ratio decreases such that the expected first-stage $F$-statistic is of constant order.
Prominent weak-instrument-robust tests include the Anderson-Rubin test \citep{anderson1951estimating}, the conditional likelihood-ratio test \citep{moreira2003conditional}, and the Lagrange multiplier test \citep{kleibergen2002pivotal}.

In some applications \citep[e.g.,][]{card1993using,tanaka2010risk,thams2024identifying} there are multiple endogenous variables.
Then, $p$-values and multivariate confidence sets for the full causal parameter are difficult to interpret.
Partial (or subvector) inference for individual coefficients of the causal parameter, similar to $t$-tests in linear regression, is more useful.
Subvector variants of the Anderson-Rubin test, the conditional likelihood-ratio test, and the Lagrange multiplier test have been proposed by \citet{guggenberger2012asymptotic,kleibergen2021efficient,londschien2024weak}.

In the following, we introduce k-class estimators, a family of instrumental variables regression estimators that includes the ordinary least-squares, two-stage least-squares, and the limited information maximum likelihood estimator.
We introduce the Wald, Anderson-Rubin, conditional likelihood-ratio, and Lagrange multiplier tests, their subvector variants, and the confidence sets resulting from their inversion.
We also introduce auxiliary tests such as the J-statistic \citep{sargan1958estimation,hansen1982large} and its LIML variant \citep{guggenberger2012asymptotic} of overidentifying restrictions, the Cragg-Donald test \citep{cragg1997inferring}, related to \citeauthor{anderson1951estimating}'s \citeyearpar{anderson1951estimating} likelihood-ratio test of reduced rank, a multivariate extension of the first-stage F-test, and \citeauthor{scheidegger2025residual}'s (\citeyear{scheidegger2025residual}) residual prediction test of misspecification.
We provide proofs or proof sketches for all results in appendix \ref{app:proofs}.
These are mostly not novel and we reference to the original literature.

All estimators and tests introduced in this manuscript are implemented in the \href{https://github.com/mlondschien/ivmodels/}{\texttt{ivmodels}} software package for Python, available on \href{https://pypi.org/project/ivmodels/}{\texttt{PyPI}} and \href{https://anaconda.org/conda-forge/ivmodels}{\texttt{conda-forge}}.
See the GitHub repository at \href{https://github.com/mlondschien/ivmodels/}{\texttt{github.com/mlondschien/ivmodels}} and the documentation at \href{https://ivmodels.readthedocs.io/}{\texttt{ivmodels.readthedocs.io}} for more details.
We use \citeauthor{card1993using}'s (\citeyear{card1993using}) application estimating the causal effect of education on log-wages as a running example throughout this manuscript and apply the estimators and tests introduced in this manuscript to their dataset.

\section{Model setup and assumptions}
\label{sec:intro_to_iv_tests:model}
\begin{figure}[!h]
    \centering
    \begin{tikzpicture}[
        node distance=2cm and 2cm,
        >=Stealth,
        every node/.style={draw, circle, minimum size=1cm, inner sep=0pt},
        dashednode/.style={draw, circle, minimum size=1cm, inner sep=0pt, dashed},
        dashedarrow/.style={->, dashed}
    ]

    \node (Z) at (0,1.5) {Z};
    \node (X) at (2,0) {X};
    \node (Y) at (5.5,0) {Y};
    \node[dashednode] (U) at (3.5,1.5) {U};

    \draw[->] (Z) -- (X) node[pos=0.3, below, draw=none] {$\Pi$};
    \draw[->] (X) -- (Y) node[pos=0.5, below, draw=none, yshift=5] {$\beta_0$};
    \draw[dashedarrow] (U) -- (X);
    \draw[dashedarrow] (U) -- (Y);

    \end{tikzpicture}
    \caption{
        \label{fig:iv_graph_X}
        Causal graph visualizing \cref{model:0}.
    }
\end{figure}
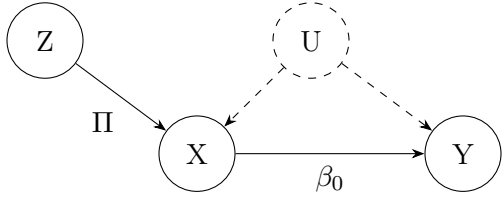

We visualize a basic instrumental variables regression model in \cref{fig:iv_graph_X}.
Here, $X$ are the endogenous treatment variables, $Z$ are instruments, $y$ is the outcome, and $U$ are unobserved confounders.
We are interested in the causal effect of $X$ on $y$, encoded by the causal parameter $\beta_0$.
The first stage parameter $\Pi$ is a nuisance parameter.

\begin{theoremEnd}[malte,restate command=modelone]{model}
    \label{model:0}
    Let $y_i = X_i^T \beta_0 + \varepsilon_i \in \BR$ with $X_i = Z_i^T \Pi + V_{X, i} \in \BR^\mX$ for random vectors $Z_i \in \BR^k, V_{X, i}\in \BR^\mX$, and $\varepsilon_i \in \BR$ for $i=1\ldots, n$ and parameters $\Pi \in \BR^{k \times \mX}$, and $\beta_0 \in \BR^\mX$.
    We call the $Z_i$ \emph{instruments}, the $X_i$ \emph{endogenous covariates}, and the $y_i$ \emph{outcomes}.
    Let $Z \in \BR^{n \times k}, X \in \BR^{n \times \mX},$ and $y \in \BR^n$ be the matrices of stacked observations.

    In \emph{strong instrument asymptotics}, we assume that $\Pi$ is fixed for all $n$ and of full column rank $\mX$.
    In \emph{weak instrument asymptotics} \citep{staiger1997instrumental}, we assume that $\sqrt{n} \Pi$ is fixed and of full column rank $m_X$.
    Thus, $\Pi = \CO(\frac{1}{\sqrt{n}})$.
    Both asymptotics imply that $k \geq \mX$.
\end{theoremEnd}%

In standard asymptotics, the expected first-stage $F$-statistic (or its multivariate extension, see \cref{sec:anderson_1951_rank_test} and \citeauthor{anderson1951estimating}, \citeyear{anderson1951estimating}, \citeauthor{cragg1997inferring}, \citeyear{cragg1997inferring}) grows with the sample size.
\citet{staiger1997instrumental} note that in many applications, even though the sample size $n$ is large, the $F$-statistic for regressing $X$ on $Z$ is small.
In these settings, standard asymptotics no longer apply, motivating weak instrument asymptotics, where the first-stage $F$-statistic is of constant order.

A setup comparable to weak instrument asymptotics is high-dimensional statistics \citep{buhlmann2011statistics}.
In standard asymptotics, the number of covariates $p$ is fixed and the sample size $n$ grows to infinity.
However, in application areas such as genomics, the number of covariates $p$ can be as large or even larger than the sample size $n$.
In these settings, standard asymptotics no longer apply, and one needs to use high-dimensional asymptotics, where $p$ grows with $n$.

We assume that a central limit theorem applies to the sums $Z^T \varepsilon$ and $Z^T V_X$.
\begin{theoremEnd}[malte,restate command=assumptionzero]{assumption}
    \label{ass:0}
    Let
    $$
    \Psi := \begin{pmatrix} \Psi_{\varepsilon} & \Psi_{V_X} \end{pmatrix} := (Z^T Z)^{-1/2} Z^T \begin{pmatrix} \varepsilon & V_X \end{pmatrix} \in \BR^{k \times (1 + \mX)}.
    $$
    Assume there exist $\Omega \in \BR^{(1+\mX) \times (1+\mX)}$ and $Q \in \BR^{k \times k}$ positive definite such that, as $n \to \infty$,
    \begin{align*}
        &\mathrm{(a)} \ \ \frac{1}{n} \begin{pmatrix}\varepsilon & V_X \end{pmatrix}^T \begin{pmatrix}\varepsilon & V_X \end{pmatrix} \toP \Omega = \begin{pmatrix}
    \sigma^2_\varepsilon & \Omega_{\varepsilon, V_X} \\
    \Omega_{V_X, \varepsilon} & \Omega_{V_X} \\
\end{pmatrix}, \\
        &\mathrm{(b)} \ \ \vecop(\Psi) \tod \CN(0, \Omega \otimes \Id_k), \text{ and }\\
        &\mathrm{(c)} \ \ \frac{1}{n} Z^T Z \toP Q,
    \end{align*}
    where $\Cov(\vecop(\Psi)) = \Omega \otimes \Id_k$ means $\Cov(\Psi_{i, j}, \Psi_{i', j'}) = 1_{i = i'} \cdot \Omega_{j, j'}$.
\end{theoremEnd}%

\noindent While standard in the literature, \cref{ass:0} might be difficult to interpret.
A sufficient condition for \cref{ass:0} is that (i) the error terms $\varepsilon_i$ and $V_{X, i}$ are homoscedastic and uncorrelated with the instruments $Z_i$ and (ii) the error terms and instruments are i.i.d.\ with finite second moments.
\begin{theoremEnd}[malte,category=model]{lemma}
    \label{lem:when_does_ass_0_hold}
    Assume the random variables $(Z_i, \varepsilon_i, V_{X, i}) \in \BR^{k +\mX + 1}$ are i.i.d.\ across $i = 1, \ldots, n$ with
    \begin{align*} 
        &\mathrm{(a)} \ \ \BE[Z_i Z_i^T] = Q \in \BR^{k \times k}, \\
        &\mathrm{(b)} \ \ \BE((\varepsilon_i, V_{X, i}) Z_i^T) = 0, \ \text{and} \\
        &\mathrm{(c)} \ \ \Cov((\varepsilon_i, V_{X, i}) \mid Z_i) = \Omega \in \BR^{(\mX+1) \times (\mX+1)}.
    \end{align*}
    Then \cref{ass:0} holds.
\end{theoremEnd}%
\begin{proofEnd}%
    Both \cref{ass:0}: a and c follow from the weak law of large numbers.
    We show \cref{ass:0}: b.
    Let $M = (m_{i, j})_{i, j} \in \BR^{k \times (\mX + 1)}$.
    Write $V := (\varepsilon, V_X) \in \BR^{n \times (\mX +1 )}$ and $\Phi := \frac{1}{\sqrt{n}} Z^T V \in \BR^{k \times (\mX + 1)}.$
    Let $\phi := M \odot \Phi = \trace( M^T \Phi )$, where $\odot$ denotes pointwise multiplication.
    Calculate
    \begin{align*}
    &\phi = \frac{1}{\sqrt{n}} \sum_{i = 1}^n \underbrace{\sum_{j=1}^k \sum_{l=1}^{m+1} m_{j, l} Z_{i, j} V_{i, l}}_{=: X_i, \ \text{i.i.d.}}, \text{ where} \ \BE[X_i] = \sum_{j, l} m_{j, l} \BE[Z_{i, j} V_{i, l}] \overset{(ii)}{=} 0, \text{ and}\\
    &\Var(X_i) = \sum_{j, l} \sum_{j', l'} m_{j, l} m_{j', l'} \BE[ Z_{i, j} Z_{i, j'} V_{i, l} V_{i, l'} ] \overset{(i), (iii)}{=} \sum_{j, l} \sum_{j', l'} m_{j, l} m_{j', l'} Q_{j, j'} \Omega_{l, l'} =: \sigma^2.
    \end{align*}
    By the central limit theorem
    $$
    \phi = \frac{1}{\sqrt{n}} \sum_{i = 1}^n X_i \tod \CN(0, \sigma^2).
    $$
    If $\vecop(Y) \sim \CN(0, \Omega \otimes Q)$, then $\Var(M \odot Y) = \sigma^2$.
    By the theorem of Cram\'er-Wold thus $\phi \tod Y$.
    As $\frac{1}{n} Z^T Z \toP Q$, by Slutsky $\vecop(\Psi) = \vecop((\frac{1}{n} Z^T Z)^{-1/2} \Phi) \tod \vecop(Q^{-1/2} Y) \sim \CN(0, \Omega \otimes \Id_k)$.    
\end{proofEnd}%

We consider \cref{model:0} and assume that \cref{ass:0} holds for the rest of this manuscript.

In practice, it is common to include additional exogenous (that is, unconfounded) covariates $C$ that enter the model equation $y = X^T \beta_0 + C \alpha_0 + \varepsilon$, as depicted in \cref{fig:iv_graph_exogenous} (left panel).
We will discuss the treatment of such included exogenous covariates later in \cref{sec:exogenous_variables}.
For now, we note that one can reduce to the model without exogenous covariates by replacing $y$, $X$, $W$, and $Z$ with their residuals after regressing out $C$.
This also allows for the inclusion of an intercept by centering the data.

\subsection{Application}
We use \citeauthor{card1993using}'s \citeyearpar{card1993using} application as a running example throughout this manuscript and apply the estimators and tests introduced in this manuscript to their data.
The data is available at \href{https://davidcard.berkeley.edu/data_sets/proximity.zip}{\texttt{davidcard.berkeley.edu/data\_sets/proximity.zip}}.
See appendix \ref{sec:omitted_code} for the omitted code to load the data.

\citet{card1993using} estimates the causal effect of length of education on hourly wages.
Their dataset is based on a 1976 subsample of the National Longitudinal Survey of Young Men and contains 3010 observations.
It includes years of education obtained by 1976 (\texttt{ed76}), log hourly wages in 1976 (\texttt{lwage76}), age in 1976 (\texttt{age76}), and indicators of whether the individual lived close to a public four-year college (\texttt{nearc4a}), a private four-year college (\texttt{nearc4b}), or a two-year college in 1966 (\texttt{nearc2}).
The dataset also contains variables or indicators on race, metropolitan area, region (summarized in \texttt{indicators}), and family background (summarized in \texttt{family}).
\citet{card1993using} defines (potential) experience \texttt{exp76} as \texttt{age76} - \texttt{ed76} - 6.

\citet{card1993using} reports estimates of the causal effect on education on log wages for multiple specifications.
We focus on their specification where experience and experience squared are included as endogenous regressors and variables or indicators on race, metropolitan area, region, and family background are included as exogenous regressors.
Whereas \citet{card1993using} uses age and age squared to instrument for experience and experience squared, we simply include these as additional instruments.
We also include all three college proximity indicators as instruments, instead of aggregating \texttt{nearc4a} and \texttt{nearc4b} into a single instrument.

\begin{jupyternotebook}
\begin{tcolorbox}[breakable, size=fbox, boxrule=1pt, pad at break*=1mm,colback=cellbackground, colframe=cellborder]
\prompt{In}{incolor}{1}{\boxspacing}
\begin{Verbatim}[commandchars=\\\{\}]
\PY{p}{[}\PY{o}{.}\PY{o}{.}\PY{o}{.}\PY{p}{]}  \PY{c+c1}{\PYZsh{} Load data. See appendix A for details.}

\PY{c+c1}{\PYZsh{} construct potential experience and its square.}
\PY{n}{df}\PY{p}{[}\PY{l+s+s2}{\PYZdq{}}\PY{l+s+s2}{exp76}\PY{l+s+s2}{\PYZdq{}}\PY{p}{]} \PY{o}{=} \PY{n}{df}\PY{p}{[}\PY{l+s+s2}{\PYZdq{}}\PY{l+s+s2}{age76}\PY{l+s+s2}{\PYZdq{}}\PY{p}{]} \PY{o}{\PYZhy{}} \PY{n}{df}\PY{p}{[}\PY{l+s+s2}{\PYZdq{}}\PY{l+s+s2}{ed76}\PY{l+s+s2}{\PYZdq{}}\PY{p}{]} \PY{o}{\PYZhy{}} \PY{l+m+mi}{6}
\PY{n}{df}\PY{p}{[}\PY{l+s+s2}{\PYZdq{}}\PY{l+s+s2}{exp762}\PY{l+s+s2}{\PYZdq{}}\PY{p}{]} \PY{o}{=} \PY{n}{df}\PY{p}{[}\PY{l+s+s2}{\PYZdq{}}\PY{l+s+s2}{exp76}\PY{l+s+s2}{\PYZdq{}}\PY{p}{]} \PY{o}{*}\PY{o}{*} \PY{l+m+mi}{2}
\PY{n}{df}\PY{p}{[}\PY{l+s+s2}{\PYZdq{}}\PY{l+s+s2}{age762}\PY{l+s+s2}{\PYZdq{}}\PY{p}{]} \PY{o}{=} \PY{n}{df}\PY{p}{[}\PY{l+s+s2}{\PYZdq{}}\PY{l+s+s2}{age76}\PY{l+s+s2}{\PYZdq{}}\PY{p}{]} \PY{o}{*}\PY{o}{*} \PY{l+m+mi}{2}

\PY{c+c1}{\PYZsh{} endogenous variables: years of education, experience, experience squared.}
\PY{n}{X} \PY{o}{=} \PY{n}{df}\PY{p}{[}\PY{p}{[}\PY{l+s+s2}{\PYZdq{}}\PY{l+s+s2}{ed76}\PY{l+s+s2}{\PYZdq{}}\PY{p}{,} \PY{l+s+s2}{\PYZdq{}}\PY{l+s+s2}{exp76}\PY{l+s+s2}{\PYZdq{}}\PY{p}{,} \PY{l+s+s2}{\PYZdq{}}\PY{l+s+s2}{exp762}\PY{l+s+s2}{\PYZdq{}}\PY{p}{]}\PY{p}{]}
\PY{n}{y} \PY{o}{=} \PY{n}{df}\PY{p}{[}\PY{l+s+s2}{\PYZdq{}}\PY{l+s+s2}{lwage76}\PY{l+s+s2}{\PYZdq{}}\PY{p}{]}  \PY{c+c1}{\PYZsh{} outcome: log wage.}
\PY{c+c1}{\PYZsh{} included exogenous variables: indicators for family background, region, and race.}
\PY{n}{C} \PY{o}{=} \PY{n}{df}\PY{p}{[}\PY{n}{family} \PY{o}{+} \PY{n}{indicators}\PY{p}{]}
\PY{c+c1}{\PYZsh{} instruments: proximity to colleges, age, and age squared.}
\PY{n}{Z} \PY{o}{=} \PY{n}{df}\PY{p}{[}\PY{p}{[}\PY{l+s+s2}{\PYZdq{}}\PY{l+s+s2}{nearc4a}\PY{l+s+s2}{\PYZdq{}}\PY{p}{,} \PY{l+s+s2}{\PYZdq{}}\PY{l+s+s2}{nearc4b}\PY{l+s+s2}{\PYZdq{}}\PY{p}{,} \PY{l+s+s2}{\PYZdq{}}\PY{l+s+s2}{nearc2}\PY{l+s+s2}{\PYZdq{}}\PY{p}{,} \PY{l+s+s2}{\PYZdq{}}\PY{l+s+s2}{age76}\PY{l+s+s2}{\PYZdq{}}\PY{p}{,} \PY{l+s+s2}{\PYZdq{}}\PY{l+s+s2}{age762}\PY{l+s+s2}{\PYZdq{}}\PY{p}{]}\PY{p}{]}

\PY{n}{pd}\PY{o}{.}\PY{n}{concat}\PY{p}{(}\PY{p}{[}\PY{n}{X}\PY{p}{,} \PY{n}{y}\PY{p}{,} \PY{n}{Z}\PY{p}{]}\PY{p}{,} \PY{n}{axis}\PY{o}{=}\PY{l+m+mi}{1}\PY{p}{)}\PY{o}{.}\PY{n}{head}\PY{p}{(}\PY{p}{)}
\end{Verbatim}
\end{tcolorbox}
\begin{tcolorbox}[breakable, size=fbox, boxrule=.5pt, pad at break*=1mm, opacityfill=0]
\prompt{Out}{outcolor}{1}{\boxspacing}
\begin{Verbatim}[commandchars=\\\{\}]
    ed76  exp76  exp762   lwage76  nearc4a  nearc4b  nearc2  age76  age762
id
2      7     16     256  6.306275        0        0       0     29     841
3     12      9      81  6.175867        0        0       0     27     729
4     12     16     256  6.580639        0        0       0     34    1156
5     11     10     100  5.521461        1        0       1     27     729
6     12     16     256  6.591674        1        0       1     34    1156
\end{Verbatim}
\end{tcolorbox}
\end{jupyternotebook}
\noindent For simplicity, we reduce to \cref{model:0} by replacing $Z$, $X$, and $y$ with their residuals after regressing out $C$.
See also \cref{sec:exogenous_variables}, where we treat the included exogenous variables explicitly.

\begin{jupyternotebook}
\begin{tcolorbox}[breakable, size=fbox, boxrule=1pt, pad at break*=1mm,colback=cellbackground, colframe=cellborder]
\prompt{In}{incolor}{2}{\boxspacing}
\begin{Verbatim}[commandchars=\\\{\}]
\PY{k+kn}{from} \PY{n+nn}{ivmodels}\PY{n+nn}{.}\PY{n+nn}{utils} \PY{k+kn}{import} \PY{n}{oproj}

\PY{c+c1}{\PYZsh{} oproj(A, B) returns B minus the projection of B onto the column space of A.}
\PY{n}{Z}\PY{p}{,} \PY{n}{X}\PY{p}{,} \PY{n}{y} \PY{o}{=} \PY{n}{oproj}\PY{p}{(}\PY{n}{C}\PY{p}{,} \PY{n}{Z}\PY{p}{,} \PY{n}{X}\PY{p}{,} \PY{n}{y}\PY{p}{)}
\end{Verbatim}
\end{tcolorbox}
\end{jupyternotebook}
\section{Estimators}
\label{sec:intro_to_iv_tests:estimators}
For any $A \in \BR^{p \times q}$ let $P_A := A (A^T A)^\dagger A^T$, where $\dagger$ denotes the Moore-Penrose inverse, be the projection matrix onto the column span of $A$.
For $B \in \BR^{p \times r}$, the projected values $P_A B$ are equal to the fitted values of the linear regression of $B$ on $A$.
Write $M_A := \Id_{p} - P_A$ for the projection onto its orthogonal complement.
Then, $M_A B$ are the residuals of the linear regression of $B$ on $A$.

An important family of estimators in instrumental variables regression are k-class estimators.
We will later see that these encompass the ordinary least squares estimator, the two-stage least-squares estimator, and the limited information maximum likelihood estimator.
\begin{definition}\label{def:kclass}
    Let $\kappa \geq 0$. The \emph{k-class estimator} with parameter $\kappa$ is
    \begin{align*}
        \hat\beta_\kclass(\kappa) :=& 
        (X^T (\kappa P_Z + (1 - \kappa) \Id) X)^{\dagger} X^T (\kappa P_Z + (1 - \kappa) \Id) y,
    \end{align*}
    where $\dagger$ denotes the Moore-Penrose pseudoinverse.
\end{definition}

\noindent An alternative definition of the k-class estimator is as the solution to an optimization problem.%
\begin{theoremEnd}[malte,category=estimators]{proposition}%
    \label{prop:k_class_well_defined}
    Let $\kappa < 1 + \lambdamin{ (X^T M_Z X)^{-1} X^T P_Z X }$.
    Then
    $$
        \beta \mapsto \kappa \| P_Z (y - X \beta) \|_2^2 + (1 - \kappa) \| y - X \beta \|_2^2
    $$
    is strictly convex with unique minimizer
    $$
        \hat\beta_\kclass(\kappa) = (X^T (\kappa P_Z + (1 - \kappa) \Id) X)^{-1} X^T (\kappa P_Z + (1 - \kappa) \Id) y.
    $$
\end{theoremEnd}%
\begin{proofEnd}%
    Calculate
    \begin{align*}
        \frac{\dd}{\dd \beta} \kappa \|  P_Z (y - X \beta) \|^2 + (1 -\kappa) \| (y - X \beta) \|^2 = - 2 X^T (\kappa P_Z + (1 -\kappa) \Id) (y - X \beta)
    \end{align*}
    and
    \begin{align*}
        \frac{\dd^2}{\dd^2 \beta}  \kappa \| P_Z (y - X \beta) \|^2 + (1 -\kappa) \| (y - X \beta) \|^2 &= 2 X^T (\kappa P_Z + (1 -\kappa) \Id) X
    \end{align*}
    By \cref{lem:kappa_pos_definite}, this Hessian is positive definite if $\kappa < 1 + \lambdamin{ (X^T M_Z X)^{-1} X^T P_Z X }$ and thus the objective is strictly convex.
    The result follows.
\end{proofEnd}%

\noindent One prominent instance of k-class estimators is the two-stage least-squares estimator.
\begin{definition}\label{def:tsls}
    The \emph{two-stage least-squares (TSLS or 2SLS)} estimator is
    \begin{align*}
        \hat\beta_\tsls := \hat \beta_\kclass(1)  =(X^T P_Z X)^{-1} X^T P_Z y =  \argmin_{\beta \in \BR^\mX} \| P_Z (y - X \beta) \|_2^2.
    \end{align*}
\end{definition}

\noindent The two-stage least-squares estimator gets its name as it results from first linearly regressing $X$ on $Z$ and then linearly regressing $y$ on the fitted values
$P_Z X$.

Why does the two-stage least-squares estimator recover $\beta_0$?
Under \cref{lem:when_does_ass_0_hold}'s assumptions, $\Cov(Z_i, V_{X,i}) = 0$ and  $\Cov(Z_i, V_{X, i} \beta_0 + \varepsilon_i) = 0$, so we can consistently estimate $\Pi$ and $\Pi \beta_0$ in $X = Z \Pi + V_X$ and $y = Z \Pi \beta_0 + V_X \beta_0 + \varepsilon$ with ordinary least-squares as $\widehat{\Pi} = (Z^T Z)^{-1} Z^T X$ and $\widehat{\Pi \beta} = (Z^T Z)^{-1}Z^T y$.
In a just-identified model, that is, when $\dim(Z) = \dim(X)$, the two-stage least-squares estimator equals $\hat\beta_\tsls = \widehat{\Pi}^{-1} \widehat{\Pi \beta}$.
In an overidentified model, that is, when $\dim(Z) > \dim(X)$, the two-stage least-squares estimator is the solution to the optimization problem $\hat\beta_\tsls = \argmin_{\beta \in \BR^{\mX}} \| \widehat{\Pi} \beta - \widehat{\Pi \beta} \|_{(Z^T Z)}^2$, where $\| x \|^2_A = x^T A x$.
As $\widehat{\Pi} \to_\BP \Pi$ and $\widehat{\Pi \beta} \to_\BP \Pi \beta_0$ under strong instrument asymptotics, this implies that $\hat\beta_\tsls \to_\BP \beta_0$ as $n \to \infty$.

The main intuition for k-class estimators should be that for $\kappa < 1$, the estimator $\hat\beta_\kclass(\kappa)$ regularizes the TSLS estimator towards the OLS estimator, and for $\kappa > 1$, it regularizes the TSLS estimator away from the OLS estimator.

For $\kappa$ close enough to 1, under strong instrument asymptotics, k-class estimators are consistent and asymptotically Gaussian.
\begin{theoremEnd}[malte,category=estimators]{proposition}%
    \label{prop:kclass_asymptotic_normality}
    Consider a sequence of k-class estimators $\hat \beta_\kclass(\kappa_n)$ with possibly random parameters $\kappa_n$.
    If $\kappa_n \to_\BP 1$ as $n \to \infty$, then, under strong instrument asymptotics, $\hat \beta_\kclass(\kappa_n)$ is a consistent estimate of $\beta_0$:
    $$
    \hat \beta_\kclass(\kappa_n) \toP \beta_0 \ \text{ as } \ n \to \infty.
    $$
    If furthermore $\sqrt{n} (\kappa_n - 1) \to_\BP 0$ as $n \to \infty$, then, under strong instrument asymptotics, $\hat\beta_\kclass(\kappa_n)$ is asymptotically normal:
    $$
    \frac{1}{\sqrt{\hat\sigma^2}} \left(X^T (\kappa_n P_Z + (1 - \kappa_n) \Id_{n}) X \right)^{1/2} \left(\hat \beta_\kclass(\kappa_n) - \beta_0 \right) \overset{d}{\to} \CN(0, \Id_{\mX}) \ \text{ as } \ n \to \infty,
    $$%
    where $\hat\sigma^2 := \frac{1}{n - m} \| y - X \hat\beta_\kclass(\kappa_n) \|^2$.
\end{theoremEnd}%
\begin{proofEnd}%
    Write $\Omega = \begin{pmatrix} \sigma_\varepsilon^2 & \Omega_{V_X, \varepsilon} \\ \Omega_{\varepsilon, V_X} & \Omega_{V_X, V_X}\end{pmatrix}$.
    Calculate
    $$\hat \beta_\kclass(\kappa_n) - \beta_0 = \left(\frac{1}{n} X^T (\kappa_n P_Z + (1 - \kappa_n) \Id) X\right)^{-1} \left(\frac{1}{n} X^T (\kappa_n P_Z + (1 - \kappa_n) \Id) \varepsilon \right).
    $$
    Under strong instrument asymptotics and \cref{ass:1},
    \begin{itemize}
        \item $\frac{1}{\sqrt{n}} (Z^T Z)^{-1/2} Z^T X = \frac{1}{\sqrt{n}} (Z^T Z)^{1/2} \Pi_X + \frac{1}{\sqrt{n}}\Psi_{V_X} \toP Q^{1/2} \Pi_X$,
        \item $(Z^T Z)^{-1/2} Z^T \varepsilon = \Psi_{\varepsilon} \tod \CN(0, \sigma_\varepsilon^2 \cdot \Id_k)$,
        \item $X^T X = \Pi_X^T Z^T Z \Pi_X + \Pi_X^T (Z^T Z)^{1/2} \Psi_{V_X} + \Psi_{V_X}^T (Z^T Z)^{1/2} \Pi_X + V_X^T V_X$ such that $\frac{1}{n} X^T X \toP \Pi_X^T Q \Pi_X + \Omega$, and
        \item $X^T \varepsilon = \Pi_X^T (Z^T Z)^{1/2} \Psi_{\varepsilon} + V_X^T \varepsilon$ such that $\frac{1}{n} X^T \varepsilon \toP \Omega_{V_X, \varepsilon}$.
    \end{itemize}
    Thus, as long as $\kappa_n \toP 1$,
    \begin{align}\label{eq:kclass_asymptotic_normality:1}
        \frac{1}{n}  X^T (\kappa_n P_Z + (1 - \kappa_n) \Id) X \toP \Pi_X^T Q \Pi_X
    \end{align}
    and
    \begin{align*}
        \frac{1}{n} X^T (\kappa_n P_Z + (1 - \kappa_n) \Id) \varepsilon \toP 0,
    \end{align*}
    proving consistency.

    To show asymptotic normality, calculate
    \begin{align}\nonumber
        \frac{1}{\sqrt{n}} X^T \left(\kappa_n P_Z + (1 - \kappa_n) \Id \right) \varepsilon &
        = \frac{\kappa_n}{\sqrt{n}} \Pi_X^T (Z^T Z)^{1/2} \Psi_{\varepsilon} + \frac{\kappa_n}{\sqrt{n}} \Psi_{V_X}^T\Psi_{\varepsilon} + \frac{\sqrt{n} (1 - \kappa_n)}{n} X^T \varepsilon \\
        &\toP \Pi_X^T Q^{1/2} \Psi_{\varepsilon} \tod \CN(0, \Pi_X^T Q \Pi_X \cdot \Omega_{\varepsilon, \varepsilon}). \label{eq:kclass_asymptotic_normality:2}
    \end{align}
    By consistency of $\hat \beta_\kclass(\kappa_n)$ and \cref{ass:1}, we have $\hat\sigma^2 \toP \Omega_{\varepsilon, \varepsilon}.$
    Combining this with Equations (\ref{eq:kclass_asymptotic_normality:1}) and (\ref{eq:kclass_asymptotic_normality:2}) yields the claim
    $$
    \frac{1}{\sqrt{\hat\sigma^2}} \left(X^T (\kappa_n P_Z + (1 - \kappa_n) \Id) X \right)^{1/2} \left(\hat \beta_\kclass(\kappa_n) - \beta_0 \right) \overset{d}{\to} \CN(0, \Id_\mX) \ \text{ as } \ n \to \infty.
    $$
\end{proofEnd}%

Next to the two-stage least-squares estimator, the limited information maximum likelihood (LIML) estimator is another important estimator in instrumental variables regression.
Sometimes introduced as a k-class estimator with data-dependent value $\kappa$, we introduce the LIML estimator as the maximum likelihood estimator of $\beta_0$ assuming i.i.d.\ Gaussian errors $(\varepsilon_i, V_{X, i}) \sim \CN( 0, \Omega) $.
In this case, the log-likelihood of observing $y$ and $X$ given $\beta \in \BR^{\mX}, \Pi \in \BR^{k \times \mX}, \Omega \succ 0$, and $Z \in \BR^{n \times k}$ is
\begin{multline*}
    \ell(y, X \mid \beta, \Pi, \Omega, Z)\\ = -\frac{n(\mX+1)}{2}\log(2\pi) - \frac{n}{2}\log( \det(\Omega) ) - \frac{1}{2}\sum_{i=1}^n \begin{pmatrix} y_i - X_i^T \beta \\ X_i - Z_i^T \Pi \end{pmatrix}^T \Omega^{-1} \begin{pmatrix} y_i - X_i^T \beta \\ X_i - Z_i^T \Pi \end{pmatrix}.
\end{multline*}
\begin{definition}
    \label{def:liml}
    The \emph{limited information maximum likelihood (LIML)} estimator of $\beta_0$ is
    \begin{align*}
        \hat\beta_\liml &:= \argmax_{\beta\in\BR^{m_x}} \max_{\Pi \in \BR^{k \times \mX}\!, \,\Omega \succ 0} \ell(y, X \mid \beta, \Pi, \Omega, Z).
    \end{align*}
\end{definition}

\begin{theoremEnd}[malte,category=estimators]{lemma}%
    \label{lem:liml_likelihood}
    Assume that $M_Z \begin{pmatrix} y & X \end{pmatrix}$ is of full column rank $\mX + 1$.
    Then,
    \begin{align*}
        \max_{\Pi \in \BR^{k \times \mX}\!,\,\Omega \succ 0} &\ell(
        y, X \mid \beta, \Pi, \Omega, Z) =
            -\frac{\mX + 1}{2}(n \log(2 \pi) + 1 - n \log(n)) \\&- \frac{n}{2}\log(1 + \frac{(y - X \beta)^T P_Z (y - X \beta)}{(y - X \beta)^T M_Z (y - X \beta)})-\frac{n}{2} \log(\det(\begin{pmatrix} y & X \end{pmatrix}^T M_Z \begin{pmatrix} y & X \end{pmatrix})).
    \end{align*}
    Also, for $\tilde X(\beta) := X - (y - X \beta) \frac{(y - X \beta)^T M_Z X}{(y - X \beta)^T M_Z (y - X \beta)}$,
    \begin{align*}
        \hat\Pi_\liml(\beta) := \argmax_{\Pi \in \BR^{k \times \mX}} \max_{\Omega \succ 0} \ell(y, X \mid \beta, \Pi, \Omega, Z) = (Z^T Z)^{-1} Z^T \tilde X(\beta).
    \end{align*}
\end{theoremEnd}%
\begin{proofEnd}%
    This proof follows \citet{morimune1993derivation}.
    Write $u = u(\beta) := y - X \beta$ and $V = V(\Pi) := X - Z \Pi$.
    The Gaussian log-likelihood $\ell(y, X \mid \beta, \Pi, \Omega, Z)$ is maximized with respect to $\Omega$ by the empirical covariance 
    $$
    \hat\Omega(\beta, \Pi) := \argmax_\Omega \ell(y, X \mid \beta, \Pi, \Omega, Z) = \frac{1}{n} \begin{pmatrix} u(\beta)^T u(\beta) & u(\beta)^T V(\Pi) \\ V(\Pi)^T u(\beta) &V(\Pi)^T V(\Pi).\end{pmatrix}
    $$
    The corresponding log-likelihood is
    \begin{equation}\label{eq:log-likelihood}
    \ell(y, X \mid  \beta, \Pi, \hat\Omega(\beta, \Pi), Z) = - \frac{n(\mX + 1)}{2}\log(2\pi) - \frac{n}{2}\log(\det(\hat\Omega(\beta, \Pi))) - \frac{\mX + 1}{2}.
    \end{equation}
    Note that $\frac{d}{dX_{i, j}} \log(\det(X)) = (X^{-1})_{j, i}$ %
    .
    Define $B := B(\beta, \Pi) := (u(\beta), V(\Pi)) \in \mathbb{R}^{n \times (\mX+1)}$ and write $\left(n \hat \Omega(\beta, \Pi)\right)^{-1} = (B^T B)^{-1} =: \begin{pmatrix} b_{1,1} & b_{1, 2} \\ b_{2, 1} & b_{2, 2} \end{pmatrix}$ with $b_{1,1} \in \BR$ and $b_{2,2} \in \BR^{\mX \times \mX}$.
    Then, the first-order condition for $\hat\Pi(\beta) := \argmin_\Pi \ell(y, X \mid \beta, \Pi, \hat\Omega(\beta, \Pi), Z)$ is
    \begin{align*}
    0 &= \frac{d}{d\Pi_{l, j}} \log( \det(n \hat\Omega(\beta, \Pi))) \\
    &= (B^T B)^{-1} \odot  \frac{d}{d B} (B^T B) \cdot \frac{d}{d\Pi_{l, j}}B\\
    & = 2 (B^T B)^{-1} \odot  B^T \cdot 
    \begin{pmatrix}
        0^{n \times j} & -Z_{\cdot, l} & 0^{n \times (\mX - j)}
    \end{pmatrix} \\
    &= - 2 (B^T B)^{-1} \odot
    \begin{pmatrix}
        0^{1 \times j} & u(\beta)^T Z_{\cdot, l} & 0^{1 \times (\mX - j)} \\
        0^{\mX \times j} & V(\Pi)^T Z_{\cdot, l} & 0^{\mX \times (\mX - j)}
    \end{pmatrix} \\
    &= - 2 (b_{1, 2})_j \cdot u(\beta)^T Z_{\cdot, l} - 2 (b_{2, 2})_{j, \cdot} \cdot V(\Pi)^T Z_{\cdot, l}\\
    \Rightarrow 0 &= \frac{d}{d \Pi} \log( \det(n \hat\Omega(\beta, \Pi)))  = -2  (b_{1, 2} u(\beta)^T Z + b_{2, 2} V(\Pi)^T Z), \numberthis \label{eq:buz}
    \end{align*}  
    where $A \odot B = \tr(A^T B)$ is the standard scalar product for matrices, $Z_{\cdot, l}$ is the $l$-th row of $Z$, and $0^{n \times l} \in \mathbb{R}^{n \times l}$ contains only zeros.

    For any matrix $A$, define $P_A := A (A^TA)^{-1} A^T$ to be the projection onto the column space of $A$ and let $M_A := \mathrm{Id} - P_A$.
    The analytic formula for blockwise inversion of matrices
    yields $b_{2,2}^{-1} b_{2, 1} = - V^T u (u^T u)^{-1}$.
    Thus, premultiplying \eqref{eq:buz} with $b_{2, 2}^{-1}$ yields
    \begin{align*}
    0 &= \frac{d}{d\Pi} \log( \det(n \hat\Omega(\beta, \Pi))) = -2(b_{1, 2} u^T Z + b_{2, 2} V^T Z)  \\
    \Rightarrow 0 &= V^Tu(u^Tu)^{-1}u^TZ - V^TZ \\
    &= V^T  u (u^T u)^{-1} u^T Z - X^T Z + \Pi^T Z^T Z \\
    \Rightarrow \Pi &= (Z^T Z)^{-1} Z^T X - (Z^TZ)^{-1} Z^T u (u^T u) u^T V \numberthis\label{eq:Pi} \\
    \Rightarrow V &= X - Z \Pi = X - P_Z X + P_Z u (u^T u)^{-1} u^T V \numberthis \label{eq:V} \\
    \Rightarrow u^TV &= u^T     M_Z X + (u^T P_Z u) (u^T u)^{-1} u^T V \\
    \Rightarrow u^T V &= \frac{u^T M_Z X}{1 - (u^T P_Z u)(u^T u)^{-1}} = u^Tu \frac{u^T M_Z X}{u^T M_Z u} \numberthis \label{eq:uV}\\
    &=  u^TM_Z X + u^T P_Zu \frac{u^T M_Z X}{u^T M_Z u}.
    \end{align*}
    This is independent of $\Pi$.
    Next, we expand equation (\ref{eq:V}) and use $M_Z P_Z = 0$ to get
    \begin{align*}
    V^T V &= (M_ZX + P_Z u (u^T u)^{-1} u^T V)^T (M_ZX + P_Z u (u^T u)^{-1} u^T V) \\
    &= X^T M_Z X + \frac{u^T P_Z u}{(u^T u)^{2}} (u^T V)^T(u^T V) \\
    &\overset{(\ref{eq:uV})}{=} X^T M_Z X + \frac{u^T P_Z u}{(u^T M_Z u)^2}(u^TM_ZX)^T (u^TM_ZX) \numberthis \label{eq:liml_likelihood:VTV}.
    \end{align*}
    Thus, we can rewrite
    \begin{equation*}
    n \hat \Omega
    = \begin{pmatrix} u^T u & u^T V \\ V^T u & V^T V \end{pmatrix}
    \overset{(\ref{eq:uV}), (\ref{eq:liml_likelihood:VTV})}{=}
    \begin{pmatrix} u^T \\ X^T\end{pmatrix} M_Z \begin{pmatrix} u & X\end{pmatrix} + \frac{u^T P_Z u}{(u^T M_Z u)^2} \begin{pmatrix} u^T M_Z u \\ X^T M_Z u \end{pmatrix} \begin{pmatrix} u^T M_Z u & u^T M_Z X \end{pmatrix}.
    \end{equation*}
    Note that
    \begin{equation}\label{eq:det0}
        M_Z \begin{pmatrix} u & X \end{pmatrix} = M_Z \begin{pmatrix} y & X \end{pmatrix} \begin{pmatrix} 1 & 0 \\ -\beta & \mathrm{Id} \end{pmatrix}.
    \end{equation}
    Thus, as $M_Z \begin{pmatrix}y & X \end{pmatrix}$ is of full column rank, so is $L := M_Z \begin{pmatrix} u & X \end{pmatrix}$ and $L^T L$ is invertible.
    Define $v := \frac{\sqrt{u^T P_Z u}}{u^T M_Z u} L^T u$ such that
    $n \hat\Omega = L^T L + v v^T$.
    The matrix determinant lemma
    states that $\det(L^T L + v v^T) = (1 + v^T (L^T L)^{-1} v)\det(L^T L)$.
    Calculate
    $$
    P_L u = P_{[M_Z u, M_Z X ]} u = P_{[M_Z u, M_Z X]} M_Z u = P_{M_Z u} M_Z u + P_{M_{M_Z u} X} M_Z u = M_Z u
    $$
    Thus,
    \begin{equation*}
        v^T (L^T L)^{-1} v = \frac{u^T P_Z u}{(u^T M_Z u)^2} u^T L (L^T L)^{-1} L^T u = \frac{u^T P_Z u}{u^T M_Z u}.
    \end{equation*}
    This implies that 
    \begin{align*}
    \det(\hat\Omega) = n^{-(\mX + 1)} \det(n\hat\Omega) &= n^{-(\mX + 1)} \left(1 + \frac{u^T P_Z u}{u^T M_Z u}\right) \det(\begin{pmatrix} u^T \\ X^T \end{pmatrix} M_Z \begin{pmatrix} u & X \end{pmatrix}) \numberthis \label{eq:detOmega} \\
    &\overset{(\ref{eq:det0})}{=} n^{-(\mX + 1)} \left(1 + \frac{u^T P_Z u}{u^T M_Z u}\right) \det(\begin{pmatrix} y^T \\ X^T \end{pmatrix} M_Z \begin{pmatrix} y & X \end{pmatrix}).
    \end{align*}
    Plugging this into \cref{eq:log-likelihood} and expanding $u = y - X \beta$ yields
    \begin{align*}
        \ell(y, X & \mid \beta, \hat\Pi, \hat\Omega, Z) = - \frac{n(\mX + 1)}{2}\log(2\pi) - \frac{n}{2}\log(\det(\hat\Omega)) - \frac{\mX + 1}{2}\\
        &=-\frac{\mX + 1}{2}(n \log(2 \pi) + 1 - n \log(n))
        \\&- \frac{n}{2}\log(1 + \frac{(y - X \beta)^T P_Z (y - X \beta)}{(y - X \beta)^T M_Z (y - X \beta)}) - \frac{n}{2}\log(\det(\begin{pmatrix} y^T \\ X^T \end{pmatrix} M_Z \begin{pmatrix} y & X \end{pmatrix})).
    \end{align*}
    The second claim follows from plugging Equation (\ref{eq:uV}) into Equation (\ref{eq:Pi}).
\end{proofEnd}%

\begin{theoremEnd}[malte]{definition}[\citeauthor{anderson1949estimation}, \citeyear{anderson1949estimation}]
    \label{def:anderson_rubin_test_statistic}
    The \emph{Anderson-Rubin test statistic} is
    \begin{equation*}
        \AR(\beta) := \frac{n - k}{k}\frac{(y - X \beta)^T P_Z (y - X \beta)}{(y - X \beta)^T M_Z (y - X \beta)}.
    \end{equation*}
\end{theoremEnd}%

\noindent We will discuss the Anderson-Rubin test in more detail in \cref{sec:intro_to_iv_tests:tests}.
If $\beta=\beta_0$, the numerator is the sum of squares of the second-stage noise terms $\varepsilon_i$ projected onto the instruments.
The denominator, divided by $n - k$, is an estimate of $\sigma_\varepsilon^2$.
For now, note the following result:
\begin{theoremEnd}[malte,category=estimators]{corollary}%
    \label{cor:liml_minimizes_ar}
    The LIML estimator minimizes the Anderson-Rubin test statistic.
\end{theoremEnd}%
\begin{proofEnd}%
    By \cref{lem:liml_likelihood},
    $$
    \max_{\Pi, \Omega \succ 0}
    \ell(y, X \mid \beta, \Pi, \Omega, Z) = 
    \mathrm{const} - \frac{n}{2}\log(
        \left(1 + \frac{k}{n-k} \AR(\beta)\right) \det(\begin{pmatrix} y^T \\ X^T \end{pmatrix} M_Z \begin{pmatrix} y & X \end{pmatrix}))
    $$
    and so
    $$
    \hat\beta_\liml = \argmax_{\beta \in \BR^\mX} \max_{\Pi, \Omega \succ 0}
    \ell(y, X \mid \beta, \Pi, \Omega, Z)
    = \argmin_{\beta \in \BR^\mX} \AR(\beta).
    $$
\end{proofEnd}%

\noindent Using \cref{cor:liml_minimizes_ar}, we can derive the k-class formulation of the LIML estimator.
\begin{theoremEnd}[malte,category=estimators]{technical_condition}
    \label{tc:liml_theorem}
    It holds that
    $$
    \lambda_1 := \lambdamin{ \left[\begin{pmatrix} y & X \end{pmatrix}^T M_Z \begin{pmatrix} y & X \end{pmatrix} \right]^{-1} \begin{pmatrix} y & X \end{pmatrix}^T P_Z \begin{pmatrix} y & X \end{pmatrix}} < \lambdamin{[X^T M_Z X]^{-1} (X^T P_Z X)}
    $$
    or, equivalently (see \cref{lem:kappa_pos_definite}),
    $$
    \lambdamin{X^T (\lambda_1 P_Z + (1 - \lambda_1) \Id) X} > 0.
    $$
\end{theoremEnd}%
\begin{theoremEnd}[malte,category=estimators]{proposition}%
    \label{thm:liml_is_kclass}
    Assume that $M_Z \begin{pmatrix} y & X \end{pmatrix}$ has full column rank $\mX + 1$ and \cref{tc:liml_theorem} holds.
    Let 
    \begin{align*}
    \hat\kappa_\liml &:= \lambda_\mathrm{min}\left( \left[\begin{pmatrix} y & X \end{pmatrix}^T M_Z \begin{pmatrix} y & X \end{pmatrix} \right]^{-1} \begin{pmatrix} y & X \end{pmatrix}^T P_Z \begin{pmatrix} y & X \end{pmatrix} \right) + 1 \geq 1.
    \end{align*}
    Then
    \begin{align*}
    \hat \beta_\liml &= \hat \beta_\kclass(\hat\kappa_\liml) \\
    &= (X^T (\hat\kappa_\liml P_Z + (1 - \hat\kappa_\liml) \Id_n) X)^{-1} X^T (\hat\kappa_\liml P_Z + (1 - \hat\kappa_\liml) \Id_n) y \\
    &= \argmin_{\beta \in \BR^\mX} \ \hat\kappa_\liml \| P_Z (y - X \beta) \|^2 + (1 - \hat\kappa_\liml) \| y - X \beta \|^2
    \end{align*}
    and
    $$\AR(\hat\beta_\liml) = \frac{n - k}{k} (\hat\kappa_\liml - 1).$$
\end{theoremEnd}%
\begin{proofEnd}%
    We first prove that $\min_\beta \frac{k}{n-k} \AR(\beta) + 1 = \hat\kappa_\liml$ and then prove that $\hat \beta_\liml = \hat\beta_\kclass(\hat\kappa_\liml)$.

    \paragraph*{Step 1: $ \min_\beta \frac{k}{n-k} \AR(\beta) + 1 = \hat\kappa_\liml$} \ \\
    
    \noindent We can rewrite the optimization problem
    $$
    \min_\beta \frac{k}{n-k} \AR(\beta) + 1 = \min_\beta \frac{(y - X\beta)^T P_Z (y - X\beta)}{(y - X\beta)^T M_Z (y - X\beta)} + 1 = \min_\beta \frac{(y - X\beta)^T (y - X\beta)}{(y - X\beta)^T M_Z (y - X\beta)}
    $$
    as
    \begin{align*}
    \min_{\kappa, \beta} \ \kappa \ \textrm{ s.t. } \ &\frac{(y - X \beta)^T (y - X \beta)}{(y - X \beta)^T M_Z (y - X \beta)} \leq \kappa \\
    &\Leftrightarrow (y - X \beta)^T (\Id - \kappa M_Z) (y - X \beta) \leq 0 \\
    &\Leftrightarrow \begin{pmatrix}1 \\ -\beta \end{pmatrix}^T \begin{pmatrix} y & X\end{pmatrix}^T (\Id - \kappa M_Z) \begin{pmatrix} y & X \end{pmatrix} \begin{pmatrix} 1 \\ -\beta \end{pmatrix} \leq 0
    \end{align*}
    Thus, setting $M(\kappa) := \begin{pmatrix} y & X\end{pmatrix}^T (\Id - \kappa M_Z) \begin{pmatrix} y & X \end{pmatrix}$, 
    \begin{align*}
        \min_\beta \frac{k}{n-k} \AR(\beta) + 1 \label{eq:min_ar_beta} \numberthis
        &= \min\{ \kappa \in \BR \mid \exists \beta \in \BR^\mX \colon \begin{pmatrix} 1 \\ -\beta \end{pmatrix}^T M(\kappa) \begin{pmatrix} 1 \\ -\beta \end{pmatrix} \leq 0 \}\\
        &= \textrm{min} \{ \kappa \in \BR \mid \exists \tilde\beta \in (\BR \backslash \{ 0 \}) \times \BR^\mX \colon \tilde\beta^T M(\kappa) \tilde\beta \leq 0 \}\\
        &\overset{TC \ref{tc:liml_theorem}}{=} \textrm{min} \{ \kappa \in \BR \mid \exists \tilde\beta \in \BR^{\mX+1} \colon \tilde\beta^T M(\kappa) \tilde\beta \leq 0 \} \\
        &= \textrm{min} \{ \kappa \in \BR \mid \lambda_\textrm{min}(M(\kappa)) \leq 0 \}
    \end{align*}
    By \cref{lem:kappa_pos_definite} applied with $X \leftarrow \begin{pmatrix} X & y \end{pmatrix}$ and continuity of $\kappa \mapsto \lambdamin{M(\kappa)}$, it follows that
    $1 + \min_\beta \frac{k}{n-k} \AR(\beta) = \hat\kappa_\liml$.

    \paragraph*{Step 2: Deriving the closed-form solution for $\hat\beta_\liml$}\ \\

    By \cref{eq:min_ar_beta}, $\hat\beta_\liml$ solves
    \begin{multline*}
        0 =  M (\hat\kappa_\liml) \begin{pmatrix} 1 \\ -\hat\beta_\liml \end{pmatrix} = \begin{pmatrix} y & X\end{pmatrix}^T \left( \hat\kappa_\liml P_Z + (1 - \hat\kappa_\liml) \Id \right) ( y - X \hat\beta_\liml ) \\
        \Rightarrow X^T (\hat\kappa_\liml P_Z + (1 - \hat\kappa_\liml) \Id) X \hat\beta_\liml = X^T ( \hat\kappa_\liml P_Z + (1 - \hat\kappa_\liml) \Id) y \\
        \Rightarrow \hat\beta_\liml = (X^T (\hat\kappa_\liml P_Z + (1 - \hat\kappa_\liml) \Id) X)^{\dagger} X^T (\hat\kappa_\liml P_Z + (1 - \hat\kappa_\liml) \Id) y = \hat\beta_\kclass(\hat\kappa_\liml).
    \end{multline*}
    Here, $\dagger$ denotes the Moore-Penrose pseudoinverse.
    Note that $(X^T (\hat\kappa_\liml P_Z + (1 - \hat\kappa_\liml) \Id) X)$ is indeed invertible by \cref{tc:liml_theorem} and \cref{lem:kappa_pos_definite}.

\end{proofEnd}%

\begin{theoremEnd}[malte,category=estimators]{corollary}%
    \label{cor:liml_tsls_if_identified}
    If $\beta_0$ is just identified, that is, $k = m_x$, then $\hat\kappa_\liml = 1$ and $\hat\beta_\liml = \hat\beta_\tsls$.
\end{theoremEnd}%
\begin{proofEnd}%
    The matrix $\begin{pmatrix} y & X \end{pmatrix}^T P_Z \begin{pmatrix} y & X \end{pmatrix}$ has rank at most $\dim(Z)=k$.
    Thus, if $k = m_x$, it is singular with eigenvalue $0$ and thus $\hat\kappa_\liml = 1$.
\end{proofEnd}%

\Cref{tc:liml_theorem} is often assumed implicitly in the literature.
It is necessary.
For
\begin{equation*}
   X = \begin{pmatrix} 0.5 & 0 \\ 0 & 1 \\ 0 & 0 \\ 1 & 0 \\ 0 & 1 \\ 0 & 0 \end{pmatrix}, \ y = \begin{pmatrix} 0 \\ 0 \\ 1 \\0 \\ 0 \\ 1\end{pmatrix}, \text{ and } Z = \begin{pmatrix} 1 & 0 & 0 \\ 0 & 1 & 0 \\ 0 & 0 & 1\\ 0 & 0 & 0 \\ 0 & 0 & 0 \\ 0 & 0 & 0 \end{pmatrix} \Rightarrow \ 
   \begin{aligned}[c]
    &\begin{pmatrix} X & y \end{pmatrix}^T M_Z \begin{pmatrix} X & y \end{pmatrix} = \Id_3 \\
    &\begin{pmatrix} X & y \end{pmatrix}^T P_Z \begin{pmatrix} X & y \end{pmatrix} = \begin{pmatrix} 0.25 & 0 & 0 \\ 0 & 1 & 0 \\ 0 & 0 & 1 \end{pmatrix}
   \end{aligned}
\end{equation*}
we have 
$$
\lambdamin{ \left[X^T M_Z X \right]^{-1} X^T P_Z X  } = \lambdamin{ \left[\begin{pmatrix} X & y \end{pmatrix}^T M_Z \begin{pmatrix} X & y \end{pmatrix} \right]^{-1} \begin{pmatrix} X & y \end{pmatrix}^T P_Z \begin{pmatrix} X & y \end{pmatrix}  } = 0.25,$$ but the minimum for $\AR(\beta)$ is not achieved.
In practice, if the noise in \cref{model:0} is absolutely continuous with respect to the Lebesgue measure, \cref{tc:liml_theorem} holds with probability one.
We assume that the \cref{tc:liml_theorem} holds throughout this manuscript.

We will see later in \cref{cor:kappa_liml_is_chi_squared} that, asymptotically, $(n - k) (\hat\kappa_\liml - 1) = k \cdot \min_\beta \AR(\beta) = k \cdot \AR(\hat\beta_\liml)$ is bounded from above by a $\chi^2(k - \mX)$ distributed random variable, implying that the LIML estimator is asymptotically Gaussian by \cref{prop:kclass_asymptotic_normality}.

If $k > \mX$ and thus $\hat\kappa_\liml > 1$, the LIML estimator $\hat\beta_\kclass(\hat\kappa_\liml)$ regularizes the two-stage least-squares estimator $\hat\beta_\tsls = \hat\beta_\kclass(1)$ away from the ordinary least-squares estimate $\hat\beta_\ols := \hat\beta_\kclass(0)$.
This mitigates a bias of $\hat\beta_\tsls$ towards $\hat\beta_\ols$ that arises when many (weak) instruments lead to overfitting in the first stage.

The following proposition shows that the LIML estimator can be seen as a two-stage estimator, using the (infeasible) maximum likelihood estimate $\hat\Pi_\liml(\hat\beta_\liml)$ from \cref{lem:liml_likelihood} for the first-stage coefficients.
\begin{theoremEnd}[malte,category=estimators]{proposition}%
    \label{prop:liml_like_tsls}
    Let $\tilde X(\beta):= X - (y - X \beta) \frac{(y - X \beta)^T M_Z X}{(y - X \beta)^T M_Z (y - X \beta)}$.
    The LIML estimator satisfies
    \begin{align*}
    \hat\beta_\liml &= \left(\tilde X(\hat\beta_\liml)^T P_Z \tilde X(\hat\beta_\liml)\right)^{-1} \tilde X(\hat\beta_\liml)^T P_Z y \\
    &= \left(\hat\Pi_\liml(\hat\beta_\liml)^T Z^T Z \hat\Pi_\liml(\hat\beta_\liml)\right)^{-1} \hat\Pi_\liml(\hat\beta_\liml)^T Z^T y.
    \end{align*}
\end{theoremEnd}%
\begin{proofEnd}%
    By \cref{cor:liml_minimizes_ar} and \cref{lemma:ar_statistic_derivative}, we have
    \begin{align*}
        \tilde X(\hat\beta_\liml)^T P_Z (y - X \hat\beta_\liml) = 0
    \end{align*}
    Calculate
    \begin{multline*}
    \tilde X(\hat\beta_\liml)^T P_Z (y - \tilde X \hat\beta_\liml) = \tilde X(\hat\beta_\liml)^T P_Z (y - X \hat\beta_\liml) \\
    + \tilde X(\hat\beta_\liml)^T P_Z (y - X \hat\beta_\liml) \frac{(y - X \hat\beta_\liml)^T M_Z X \hat\beta_\liml}{(y - X \hat\beta_\liml)^T M_Z (y - X \hat\beta_\liml)} = 0 \\
    \Rightarrow \tilde X(\hat\beta_\liml)^T P_Z \tilde X(\hat\beta_\liml) \hat\beta_\liml = \tilde X(\hat\beta_\liml)^T P_Z y \\
    \Rightarrow \hat\beta_\liml = (\tilde X(\hat\beta_\liml)^T P_Z \tilde X(\hat\beta_\liml))^{-1} \tilde X(\hat\beta_\liml)^T P_Z y.
    \end{multline*}

\end{proofEnd}%

\noindent Compare this to $\hat\beta_\tsls = \left(X^T P_Z X\right)^{-1} X^T P_Z y = \left(\hat\Pi^T Z^T Z \hat\Pi\right)^{-1} \hat\Pi^T Z^T y$ for $\hat\Pi = (Z^T Z)^{-1} Z^T X$.

Unaware of k-class estimators, \citet{rothenhausler2021anchor} proposed the anchor regression estimator $\hat b^\gamma := \hat\beta_\kclass(\frac{\gamma - 1}{\gamma})$ for $\gamma \geq 1$.
They show that anchor regression yields predictions that are robust against certain distribution shifts.
\citet{jakobsen2022distributional} discuss this relation and propose the PULSE, a k-class estimator $\hat\beta_\kclass(\hat\kappa_\mathrm{PULSE}(\alpha))$ with $\hat\kappa_\mathrm{PULSE}(\alpha) := \min \{\kappa \geq 0 \mid \AR(\hat\beta_\kclass(\hat\kappa_\mathrm{PULSE})) \leq F^{-1}_{\chi^2(k)}(1 - \alpha) / k\}$, where $F_{\chi^2(k)}$ is the cumulative distribution function of the $\chi^2(k)$ distribution.
Differently from \citet{rothenhausler2021anchor} and \citet{jakobsen2022distributional}, we focus on estimation of and inference for the causal parameter $\beta_0$ and k-class estimators with $\kappa \geq 1$.

Another important class of k-class estimators are the Fuller estimators 
$\hat\beta_\mathrm{Fuller}(\alpha) := \hat\beta_\kclass(\hat\kappa_\liml - \frac{\alpha}{n - k})$ for $\alpha > 0$ \citep{fuller1977some}.
These improve the finite-sample properties of the LIML estimator, which does not have finite moments $\| \hat\beta_\liml \|^p$ for any $p \geq 1$, whereas the Fuller estimators have finite moments $\| \hat\beta_\mathrm{Fuller}(\alpha) \|^p$ for all $\alpha > 0$ and $p \geq 1$ such that the noise terms $(V_{X, i}, \varepsilon_i)$ have finite $p$-th moments.
Furthermore, $\hat\beta_\mathrm{Fuller}(1)$ has a bias of order $O(n^{-2})$ and $\hat\beta_\mathrm{Fuller}(4)$ minimizes the mean squared error among k-class estimators under weak-instrument asymptotics \citep{fuller1977some,staiger1997instrumental}.

\subsection{Application}
We now compare the ordinary least-squares (OLS), two-stage least-squares (TSLS), and limited information maximum likelihood (LIML) estimators applied to \citeauthor{card1993using}'s \citeyearpar{card1993using} data.

\begin{jupyternotebook}
    
\begin{tcolorbox}[breakable, size=fbox, boxrule=1pt, pad at break*=1mm,colback=cellbackground, colframe=cellborder]
\prompt{In}{incolor}{3}{\boxspacing}
\begin{Verbatim}[commandchars=\\\{\}]
\PY{k+kn}{from} \PY{n+nn}{ivmodels} \PY{k+kn}{import} \PY{n}{KClass}

\PY{n}{ols} \PY{o}{=} \PY{n}{KClass}\PY{p}{(}\PY{n}{kappa}\PY{o}{=}\PY{l+s+s2}{\PYZdq{}}\PY{l+s+s2}{ols}\PY{l+s+s2}{\PYZdq{}}\PY{p}{)}\PY{o}{.}\PY{n}{fit}\PY{p}{(}\PY{n}{Z}\PY{o}{=}\PY{k+kc}{None}\PY{p}{,} \PY{n}{X}\PY{o}{=}\PY{n}{X}\PY{p}{,} \PY{n}{y}\PY{o}{=}\PY{n}{y}\PY{p}{)}
\PY{n}{ols}\PY{o}{.}\PY{n}{named\PYZus{}coef\PYZus{}}
\end{Verbatim}
\end{tcolorbox}

            \begin{tcolorbox}[breakable, size=fbox, boxrule=.5pt, pad at break*=1mm, opacityfill=0]
\prompt{Out}{outcolor}{3}{\boxspacing}
\begin{Verbatim}[commandchars=\\\{\}]
intercept    4.768683
ed76         0.072634
exp76        0.084529
exp762      -0.002290
Name: coefficients, dtype: float64
\end{Verbatim}
\end{tcolorbox}
        
    \begin{tcolorbox}[breakable, size=fbox, boxrule=1pt, pad at break*=1mm,colback=cellbackground, colframe=cellborder]
\prompt{In}{incolor}{4}{\boxspacing}
\begin{Verbatim}[commandchars=\\\{\}]
\PY{n}{tsls} \PY{o}{=} \PY{n}{KClass}\PY{p}{(}\PY{n}{kappa}\PY{o}{=}\PY{l+s+s2}{\PYZdq{}}\PY{l+s+s2}{tsls}\PY{l+s+s2}{\PYZdq{}}\PY{p}{)}\PY{o}{.}\PY{n}{fit}\PY{p}{(}\PY{n}{Z}\PY{o}{=}\PY{n}{Z}\PY{p}{,} \PY{n}{X}\PY{o}{=}\PY{n}{X}\PY{p}{,} \PY{n}{y}\PY{o}{=}\PY{n}{y}\PY{p}{)}
\PY{n}{tsls}\PY{o}{.}\PY{n}{named\PYZus{}coef\PYZus{}}
\end{Verbatim}
\end{tcolorbox}

            \begin{tcolorbox}[breakable, size=fbox, boxrule=.5pt, pad at break*=1mm, opacityfill=0]
\prompt{Out}{outcolor}{4}{\boxspacing}
\begin{Verbatim}[commandchars=\\\{\}]
intercept    3.907937
ed76         0.144954
exp76        0.061604
exp762      -0.001196
Name: coefficients, dtype: float64
\end{Verbatim}
\end{tcolorbox}
        
    \begin{tcolorbox}[breakable, size=fbox, boxrule=1pt, pad at break*=1mm,colback=cellbackground, colframe=cellborder]
\prompt{In}{incolor}{5}{\boxspacing}
\begin{Verbatim}[commandchars=\\\{\}]
\PY{n}{liml} \PY{o}{=} \PY{n}{KClass}\PY{p}{(}\PY{n}{kappa}\PY{o}{=}\PY{l+s+s2}{\PYZdq{}}\PY{l+s+s2}{liml}\PY{l+s+s2}{\PYZdq{}}\PY{p}{)}\PY{o}{.}\PY{n}{fit}\PY{p}{(}\PY{n}{Z}\PY{o}{=}\PY{n}{Z}\PY{p}{,} \PY{n}{X}\PY{o}{=}\PY{n}{X}\PY{p}{,} \PY{n}{y}\PY{o}{=}\PY{n}{y}\PY{p}{)}
\PY{n}{liml}\PY{o}{.}\PY{n}{named\PYZus{}coef\PYZus{}}
\end{Verbatim}
\end{tcolorbox}

            \begin{tcolorbox}[breakable, size=fbox, boxrule=.5pt, pad at break*=1mm, opacityfill=0]
\prompt{Out}{outcolor}{5}{\boxspacing}
\begin{Verbatim}[commandchars=\\\{\}]
intercept    3.587249
ed76         0.172352
exp76        0.051571
exp762      -0.000713
Name: coefficients, dtype: float64
\end{Verbatim}
\end{tcolorbox}
\end{jupyternotebook}
\noindent Here $3 = \mX < k = 5$ and the LIML estimator $\hat\beta_\liml$ regularizes the two-stage least-squares estimator $\hat\beta_\tsls$ away from the ordinary least-squares estimator $\hat\beta_\ols$.

We numerically validate \cref{thm:liml_is_kclass} that $\AR(\hat\beta_\mathrm{LIML}) = \frac{n-k}{k} (\hat\kappa_\mathrm{LIML} - 1)$
on our data.
\begin{jupyternotebook}
    \begin{tcolorbox}[breakable, size=fbox, boxrule=1pt, pad at break*=1mm,colback=cellbackground, colframe=cellborder]
\prompt{In}{incolor}{6}{\boxspacing}
\begin{Verbatim}[commandchars=\\\{\}]
\PY{c+c1}{\PYZsh{} Multiply with (n \PYZhy{} k \PYZhy{} 1) instead of (n \PYZhy{} k) due to the included intercept}
\PY{p}{(}\PY{n}{y}\PY{o}{.}\PY{n}{shape}\PY{p}{[}\PY{l+m+mi}{0}\PY{p}{]} \PY{o}{\PYZhy{}} \PY{n}{Z}\PY{o}{.}\PY{n}{shape}\PY{p}{[}\PY{l+m+mi}{1}\PY{p}{]} \PY{o}{\PYZhy{}} \PY{l+m+mi}{1}\PY{p}{)} \PY{o}{/} \PY{n}{Z}\PY{o}{.}\PY{n}{shape}\PY{p}{[}\PY{l+m+mi}{1}\PY{p}{]} \PY{o}{*} \PY{p}{(}\PY{n}{liml}\PY{o}{.}\PY{n}{kappa\PYZus{}} \PY{o}{\PYZhy{}} \PY{l+m+mi}{1}\PY{p}{)}
\end{Verbatim}
\end{tcolorbox}

            \begin{tcolorbox}[breakable, size=fbox, boxrule=.5pt, pad at break*=1mm, opacityfill=0]
\prompt{Out}{outcolor}{6}{\boxspacing}
\begin{Verbatim}[commandchars=\\\{\}]
0.8568537499466554
\end{Verbatim}
\end{tcolorbox}
        
    \begin{tcolorbox}[breakable, size=fbox, boxrule=1pt, pad at break*=1mm,colback=cellbackground, colframe=cellborder]
\prompt{In}{incolor}{7}{\boxspacing}
\begin{Verbatim}[commandchars=\\\{\}]
\PY{k+kn}{from} \PY{n+nn}{ivmodels}\PY{n+nn}{.}\PY{n+nn}{tests} \PY{k+kn}{import} \PY{n}{anderson\PYZus{}rubin\PYZus{}test}

\PY{c+c1}{\PYZsh{} This returns a tuple (statistic, p\PYZus{}value)}
\PY{n}{anderson\PYZus{}rubin\PYZus{}test}\PY{p}{(}\PY{n}{Z}\PY{o}{=}\PY{n}{Z}\PY{p}{,} \PY{n}{X}\PY{o}{=}\PY{n}{X}\PY{p}{,} \PY{n}{y}\PY{o}{=}\PY{n}{y}\PY{p}{,} \PY{n}{beta}\PY{o}{=}\PY{n}{liml}\PY{o}{.}\PY{n}{coef\PYZus{}}\PY{p}{)}
\end{Verbatim}
\end{tcolorbox}

            \begin{tcolorbox}[breakable, size=fbox, boxrule=.5pt, pad at break*=1mm, opacityfill=0]
\prompt{Out}{outcolor}{7}{\boxspacing}
\begin{Verbatim}[commandchars=\\\{\}]
(0.8568537499489439, 0.5092552733782495)
\end{Verbatim}
\end{tcolorbox}
\end{jupyternotebook}

\section{Tests for the causal parameter}
\label{sec:intro_to_iv_tests:tests}
We now present tests for the causal parameter.
In a typical application, one wishes to make subvector or partial inference for the causal parameter.
For example, in our application, we would like to be able to test whether the partial causal effect of education on wages is statistically significant, rather than testing the joint causal effect of education, potential experience, and its square.
To model this, we extend \cref{model:0}: We split the endogenous variables $X$ into $(X, W)$, the first-stage parameter $\Pi$ into $(\Pi_X, \Pi_W)$, and the causal parameter $\beta_0$ into $(\beta_0, \gamma_0)$.
Now $X$ includes the endogenous variables whose causal effect $\beta_0$ we are interested in and wish to make inference for.
The coefficient $\gamma_0$ corresponding to $W$ is a nuisance parameter.
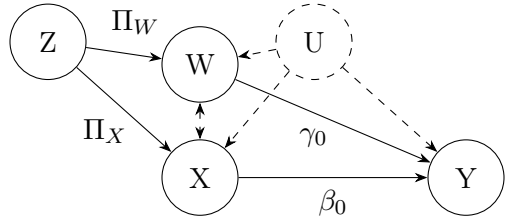
\begin{figure}[!h]
    \centering
        \begin{tikzpicture}[
            node distance=2cm and 2cm,
            >=Stealth,
            every node/.style={draw, circle, minimum size=1cm, inner sep=0pt},
            dashednode/.style={draw, circle, minimum size=1cm, inner sep=0pt, dashed},
            dashedarrow/.style={->, dashed}
        ]

        \node (Z) at (0,1.8) {Z};
        \node (W) at (2,1.5) {W};
        \node (X) at (2,0) {X};
        \node (Y) at (5.5,0) {Y};
        \node[dashednode] (U) at (3.5,1.8) {U};

        \draw[->] (Z) -- (X) node[pos=0.3, below, draw=none] {$\Pi_X$};
        \draw[->] (Z) -- (W) node[pos=0.65, above, draw=none, yshift=-2] {$\Pi_W$};
        \draw[->] (X) -- (Y) node[pos=0.5, below, draw=none, yshift=5] {$\beta_0$};
        \draw[->] (W) -- (Y) node[pos=0.4, below, draw=none, yshift=5] {$\gamma_0$};
        \draw[<->, dashed] (W) -- (X);
        \draw[dashedarrow] (U) -- (W);
        \draw[dashedarrow] (U) -- (X);
        \draw[dashedarrow] (U) -- (Y);

        \end{tikzpicture}
    \caption{
        \label{fig:iv_graph_XW}
        Causal graph visualizing \cref{model:1}.
    }
\end{figure}

\begin{theoremEnd}[malte,restate command=modelone]{model}
    \label{model:1}
    Let $y_i = X_i^T \beta_0 + W_i^T \gamma_0 + \varepsilon_i \in \BR$ with $X_i = Z_i^T \Pi_X + V_{X, i} \in \BR^\mX$ and $W_i = Z_i^T \Pi_W + V_{W, i} \in \BR^\mW$ for random vectors $Z_i \in \BR^k, V_{X, i}\in \BR^\mX, V_{W, i}\in\BR^\mW$, and $\varepsilon_i \in \BR$ for $i=1\ldots, n$ and parameters $\Pi_X \in \BR^{k \times \mX}$, $\Pi_W \in \BR^{k \times \mW}$, $\beta_0 \in \BR^\mX$, and $\gamma_0 \in \BR^\mW$.
    We call the $Z_i$ \emph{instruments}, the $X_i$ \emph{endogenous covariates of interest}, the $W_i$ \emph{endogenous covariates not of interest}, and the $y_i$ \emph{outcomes}.
    The $V_{X, i}, V_{W, i}$, and $\varepsilon_i$ are \emph{errors}.
    These need not be independent across observations.
    Let $Z \in \BR^{n \times k}, X \in \BR^{n \times \mX}, W \in \BR^{n \times \mW}$, and $y \in \BR^n$ be the matrices of stacked observations.

    In \emph{strong instrument asymptotics}, we assume that $\Pi := \begin{pmatrix} \Pi_X & \Pi_W \end{pmatrix}$ is fixed for all $n$ and of full column rank $m := \mX + \mW$.
    In \emph{weak instrument asymptotics} \citep{staiger1997instrumental}, we assume that $\sqrt{n} \, \Pi = \sqrt{n}  \begin{pmatrix} \Pi_X & \Pi_W \end{pmatrix}$ is fixed and of full column rank $m$. Thus, $\Pi = \CO(\frac{1}{\sqrt{n}})$.
    Both asymptotics imply that $k \geq m$.
\end{theoremEnd}%

\noindent We also extend \cref{ass:0}.
\begin{theoremEnd}[malte,restate command=assumptionone]{assumption}
    \label{ass:1}
    Let
    $$
    \Psi := \begin{pmatrix} \Psi_{\varepsilon} & \Psi_{V_X} & \Psi_{V_W} \end{pmatrix} := (Z^T Z)^{-1/2} Z^T \begin{pmatrix} \varepsilon & V_X & V_W \end{pmatrix} \in \BR^{k \times (1 + m)}.
    $$
    Assume there exist $\Omega \in \BR^{(1+m) \times (1+m)}$ and $Q \in \BR^{k \times k}$ positive definite such that, as $n \to \infty$,
    \begin{align*}
        &\mathrm{(a)} \ \ \frac{1}{n} \begin{pmatrix}\varepsilon & V_X & V_W\end{pmatrix}^T \begin{pmatrix}\varepsilon & V_X & V_W \end{pmatrix} \toP \Omega \\
        &\mathrm{(b)} \ \ \vecop(\Psi) \tod \CN(0, \Omega \otimes \Id_k), \text{ and }\\
        &\mathrm{(c)} \ \ \frac{1}{n} Z^T Z \toP Q,
    \end{align*}
    where $\Cov(\vecop(\Psi)) = \Omega \otimes \Id_k$ means $\Cov(\Psi_{i, j}, \Psi_{i', j'}) = 1_{i = i'} \cdot \Omega_{j, j'}$.
\end{theoremEnd}%

\Cref{lem:when_does_ass_0_hold} applies similarly by replacing $V_{X, i} \leftarrow (V_{X, i}, V_{W, i})$.
That is, if (i) the error terms $\varepsilon_i, V_{X, i}$, and $V_{W, i}$ are homoscedastic and uncorrelated with the instruments $Z_i$ and (ii) the error terms and instruments are i.i.d.\ with finite second moments, \cref{ass:1} holds.

\subsection{The Wald test}
The Wald test is a classical approach to hypothesis testing if an estimator of the parameter of interest can be shown to be asymptotically Gaussian.

\begin{definition}\label{def:wald_test_statistic}
    Let $B \in \BR^{\mX \times m}$ have ones on the diagonal and zeros elsewhere.
    Let $\kappa \geq 0$ and let $\hat \beta_\kclass(\kappa)$ be the k-class estimator of $(\beta_0^T, \gamma_0^T)^T$ using outcomes $y$, covariates $S := \begin{pmatrix} X & W \end{pmatrix}$, and instruments $Z$.
    Let $\hat\sigma^2_{\Wald}(\kappa) := \frac{1}{n - m} \| y - S \hat\beta_\kclass(\kappa) \|^2$.
    The (subvector) Wald test statistic is
    \begin{align*}
        \Wald_{\hat \beta_\kclass(\kappa)}(\beta) := \frac{1}{\hat\sigma^2_{\Wald}(\kappa)} 
        (\beta - B \hat \beta_\kclass(\kappa))^T \left( B \left[S^T (\Id_{m} - \kappa M_Z) S\right)^{-1} B^T \right]^{-1} ( \beta - B \hat \beta_\kclass(\kappa))
    \end{align*}
\end{definition}

In the following, we will use the same notation for tests on the full causal parameter and their subvector variants.
Subvector tests will always reduce to the original test if $\mW = 0$.
If a proposition only holds for $\mW = 0$, we will state this explicitly.
The following proposition follows from \cref{prop:kclass_asymptotic_normality}.

\begin{theoremEnd}[malte,category=tests]{proposition}%
    \label{prop:wald_test_chi_squared}
    Let $\kappa_n$ satisfy $\sqrt{n}(\kappa_n - 1) \toP 0$ as $n\to\infty$.
    Under the null $\beta = \beta_0$ and under strong instrument asymptotics, the (subvector) Wald test statistic $\Wald_{\hat\beta_\kclass(\kappa_n)}(\beta_0)$ is asymptotically $\chi^2(m_x)$ distributed.
\end{theoremEnd}%
\begin{proofEnd}%
    By \cref{prop:kclass_asymptotic_normality}, we have that
    $$
        \frac{1}{\sqrt{\hat\sigma^2_{\Wald}(\kappa)}} \left( B \left(S^T ( \kappa_n P_Z + (1 - \kappa_n) \Id ) S \right)^{-1} B^T \right)^{-1/2} B \left(
            \begin{pmatrix} \beta_0 \\ \gamma_0 \end{pmatrix} - \hat\beta_\kclass(\kappa)
            \right)\tod \CN(0, \Id_\mX).
    $$
\end{proofEnd}%

\begin{theoremEnd}[malte,category=tests]{corollary}%
    \label{cor:subvector_wald_test}
    Let $F_{\chi^2(\mX)}$ be the cumulative distribution function of a chi-squared random variable with $\mX$ degrees of freedom.
    Under \cref{model:1} and strong instrument asymptotics, for any sequence $\kappa_n$ such that $\sqrt{n}(\kappa_n - 1) \toP 0$ as $n\to\infty$, a test that rejects the null $H_0: \beta = \beta_0$ whenever $\Wald_{\hat\beta_\kclass(\kappa_n)}(\beta) > F^{-1}_{\chi^2(\mX)}(1 - \alpha)$ has asymptotic size $\alpha$.
\end{theoremEnd}%
\begin{proofEnd}%
    This follows directly from \cref{prop:wald_test_chi_squared}.
\end{proofEnd}%

\noindent
This applies to the two-stage least-squares estimator $\hat\beta_\tsls = \hat\beta_\kclass(1)$.
We will see in \cref{cor:kappa_liml_is_chi_squared} that this applies to the LIML estimator $\hat\beta_\liml = \hat\beta_\kclass(\hat\kappa_\liml)$ as well.

Under weak instrument asymptotics, these estimators are no longer asymptotically normal and \cref{prop:wald_test_chi_squared} and \cref{cor:subvector_wald_test} no longer apply.
In practice, when instruments are weak, $p$-values calculated based on \cref{cor:subvector_wald_test} are too small and confidence intervals are too narrow, too often not containing the true parameter of interest.

To get an intuition of why asymptotic normality fails under weak instrument asymptotics, recall the formulation $\hat\beta_\tsls = \widehat{\Pi}^{-1} \widehat{\Pi \beta}$ for $k=\mX$ and $\mW=0$ from \cref{sec:intro_to_iv_tests:estimators}.
Both ordinary least-squares estimators $\widehat{\Pi}$ and $\widehat{\Pi \beta}$ are asymptotically normal.
Under strong instrument asymptotics, the first-stage estimator $\widehat{\Pi}$ converges in probability to the non-degenerate matrix $\Pi$, and asymptotic normality of $\hat\beta_\tsls$ follows using Slutsky.
However, under weak instrument asymptotics, the first-stage estimator $\widehat{\Pi}$ converges in distribution to a multivariate Gaussian random variable with mean $\Pi$ and covariance $\Omega_V \otimes Q^{-1}$.
The two-stage least-squares estimator $\hat\beta_\tsls$ then converges in distribution to the ratio of two Gaussians, with Cauchy-like tails.

A measure of instrument strength is the first-stage $F$-statistic.
\citet{staiger1997instrumental} note that in a just-identified model with a single endogenous variable ($\mW = 0$ and $k = \mX = 1$), if $\mu^2 := n \Omega_V^{-1} \Pi^T Q \Pi = \BE[F] - 1 \geq 9$, then the Wald-based 95\%-confidence interval for $\beta_0$ has asymptotically at least 85\% coverage.
This has resulted in the ``rule of thumb'' that the first-stage $F$-statistic should exceed 10 for valid inference using the Wald test.
\citet{stock2002testing} present equivalent critical values for the (multivariate) population first-stage $F$-statistic for multiple instruments or endogenous variables ($k > 1$ or $m > 1$).
In practice, \citeauthor{staiger1997instrumental}'s \citeyearpar{staiger1997instrumental} rule of thumb value of 10 is often used even when $k > 1$ or $m > 1$.
See \cref{sec:anderson_1951_rank_test} for a discussion of the first-stage $F$-test heuristic.

We believe that a two-stage testing approach is suboptimal and recommend the usage of tests that are valid under both weak and strong instrument asymptotics.
We present three such tests, the Anderson-Rubin test \citep{anderson1949estimation,guggenberger2012asymptotic}, the conditional likelihood-ratio test \citep{moreira2003conditional,kleibergen2021efficient}, and the Lagrange multiplier test \citep{kleibergen2002pivotal,londschien2024weak} in the following.

\subsection{The (subvector) Anderson-Rubin test}
\label{sec:anderson_rubin_test}
We have already seen the Anderson-Rubin test statistic in \cref{def:anderson_rubin_test_statistic}.
We first present the following result for $\mW = 0$.
\begin{theoremEnd}[malte,category=tests]{proposition}[\citeauthor{anderson1949estimation}, \citeyear{anderson1949estimation}]%
    \label{prop:anderson_rubin_test}
    Under the null $H_0: \beta = \beta_0$ and both strong and weak instrument asymptotics, the Anderson-Rubin test statistic is asymptotically $\chi^2(k) / k$ distributed.
\end{theoremEnd}%
\begin{proofEnd}%
    By \cref{ass:1} (b),
    $$
    (y - X \beta_0)^T P_Z (y - X \beta_0) = \Psi_{\varepsilon}^T \Psi_{\varepsilon} \tod \sigma_\varepsilon^2 \cdot \chi^2(k).
    $$
    The result then follows from $\frac{1}{n-k}(y - X \beta_0)^T M_Z (y - X \beta_0) \toP \sigma_\varepsilon^2$ (\cref{ass:1}, a).
\end{proofEnd}%
 
The proof follows from \cref{ass:0}.
Recall the estimators $\widehat \Pi = (Z^T Z)^{-1} Z^T X$ and $\widehat{\Pi \beta} = (Z^T Z)^{-1} Z^T y$ from \cref{sec:intro_to_iv_tests:estimators}.
We present a more constructive proof based on the assumptions of \cref{lem:when_does_ass_0_hold}, which imply that $\widehat \Pi$ and $\widehat{\Pi \beta}$ are asymptotically jointly Gaussian after subtracting their means:
$$
\sqrt{n}
\left(
    \begin{pmatrix}
        \widehat{\Pi \beta} &  \widehat \Pi
    \end{pmatrix}
- 
    \begin{pmatrix}
        \Pi \beta_0 & \Pi
    \end{pmatrix}
\right)
\tod \CN\left(0, \Omega_{y X} \otimes Q \right), \text{ where } \Omega_{yX} := \begin{pmatrix} 1 & 0 \\ \beta_0 & \Id_{\mX} \end{pmatrix}^T \!\! \Omega \begin{pmatrix} 1 & 0 \\ \beta_0 & \Id_{\mX} \end{pmatrix}.
$$
Consequently, for any $\beta \in \BR^\mX$,
$$
    \sqrt{n} \left(\begin{pmatrix}
        \widehat{\Pi \beta} & \widehat \Pi
    \end{pmatrix}
    \begin{pmatrix}
        1 \\ -\beta
    \end{pmatrix} - \Pi (\beta_0 - \beta) \right) \tod \CN \Big(0,  Q \cdot \underbrace{\begin{pmatrix}
        1 \\ -\beta
    \end{pmatrix}^T \Omega_{y X} \begin{pmatrix}
        1 \\ -\beta
    \end{pmatrix}}_{=: \sigma^2_{y - X \beta}} \Big).$$
The covariance terms $Q$, $\Omega_{y X}$, and $\sigma^2_{y - X \beta}$ can be estimated consistently as $\hat Q := \frac{1}{n} Z^T Z$, $\widehat\Omega_{y X} := \frac{1}{n - k} \begin{pmatrix} y & X \end{pmatrix}^T M_Z \begin{pmatrix} y & X \end{pmatrix}$, and $\hat\sigma^2_{y - X \beta} := \frac{1}{n - k} (y - X \beta)^T M_Z (y - X \beta)$.
Define
$$ T(\beta) := \frac{1}{\hat \sigma_{y - X \beta}} \hat Q^{-1/2} \begin{pmatrix}
        \widehat{\Pi \beta} & \widehat \Pi
    \end{pmatrix}
    \begin{pmatrix}
        1 \\ -\beta
    \end{pmatrix},
$$
such that $\sqrt{n} \, (T(\beta) - \Pi (\beta_0 - \beta)) \tod \CN(0, \Id_k)$.
The Anderson-Rubin test statistic is equal to $\AR(\beta) = n \, T(\beta)^T T(\beta) / k$.

We also define a subvector extension.
\begin{theoremEnd}[malte, restate command=subvectorandersonrubinteststatistic]{definition}[\citeauthor{guggenberger2012asymptotic}, \citeyear{guggenberger2012asymptotic}]
    \label{def:subvector_anderson_rubin_test_statistic}
    The subvector Anderson-Rubin test statistic is
    \begin{align*}
        \AR(\beta) &:=  \min_{\gamma \in \BR^\mW} \frac{n - k}{k - \mW} \frac{(y - X \beta - W \gamma)^T P_Z (y - X \beta - W \gamma)}{(y - X \beta - W \gamma)^T M_Z (y - X \beta - W \gamma)} \\
        &= \frac{n - k}{k - \mW} \frac{(y - X \beta - W \hat\gamma_\liml)^T P_Z (y - X \beta - W \hat\gamma_\liml)}{(y - X \beta - W \hat\gamma_\liml)^T M_Z (y - X \beta - W \hat\gamma_\liml)} \\
        &= \frac{n - k}{k - \mW} \lambdamin{ \kern-1pt  \left[  \setlength\arraycolsep{3pt} \begin{pmatrix} y \kern-1pt - \kern-1pt X \beta & W \end{pmatrix}^T  \kern-3pt M_Z \kern-1pt \begin{pmatrix} y\kern-1pt - \kern-1pt X\beta & W \end{pmatrix} \right]^{-1}  \kern-2pt  \setlength\arraycolsep{3pt} \begin{pmatrix} y\kern-1pt - \kern-1pt X\beta & W \end{pmatrix}^T  \kern-3pt M_Z \kern-1pt \begin{pmatrix} y\kern-1pt - \kern-1pt X\beta & W \end{pmatrix} \kern-1pt  },
    \end{align*}
    where $\hat\gamma_\liml$ is the LIML estimator using outcomes $y - X \beta$, covariates $W$, and instruments $Z$.
\end{theoremEnd}%

\begin{theoremEnd}[malte,category=tests]{proposition}[\citeauthor{guggenberger2012asymptotic}, \citeyear{guggenberger2012asymptotic}]%
    \label{prop:subvector_anderson_rubin_test_statistic}
    Under the null $H_0: \beta = \beta_0$, for both strong and weak instrument asymptotics, the subvector Anderson-Rubin test statistic is asymptotically bounded from above by a $\chi^2(k-\mW) / (k - \mW)$ distributed random variable.
\end{theoremEnd}%
\begin{proofEnd}%
    This proof follows \citet{guggenberger2012asymptotic}.
    It has four steps:
    \begin{itemize}
        \item First, we construct $\tilde W$, $\eta$, and $\gamma^\star$ such that $P_Z (y - X \beta_0 - W \gamma^\star) = \frac{1}{1 + \eta} M_{P_Z \tilde W} P_Z \varepsilon$.
        \item Then, we show that $\varepsilon^T P_Z M_{P_Z \tilde W} P_Z \varepsilon \tod \sigma_\varepsilon^2 \cdot \chi^2(k-\mW)$.
        \item Next, we show that $\plim \frac{(1 + \eta)^2}{n - k} (y - X \beta_0 - W \gamma^\star)^T M_Z (y - X \beta_0 - W \gamma^\star) \geq \sigma_\varepsilon^2$.
        \item Finally, as $\AR(\beta) \leq (n-k) \frac{(y - X \beta_0 - W \gamma^\star)^T P_Z (y - X \beta_0 - W \gamma^\star)}{(y - X \beta_0 - W \gamma^\star)^T M_Z (y - X \beta_0 - W \gamma^\star)}$, this concludes the proof.
    \end{itemize}
    \paragraph{Step 1:}
    Write 
    $\Omega = \begin{pmatrix}
        \sigma_\varepsilon^2 & \Omega_{\varepsilon, V_X} & \Omega_{\varepsilon, V_W} \\
        \Omega_{V_X, \varepsilon} & \Omega_{V_X} & \Omega_{V_X, V_W} \\ \Omega_{V_W, \varepsilon} & \Omega_{V_W, V_X} & \Omega_{V_W} \end{pmatrix}$.
    Let $\tilde W := W - \varepsilon \frac{\Omega_{\varepsilon, V_W}}{\sigma_\varepsilon^2} \in \BR^{n \times \mW}$ and $\eta := \frac{\Omega_{\varepsilon, V_W}}{\sigma_\varepsilon^2} (\tilde W^T P_Z \tilde W)^{-1} \tilde W^T P_Z \varepsilon \in \BR$
    and define
    \begin{equation*}
        \gamma^\star = \frac{(\tilde W^T P_Z \tilde W)^{-1} \tilde W^T P_Z \varepsilon}{1 + \eta} + \gamma_0 \in \BR^\mW.
    \end{equation*}
    We calculate
    \begin{align*}
        (1 + \eta)  (y - X \beta_0 - W \gamma^\star) &= (1 + \eta)(\varepsilon + W (\gamma_0 - \gamma^\star)) \\
        &= (1 + \eta)\varepsilon - W (\tilde W^T P_Z \tilde W)^{-1} \tilde W^T P_Z \varepsilon \\
        &= \varepsilon - \tilde W (\tilde W^T P_Z \tilde W)^{-1} \tilde W^T P_Z \varepsilon,
    \end{align*}
    and thus
    \begin{align*}
        (1 + \eta) &P_Z (y - X \beta_0 - W \gamma^\star) =
        P_Z \varepsilon - P_Z \tilde W (\tilde W^T P_Z \tilde W)^{-1} \tilde W^T P_Z \varepsilon =
        P_Z M_{P_Z \tilde W} \varepsilon \\
        &\Rightarrow
        (y - X \beta - W \gamma^\star)^T P_Z (y - X \beta - W \gamma^\star) = \frac{\varepsilon  M_{P_Z \tilde W} P_Z M_{P_Z \tilde W} \varepsilon}{(1 + \eta)^2}.
    \end{align*}
    The projections $P_Z$ and $M_{P_Z \tilde W}$ commute, concluding Step 1.
    \paragraph*{Step 2:}
    Let 
    \begin{align*}
    \Psi_{\tilde W} := (Z^T Z)^{-1/2} Z^T \tilde W &= (Z^T Z)^{-1/2} Z^T (Z \Pi_W + V_W - \varepsilon \frac{\Omega_{\varepsilon, V_W}}{\sigma_\varepsilon^2}) \\
    &= (Z^T Z)^{1/2} \Pi_W + \Psi_{V_W} - \Psi_{\varepsilon} \frac{\Omega_{\varepsilon, V_W}}{\sigma_\varepsilon^2}
    \end{align*}
    such that
    \begin{align*}
        \varepsilon^T &M_{P_Z \tilde W} P_Z M_{P_Z \tilde W} \varepsilon = \varepsilon^T P_Z M_{P_Z \tilde W} P_Z \varepsilon \\
        &= \Psi_{\varepsilon}^T (Z^T Z)^{-1/2} Z^T (\Id_n - Z (Z^T Z)^{-1/2} \Psi_{\tilde W} (\Psi_{\tilde W}^T \Psi_{\tilde W})^{-1} \Psi_{\tilde W}^T (Z^T Z)^{-1/2} Z) Z (Z^T Z)^{-1/2} \Psi_{\varepsilon} \\
        &= \Psi_{\varepsilon}^T (\Id_k - \Psi_{\tilde W} (\Psi_{\tilde W}^T \Psi_{\tilde W})^{-1} \Psi_{\tilde W}^T) \Psi_{\varepsilon} = \Psi_{\varepsilon}^T M_{\Psi_{\tilde W}} \Psi_{\varepsilon}.
    \end{align*}
    We show that this is asymptotically $\chi^2$-distributed separately for strong and weak instruments asymptotics.

    \textbf{Assuming strong instruments}, that is, $\Pi_W$ is constant and of full rank, then, by \cref{ass:1}, $\frac{1}{\sqrt{n}} \Psi_{\tilde W} \toP Q^{1/2}\Pi_W$ and thus
    $$
    \Psi_{\varepsilon}^T M_{\Psi_{\tilde W}} \Psi_{\varepsilon} = \Psi_{\varepsilon}^T M_{\frac{1}{\sqrt{n}}\Psi_{\tilde W}} \Psi_{\varepsilon} \overset{\BP}{\to} \Psi_{\varepsilon}^T M_{Q^{1/2} \Pi_W} \Psi_{\varepsilon} \tod \sigma_\varepsilon^2 \cdot \chi^2(k - \mW),
    $$
    as $\Psi_{\varepsilon} \tod \CN(0, \sigma_\varepsilon^2 \cdot \Id_k)$ and $Q^{1/2}\Pi_W$ has rank $\mW$.

    \textbf{Assuming weak instruments}, that is, $\Pi_W = \frac{1}{\sqrt{n}} C_W$ for some constant $C_W$, then
    $$
    \Psi_{\tilde W} \toP Q^{1/2} C_W + \Psi_{W} - \Psi_{\varepsilon} \frac{\Omega_{\varepsilon, V_W}}{\sigma_\varepsilon^2}
    $$
    is asymptotically Gaussian by \cref{ass:1}.
    We calculate
    \begin{align*}
        \Cov(\Psi_\varepsilon, \Psi_{\tilde W}) & = \Cov(\Psi_{\varepsilon}, \Psi_{V_W} - \Psi_{\varepsilon} \frac{\Omega_{\varepsilon, V_W}}{\sigma_\varepsilon^2}) \\
        &=
        \Cov(\Psi_{\varepsilon}, \Psi_{V_W}) - \Cov(\Psi_{\varepsilon}, \Psi_{\varepsilon}) \frac{\Omega_{\varepsilon, V_W}}{\sigma_\varepsilon^2} \\
        &\overset{\cref{ass:1} b}{\toP} \Omega_{\varepsilon, V_W} - \sigma_\varepsilon^2 \frac{\Omega_{\varepsilon, V_W}}{\sigma^2_\varepsilon} = 0.
    \end{align*}
    Thus, $\Psi_{\tilde W}$ and $\Psi_{\varepsilon}$ are asymptotically jointly Gaussian and uncorrelated, and thus asymptotically independent.
    Conditional on $\Psi_{\tilde W}$, the random variable
    $\Psi_{\varepsilon}^T M_{\Psi_{\tilde W}} \Psi_{\varepsilon}
    $
    is asymptotically $\sigma_\varepsilon^2 \cdot \chi^2(k - \mW)$ distributed, as $\rank(M_{\Psi_{\tilde W}}) = k-\mW$ almost surely.
    By asymptotic independence of $\Psi_{\varepsilon}$ and $\Psi_{\tilde W}$, asymptotically this also holds unconditionally.

    \paragraph*{Step 3:}
    Write $\sigma^2_{\varepsilon + V_W (\gamma_0 - \gamma^\star)} := \begin{pmatrix} 1 \\ \gamma_0 - \gamma^\star \end{pmatrix}^T \begin{pmatrix} \sigma^2_\varepsilon & \Omega_{\varepsilon, V_W} \\ \Omega_{\varepsilon, V_W}^T & \Omega_{V_W} \end{pmatrix} \begin{pmatrix} 1 \\ \gamma_0 - \gamma^\star \end{pmatrix}$.
    By \cref{ass:1} a, the denominator
    $$
    \frac{1}{n-k}(y - X \beta_0 - W \gamma^\star)^T M_Z (y - X \beta_0 - W \gamma^\star) \toP \sigma^2_{\varepsilon + V_W (\gamma_0 - \gamma^\star)}.
    $$
    By definition of $\eta = \frac{\Omega_{\varepsilon, V_W}}{\sigma_\varepsilon^2} (\tilde W^T P_Z \tilde W)^{-1} \tilde W P_Z \varepsilon$ and $\gamma^\star = \gamma_0 + \frac{1}{1 + \eta}(\tilde W^T P_Z \tilde W)^{-1} \tilde W P_Z \varepsilon$, we have
    $$
    \frac{\Omega_{\varepsilon, V_W}}{\sigma_\varepsilon^2} (\gamma_0 - \gamma^\star) = -\frac{\eta}{1 + \eta}
    $$
    and thus
    $$
    \frac{\sigma_\varepsilon^2 + \Omega_{\varepsilon, V_W} (\gamma_0 - \gamma^\star)}{\sigma_\varepsilon^2} = 1 - \frac{\eta}{1 + \eta} = \frac{1}{1 + \eta}.
    $$
    Calculate
    \begin{align*}
    \frac{\sigma_\varepsilon^2}{1 + \eta} &=
    \sigma_\varepsilon^2 + \Omega_{\varepsilon, V_W} (\gamma_0 - \gamma^\star)
    = \begin{pmatrix} 1 \\ 0 \end{pmatrix}^T \begin{pmatrix} \sigma^2_\varepsilon & \Omega_{\varepsilon, V_W} \\ \Omega_{\varepsilon, V_W}^T & \Omega_{V_W} \end{pmatrix} \begin{pmatrix} 1 \\ \gamma_0 - \gamma^\star \end{pmatrix} \\
    &\overset{\text{Cauchy Schwarz}}{\leq} \sqrt{\sigma_\varepsilon^2} \sqrt{\sigma^2_{\varepsilon + V_W (\gamma_0 - \gamma^\star)}}
    \end{align*}
    Thus
    $$
    \plim \frac{(1 + \eta)^2}{n - k}  (y - X \beta_0 - W \gamma^\star)^T M_Z (y - X \beta_0 - W \gamma^\star) \geq \sigma_\varepsilon^2.
    $$
    \paragraph*{Step 4:}
    Steps 1, 2, and 3 combined imply that, both under strong and weak instrument asymptotics, the random variable
    $$
    (n - k) \frac{(y - X\beta_0 - W \gamma^\star) P_Z (y - X\beta_0 - W \gamma^\star)}{(y - X\beta_0 - W \gamma^\star) M_Z (y - X\beta_0 - W \gamma^\star)}
    $$
    is asymptotically upper bounded by a $\chi^2(k - \mW)$ random variable.
    As
    \begin{align*}
    \AR(\beta_0) &= \min_\gamma \frac{(y - X \beta_0 - W \gamma)^T P_Z (y - X \beta_0 - W \gamma)}{(y - X \beta_0 - W \gamma)^T M_Z (y - X \beta_0 - W \gamma)} \\
    &\leq \frac{(y - X \beta_0 - W \gamma^\star)^T P_Z (y - X \beta_0 - W \gamma^\star)}{(y - X \beta_0 - W \gamma^\star)^T M_Z (y - X \beta_0 - W \gamma^\star)},
    \end{align*}
    this concludes the proof.
\end{proofEnd}%

That is, by marginalizing over the nuisance parameter $\gamma_0$, we recover $\mW$ degrees of freedom for the limiting chi-squared distribution.
The special case $\mX = 0$ yields the following corollary.%
\begin{theoremEnd}[malte,category=tests]{corollary}
    \label{cor:kappa_liml_is_chi_squared}
    Both for strong and weak instrument asymptotics, $(n - k) (\hat\kappa_\liml - 1)$ is asymptotically bounded from above by a $\chi^2(k - m)$ distributed random variable.
\end{theoremEnd}%
\begin{proofEnd}%
    This follows directly from \cref{thm:liml_is_kclass} and \cref{prop:subvector_anderson_rubin_test_statistic}.
\end{proofEnd}%
\noindent Thus $\sqrt{n}(\hat\kappa_\liml - 1) \toP 0$ as $n \to \infty$ and by \cref{prop:kclass_asymptotic_normality} the LIML estimator is, under strong instrument asymptotics, asymptotically Gaussian.
\Cref{prop:anderson_rubin_test,prop:subvector_anderson_rubin_test_statistic} motivate the following test.
\begin{theoremEnd}[malte,category=tests]{corollary}
    \label{cor:subvector_anderson_rubin_test}
    Let $F_{\chi^2(k - \mW)}$ be the cumulative distribution function of a chi-squared random variable with $k - \mW$ degrees of freedom.
    Under \cref{model:1}, for both strong and weak instrument asymptotics, a test that rejects the null $H_0 : \beta = \beta_0$ whenever $\AR(\beta) > \frac{1}{k - \mW} F^{-1}_{\chi^2(k - \mW)}(1 - \alpha)$ has asymptotic size at most $\alpha$.
    If  $\mW = 0$, then the test has asymptotic size equal to $\alpha$.
\end{theoremEnd}%
\begin{proofEnd}%
    This follows directly from \cref{prop:subvector_anderson_rubin_test_statistic}.
\end{proofEnd}%

Thus, the (subvector) Anderson-Rubin test is a weak-instrument-robust test and it allows the construction of weak-instrument-robust confidence sets for components of the causal parameter, see \cref{def:confidence_sets}.
However, compared to the Wald test, the Anderson-Rubin test has two disadvantages:
First, the number of degrees of freedom of the limiting chi-squared distribution increases with the number of instruments.
Consequently, additional instruments possibly increase the size of the confidence sets.
Second, the Anderson-Rubin test does not only test for the ``goodness of fit'' of the parameter $\beta$, but it also tests so-called overidentifying restrictions, that is, whether the rest of the model assumptions are met.

Essentially, the Anderson-Rubin test statistic can be decomposed as
$$
\AR(\beta) = \underbrace{\left( \AR(\beta) - \min_b \AR(b) \right)}_{\LR(\beta) / (k - \mW) } + \underbrace{\min_b \AR(b)}_{J_\liml  / (k - \mW)}.
$$
Here, the likelihood-ratio test statistic $\LR(\beta)$ (see \cref{def:lr_test_statistic} below) tests the goodness of fit of $\beta$ and the LIML variant of the J-statistic $J_\liml$ (see \cref{def:j_liml_statistic} in \cref{sec:j_statistic}) tests misspecification.
Consequently, a little bit of misspecification improves the power of the Anderson-Rubin test and results in narrower conﬁdence intervals by increasing $J_\liml$.
However, too much misspecification, that is, if $\min_b \AR(b) = J_\liml / (k - \mW) > F^{-1}_{\chi^2(\mX)}(1 - \alpha) / (k - \mW)$, will lead to the Anderson-Rubin test rejecting for all $\beta$ and thus yield an empty confidence set.

\subsection{The (conditional) likelihood ratio test}
The likelihood-ratio test is another classical approach to hypothesis testing.
\begin{definition}
    \label{def:lr_test_statistic}
    The \emph{likelihood-ratio (LR)} test statistic is defined as
    $$
    \LR(\beta) := (k - \mW) \left( \AR(\beta) - \min_{b \in \BR^\mX} \AR(b) \right) = (k - \mW) \left( \AR(\beta) - \AR(\hat\beta_\liml) \right).
    $$
\end{definition}

\noindent Here, $\AR(\beta)$ is the subvector Anderson-Rubin test statistic from \cref{def:subvector_anderson_rubin_test_statistic} if $\mW > 0$.
Note that $\max_{\Pi, \Omega, \gamma} \ell(y, (X, W) \mid (\beta, \gamma), \Pi, \Omega, Z) = \mathrm{const} - \frac{n}{2} \log(1 + \frac{k - \mW}{n-k} \AR(\beta))$ by \cref{lem:liml_likelihood} and thus
$$
\LR(\beta) \toP  2 \max_b \max_{\Pi, \Omega, \gamma} \ell(y, (X, W) \mid (b, \gamma), \Pi, \Omega, Z) - 2 \max_{\Pi, \Omega, \gamma} \ell(y, (X, W) \mid (\beta, \gamma), \Pi, \Omega, Z) \ \text{ as } n \to \infty
$$
as $\log(1 + x) \approx x$ as $x \to 0$.
\begin{theoremEnd}[malte,category=tests]{proposition}%
    \label{prop:lr_is_chi_squared}
    Under the null $H_0: \beta = \beta_0$ and strong instrument asymptotics, the likelihood-ratio test statistic is asymptotically $\chi^2(\mX)$ distributed.
\end{theoremEnd}%
\begin{proofEnd}%
    This is Wilk's theorem.
\end{proofEnd}%

\begin{theoremEnd}[malte,category=tests]{corollary}
    \label{cor:subvector_likelihood_ratio_test}
    Let $F_{\chi^2(\mX)}$ be the cumulative distribution function of a chi-squared random variable with $\mX$ degrees of freedom.
    Under \cref{model:1} and strong instrument asymptotics, a test that rejects the null $H_0: \beta = \beta_0$ whenever $\LR(\beta) > F^{-1}_{\chi^2(\mX)}(1 - \alpha)$ has asymptotic size $\alpha$.
\end{theoremEnd}%
\begin{proofEnd}%
    This follows from \cref{prop:lr_is_chi_squared}.
\end{proofEnd}%

\citet{sijpe2023power} note the following connection between the likelihood-ratio test and the Wald test using the LIML estimator.
\begin{theoremEnd}[malte,category=tests]{proposition}[\citeauthor{sijpe2023power}, \citeyear{sijpe2023power}]%
    \label{prop:wald_is_lr}
    Assume $\mW = 0$. Define $\hat\sigma(\beta)^2 := \frac{1}{n-k} \| M_Z (y - X \beta) \|^2$ and $\hat\sigma^2_\liml := \frac{1}{n - \mX}  \| y - X \hat\beta_\liml \|^2 $.
    Then,
    \begin{align*}
        \LR(\beta) = \frac{\hat\sigma^2_\liml}{\hat\sigma(\beta)^2} \Wald_{\hat\beta_\liml}(\beta).
    \end{align*}
\end{theoremEnd}%
\begin{proofEnd}%
    This follows \citet[Proposition 1]{sijpe2023power}.
    In the following, let $u := u(\beta) := y - X \beta$ and $B(\beta) := \frac{k}{n - k} \AR(\beta) = \frac{u^T P_Z u}{u^T M_Z u}$ such that $B(\hat\beta_\liml) = \hat\kappa_\liml - 1$.
    Furthermore, write $\hat u_\liml := y - X \hat\beta_\liml$.
    We wish to show that $\LR(\beta) = (n-k) (B(\beta) - B(\hat\beta_\liml)) = \frac{\hat\sigma^2_\liml}{\hat\sigma(\beta)^2} \Wald_{\hat\beta_\liml}(\beta)$.
    For this, we expand $B(\beta)$.
    By \cref{lemma:ar_statistic_derivative},
    $$
    \frac{\dd}{\dd \beta} \AR(\beta) = - 2 (n - k) \frac{u^T P_Z X - u^T P_Z u \frac{u^T M_Z X}{u^T M_Z u}}{u^T M_Z u}
    $$
    such that, as $\hat\beta_\liml = \argmin_\beta \AR(\beta)$ and thus $\frac{\dd}{\dd \beta} \AR(\beta) |_{\beta = \hat\beta_\liml} = 0$,
    \begin{align*}
    \hat u_\liml^T P_Z X &= \hat u_\liml^T P_Z \hat u_\liml \frac{\hat u_\liml^T M_Z X}{\hat u_\liml^T M_Z \hat u_\liml} \\
    &=  B(\hat\beta_\liml) \hat u_\liml^T M_Z X. \numberthis\label{eq:wald_is_lr:1}
    \end{align*}
    Using $y - X \beta = y - X \hat\beta_\liml + X (\hat\beta_\liml - \beta)$, rewrite
    \begin{align*}
    u^T P_Z u &= \hat u_\liml^T P_Z \hat u_\liml + (\hat\beta_\liml - \beta)^T X^T P_Z X (\hat\beta_\liml - \beta) \\
    &+ \numberthis\label{eq:wald_is_lr:2}
    \underbrace{
        \hat u_\liml^T P_Z X (\hat\beta_\liml - \beta)
     }_{ \overset{(\ref{eq:wald_is_lr:1})}{=} B(\hat\beta_\liml)\hat u_\liml^T M_Z X (\hat\beta_\liml - \beta)} +
    \underbrace{
        (\hat\beta_\liml - \beta)^T X^T P_Z \hat u_\liml
    }_{ \overset{(\ref{eq:wald_is_lr:1})}{=} B(\hat\beta_\liml) (\hat\beta_\liml - \beta)^T X^T M_Z \hat u_\liml}.
    \end{align*}
    Furthermore, rewrite
    \begin{align}\label{eq:wald_is_lr:3}    
        \frac{\hat u_\liml^T P_Z \hat u_\liml}{u^T M_Z u} &= \frac{B(\hat\beta_\liml)}{u^T M_Z u} \left( \hat u_\liml^T M_Z \hat u_\liml - u^T M_Z u \right),
    \end{align}
    where, again using $y - X \beta = y - X \hat\beta_\liml + X (\hat\beta_\liml - \beta)$,
    \begin{align*}
        \hat u_\liml^T M_Z \hat u_\liml - u^T M_Z u = -\hat u_\liml M_Z & X (\hat\beta_\liml - \beta) +  (\hat\beta_\liml - \beta)^T X M_Z \hat u_\liml \\
        & -  (\hat\beta_\liml - \beta)^T X M_Z  X (\hat\beta_\liml - \beta). \numberthis\label{eq:wald_is_lr:4}
    \end{align*}
    Combining Equations (\ref{eq:wald_is_lr:2}), (\ref{eq:wald_is_lr:3}), and (\ref{eq:wald_is_lr:4}), yields
    \begin{align*}
    B(\beta) - B(\hat\beta_\liml)  &= \frac{(\hat\beta_\liml - \beta)^T X^T P_Z X (\hat\beta_\liml - \beta) - B(\hat\beta_\liml)  (\hat\beta_\liml - \beta)^T X M_Z  X (\hat\beta_\liml - \beta)}{u^T M_Z u} \\
    &= \frac{(\hat\beta_\liml - \beta)^T X^T (\hat\kappa_\liml P_Z - (\hat\kappa_\liml - 1) \Id ) X (\hat\beta_\liml - \beta) }{u^T M_Z u} \\
    &= \frac{1}{n-k} \frac{\hat\sigma^2_\liml}{\hat\sigma(\beta)^2} \Wald_{\hat\beta_\liml}(\beta).
    \end{align*}

\end{proofEnd}%

That is, the likelihood-ratio and Wald test statistics using the LIML (the maximum likelihood estimator assuming i.i.d.\ Gaussian noise) differ only in the estimation of the second-stage noise variance.
The Wald test statistic's noise variance estimate uses the LIML, while the likelihood-ratio test statistic's noise variance uses the parameter under test.

The proof of Wilk's theorem, and thus \cref{prop:lr_is_chi_squared}, requires that the tuple $(\beta_0, \gamma_0, \Pi, \Omega)$ does not lie on the boundary of the parameter space.
This is not the case under weak instrument asymptotics where $\Pi \to 0$ and thus the likelihood-ratio test is not robust to weak instruments when using $\chi^2(\mX)$ critical values.
However, as shown by \citet{moreira2003conditional}, if $\mW = 0$, a variant of the likelihood-ratio test using a different critical value function is.

\begin{theoremEnd}[malte,category=tests]{proposition}[\citeauthor{moreira2003conditional}, \citeyear{moreira2003conditional}]%
    \label{prop:clr_test_statistic}
    Assume $\mW = 0$.
    Let $\tilde X(\beta) := X - (y - X \beta) \frac{(y - X \beta )^T M_Z X}{(y - X \beta)^T M_Z (y - X \beta)}$ as in 
    \cref{lem:liml_likelihood} and let 
    $$
    s_\mathrm{min}(\beta) := (n - k) \ \lambdamin{ \left(\tilde X(\beta)^T M_Z \tilde X(\beta) \right)^{-1} \tilde X(\beta)^T P_Z \tilde X(\beta)}.
    $$
    Then, under the null $H_0: \beta = \beta_0$ and both strong and weak instrument asymptotics, conditionally on $s_\mathrm{min}(\beta)$, the statistic $\LR(\beta)$ is asymptotically bounded from above by a random variable with distribution
    \begin{multline*}
    \Gamma( k-\mX, \mX, s_\mathrm{min}(\beta_0) ) := \\
    \frac{1}{2} \left(Q_{k - \mX} + Q_\mX - s_\mathrm{min}(\beta_0) + \sqrt{(Q_{k - \mX} + Q_\mX + s_\mathrm{min}(\beta_0))^2 - 4 Q_{k - \mX} s_\mathrm{min}(\beta_0)} \right) \\
    = \frac{1}{2} \left(Q_{k - \mX} + Q_\mX - s_\mathrm{min}(\beta_0) + \sqrt{(Q_{k - \mX} + Q_\mX - s_\mathrm{min}(\beta_0))^2 + 4 Q_{\mX} s_\mathrm{min}(\beta_0)} \right)
    \end{multline*}
    where $Q_{k - \mX} \sim \chi^2(k - \mX)$ and $Q_\mX \sim \chi^2(\mX)$ are independent.
\end{theoremEnd}%
\begin{proofEnd}%
    This proof follows \citet{kleibergen2007generalizing}.
    It proceeds in 5 steps.
    \paragraph*{Step 1:}
    Write $\tilde X = \tilde X(\beta_0)$.
    Then,
    \begin{equation} \label{eq:clr1}
        \begin{pmatrix} y & X \end{pmatrix} \begin{pmatrix} 1 & 0 \\ - \beta_0 & \Id_\mX \end{pmatrix} 
        \begin{pmatrix}
             1 & -\frac{\varepsilon^T M_Z X}{\varepsilon^T M_Z \varepsilon} \\ 0 & \Id_\mX
         \end{pmatrix}
        =
        \begin{pmatrix} \varepsilon & X \end{pmatrix} 
        \begin{pmatrix}
            1 & -\frac{\varepsilon^T M_Z X}{\varepsilon^T M_Z \varepsilon} \\ 0 & \Id_\mX
        \end{pmatrix} =
        \begin{pmatrix} \varepsilon & \tilde X \end{pmatrix}
    \end{equation}
    Note that $\varepsilon^T M_Z \tilde X = 0$ and thus
    \begin{equation}\label{eq:clr2}
        \tilde \Omega := 
        \begin{pmatrix} \varepsilon & \tilde X \end{pmatrix}^T M_Z \begin{pmatrix} \varepsilon & \tilde X \end{pmatrix} = 
        \begin{pmatrix}
            \varepsilon^T M_Z \varepsilon & 0 \\
            0 & \tilde X^T M_Z \tilde X
        \end{pmatrix},
    \end{equation}
    We can calculate
    \begin{align*}
    \min_\beta & \ \frac{n - k}{k} \AR(\beta) \overset{\text{Theorem \ref{thm:liml_is_kclass}}}{=} \lambda_\text{min}\left(\left(\begin{pmatrix} X & y \end{pmatrix}^T M_Z \begin{pmatrix} X & y \end{pmatrix}\right)^{-1} \begin{pmatrix} X & y \end{pmatrix}^T P_Z \begin{pmatrix} X & y \end{pmatrix}\right) \\
    &= \min \{\mu \in \BR \colon \det( 
        \mu \cdot \Id_{\mX+1} - \left(\begin{pmatrix} y & X \end{pmatrix}^T M_Z \begin{pmatrix} y & X \end{pmatrix}\right)^{-1} \begin{pmatrix} y & X \end{pmatrix}^T P_Z \begin{pmatrix} y & X \end{pmatrix}
    ) = 0 \}\\
    &=
    \min \{ \mu \in \BR \colon \det( 
        \mu \cdot \begin{pmatrix} y & X \end{pmatrix}^T M_Z \begin{pmatrix} y & X \end{pmatrix} -  \begin{pmatrix} y & X \end{pmatrix}^T P_Z \begin{pmatrix} y & X \end{pmatrix}
    ) = 0\}\\
    &\overset{\text{(\ref{eq:clr1}, \ref{eq:clr2})}}{=}
    \min \{ \mu \in \BR \colon \det( 
        \mu \cdot \Id_{\mX+1}
        - \tilde \Omega^{-1/2, T} \begin{pmatrix} \varepsilon & \tilde X \end{pmatrix}^T P_Z \begin{pmatrix} \varepsilon & \tilde X \end{pmatrix}  \tilde \Omega^{-1/2}
    ) = 0\}.
    \end{align*}

    \paragraph*{Step 2:}
    Let $U D V = P_Z \tilde X (\tilde X^T M_Z \tilde X)^{-1/2}$ be a singular value decomposition with $0 \leq d_1,\ldots, d_\mX$, the diagonal entries of $D^2 \in \BR^{\mX \times \mX}$, and thus the eigenvalues of $ (\tilde X M_Z \tilde X)^{-1} \tilde X P_Z \tilde X$, sorted ascending.
    Then $P_{P_Z \tilde X} = U U^T$ and
    \begin{align*}
       \Sigma &:= \tilde \Omega^{-1/2, T} \begin{pmatrix} \varepsilon & \tilde X \end{pmatrix}^T P_Z \begin{pmatrix} \varepsilon & \tilde X \end{pmatrix} \tilde \Omega^{-1/2} \numberthis\label{eq:prop_clr_diagonal} \\
       &= 
        \begin{pmatrix} 1 & 0 \\ 0 & V^T \end{pmatrix}
        \begin{pmatrix} \varepsilon^T P_Z \varepsilon / \varepsilon^T M_Z \varepsilon & \varepsilon^T P_Z U D / \sqrt{\varepsilon^T M_Z \varepsilon} \\ D U^T P_Z \varepsilon / \sqrt{\varepsilon^T M_Z \varepsilon} & D^2 \end{pmatrix}
        \begin{pmatrix} 1 & 0 \\ 0 & V \end{pmatrix}
    \end{align*}
    such that, for $\mu \notin \{d_1, \ldots, d_\mX\}$,
    \begin{align*}
        \phi_\Sigma(\mu) &:= \det(\mu \cdot \Id_{\mX+1} - \Sigma) 
        = \det \begin{pmatrix} \mu - \varepsilon^T P_Z \varepsilon / \varepsilon^T M_Z \varepsilon & - \varepsilon^T P_Z U D / \sqrt{\varepsilon^T M_Z \varepsilon} \\ - D U^T P_Z \varepsilon /\sqrt{\varepsilon^T M_Z \varepsilon} & \mu \cdot \Id_\mX - D^2 \end{pmatrix} \\
        &= \det(\mu \cdot \Id_\mX - D^2) \cdot \left(\mu - \frac{\varepsilon^T P_Z \varepsilon}{\varepsilon^T M_Z \varepsilon} - \frac{\varepsilon^T P_Z U D (\mu \cdot \Id_\mX - D^2)^{-1} D U^T P_Z \varepsilon}{\varepsilon^T M_Z \varepsilon} \right).
    \end{align*}
    Here we used the block matrix equality 
    $$
    \begin{pmatrix} A & B \\ C & D \end{pmatrix} \begin{pmatrix}\Id & 0 \\ -D^{-1} C & \Id \end{pmatrix} = \begin{pmatrix} A - B D^{-1} C & 0 \\ 0 & D \end{pmatrix}
    \Rightarrow \det \begin{pmatrix} A & B \\ C & D \end{pmatrix} = \det(D) \det( A - B D^{-1} C) 
    $$
    for invertible $D$.
    \paragraph*{Step 3:} Let $\tilde D := d_1 \cdot \Id_\mX$ and $Q := \begin{pmatrix} 1 & 0 \\ 0 & V^{-1} D^{-1} \tilde D V \end{pmatrix}$ and let $\tilde \Sigma := Q^T \Sigma Q$.
    As all nonzero entries of $D^{-1} \cdot \tilde D$ are bounded from above by 1 and $V$ is orthogonal,
    $$
    \lambda_\mathrm{min}(\tilde \Sigma) = \min_{x \neq 0} \frac{x^T \tilde\Sigma x}{x^T x} = \min_{x \neq 0} \frac{x^T Q^T \Sigma Q x}{x^T x} = \min_{x \neq 0} \frac{x^T \Sigma x}{x^T Q^T Q x} \geq \min_{x \neq 0} \frac{x^T \Sigma x}{x^T x} = \lambda_\mathrm{min}(\Sigma).
    $$
    \paragraph*{Step 4:}
    Compute
    $$
    \phi_{\tilde \Sigma}(\mu) = (\mu - d)^\mX \cdot (\mu - \frac{\varepsilon^T P_Z \varepsilon}{\varepsilon^T M_Z \varepsilon} - \frac{d}{\mu - d} \frac{\varepsilon^T P_{P_Z \tilde X} \varepsilon}{\varepsilon^T M_Z \varepsilon}),
    $$
    with 
    $$
    \phi_{\tilde \Sigma}(\mu) = 0 \Rightarrow (\mu = d) \  \vee \ (\mu \cdot (\mu - d) - (\mu - d) \cdot \frac{\varepsilon^T P_Z \varepsilon}{\varepsilon^T M_Z \varepsilon} - d \cdot \frac{\varepsilon^T P_{P_Z \tilde X} \varepsilon}{\varepsilon^T M_Z \varepsilon} = 0).
    $$
    The latter polynomial equation has solutions
    \begin{align*}
    \mu_\pm &= \frac{1}{2} \left(d + \frac{\varepsilon^T P_Z \varepsilon}{\varepsilon^T M_Z \varepsilon} \pm \sqrt{ \left(d + \frac{\varepsilon^T P_Z \varepsilon}{\varepsilon^T M_Z \varepsilon} \right)^2 - 4 d \frac{\varepsilon^T (P_Z - P_{P_Z \tilde X}) \varepsilon}{\varepsilon^T M_Z \varepsilon} }\right) \\
    &= \frac{1}{2} \left(d + \frac{\varepsilon^T P_Z \varepsilon}{\varepsilon^T M_Z \varepsilon} \pm \sqrt{\left(d + \frac{\varepsilon^T P_Z \varepsilon}{\varepsilon^T M_Z \varepsilon} - 2 \frac{\varepsilon^T P_Z \varepsilon}{\varepsilon^T M_Z \varepsilon} \right)^2 + 4 d \frac{\varepsilon^T P_{P_Z \tilde X} \varepsilon}{\varepsilon^T M_Z \varepsilon}} \right).
    \end{align*}
    If $d \leq \frac{\varepsilon^T P_Z \varepsilon}{\varepsilon^T M_Z \varepsilon}$, then
    $$
    \mu_{-} \leq \frac{1}{2} \left(d + \frac{\varepsilon^T P_Z \varepsilon}{\varepsilon^T M_Z \varepsilon} - \sqrt{ \left(d - \frac{\varepsilon^T P_Z \varepsilon}{\varepsilon^T M_Z \varepsilon} \right)^2}\right) = \frac{1}{2} \left(d + \frac{\varepsilon^T P_Z \varepsilon}{\varepsilon^T M_Z \varepsilon} - (\frac{\varepsilon^T P_Z \varepsilon}{\varepsilon^T M_Z \varepsilon} - d) \right) = d.
    $$
    If $d > \frac{\varepsilon^T P_Z \varepsilon}{\varepsilon^T M_Z \varepsilon}$, then
    $$
    \mu_{-} \leq \frac{1}{2} \left(d + \frac{\varepsilon^T P_Z \varepsilon}{\varepsilon^T M_Z \varepsilon} - \sqrt{(d - \frac{\varepsilon^T P_Z \varepsilon}{\varepsilon^T M_Z \varepsilon})^2}\right) = \frac{1}{2} \left( d + \frac{\varepsilon^T P_Z \varepsilon}{\varepsilon^T M_Z \varepsilon} - (d - \frac{\varepsilon^T P_Z \varepsilon}{\varepsilon^T M_Z \varepsilon}) \right) = \frac{\varepsilon^T P_Z \varepsilon}{\varepsilon^T M_Z \varepsilon} < d.
    $$
    Thus, $\mu_{-} \leq d$ and thus $\lambda_\text{min}(\tilde \Sigma) = \mu_{-}$.

    \paragraph*{Step 5:}
    Putting everything together, we have
    \begin{align*}
        \LR(\beta_0) &= (n - k) \left(\frac{\varepsilon^T P_Z \varepsilon}{\varepsilon^T M_Z \varepsilon} - \lambda_\text{min}( \Sigma) \right)
        \leq(n - k) \left(\frac{\varepsilon^T P_Z \varepsilon}{\varepsilon^T M_Z \varepsilon} - \lambda_\text{min}(\tilde \Sigma) \right) \\
        &=  \frac{n - k}{2}\left(\frac{\varepsilon^T P_Z \varepsilon}{\varepsilon^T M_Z \varepsilon} - d + \sqrt{\left(d + \frac{\varepsilon^T P_Z \varepsilon}{\varepsilon^T M_Z \varepsilon} \right)^2 - 4 d\frac{\varepsilon^T (P_Z - P_{P_Z \tilde X}) \varepsilon}{\varepsilon^T M_Z \varepsilon}}\right)
    \end{align*}
    Here 
    $$Q_{\mX} := (n - k) \frac{\varepsilon^T P_{P_Z \tilde X} \varepsilon}{\varepsilon^T M_Z \varepsilon} \tod \chi^2(\mX) \text{ and }
    Q_{k - \mX} := (n - k) \frac{\varepsilon^T (P_Z - P_{P_Z \tilde X}) \varepsilon}{\varepsilon^T M_Z \varepsilon} \tod \chi^2(k - \mX)$$
    are asymptotically independent and asymptotically independent of $(Z^T Z)^{-1/2} Z^T \tilde X(\beta_0)$.

\end{proofEnd}%

\begin{theoremEnd}[malte,category=tests]{corollary}%
    \label{cor:subvector_conditional_likelihood_ratio_test}
    Let $F_{\Gamma(k - \mX, \mX, s)}$ be the cumulative distribution function of the random variable $\Gamma(k - \mX, \mX, s)$.
    Under \cref{model:1}, with $\mW = 0$, under both strong and weak instrument asymptotics, a test that rejects the null $H_0: \beta = \beta_0$ whenever $\LR(\beta) > F^{-1}_{\Gamma(k - \mX, \mX, s_\mathrm{min}(\beta))}(1 - \alpha)$ has asymptotic size at most $\alpha$.
\end{theoremEnd}%
\begin{proofEnd}%
    This follows from \cref{prop:clr_test_statistic}.
\end{proofEnd}%

The test based on \cref{cor:subvector_conditional_likelihood_ratio_test} is called the \emph{conditional likelihood-ratio test}.
The statistic $s_\mathrm{min}(\beta_0)$ is a measure of instrument strength.
It is equal to the Cragg-Donald test statistic with null hypothesis that $\Pi_X$ is of full rank in the model $\tilde X(\beta_0) = Z \Pi_X + V_X - \varepsilon \frac{\varepsilon^T M_Z V_X}{\varepsilon^T M_Z \varepsilon}$, where $\frac{\varepsilon^T M_Z V_X}{\varepsilon^T M_Z \varepsilon} \to_\BP \Omega_{\varepsilon, V_X} / \sigma^2_\varepsilon$.
The distribution of $\Gamma(k - \mX, \mX, s)$ interpolates between those of the $\chi^2(\mX) = \plim_{s \to \infty} \Gamma(k - \mX, \mX, s)$ and the $\chi^2(k) = \plim_{s \to 0} \Gamma(k - \mX, \mX, s)$ distributions.
Thus, if identification is strong, the conditional likelihood-ratio test is equivalent to the likelihood-ratio test.
If identification is very weak, as $\AR(\hat\beta_\liml) \leq s_\mathrm{min}(\beta_0) / (k - \mW)$ using \cref{thm:liml_is_kclass}, the conditional likelihood-ratio test is equivalent to the Anderson-Rubin test.

\Cref{prop:clr_test_statistic} does not apply if $\mW > 0$ and thus does not allow testing individual coefficients of the causal parameter.
\citet{kleibergen2021efficient} conjectures the following.
\begin{theoremEnd}[malte,category=tests,text proof={Proof sketch},restate command=subvectorclr,one big link translated={\hspace{-0.71cm} See proof sketch on page}]{conjecture}[\citeauthor{kleibergen2021efficient}, \citeyear{kleibergen2021efficient}]%
    \label{con:subvector_clr}
    Let $\lambda_1, \lambda_2$ be the smallest and second smallest eigenvalues of
    \begin{equation*}
        (n - k) \left[\begin{pmatrix}X & W & y\end{pmatrix}^T M_Z \begin{pmatrix}X & W & y\end{pmatrix}\right]^{-1} \begin{pmatrix}X & W & y\end{pmatrix}^T P_Z \begin{pmatrix}X & W & y\end{pmatrix}
    \end{equation*} and 
    $$
    \mu(\beta) := (n - k) \, \lambda_\mathrm{min}\left(\left[\begin{pmatrix}W \!\! & y\!-\!X \beta\end{pmatrix}^T M_Z \begin{pmatrix}W \!\! & y\!-\!X \beta\end{pmatrix}\right]^{-1} \begin{pmatrix}W \!\! & y\!-\!X \beta \end{pmatrix}^T P_Z \begin{pmatrix}W \!\! & y\!-\!X \beta\end{pmatrix}\right).
    $$
    Write $\tilde s_\mathrm{min}(\beta) := \lambda_1 + \lambda_2 - \mu(\beta)$.
    Then, under the null $\beta = \beta_0$ and both strong and weak instrument asymptotics, the likelihood-ratio test statistic $\LR(\beta_0)$ is asymptotically bounded from above by a random variable with distribution $\Gamma(k - m, \mX, \tilde s_\mathrm{min}(\beta_0))$.
\end{theoremEnd}%
\begin{proofEnd}%
\citet{kleibergen2021efficient} shows the following arguments first for $\mX = 1$ and then argues how to extend to $\mX > 1$.

When $\mX = 1$ and $\mW = 0$, then, by Equation \eqref{eq:prop_clr_diagonal}, 
$$
    \tr( \left( \begin{pmatrix} X & y \end{pmatrix}^T M_Z \begin{pmatrix} X & y \end{pmatrix} \right)^{-1} \begin{pmatrix} X & y \end{pmatrix}^T P_Z \begin{pmatrix} X & y \end{pmatrix} ) = \AR(\beta_0) + s_\mathrm{min}(\beta_0).
$$
At the same time, the trace is the sum of eigenvalues $\lambda_1 + \lambda_2$, implying that $s_\mathrm{min}(\beta_0) = \lambda_1 + \lambda_2 - \AR(\beta_0)$.
\end{proofEnd}%

Note that, using \cref{thm:liml_is_kclass}, we have $\lambda_1 = (k - \mW) \cdot \min_b \AR(b)$ and $\mu(\beta) = (k - \mW) \cdot \AR(\beta)$ such that $\tilde s_\mathrm{min}(\beta) = \lambda_2 - \LR(\beta)$.

Implementing the conditional likelihood ratio test and its subvector variants requires the approximation of the cumulative distribution function of the limiting distribution $\Gamma(k - \mX, \mX, s)$, that is, the critical value function of the corresponding test.
We discuss approaches to do so in appendix \ref{sec:approximate_clr_critical_value_function}.

\subsection{The Lagrange multiplier test}
As discussed in \cref{sec:anderson_rubin_test}, the Anderson-Rubin test tests overidentification restrictions additionally to the goodness of fit of the parameter $\beta$.
Consequently, its limiting distribution has $k - m$ excess degrees of freedom, equal to the rate of overidentification.
The (conditional) likelihood-ratio removes (a part of) these excess degrees of freedom by subtracting the LIML variant of the J-statistic (see \cref{def:j_liml_statistic}) from the Anderson-Rubin test statistic.

An alternative approach is to consider only $P_Z X = Z \hat\Pi$, the part of the instruments that is relevant for identification, as instruments.
This, but replacing $\hat\Pi$ with $\hat\Pi_\liml$ from \cref{lem:liml_likelihood} to make the test robust to weak instruments, is the idea behind \citeauthor{kleibergen2002pivotal}'s (\citeyear{kleibergen2002pivotal}) Lagrange multiplier test statistic.
\begin{theoremEnd}[malte]{definition}[\citeauthor{kleibergen2002pivotal}, \citeyear{kleibergen2002pivotal}]\label{def:klm_test_statistic}
    Assume $\mW = 0$.
    Define
    $$
    \tilde X(\beta) := X - (y - X \beta) \frac{(y - X \beta)^T M_Z X}{
        (y - X \beta)^T M_Z (y - X \beta)}.
    $$
    The Lagrange multiplier test statistic is
    $$
    \LM(\beta) = (n - k) \frac{(y - X \beta)^T P_{P_Z \tilde X(\beta)} (y - X \beta)}{(y - X \beta)^T M_Z (y - X \beta)}.
    $$
\end{theoremEnd}%

\noindent Recall from \cref{lem:liml_likelihood} that $\hat \Pi_\liml(\beta) = (Z^T Z)^{-1} Z^T \tilde X(\beta)$ such that $P_Z \tilde X(\beta) = Z \hat\Pi_\liml(\beta)$.
\begin{theoremEnd}[malte,category=tests]{proposition}[\citeauthor{kleibergen2002pivotal}, \citeyear{kleibergen2002pivotal}]%
    \label{prop:klm_test_statistic_chi_squared}
    Under both strong and weak instrument asymptotics, $\LM(\beta_0) \overset{d}{\to} \chi^2(\mX)$ as $n \to \infty$.
\end{theoremEnd}%
\begin{proofEnd}%
    This proof is based on \citep{kleibergen2002pivotal}. It proceeds in two steps:
    \begin{itemize}
        \item First, we show, separately for weak and strong instrument asymptotics, that 
        $$(y - X \beta_0)^T P_{P_Z \tilde X(\beta_0)} (y - X \beta_0) \tod \sigma^2_\varepsilon \cdot \chi^2(\mX).$$
        \item Second, using Slutsky's Lemma, we conclude that $\LM(\beta_0) \overset{d}{\to} \chi^2(\mX)$.
    \end{itemize}
    \paragraph*{Step 1:}
    First, using $y - X \beta_0 = \varepsilon$, expand
    \begin{align*}
        (y - &X \beta_0)^T P_{P_Z \tilde X(\beta_0)} (y - X \beta_0) =
        \Psi_{\varepsilon}^T (Z^T Z)^{-1/2} Z^T \tilde X(\beta_0) (\tilde X (\beta_0)^T Z (Z^T Z)^{-1} Z^T \tilde X(\beta_0))^{-1} \\&\hskip 2cm \cdot \tilde X(\beta_0)^T Z (Z^T Z)^{-1/2} Z^T \Psi_{\varepsilon} = \Psi_{\varepsilon}^T P_{ (Z^T Z)^{-1/2} Z^T \tilde X(\beta_0)} \Psi_{\varepsilon}.
    \end{align*}
    Let $\Psi_{\tilde X} := (Z^T Z)^{-1/2} Z^T \tilde X(\beta_0)$.
    \paragraph*{Assuming strong instruments,} that is, $\Pi_X$ is constant and of full rank, we calculate
    \begin{align*}
        \frac{1}{\sqrt{n}}\Psi_{\tilde X} &= \frac{1}{\sqrt{n}} (Z^T Z)^{1/2} \Pi_X + \frac{1}{\sqrt{n}} \Psi_{V_X} - \frac{1}{\sqrt{n}} \Psi_{\varepsilon} \frac{\varepsilon^T M_Z X}{\varepsilon^T M_Z \varepsilon} \\
        &\overset{\BP}{\to} Q^{1/2} \Pi_X.
    \end{align*}
    Thus,
    \begin{align*}
    (y - X &\beta_0)^T P_{P_Z \tilde X(\beta_0)} (y - X \beta_0)  = \Psi_{\varepsilon}^T P_{\frac{1}{\sqrt{n}} (Z^T Z)^{-1/2} Z^T \tilde X(\beta_0)} \Psi_{\varepsilon} \\
    &\toP
    \Psi_{\varepsilon}^T P_{Q^{1/2} \Pi_X} \Psi_{\varepsilon} \tod \sigma^2_\varepsilon \cdot \chi^2(\mX)
    \end{align*}
    as $\rank(Q^{1/2} \Pi_X) = \mX$

    \paragraph*{Assuming weak instruments,} that is,
    $\Pi_X = \frac{1}{\sqrt{n}} C_X$ for some fixed $C_X$ of full rank, we calculate
    \begin{align*}
       \Psi_{\tilde X} &= \frac{1}{\sqrt{n}} (Z^T Z)^{1/2} C_X + \Psi_{V_X} - \Psi_{\varepsilon} \frac{\varepsilon^T M_Z X}{\varepsilon^T M_Z \varepsilon} \\
        &\toP Q^{1/2} C_X + \Psi_{V_X} - \Psi_{\varepsilon} \frac{\Omega_{\varepsilon, V_X}}{\sigma^2_\varepsilon},
    \end{align*}
    as $\frac{1}{n-k} \varepsilon^T M_Z X = \frac{1}{n-k} \varepsilon^T M_Z V_X = \frac{1}{n-k} \varepsilon^T V_X - \frac{1}{n-k} \Psi_\varepsilon^T \Psi_{V_X} \overset{\BP}{\to} \Omega_{\varepsilon, V_X}$ by \cref{ass:1} (a) and similarly
    $\frac{1}{n-k} \varepsilon^T M_Z \varepsilon \overset{\BP}{\to} \sigma^2_\varepsilon$.
    By Slutsky's Lemma and \cref{ass:1} (b), as $\Psi_{\varepsilon}, \Psi_{V_X}$ are asymptotically jointly normal, so are $\Psi_{\varepsilon}, \Psi_{\tilde X}$.
    We calculate
    \begin{align*}
    \Cov(\Psi_{\tilde X}, \Psi_{\varepsilon}) &= \Cov(\Psi_{V_X}, \Psi_{\varepsilon}) - \Cov(\Psi_{\varepsilon} \frac{\varepsilon^T M_Z X}{\varepsilon^T M_Z \varepsilon}, \Psi_{\varepsilon}) \\
    &\overset{\BP, \text{\ by \cref{ass:1} a, b}}{\to} \Omega_{V_X, \varepsilon} - \sigma^2_\varepsilon \frac{\Omega_{V_X, \varepsilon}}{\sigma^2_\varepsilon} = 0.
    \end{align*}
    Thus, $\Psi_{\tilde X}$ and $\Psi_{\varepsilon}$ are asymptotically uncorrelated and by their asymptotic normality thus asymptotically independent. The random variable
    \begin{align*}
    (y - X \beta_0)^T P_{P_Z \tilde X(\beta_0)} (y - X \beta_0) &= \Psi_{\varepsilon}^T P_{\frac{1}{\sqrt{n}} \Psi_{\tilde X}} \Psi_{\varepsilon}
    \end{align*}
    is, conditionally on $\Psi_{\tilde X}$, asymptotically $\sigma^2_\varepsilon \cdot \chi^2(\mX)$-distributed.
    By asymptotic independence of $\Psi_{\tilde X}$ and $\Psi_{\varepsilon}$, this also holds unconditionally.
    \paragraph*{Step 2:}
    Finally,
    $$
    \frac{1}{n - k} (y - X \beta_0)^T M_Z (y - X\beta_0) \toP \sigma^2_\varepsilon
    $$
    and thus
    $$
    \LM(\beta_0) = (n-k) \frac{(y - X \beta)^T P_{P_Z \tilde X(\beta)} (y - X \beta)}{(y - X \beta)^T M_Z (y - X \beta)} \overset{\BP}{\to} \chi^2(\mX)
    $$
    by Slutsky's Lemma.
\end{proofEnd}%

How does this work?
Let $\Psi_X := (Z^T Z)^{-1/2} Z^T X = (Z^T Z)^{1/2} \Pi_X + \Psi_{V_X}$.
Under \cref{ass:0} with $\mW = 0$, $(y - X \beta_0)^T P_{P_Z X} (y - X \beta_0) = \Psi_\varepsilon^T P_{\Psi_X} \Psi_\varepsilon$ converges in distribution to the square of the projection of the $k$ i.i.d.\ Gaussian random variables in $\Psi_\varepsilon$ onto the column span of $\Psi_{X}$.
Under strong instrument asymptotics, $\frac{1}{\sqrt n}\Psi_{X} \to_\BP Q^{1/2} \Pi_X$ and ${(y - X \beta_0)^T P_{P_Z X} (y - X \beta_0) \to_d \sigma_\varepsilon^2 \chi^2(\mX)}$.
However, under weak instrument asymptotics, ${\Psi_{X} \to_\BP Q^{1/2} \Pi + \Psi_{V_X}}$ and $\Psi_\varepsilon$ are asymptotically jointly normal.
They are correlated due to the correlation between the $V_{X, i}$ and $\varepsilon_i$ and thus ${\Psi_\varepsilon^T P_{\Psi_X} \Psi_\varepsilon}$ is not asymptotically $\chi^2(\mX)$ distributed.
The decorrelation trick, replacing $X$ with $\tilde X(\beta_0)$, yields ${\Psi_{\tilde X(\beta_0)} = \Psi_X - \Psi_\varepsilon \frac{\varepsilon^T M_Z X}{\varepsilon^T M_Z \varepsilon}}$ ${\to_\BP \Psi_X - \Psi_\varepsilon \Omega_{V_X, \varepsilon} / \sigma^2_\varepsilon}$ that is asymptotically uncorrelated with $\Psi_\varepsilon$.
By joint Gaussianity of $\Psi_\varepsilon$ and $\Psi_{\tilde X(\beta_0)}$, they are also asymptotically independent and thus $\Psi_\varepsilon^T P_{\Psi_{\tilde X(\beta)}} \Psi_\varepsilon$ is asymptotically $\sigma_\varepsilon^2 \chi^2(\mX)$ distributed.
The denominator $(y - X \beta)^T M_Z (y - X \beta) / (n - k) \to_\BP \sigma^2_\varepsilon$ estimates the noise variance $\sigma^2_\varepsilon$.

The Lagrange multiplier test is also called \emph{score test}, as it is a quadratic form of the score
$$
    \frac{\dd}{\dd \beta} k \AR(\beta) = - 2 (n - k) \frac{\tilde X(\beta) P_Z(y - X \beta)}{(y - X \beta)^T M_Z (y - X \beta)},
$$
see \cref{lemma:ar_statistic_derivative}.
However, as the Fisher information $I(\beta) \neq 4 (n - k)^2 \frac{\tilde X(\beta)^T P_Z \tilde X(\beta)}{((y - X \beta)^T M_Z (y - X \beta))^2}$, it is not equal to the classical score test, which would not be robust to weak instruments.

The Lagrange multiplier test statistic overcomes the problem of excess degrees of freedom of the Anderson-Rubin test.
However, it does not allow for testing of individual components of the causal parameter.
\citet{londschien2024weak} propose a subvector extension of the Lagrange multiplier test statistic.
\begin{definition}[\citeauthor{londschien2024weak}, \citeyear{londschien2024weak}]
    \label{def:subvector_klm_test_statistic}
    Let 
    $$
    \tilde S(\beta, \gamma) := \begin{pmatrix} X & W \end{pmatrix} - (y - X \beta - W \gamma) \frac{(y - X \beta - W \gamma)^T M_Z \begin{pmatrix} X & W \end{pmatrix}}{(y - X \beta - W \gamma)^T M_Z (y - X \beta - W \gamma)}.
    $$
    The subvector Lagrange multiplier test statistic is
    \begin{equation*}
        \label{eq:subvector_klm_test_statistic}
        \LM(\beta) := (n - k) \min_{\gamma \in \BR^\mW} \frac{(y - X \beta - W \gamma)^T P_{P_Z \tilde S(\beta, \gamma)} (y - X \beta - W \gamma)}{(y - X \beta - W \gamma)^T M_Z (y - X \beta - W \gamma)}.
    \end{equation*}
\end{definition}

\begin{theoremEnd}[malte,category=tests]{technical_condition}
    \label{tc:subvector_klm}
    Assume there exists a $\gamma^\star \in \BR^\mW$ such that 
    $$
        \gamma^\star = \gamma_0 + (\Pi_W^T Z^T P_{P_Z \tilde S(\beta_0, \gamma^\star)} W)^{-1} \Pi_W^T Z^T P_{P_Z \tilde S(\beta_0, \gamma^\star)} \varepsilon,
    $$
    or, equivalently,
    $$
    \Pi_W^T Z^T P_{P_Z \tilde S(\beta_0, \gamma^\star)} (\varepsilon + W(\gamma_0 - \gamma^\star)) = 0.
    $$    
\end{theoremEnd}%
\begin{theoremEnd}[malte,category=tests]{proposition}[\citeauthor{londschien2024weak}, \citeyear{londschien2024weak}]
    \label{prop:subvector_klm_test_statistic_chi_squared}
    Consider \cref{model:1} and assume \cref{ass:1} and \cref{tc:subvector_klm} hold.
    Under the null $\beta = \beta_0$, under both strong and weak instrument asymptotics, the subvector Lagrange multiplier test statistic is bounded from above by a random variable that is asymptotically $\chi^2(\mX)$ distributed.
\end{theoremEnd}%
\begin{proofEnd}%
    We proceed in four steps:
    \begin{itemize}
        \item First, we show that $\gamma^\star$ from \cref{tc:subvector_klm} satisfies
        $$(y - X\beta_0 - W \gamma^\star)^T P_{P_Z \tilde S(\beta_0, \gamma^\star)} (y - X\beta_0 - W \gamma^\star) = \Psi_{\varepsilon^\star}^T (P_{\Psi_{\tilde S}} - P_{P_{\Psi_{\tilde S}} Q^{1/2} \Pi_W}) \Psi_{\varepsilon^\star},$$
        where $\Psi_{\varepsilon^\star} \in \BR^k$ and $\Psi_{\tilde S} \in \BR^{k \times m}$ are random variables.
        \item Second, we show that under \cref{ass:1}, the random variables' $\Psi_{\varepsilon^\star}$ and $\Psi_{\tilde V} := \Psi_{\tilde S} - (Z^T Z)^{1/2} (\Pi_X \ \Pi_W)$ rows are asymptotically centered Gaussian, asymptotically uncorrelated, and thus asymptotically independent.
        \item Third, we argue, both for weak and strong instrument asymptotics, that
        $$\Psi_{\varepsilon^\star}^T (P_{\Psi_{\tilde V}} - P_{P_{\Psi_{\tilde V}} Q^{1/2} \Pi_W}) \Psi_{\varepsilon^\star} \tod \Var(\Psi_{\varepsilon^\star}) \chi^2(\mX).$$
        \item We conclude as
        $$\LM(\beta_0) \leq (n - k) \frac{(y - X \beta_0 - W \gamma^\star)^T P_{P_Z \tilde S(\beta_0, \gamma^\star)} (y - X \beta_0 - W \gamma^\star)}{(y - X \beta_0 - W \gamma^\star)^T M_Z (y - X \beta_0 - W \gamma^\star)}
        $$
        and $\frac{1}{n-k} (y - X \beta_0 - W \gamma^\star)^T M_Z (y - X \beta_0 - W \gamma^\star) \toP \Var(\Psi_{\varepsilon^\star})$.
    \end{itemize}
    \paragraph*{Step 1:}
    By \cref{tc:subvector_klm}, there exists some $\gamma^\star \in \BR^{\mW}$ such that $\Pi_W^T Z^T P_{P_Z \tilde S(\beta_0, \gamma^\star)} (\varepsilon + W(\gamma_0 - \gamma^\star)) = 0$.
    Then
    \begin{align*}
        &\gamma^\star - \gamma_0 = 
        - (\Pi_W^T Z^T P_{P_Z \tilde S(\beta_0, \gamma^\star)} Z \Pi_W)^{-1} \Pi_W^T Z^T P_{P_Z \tilde S(\beta_0, \gamma^\star)} Z \Pi_W (\gamma_0 - \gamma^\star) \\
        &\overset{\text{\cref{tc:subvector_klm}}}{=} (\Pi_W^T Z^T P_{P_Z \tilde S(\beta_0, \gamma^\star)} Z \Pi_W)^{-1} \Pi_W^T Z^T P_{P_Z \tilde S(\beta_0, \gamma^\star)} (\varepsilon + V_W (\gamma_0 - \gamma^\star)) \\
        &\ \Rightarrow P_{P_Z \tilde S(\beta_0, \gamma^\star)} Z \Pi_W (\gamma_0 - \gamma^\star) = - P_{P_{P_Z \tilde S(\beta_0, \gamma^\star)} Z \Pi_W} (\varepsilon + V_W (\gamma_0 - \gamma^\star))
    \end{align*}
    and thus
    \begin{align*}
    P_{P_Z \tilde S(\beta_0, \gamma^\star)} (y - X\beta_0 - W \gamma^\star) &= P_{P_Z \tilde S(\beta_0, \gamma^\star)} (\varepsilon + V_W (\gamma_0 - \gamma^\star) + Z \Pi_W (\gamma_0 - \gamma^\star)) \\
    &= P_{P_Z \tilde S(\beta_0, \gamma^\star)} M_{P_{P_Z \tilde S(\beta_0, \gamma^\star)} Z \Pi_W} (\varepsilon + V_W (\gamma_0 - \gamma^\star))\\
    &=   M_{P_{P_Z \tilde S(\beta_0, \gamma^\star)} Z \Pi_W} P_{P_Z \tilde S(\beta_0, \gamma^\star)} (\varepsilon + V_W (\gamma_0 - \gamma^\star))  \label{eq:1}\numberthis
    \end{align*}
    as $P_{P_Z \tilde S(\beta_0, \gamma^\star)}$ and $M_{P_{P_Z \tilde S(\beta_0, \gamma^\star)} Z \Pi_W} $ commute.
    Define $\varepsilon^\star := \varepsilon + V_W (\gamma_0 - \gamma^\star)$.
    Write
    \begin{align*}
        &\Psi_{\varepsilon^\star} := \Psi_{\varepsilon} + \Psi_{V_W} (\gamma_0 - \gamma^\star) = (Z^T Z)^{-1/2} Z^T \varepsilon^\star \\
        &\Psi_{\tilde V} := \begin{pmatrix} \Psi_{V_X} & \Psi_{V_W} \end{pmatrix} - \Psi_{\varepsilon^\star}\frac{{\varepsilon^\star}^T M_Z \begin{pmatrix}X & W \end{pmatrix}}{{\varepsilon^\star}^T M_Z \varepsilon^\star}, \text{ and} \\
        &\Psi_{\tilde S} := \Psi_{\tilde V} + (Z^T Z)^{1/2} \begin{pmatrix} \Pi_X & \Pi_W \end{pmatrix} = (Z^T Z)^{-1/2} Z^T \tilde S(\beta_0, \gamma^\star)
    \end{align*}
    Expand
    $$
    P_{P_Z \tilde S(\beta_0, \gamma^\star)} =
    Z (Z^T Z)^{-1/2} \Psi_{\tilde S} (\Psi_{\tilde S}^T \Psi_{\tilde S})^{-1} \Psi_{\tilde S}^T (Z^T Z)^{-1/2} Z^T = Z (Z^T Z)^{-1/2} P_{\Psi_{\tilde S}} (Z^T Z)^{-1/2} Z^T
    $$
    such that
    $$
    P_{P_Z \tilde S(\beta_0, \gamma^\star)} Z \Pi_W = Z (Z^T Z)^{-1/2} P_{\Psi_{\tilde S}} (Z^T Z)^{1/2} \Pi_W
    $$
    and
    \begin{align*}
    &M_{P_{P_Z \tilde S(\beta_0, \gamma^\star)} Z \Pi_W} = (\Id_n - P_{P_{P_Z \tilde S(\beta_0, \gamma^\star)} Z \Pi_W
    }) \\
    &\hskip 0.5cm = (\Id_n - Z (Z^T Z)^{-1/2} P_{\Psi_{\tilde S}} (Z^T Z)^{1/2} \Pi_W (\Pi_W^T (Z^T Z)^{1/2} P_{\Psi_{\tilde S}}^T P_{\Psi_{\tilde S}} (Z^T Z)^{1/2} \Pi_W)^{-1} \\
    &\hskip 1cm \Pi_W^T (Z^T Z)^{1/2} P_{\Psi_{\tilde S}}^T (Z^T Z)^{-1/2} Z^T)
    = \Id_n - Z (Z^T Z)^{-1/2} P_{P_{\Psi_{\tilde S}} (Z^T Z)^{1/2} \Pi_W} (Z^T Z)^{-1/2} Z^T.
    \end{align*}
    We combine this to get
    \begin{align*}
        (y - &X \beta_0 - W \gamma^\star)^T P_{P_Z \tilde S(\beta_0, \gamma^\star)} (y - X \beta_0 - W \gamma^\star)        
        &\overset{(\ref{eq:1})}{=} {\varepsilon^\star}^T P_{P_Z \tilde S(\beta_0, \gamma^\star)} M_{P_{P_Z \tilde S(\beta_0, \gamma^\star)}Z \Pi_W} P_{P_Z \tilde S(\beta_0, \gamma^\star)}  \varepsilon^\star \\
        &= \Psi_{\varepsilon^\star}^T 
        (P_{\Psi_{\tilde S}} - P_{P_{\Psi_{\tilde S}} (Z^T Z)^{1/2} \Pi_W}) \Psi_{\varepsilon^\star}.
        \end{align*}
    \paragraph*{Step 2:}
    By \cref{ass:1}
    $$
    \vecop\begin{pmatrix} \Psi_{\varepsilon} & \Psi_{V_X} & \Psi_{V_W} \end{pmatrix} \tod \CN(0, \Omega \otimes \Id_k)
    $$
    and thus
    $$
    \vecop\begin{pmatrix} \Psi_{\varepsilon^\star} & \Psi_{V_X} & \Psi_{V_W} \end{pmatrix} \tod \CN(0, \underbrace{\begin{pmatrix} 1 & 0 & 0 \\ 0 & \Id_\mX & 0 \\ \gamma_0 - \gamma & 0 & \Id_\mW \end{pmatrix}^T \Omega \begin{pmatrix} 1 & 0 & 0 \\ 0 & \Id_\mX & 0 \\ \gamma_0 - \gamma & 0 & \Id_\mW \end{pmatrix}}_{=: \Omega^\star} \otimes \Id_k).
    $$
    By \cref{ass:1} (a), $\frac{{\varepsilon^\star}^T M_Z (X \ W)}{{\varepsilon^\star}^T M_Z \varepsilon^\star} \toP \frac{\Omega^\star_{\varepsilon^\star, (V_X \ V_W)}}{\Omega^\star_{\varepsilon^\star}}$, implying that
    \begin{align*}
    \Cov(\Psi_{\varepsilon^\star}, \Psi_{\tilde V}) &= \Cov( \Psi_{\varepsilon^\star}, \begin{pmatrix} \Psi_{V_X} & \Psi_{V_W} \end{pmatrix} - \Psi_{\varepsilon^\star} \frac{{\varepsilon^\star}^T M_Z \begin{pmatrix} X & W \end{pmatrix}}{{\varepsilon^\star}^T M_Z \varepsilon^\star}) \\
    &\overset{\BP}{\to}
    \Cov(\Psi_{\varepsilon^\star}, \begin{pmatrix} \Psi_{V_X} & \Psi_{V_W} \end{pmatrix} - \Psi_{\varepsilon^\star} \frac{\Omega^\star_{\varepsilon^\star, (V_X \ V_W)}}{\Omega^\star_{\varepsilon^\star}}) \\
    &\toP \Omega^\star_{\varepsilon^\star, (V_X \ V_W)} - \Omega^\star_{\varepsilon^\star} \frac{\Omega^\star_{\varepsilon^\star, (V_X \ V_W)}}{\Omega^\star_{\varepsilon^\star}} = 0.
    \end{align*}
    Thus $\Psi_{\varepsilon^\star}$ and $\Psi_{\tilde V}$ are asymptotically uncorrelated and by their asymptotic normality thus asymptotically independent.

    \paragraph*{Step 3 for strong instrument asymptotics:} Here, $\Pi_W$ is constant and of full rank.
    We calculate 
    $$
    \plim \frac{1}{\sqrt{n}} \Psi_{\tilde S} = \plim \frac{1}{\sqrt{n}} (Z^T Z)^{1/2} \Pi_W + \frac{1}{\sqrt{n}} \Psi_{\tilde V} = Q^{1/2} \Pi_W,
    $$
    as $\Psi_{\tilde V}$ is asymptotically Gaussian and thus $\frac{1}{\sqrt{n}} \Psi_{\tilde V} \toP 0$.
    Then,
    \begin{align*}
    \Psi_{\varepsilon^\star}^T (P_{\Psi_{\tilde S}} -& P_{P_{\Psi_{\tilde S}} (Z^T Z)^{1/2} \Pi_W}) \Psi_{\varepsilon^\star} = \Psi_{\varepsilon^\star}^T (P_{\frac{1}{\sqrt{n}}\Psi_{\tilde S}} - P_{P_{\frac{1}{\sqrt{n}}\Psi_{\tilde S}} \frac{1}{\sqrt{n}}(Z^T Z)^{1/2} \Pi_W}) \Psi_{\varepsilon^\star} \\
    &\toP \Psi_{\varepsilon^\star}^T (P_{Q^{1/2} (\Pi_X \ \Pi_W)} - P_{Q^{1/2} \Pi_W}) \Psi_{\varepsilon^\star}\\
    &=
    \Psi_{\varepsilon^\star}^T P_{M_{Q^{1/2} \Pi_W} Q^{1/2} \Pi_X} \Psi_{\varepsilon^\star} \tod \Var(\Psi_{\varepsilon^\star}) \cdot \chi^2(\mX),
    \end{align*}
    as $\Psi_{\varepsilon^\star} \tod \CN(0, \Omega^\star_{\varepsilon^\star} \cdot \Id_k)$ and $\rank(M_{Q^{1/2} \Pi_W} Q^{1/2} \Pi_X) = \rank(Q^{1/2} M_{ \Pi_W} \Pi_X) = \mX$.

    \paragraph*{Step 3 for weak instrument asymptotics:} Here, $\Pi_X = \frac{1}{\sqrt{n}} C_X$ for some fixed $C_X$ of full rank.
    We calculate 
    $$
    \plim \Psi_{\tilde S} = \plim \frac{1}{\sqrt{n}} (Z^T Z)^{1/2} C_W + \Psi_{\tilde V} = Q^{1/2} C_W + \Psi_{\tilde V}
    $$
    and
    \begin{align*}
        \Psi_{\varepsilon^\star}^T (P_{\Psi_{\tilde S}} -& P_{P_{\Psi_{\tilde S}} (Z^T Z)^{1/2} \Pi_W}) \Psi_{\varepsilon^\star}  \\
        &\toP \Psi_{\varepsilon^\star}^T (P_{Q^{1/2}(C_X \ C_W) + \Psi_{\tilde V}} - P_{P_{Q^{1/2}(C_X \ C_W) + \Psi_{\tilde V}}Q^{1/2} C_W})  \Psi_{\varepsilon^\star}
    \end{align*}
    We have $\Psi_{\varepsilon^\star} \tod \CN(0, \Omega_{\varepsilon^\star}^\star \cdot \Id_k)$, 
    $$\rank(P_{Q^{1/2}(C_X \ C_W) + \Psi_{\tilde V}} - P_{P_{Q^{1/2}(C_X \ C_W) + \Psi_{\tilde V}}Q^{1/2} \Pi_W}) = \mX,$$
    and that, by Step 2, $\Psi_{\tilde V}$ and $\Psi_{\varepsilon^\star}$ are asymptotically independent.
    Thus, the above is asymptotically $\Omega^\star_{\varepsilon^\star} \cdot \chi^2(\mX)$-distributed.

    \paragraph*{Step 4:}
    Note that 
    $$
    \frac{1}{n-k} (y - X \beta_0 - W \gamma^\star)^T M_Z (y - X \beta_0 - W \gamma^\star) = \frac{1}{n-k} {\varepsilon^\star}^T M_Z \varepsilon^\star \overset{\BP}{\to} \Omega^\star_{\varepsilon^\star}.
    $$
    Thus
    \begin{align*}
    \LM(\beta_0) &= (n-k) \min_{\gamma \in \BR^\mW} \frac{(y - X \beta_0 - W \gamma )^T P_{P_Z \tilde S(\beta_0, \gamma)} (y - X \beta_0 - W \gamma)}{(y - X \beta_0 - W \gamma)^T M_Z (y - X \beta_0 - W \gamma)} \\
    & \leq
    (n-k) \frac{(y - X \beta_0 - W \gamma^\star)^T P_{P_Z \tilde S(\beta_0, \gamma^\star)} (y - X \beta_0 - W \gamma^\star)}{(y - X \beta_0 - W \gamma^\star)^T M_Z (y - X \beta_0 - W \gamma^\star)} \tod \chi^2(\mX)
    \end{align*}
    by Slutsky's Lemma.
\end{proofEnd}%

\noindent This directly implies the following corollary.
\begin{theoremEnd}[category=tests,malte,restate command=corsubvectorlagrangemultipliertest]{corollary}%
    \label{cor:subvector_lagrange_multiplier_test}
    Consider \cref{model:1} and assume \cref{ass:1} and \cref{tc:subvector_klm} hold.
    Let $F_{\chi^2(\mX)}$ be the cumulative distribution function of a chi-squared random variable with $\mX$ degrees of freedom.
    Under \cref{model:1} and both strong and weak instrument asymptotics, a test that rejects the null $H_0: \beta = \beta_0$ whenever $\LM(\beta) > F^{-1}_{\chi^2(\mX)}(1 - \alpha)$ has asymptotic size at most $\alpha$.
\end{theoremEnd}%
\begin{proofEnd}%
    This follows directly from \cref{prop:subvector_klm_test_statistic_chi_squared}.
\end{proofEnd}%

Thus, under \cref{tc:subvector_klm}, the weak-instrument-robust subvector Lagrange multiplier test achieves the same degrees of freedom as the non-robust but commonly used Wald test.
\subsection{Application}

We test the null hypothesis
$$H_0 \colon \text{ The causal effect of education on log wages } \beta_0 = 0$$
using the subvector Wald, Anderson-Rubin, (conditional) likelihood-ratio, and lagrange multiplier tests.
\begin{jupyternotebook}
\begin{tcolorbox}[breakable, size=fbox, boxrule=1pt, pad at break*=1mm,colback=cellbackground, colframe=cellborder]
\prompt{In}{incolor}{8}{\boxspacing}
\begin{Verbatim}[commandchars=\\\{\}]
\PY{k+kn}{from} \PY{n+nn}{functools} \PY{k+kn}{import} \PY{n}{partial}
\PY{k+kn}{from} \PY{n+nn}{ivmodels}\PY{n+nn}{.}\PY{n+nn}{tests} \PY{k+kn}{import} \PY{p}{(}
    \PY{n}{wald\PYZus{}test}\PY{p}{,}
    \PY{n}{likelihood\PYZus{}ratio\PYZus{}test}\PY{p}{,}
    \PY{n}{conditional\PYZus{}likelihood\PYZus{}ratio\PYZus{}test}\PY{p}{,}
    \PY{n}{lagrange\PYZus{}multiplier\PYZus{}test}
\PY{p}{)}

\PY{c+c1}{\PYZsh{} Split X into (X, W) according to model 2.}
\PY{n}{X}\PY{p}{,} \PY{n}{W} \PY{o}{=} \PY{n}{X}\PY{p}{[}\PY{p}{[}\PY{l+s+s2}{\PYZdq{}}\PY{l+s+s2}{ed76}\PY{l+s+s2}{\PYZdq{}}\PY{p}{]}\PY{p}{]}\PY{p}{,} \PY{n}{X}\PY{p}{[}\PY{p}{[}\PY{l+s+s2}{\PYZdq{}}\PY{l+s+s2}{exp76}\PY{l+s+s2}{\PYZdq{}}\PY{p}{,} \PY{l+s+s2}{\PYZdq{}}\PY{l+s+s2}{exp762}\PY{l+s+s2}{\PYZdq{}}\PY{p}{]}\PY{p}{]}

\PY{k}{for} \PY{n}{test}\PY{p}{,} \PY{n}{name} \PY{o+ow}{in} \PY{p}{[}
    \PY{p}{(}\PY{n}{partial}\PY{p}{(}\PY{n}{wald\PYZus{}test}\PY{p}{,} \PY{n}{estimator}\PY{o}{=}\PY{l+s+s2}{\PYZdq{}}\PY{l+s+s2}{tsls}\PY{l+s+s2}{\PYZdq{}}\PY{p}{)}\PY{p}{,} \PY{l+s+s2}{\PYZdq{}}\PY{l+s+s2}{Wald (TSLS)}\PY{l+s+s2}{\PYZdq{}}\PY{p}{)}\PY{p}{,}
    \PY{p}{(}\PY{n}{partial}\PY{p}{(}\PY{n}{wald\PYZus{}test}\PY{p}{,} \PY{n}{estimator}\PY{o}{=}\PY{l+s+s2}{\PYZdq{}}\PY{l+s+s2}{liml}\PY{l+s+s2}{\PYZdq{}}\PY{p}{)}\PY{p}{,} \PY{l+s+s2}{\PYZdq{}}\PY{l+s+s2}{Wald (LIML)}\PY{l+s+s2}{\PYZdq{}}\PY{p}{)}\PY{p}{,}
    \PY{p}{(}\PY{n}{anderson\PYZus{}rubin\PYZus{}test}\PY{p}{,} \PY{l+s+s2}{\PYZdq{}}\PY{l+s+s2}{AR}\PY{l+s+s2}{\PYZdq{}}\PY{p}{)}\PY{p}{,}
    \PY{p}{(}\PY{n}{likelihood\PYZus{}ratio\PYZus{}test}\PY{p}{,} \PY{l+s+s2}{\PYZdq{}}\PY{l+s+s2}{LR}\PY{l+s+s2}{\PYZdq{}}\PY{p}{)}\PY{p}{,}
    \PY{p}{(}\PY{n}{conditional\PYZus{}likelihood\PYZus{}ratio\PYZus{}test}\PY{p}{,} \PY{l+s+s2}{\PYZdq{}}\PY{l+s+s2}{CLR}\PY{l+s+s2}{\PYZdq{}}\PY{p}{)}\PY{p}{,}
    \PY{p}{(}\PY{n}{lagrange\PYZus{}multiplier\PYZus{}test}\PY{p}{,} \PY{l+s+s2}{\PYZdq{}}\PY{l+s+s2}{LM}\PY{l+s+s2}{\PYZdq{}}\PY{p}{)}\PY{p}{,}
\PY{p}{]}\PY{p}{:}
    \PY{n}{stat}\PY{p}{,} \PY{n}{pval} \PY{o}{=} \PY{n}{test}\PY{p}{(}\PY{n}{Z}\PY{o}{=}\PY{n}{Z}\PY{p}{,} \PY{n}{X}\PY{o}{=}\PY{n}{X}\PY{p}{,} \PY{n}{W}\PY{o}{=}\PY{n}{W}\PY{p}{,} \PY{n}{y}\PY{o}{=}\PY{n}{y}\PY{p}{,} \PY{n}{beta}\PY{o}{=}\PY{n}{np}\PY{o}{.}\PY{n}{array}\PY{p}{(}\PY{p}{[}\PY{l+m+mf}{0.}\PY{p}{]}\PY{p}{)}\PY{p}{)}
    \PY{n+nb}{print}\PY{p}{(}\PY{l+s+sa}{f}\PY{l+s+s2}{\PYZdq{}}\PY{l+s+si}{\PYZob{}}\PY{n}{name}\PY{l+s+si}{:}\PY{l+s+s2}{\PYZlt{}11}\PY{l+s+si}{\PYZcb{}}\PY{l+s+s2}{: statistic=}\PY{l+s+si}{\PYZob{}}\PY{n}{stat}\PY{l+s+si}{:}\PY{l+s+s2}{5.2f}\PY{l+s+si}{\PYZcb{}}\PY{l+s+s2}{, p\PYZhy{}value=}\PY{l+s+si}{\PYZob{}}\PY{n}{pval}\PY{l+s+si}{:}\PY{l+s+s2}{.4f}\PY{l+s+si}{\PYZcb{}}\PY{l+s+s2}{\PYZdq{}}\PY{p}{)}
\end{Verbatim}
\end{tcolorbox}
\vspace{-0.3cm}
\begin{Verbatim}[commandchars=\\\{\}]
Wald (TSLS): statistic=10.62, p-value=0.0011
Wald (LIML): statistic= 9.46, p-value=0.0021
AR         : statistic= 5.07, p-value=0.0016
LR         : statistic=10.93, p-value=0.0009
CLR        : statistic=10.93, p-value=0.0024
LM         : statistic= 5.79, p-value=0.0161
\end{Verbatim}
\end{jupyternotebook}

\section{Testing the model assumptions}
\label{sec:intro_to_iv_tests:auxiliary_tests}
We present auxiliary tests to validate \cref{model:0} and \cref{ass:0}.
Assume $\mW = 0$.

\subsection{Sargan's $J$-statistic and its LIML variant}
\citet{sargan1958estimation} proposed the Sargan J-test to test the validity of the overidentifying restrictions in \cref{model:0}.
If there are more instruments than endogenous variables, that is, if $k > m$, the Sargan J-test checks whether these ``additional'' instruments are uncorrelated with the error term $\varepsilon$.
A rejection of the J-test suggests that there is at least one invalid instrument, casting doubt on the instrumental variable regression estimates and inference.

\citet{hansen1982large} extended the Sargan J-test to generalized method of moments (GMM) estimators and the resulting test is sometimes also called the Sargan-Hansen test.

\label{sec:j_statistic}
\begin{definition}[\citeauthor{sargan1958estimation}, \citeyear{sargan1958estimation} and \citeauthor{hansen1982large}, \citeyear{hansen1982large}]
    The J-statistic is
    $$
    J := k \AR(\hat\beta_\tsls) = (n - k) \frac{(y - X \hat\beta_\tsls) P_Z (Y - X \hat\beta_\tsls)}{(y - X \hat\beta_\tsls)^T M_Z (y - X \hat\beta_\tsls)}.
    $$
\end{definition}

\begin{theoremEnd}[malte,category=tests2]{proposition}[\citeauthor{sargan1958estimation}, \citeyear{sargan1958estimation} and \citeauthor{hansen1982large}, \citeyear{hansen1982large}]
    \label{prop:j_statistic_chi_squared}
    Under strong instrument asymptotics, the J statistic is asymptotically $\chi^2(k-m)$ distributed.
\end{theoremEnd}%
\begin{proofEnd}%
    Recall that
    \begin{equation*}
        \hat \beta_\tsls = (X^T P_Z X)^{-1} X^T P_Z y \Rightarrow P_Z X (\beta_0 - \hat\beta_\tsls) = - P_{P_Z X} \varepsilon.
    \end{equation*}
    Hence
    \begin{align*}
        \AR(\hat\beta_\tsls) &= \frac{n - k}{k} \frac{ (y - X \hat\beta_\tsls) P_Z (y - X \hat\beta_\tsls) }{ (y - X \hat\beta_\tsls) M_Z (y - X \hat\beta_\tsls) } \\
        &= \frac{n - k}{k} \frac{ (\varepsilon + X (\beta_0 - \hat\beta_\tsls))^T P_Z (\varepsilon + X (\beta_0 - \hat\beta_\tsls)) }{ (\varepsilon + X (\beta_0 - \hat\beta_\tsls))^T M_Z (\varepsilon + X (\beta_0 - \hat\beta_\tsls)) } \\
        &= \frac{n - k}{k} \frac{ \varepsilon^T M_{P_Z X} P_Z M_{P_Z X} \varepsilon }{ (\varepsilon^T + V_X (\beta_0 - \hat\beta_\tsls))^T (\varepsilon^T + V_X (\beta_0 - \hat\beta_\tsls)) }.
    \end{align*}
    Under strong instrument asymptotics,
    $$
    \varepsilon^T M_{P_Z X} P_Z M_{P_Z X} \varepsilon \toP \Psi_{\varepsilon} M_{Q \Pi_X} \Psi_{\varepsilon} \tod \Var(\varepsilon) \cdot \chi^2(k - m)
    $$
    and, by consistency of $\hat\beta_\tsls$,
    $$\frac{1}{n-k} (\varepsilon^T + V_X (\beta_0 - \hat\beta_\tsls))^T (\varepsilon^T + V_X (\beta_0 - \hat\beta_\tsls)) \toP \Var(\varepsilon).$$
    This implies the claim.
    
\end{proofEnd}%

The resulting test rejects the null hypothesis of model specification as in \cref{model:0} and \cref{ass:1} if $J > F^{-1}_{\chi^2(k - m)}(1 - \alpha)$,
where $F_{\chi^2(k - m)}$ is the cumulative distribution function of the chi-squared distribution with $k - m$ degrees of freedom.

This test is not robust to weak instruments.
However, the following variant based on plugging in the LIML estimator instead of the TSLS estimator is.
\begin{definition}[\citeauthor{guggenberger2012asymptotic}, \citeyear{guggenberger2012asymptotic}]
    \label{def:j_liml_statistic}
    The LIML variant of the J-statistic is
    $$
    J_\liml := k \min_{b \in \BR^\mX} \AR(b) \overset{\cref{thm:liml_is_kclass}}{=} (n - k) \frac{(y - X \hat\beta_\liml) P_Z (Y - X \hat\beta_\liml)}{(y - X \hat\beta_\liml)^T M_Z (y - X \hat\beta_\liml)}
    $$
\end{definition}

\noindent The following is a direct application of \cref{prop:subvector_anderson_rubin_test_statistic} with $\mX = 0$.
\begin{theoremEnd}[malte,category=tests2]{proposition}[\citeauthor{guggenberger2012asymptotic}, \citeyear{guggenberger2012asymptotic}]
    \label{prop:j_liml_statistic_chi_squared}
    Under both strong and weak instrument asymptotics, the LIML variant of the J-statistic is asymptotically bounded from above by a $\chi^2(k - m)$ distributed random variable.
\end{theoremEnd}%

\subsection{The Cragg-Donald test statistic and Anderson's test of reduced rank}
\label{sec:anderson_1951_rank_test}
\citet{staiger1997instrumental} note that the convergence of the TSLS and LIML estimators is a function of the concentration parameter $\mu^2 / k := n \Omega_{V}^{-1} \Pi^T Q \Pi /  k$, which acts as an effective sample size.
For $m = 1$, the concentration parameter can be estimated using the $F$-test for the first stage, with $\BE[F] = 1 + \mu^2 / k$.
If $k=1$ and $\mu^2 / k \geq 9$, \citet{staiger1997instrumental} prove that the TSLS estimator has a bias of at most 10\% of the bias of the OLS estimator, that is, $| \BE[\hat\beta_\tsls] - \beta_0 |  \leq 0.1 \cdot | \BE[\hat\beta_\ols]  - \beta_0 |$, and that the Wald-based 95\%-confidence interval around the TSLS estimator has at least 85\% coverage.
This motivated the ``rule of thumb'' that instruments are considered weak if the first-stage $F$-statistic is below 10.

As noted by \citet{stock2002testing}, this rule of thumb is misleading if not $m = k = 1$.
If $m > 1$, underidentification can be tested with the Cragg-Donald test statistic.
\begin{definition}[\citeauthor{cragg1997inferring}, \citeyear{cragg1997inferring}]
    \label{def:cragg_donald_test_statistic}
    The Cragg-Donald test statistic is
    $$
    \mathrm{CD} := (n - k) \cdot \lambda_\mathrm{min} \left( \left[X ^T M_Z  X  \right]^{-1}  X ^T P_Z X  \right).
    $$
\end{definition}

\noindent This is the Wald test statistic of the first-stage regression parameter $\Pi$ in $X = Z\Pi + V_X$ being of reduced rank.
This is asymptotically equivalent to \citeauthor{anderson1951estimating}'s (\citeyear{anderson1951estimating}) likelihood-ratio test statistic that $\rank(\Pi) \leq r$ for $r = m-1$.
\begin{theoremEnd}[malte,category=tests2]{proposition}[\citeauthor{anderson1951estimating}, \citeyear{anderson1951estimating}]
    Let $\Pi \in \BR^{k \times m}$ for $k \geq m$, let $Z_i \in \BR^k$ and $V_i \sim \CN(0, \Omega)$ for $i = 1, \ldots, n$ be i.i.d.\ 
    and let $X_i = Z_i^T \Pi + V_i$.
    Write $Z \in \BR^{n \times k}$ and $X \in \BR^{n \times m}$ for the matrices of stacked observations.
    The log-likelihood of observing $X$ given $\Pi$, $\Omega$, and $Z$ is
    $$
        \ell(X \mid Z, \Pi, \Omega) = - \frac{m}{2} \log(2 \pi) - \frac{n}{2} \log(\det(\Omega)) - \frac{1}{2} \trace((X - Z \Pi) \Omega^{-1} (X - Z \Pi)^T).
    $$
    Let $r \geq 0$ and write $\tilde d_i$ for the $i$-th smallest eigenvalue of $(X^T M_Z X)^{-1} X^T P_Z X$.
    Then,
    \begin{multline*}
        \max_{\substack{{\Omega \succ 0, \, \Pi \in \BR^{k \times m}} \\ {\rank(\Pi) \leq m-r}}} \ell(X \mid Z, \Pi, \Omega) = -\frac{m}{2} (\log(2\pi) + n \log(n) + 1)+  \frac{n}{2} \log(\det(X^T M_Z X)) \\ 
        - \frac{n}{2} \sum_{i=1}^{r} \log(1 + \tilde d_i).
        \end{multline*}
    Consequently, the likelihood ratio test statistic for the null hypothesis $\rank(\Pi) \leq m - r$ is
    \begin{align*}
        \lambda := 2 \max_{\substack{{\Omega \succ 0, \, \Pi \in \BR^{k \times m}} \\ {\rank(\Pi) \leq m-r}}} \ell(X \mid Z, \Pi, \Omega) - 2 \max_{\substack{{\Omega \succ 0, \, \Pi \in \BR^{k \times m}}}} \ell(X \mid Z, \Pi, \Omega)
         = n \sum_{i = 1}^r \log(1 + \tilde d_i).
    \end{align*}
\end{theoremEnd}%
\begin{proofEnd}%
    For some matrix $A$ write $P_A := A (A^T A)^{-1} A^T$ for the projection onto the column space of $A$ and $M_A := \Id_n - P_A$ for the projection onto the orthogonal complement.
    We calculate
    $$
    \max_{\substack{{\Omega \succ 0, \Pi \in \BR^{k \times m}} \\ {\rank(\Pi) \leq m-r}}} \ell(X \mid Z, \Pi, \Omega)
    = \max_{\substack{{\Omega \succ 0, \Pi \in \BR^{k \times m}}, \Gamma \in \BR^{m \times r} \\ \Pi \Gamma = 0, \Gamma^T \Omega \Gamma = \Id_{r}}} \ell(X \mid Z, \Pi, \Omega).
    $$
    The normalization $\Gamma^T \Omega \Gamma = \Id$ is arbitrary to simplify calculations.
    Using Lagrange multipliers $\Phi \in \BR^{k \times r}$ and $\Psi \in \BR^{r \times r},$ the Lagrangian is
    $$
    L(\Pi, \Omega, \Gamma,  \Phi, \Psi) = \ell(X \mid Z, \Pi, \Omega) + \trace{\Phi^T \Pi \Gamma} + \frac{1}{2} \trace(\Psi^T (\Gamma^T \Omega \Gamma - \Id_r)).
    $$
    The first-order conditions are
    \begin{align}
        0 &= \frac{\partial}{\partial \Pi} L(\Pi, \Omega, \Gamma, \Phi, \Psi) = Z^T (X - Z \Pi) \Omega^{-1} + \Phi \Gamma^T\label{eq:rank-pi},\\
        0 &= \frac{\partial}{\partial \Omega^{-1}} L(\Pi, \Omega, \Gamma, \Phi, \Psi) = - \frac{n}{2} \Omega^{-1} + \frac{1}{2} (X - Z \Pi)^T (X - Z \Pi) + \frac{1}{2} \Gamma \Gamma^T \Psi \label{eq:rank-omega},\\
        0 &= \frac{\partial}{\partial \Gamma} L(\Pi, \Omega, \Gamma, \Phi, \Psi) = \Phi^T \Pi + \Psi \Gamma^T \Omega \label{eq:rank-gamma},\\
        0 &= \frac{\partial}{\partial \Phi} L(\Pi, \Omega, \Gamma, \Phi, \Psi) = \Pi \Gamma \label{eq:rank-psi}, \ \text{and}\\
        0 &= \frac{\partial}{\partial \Psi} L(\Pi, \Omega, \Gamma, \Phi, \Psi) = \Gamma^T \Omega \Gamma - \Id_{r} \label{eq:rank-phi},
    \end{align}
    where we used $\frac{\partial}{\partial A}\trace(A B A^T) = A (B + B^T)$ in Equation (\ref{eq:rank-pi}).
    Multiplying equation (\ref{eq:rank-gamma}) by $\Gamma$ from the right yields
    $$
    0 = \Phi^T \Pi \Gamma + \Psi \Gamma^T \Omega \Gamma \Rightarrow \Psi = 0.
    $$
    Plugging this into equation (\ref{eq:rank-omega}) yields
    \begin{equation}\label{eq:rank-omega-2}
    \Omega = \frac{1}{n} (X - Z \Pi)^T (X - Z \Pi).
    \end{equation}
    Equation (\ref{eq:rank-pi}) yields
    $$
    \Pi = (Z^T Z)^{-1} Z^T X + (Z^T Z)^{-1} \Phi \Gamma^T \Omega.
    $$
    Multiplying this from the right by $\Gamma$ yields
    $$
    0 = (Z^T Z)^{-1} Z^T X \Gamma + (Z^T Z)^{-1} \Phi \Gamma^T \Omega \Gamma \Rightarrow \Phi = - Z^T X \Gamma.
    $$
    and thus
    $$
    \Pi = (Z^T Z)^{-1} Z^T X (\Id_p - \Gamma \Gamma^T \Omega).
    $$
    Plugging the last two equations into (\ref{eq:rank-gamma}), using $\Psi = 0$, yields
    \begin{align*}
        \Gamma^T X^T Z \Pi = \Gamma^T X P_Z X (\Id_p - \Gamma \Gamma^T \Omega) = 0 \\
        \Rightarrow \Gamma^T X^T P_Z X = \Gamma^T X^T P_Z X \Gamma \Gamma^T \Omega \numberthis\label{eq:rank-gamma-2}.
    \end{align*}
    Next, we expand equation (\ref{eq:rank-omega-2}).
    This yields
    \begin{align*}
    n \Omega &= (X - Z \Pi)^T (X - Z \Pi) \\
    &= (X - P_Z X - P_Z X \Gamma \Gamma^T \Omega)^T (X - P_Z X - P_Z X \Gamma \Gamma^T \Omega) \\
    &= (X - P_Z X)^T (X - P_Z X) + (X - P_Z X)^T P_Z X \Gamma \Gamma^T \Omega + \Omega \Gamma \Gamma^T X^T P_Z X \Gamma \Gamma^T \Omega \\
    &= X^T M_Z X + \Omega \Gamma \Gamma^T X^T P_Z X \Gamma \Gamma^T \Omega \numberthis \label{eq:rank-omega-4} %
    \end{align*}

    Let $O \in \BR^{r \times r}$ be orthonormal such that $\frac{1}{n} O^T \Gamma^T X^T P_Z X \Gamma O =: \diag(d_1, \ldots, d_{r}) =: D$ is diagonal.
    Write $\tilde \Gamma = \Gamma O$ and note that $\tilde \Gamma^T \Omega \tilde \Gamma = O^T \Id_{r} O = \Id_{r}$.
    Then, from Equation (\ref{eq:rank-omega-4}), we have
    \begin{align*}
    &\phantom{\Rightarrow} n \Omega = X^T M_Z X + n \Omega \tilde \Gamma D \tilde \Gamma^T \Omega \numberthis \label{eq:rank-omega-3-1}\\
    &\Rightarrow n \Omega \tilde\Gamma = X^T M_Z X \tilde \Gamma + n \Omega \tilde \Gamma D \numberthis \label{eq:rank-omega-3}.
    \end{align*}
    Recall Equation (\ref{eq:rank-gamma-2}), which we transpose and rewrite as
    \begin{equation}
    X^T P_Z X \tilde \Gamma = n \Omega \tilde \Gamma D. \label{eq:rank-gamma-3}
    \end{equation}
    Define $\tilde D := D (\Id_r - D)^{\dagger}$ with entries $\tilde d_i$ on the diagonal.
    If $d_i = 0$ (that is, $X^T P_Z X$ is singular), then $\tilde d_i = 0$.
    Plugging equation (\ref{eq:rank-gamma-3}) into equation (\ref{eq:rank-omega-3}) yields
    \begin{align*}
    n \Omega \tilde \Gamma = X^T X \tilde \Gamma  &\overset{(\ref{eq:rank-gamma-3})}{\Rightarrow}
    X^T P_Z X \tilde \Gamma = X^T P_Z X \tilde \Gamma D + X^T M_Z X \tilde \Gamma D \\
    &\Rightarrow (X^T M_Z X)^{-1} X^T P_Z X = \tilde \Gamma \tilde D \numberthis \label{eq:rank-gamma-eigenvalue}
    \end{align*}
    That is, the columns $\gamma_i$ of $\tilde\Gamma$ are eigenvectors of $(X^T M_Z X)^{-1} X^T P_Z X$ with eigenvalues $\tilde d_i$ for $i = 1, \ldots, r$.

    We multiply equation (\ref{eq:rank-omega-3-1}) from the left by $(X^T M_Z X)^{-1}$ and plug in equations (\ref{eq:rank-gamma-3}) and (\ref{eq:rank-gamma-eigenvalue}):
    $$
        (X^T M_Z X)^{-1} n \Omega = \Id_m + \tilde \Gamma \tilde D \tilde \Gamma^T \Omega.
    $$
    Using the matrix determinant lemma $\det(A + U V^T) = \det(\Id + V^T A^{-1} U) \det(A)$ with $A = \Id, U = \tilde \Gamma$, and $V =  \Omega \tilde \Gamma \tilde D$ yields
    $$
        \det(X^T M_Z X)^{-1} \det(n \Omega) = \det(\Id_r + \tilde D \Gamma^T \Omega \tilde \Gamma) = \prod_{i=1}^{r} (1 + \tilde d_i).
    $$
    This is minimized if the $\tilde d_i$ are the minimal $r$ eigenvalues of $(X^T M_Z X)^{-1} X^T P_Z X$ and
    $$
    \log(\det(\Omega)) = - p \log(n) - \log(\det(X^T M_Z X)) + \sum_{i=1}^{r} \log(1 + \tilde d_i).
    $$
    such that
    \begin{multline*}
    \max_{\substack{{\Omega \succ 0, \Pi \in \BR^{k \times m}} \\ {\rank(\Pi) \leq m-r}}} \ell(X \mid Z, \Pi, \Omega) = -\frac{m}{2} (\log(2\pi) + n \log(n) + 1)+  \frac{n}{2} \log(\det(X^T M_Z X)) \\ 
    - \frac{n}{2} \sum_{i=1}^{r} \log(1 + \tilde d_i).
    \end{multline*}
\end{proofEnd}%

\begin{theoremEnd}[malte,category=tests2]{corollary}\label{cor:anderson_1951_rank_test}
Let $\Pi \in \BR^{k \times m}$ for $k \geq m$, let $Z_i \in \BR^k$ and $V_i \sim \CN(0, \Omega)$ for $i = 1, \ldots, n$ be i.d.d.\ and let $X_i = Z_i^T \Pi + V_i$.
Write $Z \in \BR^{n \times k}$ and $X \in \BR^{n \times m}$ for the matrices of stacked observations.
Let $\lambda_i$ be the $i$-th smallest eigenvalue of $(n - k) (X^T M_Z X)^{-1} X^T P_Z X$ and let $F_{\chi^2(r \cdot (k - m + r))}$ be the cumulative distribution function of the $\chi^2(r \cdot (k - m + r))$ distribution.
A test that rejects the null $H_0: \rank(\Pi) \leq m - r$ whenever $\sum_{i=1}^r \lambda_i > F^{-1}_{\chi^2(r \cdot (k - m + r))}(1 - \alpha)$ has asymptotic size $\alpha$.
\end{theoremEnd}%
\begin{proofEnd}%
    This is equivalent to the likelihood-ratio test, with
    \begin{multline*}
        -2 \left( \max_{\substack{{\Omega \succ 0, \Pi \in \BR^{k \times m}} \\ {\rank(\Pi) \leq m-r}}} \ell(X \mid Z, \Pi, \Omega) - \max_{\Omega \succ 0, \Pi \in \BR^{k \times m}} \ell(X \mid Z, \Pi, \Omega) \right) \\
        = n \sum_{i=1}^r \log(1 + \tilde d_i) = n \sum_{i=1}^r \log(1 + \frac{1}{n - k} \lambda_i) \to \sum_{i=1}^r \lambda_i
    \end{multline*}
    and the result follows from Wilk's Theorem.
    The degrees of freedom result from counting.
    In the full model, we have $k \cdot m$ parameters.
    In the restricted model, we have $k \cdot (m - r) + r \cdot (m - r)$ parameters, first for the unrestricted entries and then those which are a linear combination of the unrestricted entries.
    The difference is $k m - k (m - r) - r (m - r) = k r - r (m - r) = r \cdot (k - m + r)$.
\end{proofEnd}%

\noindent We are most interested in the case $r = 1$.
If $\rank(\Pi) \leq m - 1 < m$, then \citeauthor{anderson1951estimating}'s \citeyearpar{anderson1951estimating} likelihood-ratio test statistic $\lambda$ and the \citet{cragg1997inferring} test statistic are asymptotically $\chi^2(k - m + 1)$ distributed.

Finally, note that if $m = 1$, then
$$
\lambda / k = \frac{n - k}{k} \lambdamin{(X^T M_Z X)^{-1} X^T P_Z X} =\frac{n - k}{k}  \frac{X^T X - X^T M_Z X}{X^T M_Z X} = \frac{n-k}{k} \frac{\mathrm{RSS}_\mathrm{empty} - \mathrm{RSS}_\mathrm{full}}{\mathrm{RSS}_\mathrm{full}}
$$
is the $F$-statistic of the first-stage regression of $X$ on $Z$ using $F_{k, n-k}$ critical values.
That is, the Cragg-Donald test of reduced rank reduces to the $F$-test of the first-stage regression if $m = 1$.

\subsection{\citeauthor{scheidegger2025residual}'s (\citeyear{scheidegger2025residual}) residual-prediction test of well-specification}
\label{sec:cyrill}
Consistent estimation and inference in instrumental variables regression requires three main assumptions: (i)~relevance, (ii)~linearity, and (iii)~exogeneity. (i)~Relevance requires that $\Pi$ has full column rank, (ii)~linearity requires that $X_i$ affects the outcome $y_i$ linearly, and (iii)~exogeneity requires that $\BE[(\varepsilon_i, V_{X, i}) Z_i^T] = 0$, or in terms of \cref{ass:0}, that $\BE[\Psi_\varepsilon] \to_\BP 0$ and $\BE[\Psi_{V_X}] \to_\BP 0$.
(i)~Relevance can be tested with the first-stage F-test or its multivariate extension \cref{cor:anderson_1951_rank_test}.
In an overidentified model, (ii)~linearity and (iii)~exogeneity can be jointly tested using Sargan's $J$ test or its LIML variant.
However, these $J$-tests have low power if identification is low and are not applicable in the common just-identified settings.

\citet{scheidegger2025residual} note that, if practitioners argue for the validity of (ii) and (iii), they often talk about the instruments being \emph{unrelated} to the error terms, not simply \emph{uncorrelated}.
\citet{scheidegger2025residual} thus propose to test mean independence
$$
H_0 \colon \exists \beta \in \BR^m \text{ such that } \BE[Y_i - X_i \beta  \mid Z_i] = 0 \text{ almost surely.}
$$
The mean independence $\BE[Y_i - X_i \beta \mid Z_i] = 0$ implies that $\BE[(Y_i - X_i \beta) w(Z_i)] = 0$, that is $w(Z_i)$ satisfies exogeneity, for all measurable functions $w$.

A good choice of $w$ to detect violations of $H_0$ would be $w_\mathrm{opt}(Z_i) = \BE[\varepsilon_i \mid Z_i]$.
\citet{scheidegger2025residual} propose the following procedure to estimate a version of $w_\mathrm{opt}$ while maintaining type 1 error control.
\begin{itemize}
    \item[1.] Split the data into a train and a test set $\CD_\mathrm{train}$ and $\CD_\mathrm{test}$.
    \item[2.] Compute two-stage least-squares estimator $\hat\beta_\tsls^\mathrm{train}$ using the train set $\CD_\mathrm{train}$.
    \item[3.] Nonlinearly regress $\hat w_0 \colon y_i - X_i \hat\beta_\tsls \sim Z_i$ using the train set.
    \item[4.] Let $K$ be some $1-\eta$ quantile of the $| \hat w_0(Z_i) |$ on the train set. Rescale and clip $\hat w_0$ to obtain $\hat w(z) := \min(1, \max(-1, \hat w_0(z) / K)) \in [-1, 1]$.
\end{itemize}
\begin{theoremEnd}{definition}[\citeauthor{scheidegger2025residual}, \citeyear{scheidegger2025residual}]
Let $\gamma > 0$. Define
$$
\hat\sigma := \frac{1}{\sqrt{n_\mathrm{test}}} \| \hat w(Z_\mathrm{test}) - P_{Z_\mathrm{test}} X_\mathrm{test} (X_\mathrm{test}^T P_{Z_\mathrm{test}} X_\mathrm{test})^{-1} X_\mathrm{test}^T \hat w({Z_\mathrm{test}}) \| \cdot \| y_\mathrm{test} - X_\mathrm{test} \hat\beta_\tsls^\mathrm{test}\|,
$$
where $\hat\beta_\tsls^\mathrm{test}$ is the two-stage least-squares estimator using the test set $\CD_\mathrm{test}$.
The \emph{residual-prediction test} statistic for well-specification of the linear instrumental variables regression model (\cref{model:0}) is
$$
\frac{1}{\sqrt{n_\mathrm{test}}}\frac{1}{\max(\gamma, \hat\sigma)}\sum_{i \in \CD_\mathrm{test}} \hat w(Z_i) \cdot (y_i - X_i \hat\beta_\tsls^\mathrm{test}) .
$$
\end{theoremEnd}%

\begin{theoremEnd}[malte,category=cyrill,text proof={Proof sketch},one big link translated={\hspace{-0.71cm} See proof sketch on page}]{proposition}[\citeauthor{scheidegger2025residual}, \citeyear{scheidegger2025residual}]
    \label{prop:residual_prediction}
    Let $0 \leq \alpha \leq 0.5$ and $\gamma > 0$.
    Under strong instrument asymptotics, a test that reject the null $H_0: \exists \beta \in \BR^{\mX}: \BE[y_i - X_i \beta \mid Z_i] = 0$ if $T > F^{-1}_{\CN(0, 1)}(1 - \alpha)$, where $F_{\CN(0, 1)}(1 - \alpha)$ is the cumulative distribution function of a standard Gaussian, has asymptotic size at most $\alpha$.
\end{theoremEnd}%
\begin{proofEnd}%
    Let $\beta^\star$ be the population variant of $\hat\beta_\tsls^\mathrm{train}$.
    This minimizes $\|\Cov( Z_i, y_i - X_i \beta^\star)\|_{\Cov(Z_i)}^2$, were $\|x\|_A = x^T A x$.
    The parameter $\beta$ satisfying $\BE[y_i - X_i \beta \mid Z_i] = 0$ under $H_0$ satisfies $\Cov( Z_i, y_i - X_i \beta) = 0$, and thus, so does $\beta^\star$.
    Assuming relevance, that is, $\Cov(Z_i, X_i)$ is of full rank, $\beta^\star$ is unique and thus $\beta = \beta_\star$.
    
    The $\hat w (Z_i) \cdot (y_i - X_i \hat\beta_\tsls^\mathrm{test})$ are i.i.d.\ random variables.
    Under strong instrument asymptotics, $\hat\beta_\tsls^\mathrm{test} \to_\BP \beta^\star$ and thus, under $H_0$, they are asymptotically centered.
    If the $X_i$, $Z_i$, and $\varepsilon_i$ have finite moments of order $2 + \eta$ for some $\eta > 0$, then we can apply the central limit theorem to get uniform convergence in distribution of $n^{-1/2} \sum_i \hat w (Z_i) \cdot (y_i - X_i \hat\beta_\tsls^\mathrm{test})$ over all bounded functions $\hat w$.

    What is the limiting variance?
    Calculate
    $$
    y_\mathrm{test} - X_\mathrm{test} \hat\beta_\tsls^{\mathrm{train}} = \varepsilon - X_\mathrm{test} (X_\mathrm{test} P_{Z_\mathrm{test}} X_\mathrm{test})^{-1} X_\mathrm{test}^T P_{Z_\mathrm{test}} \varepsilon.
    $$
    and thus
    $$(y_\mathrm{test} - X_\mathrm{test} \hat\beta^\mathrm{test}_\tsls)^T \hat w(Z) = \varepsilon^T \hat w(Z) - 
    P_{Z_\mathrm{test}} X_\mathrm{test} (X_\mathrm{test}^T P_{Z_\mathrm{test}} X_\mathrm{test})^{-1} X_\mathrm{test}^T \hat w({Z_\mathrm{test}}).
    $$
    If the noise is homoscedastic and $\hat w$ well-behaved, we can estimate the variance of the individual $(y_i - X_i \beta_\tsls^\mathrm{test})$ by $\hat \sigma^2$.
    If the noise is not homoscedastic, we can use a different estimator for the variance
    \begin{multline*}        
        \hat\sigma_\mathrm{robust}^2 := \frac{1}{n} \sum_{i \in \CD_\mathrm{test}} (\hat w(Z_i) - Z_i (Z_\mathrm{test}^T Z_\mathrm{test})^{-1} Z_\mathrm{test}^T X_\mathrm{test} (X_\mathrm{test}^T P_{Z_\mathrm{test}} X_\mathrm{test})^{-1} X_\mathrm{test}^T \hat w(Z_\mathrm{test}))^2 \\
        \cdot (y_i - X_i \hat\beta_\tsls^\mathrm{train})^2.
    \end{multline*}
    
    The function $\hat w$ is random and changes with $n$.
    In a worst-case scenario, it assigns a weight of $1$ to the first observation, and weights $0$ to all other.
    To avoid restricting the nonlinear machine learner that produces $\hat w$, we lower bound the variance estimate by the fixed constant $\gamma$.
\end{proofEnd}%

As the residual-prediction test uses the two-stage least-squares estimator, it is not robust to weak instruments.
To our knowledge, no weak-instrument-robust version exists.

\subsection{Application}
We compute the LIML variant of the Sargan J-statistic and the Cragg-Donald test statistic of reduced rank.
\begin{jupyternotebook}
\begin{tcolorbox}[breakable, size=fbox, boxrule=1pt, pad at break*=1mm,colback=cellbackground, colframe=cellborder]
\prompt{In}{incolor}{9}{\boxspacing}
\begin{Verbatim}[commandchars=\\\{\}]
\PY{k+kn}{from} \PY{n+nn}{ivmodels}\PY{n+nn}{.}\PY{n+nn}{tests} \PY{k+kn}{import} \PY{n}{j\PYZus{}test}\PY{p}{,} \PY{n}{rank\PYZus{}test}

\PY{n}{XW} \PY{o}{=} \PY{n}{pd}\PY{o}{.}\PY{n}{concat}\PY{p}{(}\PY{p}{[}\PY{n}{X}\PY{p}{,} \PY{n}{W}\PY{p}{]}\PY{p}{,} \PY{n}{axis}\PY{o}{=}\PY{l+m+mi}{1}\PY{p}{)}

\PY{n}{j\PYZus{}stat}\PY{p}{,} \PY{n}{j\PYZus{}pval} \PY{o}{=} \PY{n}{j\PYZus{}test}\PY{p}{(}\PY{n}{Z}\PY{o}{=}\PY{n}{Z}\PY{p}{,} \PY{n}{X}\PY{o}{=}\PY{n}{XW}\PY{p}{,} \PY{n}{y}\PY{o}{=}\PY{n}{y}\PY{p}{,} \PY{n}{estimator}\PY{o}{=}\PY{l+s+s2}{\PYZdq{}}\PY{l+s+s2}{liml}\PY{l+s+s2}{\PYZdq{}}\PY{p}{)}
\PY{n+nb}{print}\PY{p}{(}\PY{l+s+sa}{f}\PY{l+s+s2}{\PYZdq{}}\PY{l+s+s2}{J\PYZhy{}statistic   : }\PY{l+s+si}{\PYZob{}}\PY{n}{j\PYZus{}stat}\PY{l+s+si}{:}\PY{l+s+s2}{6.3f}\PY{l+s+si}{\PYZcb{}}\PY{l+s+s2}{, p\PYZhy{}value: }\PY{l+s+si}{\PYZob{}}\PY{n}{j\PYZus{}pval}\PY{l+s+si}{:}\PY{l+s+s2}{.4f}\PY{l+s+si}{\PYZcb{}}\PY{l+s+s2}{\PYZdq{}}\PY{p}{)}
\PY{n}{rank\PYZus{}stat}\PY{p}{,} \PY{n}{rank\PYZus{}pval} \PY{o}{=} \PY{n}{rank\PYZus{}test}\PY{p}{(}\PY{n}{Z}\PY{o}{=}\PY{n}{Z}\PY{p}{,} \PY{n}{X}\PY{o}{=}\PY{n}{XW}\PY{p}{)}
\PY{n+nb}{print}\PY{p}{(}\PY{l+s+sa}{f}\PY{l+s+s2}{\PYZdq{}}\PY{l+s+s2}{Rank statistic: }\PY{l+s+si}{\PYZob{}}\PY{n}{rank\PYZus{}stat}\PY{l+s+si}{:}\PY{l+s+s2}{6.3f}\PY{l+s+si}{\PYZcb{}}\PY{l+s+s2}{, p\PYZhy{}value: }\PY{l+s+si}{\PYZob{}}\PY{n}{rank\PYZus{}pval}\PY{l+s+si}{:}\PY{l+s+s2}{.4f}\PY{l+s+si}{\PYZcb{}}\PY{l+s+s2}{\PYZdq{}}\PY{p}{)}
\end{Verbatim}
\end{tcolorbox}
\vspace{-0.3cm}
\begin{Verbatim}[commandchars=\\\{\}]
J-statistic   :  4.284, p-value: 0.1174
Rank statistic: 15.613, p-value: 0.0014
\end{Verbatim}
\end{jupyternotebook}
The (LIML variant of the) J-test does not reject at \(\alpha=0.05\).

\citet{stock2002testing} do not tabulate critical values for the Cragg-Donald statistic of reduced rank for Wald tests to have acceptable size for $m = 3$ endogenous regressors.
For the TSLS centered Wald test at significance \(\alpha = 0.05\) to have an expected size not exceeding $r = 0.1$ for $k=5$ and $m=1, 2$ these critical values would be 26.87 and 19.45.
These values are close to the observed value of 15.6.

We apply \citeauthor{scheidegger2025residual}'s \citeyearpar{scheidegger2025residual} residual prediction test of model misspecification to our dataset. This tests
$$
H_0 \colon \exists \beta \in \mathbb{R}^m \text{ such that } \mathbb{E}[ y_i - X_i \beta \mid Z_i ] = 0.
$$
Here, we need to explicitly include the excluded exogenous regressors.

\begin{jupyternotebook}    

    \begin{tcolorbox}[breakable, size=fbox, boxrule=1pt, pad at break*=1mm,colback=cellbackground, colframe=cellborder]
\prompt{In}{incolor}{10}{\boxspacing}
\begin{Verbatim}[commandchars=\\\{\}]
\PY{k+kn}{from} \PY{n+nn}{ivmodels}\PY{n+nn}{.}\PY{n+nn}{tests} \PY{k+kn}{import} \PY{n}{residual\PYZus{}prediction\PYZus{}test}

\PY{n}{residual\PYZus{}prediction\PYZus{}test}\PY{p}{(}
    \PY{n}{Z}\PY{o}{=}\PY{n}{df}\PY{p}{[}\PY{p}{[}\PY{l+s+s2}{\PYZdq{}}\PY{l+s+s2}{nearc4a}\PY{l+s+s2}{\PYZdq{}}\PY{p}{,} \PY{l+s+s2}{\PYZdq{}}\PY{l+s+s2}{nearc4b}\PY{l+s+s2}{\PYZdq{}}\PY{p}{,} \PY{l+s+s2}{\PYZdq{}}\PY{l+s+s2}{nearc2}\PY{l+s+s2}{\PYZdq{}}\PY{p}{,} \PY{l+s+s2}{\PYZdq{}}\PY{l+s+s2}{age76}\PY{l+s+s2}{\PYZdq{}}\PY{p}{,} \PY{l+s+s2}{\PYZdq{}}\PY{l+s+s2}{age762}\PY{l+s+s2}{\PYZdq{}}\PY{p}{]}\PY{p}{]}\PY{p}{,}
    \PY{n}{X}\PY{o}{=}\PY{n}{df}\PY{p}{[}\PY{p}{[}\PY{l+s+s2}{\PYZdq{}}\PY{l+s+s2}{ed76}\PY{l+s+s2}{\PYZdq{}}\PY{p}{,} \PY{l+s+s2}{\PYZdq{}}\PY{l+s+s2}{exp76}\PY{l+s+s2}{\PYZdq{}}\PY{p}{,} \PY{l+s+s2}{\PYZdq{}}\PY{l+s+s2}{exp762}\PY{l+s+s2}{\PYZdq{}}\PY{p}{]}\PY{p}{]}\PY{p}{,}
    \PY{n}{y}\PY{o}{=}\PY{n}{df}\PY{p}{[}\PY{l+s+s2}{\PYZdq{}}\PY{l+s+s2}{lwage76}\PY{l+s+s2}{\PYZdq{}}\PY{p}{]}\PY{p}{,}
    \PY{n}{C}\PY{o}{=}\PY{n}{df}\PY{p}{[}\PY{n}{family} \PY{o}{+} \PY{n}{indicators}\PY{p}{]}\PY{p}{,}
    \PY{n}{seed}\PY{o}{=}\PY{l+m+mi}{0}\PY{p}{,}
\PY{p}{)}
\end{Verbatim}
\end{tcolorbox}

            \begin{tcolorbox}[breakable, size=fbox, boxrule=.5pt, pad at break*=1mm, opacityfill=0]
\prompt{Out}{outcolor}{10}{\boxspacing}
\begin{Verbatim}[commandchars=\\\{\}]
(1.6869718742810358, 0.045804380122060895)
\end{Verbatim}
\end{tcolorbox}
\end{jupyternotebook}
\noindent The result is just barely significant at 5\%.
To avoid the $p$-value lottery due to the random train, test split used in the residual prediction test, \citet{scheidegger2025residual} suggest aggregating the $p$-values from 50 random splits by taking 2 times the median.
This results in a conservative $p$-value \citep{meinshausen2009p}.

\begin{jupyternotebook}
    \begin{tcolorbox}[breakable, size=fbox, boxrule=1pt, pad at break*=1mm,colback=cellbackground, colframe=cellborder]
\prompt{In}{incolor}{11}{\boxspacing}
\begin{Verbatim}[commandchars=\\\{\}]
\PY{n}{ps} \PY{o}{=} \PY{n}{np}\PY{o}{.}\PY{n}{zeros}\PY{p}{(}\PY{l+m+mi}{50}\PY{p}{)}
\PY{k}{for} \PY{n}{i} \PY{o+ow}{in} \PY{n+nb}{range}\PY{p}{(}\PY{l+m+mi}{50}\PY{p}{)}\PY{p}{:}
    \PY{n}{\PYZus{}}\PY{p}{,} \PY{n}{ps}\PY{p}{[}\PY{n}{i}\PY{p}{]} \PY{o}{=} \PY{n}{residual\PYZus{}prediction\PYZus{}test}\PY{p}{(}
        \PY{n}{Z}\PY{o}{=}\PY{n}{df}\PY{p}{[}\PY{p}{[}\PY{l+s+s2}{\PYZdq{}}\PY{l+s+s2}{nearc4a}\PY{l+s+s2}{\PYZdq{}}\PY{p}{,} \PY{l+s+s2}{\PYZdq{}}\PY{l+s+s2}{nearc4b}\PY{l+s+s2}{\PYZdq{}}\PY{p}{,} \PY{l+s+s2}{\PYZdq{}}\PY{l+s+s2}{nearc2}\PY{l+s+s2}{\PYZdq{}}\PY{p}{,} \PY{l+s+s2}{\PYZdq{}}\PY{l+s+s2}{age76}\PY{l+s+s2}{\PYZdq{}}\PY{p}{,} \PY{l+s+s2}{\PYZdq{}}\PY{l+s+s2}{age762}\PY{l+s+s2}{\PYZdq{}}\PY{p}{]}\PY{p}{]}\PY{p}{,}
        \PY{n}{X}\PY{o}{=}\PY{n}{df}\PY{p}{[}\PY{p}{[}\PY{l+s+s2}{\PYZdq{}}\PY{l+s+s2}{ed76}\PY{l+s+s2}{\PYZdq{}}\PY{p}{,} \PY{l+s+s2}{\PYZdq{}}\PY{l+s+s2}{exp76}\PY{l+s+s2}{\PYZdq{}}\PY{p}{,} \PY{l+s+s2}{\PYZdq{}}\PY{l+s+s2}{exp762}\PY{l+s+s2}{\PYZdq{}}\PY{p}{]}\PY{p}{]}\PY{p}{,}
        \PY{n}{y}\PY{o}{=}\PY{n}{df}\PY{p}{[}\PY{l+s+s2}{\PYZdq{}}\PY{l+s+s2}{lwage76}\PY{l+s+s2}{\PYZdq{}}\PY{p}{]}\PY{p}{,}
        \PY{n}{C}\PY{o}{=}\PY{n}{df}\PY{p}{[}\PY{n}{family} \PY{o}{+} \PY{n}{indicators}\PY{p}{]}\PY{p}{,}
        \PY{n}{seed}\PY{o}{=}\PY{n}{i}
    \PY{p}{)}

\PY{l+m+mi}{2} \PY{o}{*} \PY{n}{np}\PY{o}{.}\PY{n}{median}\PY{p}{(}\PY{n}{ps}\PY{p}{)}
\end{Verbatim}
\end{tcolorbox}
            \begin{tcolorbox}[breakable, size=fbox, boxrule=.5pt, pad at break*=1mm, opacityfill=0]
\prompt{Out}{outcolor}{11}{\boxspacing}
\begin{Verbatim}[commandchars=\\\{\}]
0.10588417814375828
\end{Verbatim}
\end{tcolorbox}
\end{jupyternotebook}
\noindent This is not significant at 5\%.

Is it necessary to treat experience and experience squared as endogenous variables in \citeauthor{card1993using}'s \citeyearpar{card1993using} analysis?
Treating them as exogenous would simplify the analysis by negating the need for subvector tests.
Using \citeauthor{scheidegger2025residual}'s \citeyearpar{scheidegger2025residual} residual prediction test we can verify that instruments are likely not valid if we treat experience and its square as exogenous.

\begin{jupyternotebook}
    \begin{tcolorbox}[breakable, size=fbox, boxrule=1pt, pad at break*=1mm,colback=cellbackground, colframe=cellborder]
\prompt{In}{incolor}{12}{\boxspacing}
\begin{Verbatim}[commandchars=\\\{\}]
\PY{k}{for} \PY{n}{i} \PY{o+ow}{in} \PY{n+nb}{range}\PY{p}{(}\PY{l+m+mi}{50}\PY{p}{)}\PY{p}{:}
    \PY{n}{\PYZus{}}\PY{p}{,} \PY{n}{ps}\PY{p}{[}\PY{n}{i}\PY{p}{]} \PY{o}{=} \PY{n}{residual\PYZus{}prediction\PYZus{}test}\PY{p}{(}
        \PY{n}{Z}\PY{o}{=}\PY{n}{df}\PY{p}{[}\PY{p}{[}\PY{l+s+s2}{\PYZdq{}}\PY{l+s+s2}{nearc4a}\PY{l+s+s2}{\PYZdq{}}\PY{p}{,} \PY{l+s+s2}{\PYZdq{}}\PY{l+s+s2}{nearc4b}\PY{l+s+s2}{\PYZdq{}}\PY{p}{,} \PY{l+s+s2}{\PYZdq{}}\PY{l+s+s2}{nearc2}\PY{l+s+s2}{\PYZdq{}}\PY{p}{,} \PY{l+s+s2}{\PYZdq{}}\PY{l+s+s2}{age76}\PY{l+s+s2}{\PYZdq{}}\PY{p}{,} \PY{l+s+s2}{\PYZdq{}}\PY{l+s+s2}{age762}\PY{l+s+s2}{\PYZdq{}}\PY{p}{]}\PY{p}{]}\PY{p}{,}
        \PY{n}{X}\PY{o}{=}\PY{n}{df}\PY{p}{[}\PY{p}{[}\PY{l+s+s2}{\PYZdq{}}\PY{l+s+s2}{ed76}\PY{l+s+s2}{\PYZdq{}}\PY{p}{]}\PY{p}{]}\PY{p}{,}
        \PY{n}{y}\PY{o}{=}\PY{n}{df}\PY{p}{[}\PY{l+s+s2}{\PYZdq{}}\PY{l+s+s2}{lwage76}\PY{l+s+s2}{\PYZdq{}}\PY{p}{]}\PY{p}{,}
        \PY{n}{C}\PY{o}{=}\PY{n}{df}\PY{p}{[}\PY{n}{family} \PY{o}{+} \PY{n}{indicators} \PY{o}{+} \PY{p}{[}\PY{l+s+s2}{\PYZdq{}}\PY{l+s+s2}{exp76}\PY{l+s+s2}{\PYZdq{}}\PY{p}{,} \PY{l+s+s2}{\PYZdq{}}\PY{l+s+s2}{exp762}\PY{l+s+s2}{\PYZdq{}}\PY{p}{]}\PY{p}{]}\PY{p}{,}
        \PY{n}{seed}\PY{o}{=}\PY{n}{i}
    \PY{p}{)}

\PY{l+m+mi}{2} \PY{o}{*} \PY{n}{np}\PY{o}{.}\PY{n}{median}\PY{p}{(}\PY{n}{ps}\PY{p}{)}
\end{Verbatim}
\end{tcolorbox}

\begin{tcolorbox}[breakable, size=fbox, boxrule=.5pt, pad at break*=1mm, opacityfill=0]
\prompt{Out}{outcolor}{12}{\boxspacing}
\begin{Verbatim}[commandchars=\\\{\}]
0.003990876020777634
\end{Verbatim}
\end{tcolorbox}
\end{jupyternotebook}

\section{Confidence sets}
\label{sec:intro_to_iv_tests:confidence_sets}
The Wald, Anderson-Rubin, (conditional) likelihood-ratio, Lagrange multiplier test, and their subvector variants can be inverted to yield confidence sets for the causal parameter of interest $\beta_0$ in \cref{model:1}.
\begin{definition}\label{def:confidence_sets}
    Assume \cref{model:1}. Let $\alpha > 0$.
    \begin{itemize}
        \item The \emph{(subvector) inverse Wald test confidence set} using estimator $\hat\beta$ of confidence level $1 - \alpha$ is
        \begin{align*}
            \CI_{\Wald_{\hat\beta}}(1 - \alpha) := \left\{ \beta \in \BR^{\mX} \mid \Wald_{\hat\beta}(\beta) \leq F^{-1}_{\chi^2(\mX)}(1 - \alpha) \right\}.
        \end{align*}
        \item The \emph{(subvector) inverse Anderson-Rubin test confidence set} of confidence level $1 - \alpha$ is
        \begin{align*}
            \CI_{\AR}(1 - \alpha) := \left\{ \beta \in \BR^{\mX} \mid (k - \mW) \cdot  \AR(\beta) \leq F^{-1}_{\chi^2(k - \mW)}(1 - \alpha) \right\}.
        \end{align*}
        \item The \emph{(subvector) inverse likelihood-ratio test confidence set} of confidence level $1 - \alpha$ is
        \begin{align*}
            \CI_{\LR}(1 - \alpha) := \left\{ \beta \in \BR^{\mX} \mid \LR(\beta) \leq F^{-1}_{\chi^2(\mX)}(1 - \alpha) \right\}.
        \end{align*}
        \item The \emph{(subvector) inverse conditional likelihood-ratio test confidence set} of confidence level $1 - \alpha$ is
        \begin{align*}
            \CI_{\CLR}(1 - \alpha) := \left\{ \beta \in \BR^{\mX} \mid \LR(\beta) \leq F^{-1}_{\Gamma(k - m, \mX, s_\mathrm{min}(\beta))} (1 - \alpha) \right\},
        \end{align*}
        where $\Gamma$ and $s_\mathrm{min}$ are defined in \cref{prop:clr_test_statistic} (\cref{con:subvector_clr}).
        \item The \emph{(subvector) inverse Lagrange multiplier test confidence set} of confidence level $1 - \alpha$ is
        \begin{align*}
            \CI_{\LM}(1 - \alpha) := \left\{ \beta \in \BR^{\mX} \mid \LM(\beta) \leq F^{-1}_{\chi^2(\mX)}(1 - \alpha) \right\}.
        \end{align*}
    \end{itemize}   
\end{definition}

\begin{theoremEnd}[malte,category=confsets]{corollary}%
    \label{cor:confidence_sets_have_correct_size}
    Assume \cref{model:1}. Let $\alpha > 0$.
    \begin{itemize}
        \item Under strong instrument asymptotics, for any sequence $\kappa_n$ such that $\sqrt{n} (\kappa_n - 1) \toP 0$, the inverse Wald confidence set using estimator $\hat\beta_\kclass(\kappa_n)$ of confidence level $1 - \alpha$ has asymptotic coverage $1 - \alpha$.
        \item Under strong and weak instrument asymptotics, the inverse (subvector) Anderson-Rubin test confidence set of confidence level $1 - \alpha$ has asymptotic coverage at least $1 - \alpha$.
        \item Under strong instrument asymptotics, the inverse (subvector) likelihood ratio test confidence set of confidence level $1 - \alpha$ has asymptotic coverage at least $1 - \alpha$.
        \item Assume $\mW = 0$. Under strong and weak instrument asymptotics, the inverse conditional likelihood ratio test confidence set of confidence level $1 - \alpha$ has asymptotic coverage at least $1 - \alpha$.
        \item Under strong and weak instrument asymptotics and if \cref{tc:subvector_klm} holds, the inverse Lagrange multiplier confidence set of confidence level $1 - \alpha$ has asymptotic coverage at least $1 - \alpha$.
    \end{itemize}
\end{theoremEnd}%
\begin{proofEnd}%
    This follows from \cref{cor:subvector_wald_test}, \cref{cor:subvector_anderson_rubin_test}, \cref{cor:subvector_likelihood_ratio_test}, \cref{cor:subvector_conditional_likelihood_ratio_test}, and \cref{cor:subvector_lagrange_multiplier_test}.
\end{proofEnd}%

\noindent All the above confidence sets can be numerically approximated by constructing a grid of values for $\beta$ and computing the corresponding test statistic for each value.
\citet{dufour2005projection} noted that the confidence sets obtained by inversion of the (full) Anderson-Rubin test are quadrics, and that thus conservative confidence sets can be obtained analytically by projection.
\citet{londschien2024weak} extend this: They use the more powerful $\chi^2(k - \mW)$ critical values by \citet{guggenberger2012asymptotic} and give the closed-form solution for the inverse subvector Anderson-Rubin test confidence sets.
These can then be compared to the inverse Wald and inverse likelihood-ratio test confidence sets, which are also quadrics with closed-form solutions. %
\begin{theoremEnd}[malte,category=confsets]{proposition}[\citeauthor{londschien2024weak}, \citeyear{londschien2024weak}]%
    \label{prop:cis_closed_forms}
    The inverse Wald, likelihood-ratio, and Anderson-Rubin test confidence sets describe quadrics in $\BR^{\mX}$.
    Let $S := \begin{pmatrix} X & W \end{pmatrix}$ and $B \in \BR^{\mX \times m}$ have ones on the diagonal and zeros elsewhere, such that $B S = X$.
    Let
    $$\kappa_{\AR}(\alpha) = 1 + F^{-1}_{\chi^2(k - \mW)}(1 - \alpha) / (n - k) \ \text{ and } \ \kappa_{\LR}(\alpha) = \hat\kappa_\liml + F^{-1}_{\chi^2(\mX)}(1 - \alpha) / (n - k).$$
    Let
    \begin{align*}
    A(\kappa) &:= \left[ B \left(S^T (\Id_n - \kappa M_Z) S \right)^{-1} B^T \right]^{-1} =
    X^T (\Id_n - \kappa M_Z) X \\ & \hspace{1.5cm} - X^T (\Id_n - \kappa M_Z) W \left[W^T (\Id_n - \kappa M_Z) W \right]^{-1} W^T (\Id_n - \kappa M_Z) X.
    \end{align*}
    Let $\hat\sigma^2_\mathrm{Wald}(\kappa) := \frac{1}{n - m} \| y - S \hat\beta_\kclass(\kappa) \|^2$ and $\hat\sigma^2(\kappa) := \frac{1}{n-k} \| M_Z (y - S \hat\beta_\kclass(\kappa)) \|^2$.
    If $\mW > 0$, let $\kappa_\mathrm{max} := \lambdamin{(W^T M_Z W)^{-1} W^T P_Z W} + 1$. Else $\kappa_{\mathrm{max}} := \infty$.
    \begin{itemize}
    \item If $\kappa \leq \kappa_\mathrm{max}$, then
    \begin{multline*}
        \CI_{\mathrm{Wald}_{\hat\beta_\kclass(\kappa)}}(1 - \alpha) = \Big\{ \beta \in \BR^{\mX}  \mid \left(\beta - B \hat\beta_\kclass(\kappa) 
        \right)^T A(\kappa) \left(\beta - B\hat\beta_\kclass(\kappa) \right) \\
        \leq \hat\sigma^2_\mathrm{Wald}(\kappa) \cdot F^{-1}_{\chi^2(\mX)}(1 - \alpha)\Big\}.
    \end{multline*}
    Else, $\CI_{\mathrm{Wald}_{\hat\beta_\kclass(\kappa)}}(1 - \alpha) = \BR^{\mX}$.
    \item  If $\kappa_{\AR}(\alpha) \leq \kappa_\mathrm{max}$, then
    \begin{multline*}
        \CI_{\AR}(1 - \alpha) = \Big\{ \beta \in \BR^{\mX} \mid \left(\beta - B \hat\beta_\kclass(\kappa(\alpha))\right)^T A(\kappa_{\AR}(\alpha)) \left(\beta - B\hat\beta_\kclass(\kappa_{\AR}(\alpha)) \right) \\
        \leq \hat\sigma^2(\kappa_{\AR}(\alpha)) \cdot (F^{-1}_{\chi^2(k - \mW)}(1 - \alpha) - k \AR(\hat\beta_\kclass(\kappa_{\AR}(\alpha))))\Big\}.
    \end{multline*}
    Else, $\CI_{\AR}(1 - \alpha) = \BR^{\mX}$.
    \item If $\kappa_{\LR}(\alpha) \leq \kappa_\mathrm{max}$, then
    \begin{multline*}
        \CI_{\LR}(1 - \alpha) = \Big\{ \beta \in \BR^{\mX} \mid \left(\beta - B \hat\beta_\kclass(\kappa_{\LR}(\alpha)) \right)^T A(\kappa_{\LR}(\alpha))  \left(\beta - B \hat\beta_\kclass(\kappa_{\LR}(\alpha)) \right) \\
        \leq \hat\sigma^2(\kappa_{\LR}(\alpha)) \cdot (F^{-1}_{\chi^2(\mX)}(1 - \alpha) + k \AR(\hat\beta_\liml) - k \AR(\hat\beta_\kclass(\kappa_{\LR}(\alpha)))) \Big\}.
    \end{multline*}
    Else, $\CI_{\LR}(1 - \alpha) = \BR^{\mX}$.
\end{itemize}
\end{theoremEnd}%
\begin{proofEnd}%
    Let $\mW = 0$.
    By definition
    \begin{multline*}
        \mathrm{Wald}_{\hat \beta_\kclass(\kappa)}(\beta) \leq F^{-1}_{\chi^2(\mX)}(1 - \alpha) \Leftrightarrow (\beta - \hat\beta_\kclass(\kappa))^T \left( X^T (P_Z + (1 - \kappa) M_Z) X \right) (\beta - \hat\beta_\kclass(\kappa)) \\   
        \leq \hat\sigma^2_\mathrm{Wald}(\kappa) \cdot F^{-1}_{\chi^2(\mX)}(1 - \alpha). \numberthis \label{eq:ci_closed_form:wald_1}    
    \end{multline*}
    Let $\tilde \AR(\beta) := (n - k) \frac{(y - X \beta)^T P_Z (y - X \beta)}{(y - X \beta)^T M_Z (y - X \beta)}$.
    We prove that
    \begin{align*}
        \tilde\AR(\beta) \leq (n - k)(\kappa - 1) \Leftrightarrow (\beta - \hat\beta_\kclass(\kappa))^T \left( X^T (\kappa P_Z + (1 - \kappa) \Id) X \right) (\beta -  \hat\beta_\kclass(\kappa)) \\
        \leq \hat\sigma^2(\kappa) \cdot \left((n - k) (\kappa - 1)- \tilde\AR(\hat\beta_\kclass(\kappa)) \right). \numberthis \label{eq:ci_closed_form:1}
    \end{align*}
    Calculate
    \begin{align*}
        \tilde\AR&(\beta) = (n - k) \frac{(y - X \beta)^T P_Z (y - X \beta)}{(y - X \beta)^T M_Z (y - X \beta)} \leq (n - k) (\kappa - 1) \\
        &\Leftrightarrow (y - X \beta)^T P_Z (y - X \beta) \leq (\kappa - 1) (y - X \beta)^T M_Z (y - X \beta)\\
        &\Leftrightarrow (y - X \beta)^T (P_Z + (1 - \kappa) M_Z) (y - X \beta) \leq 0.
    \end{align*}
    Here,
    \begin{align*}
        &(y - X \beta)^T (P_Z + (1 - \kappa) M_Z) (y - X \beta)  \\
        &= \beta^T \left( X^T ( P_Z + (1 - \kappa) M_Z) X \right) \beta - 2 \beta^T \underbrace{X^T (P_Z + (1 - \kappa) M_Z) y}_{\mathclap{=(X^T (P_Z + (1 - \kappa) M_Z) X) \hat\beta_\kclass(\kappa)}} + y^T (P_Z + (1 - \kappa) M_Z) y \\
        &= (\beta -  \hat\beta_\kclass(\kappa))^T \left( X^T (P_Z + (1 - \kappa) M_Z) X \right) (\beta -  \hat\beta_\kclass(\kappa))
        + y^T (P_Z + (1 - \kappa) M_Z) y \\
        & \hspace{3cm} - \underbrace{\hat\beta_\kclass(\kappa)^T X^T (P_Z + (1 - \kappa) M_Z) X \hat\beta_\kclass(\kappa)}_{\mathclap{= \hat\beta_\kclass(\kappa)^T X^T (P_Z + (1 - \kappa) M_Z) y}}\\
        &= (\beta -  \hat\beta_\kclass(\kappa))^T \left( X^T (P_Z + (1 - \kappa) M_Z) X \right) (\beta - \hat\beta_\kclass(\kappa)) \\
        &\hspace{3cm} +  (y -  X \hat\beta_\kclass(\kappa))^T  (P_Z + (1 - \kappa) M_Z) (y - X \hat\beta_\kclass(\kappa)),
    \end{align*}
    as $(y - X \hat\beta_\kclass(\kappa))^T (P_Z + (1 - \kappa) M_Z) X \hat\beta_\kclass(\kappa) = 0$. Now, rewrite
    \begin{align*}
        (y& - X \hat\beta_\kclass(\kappa))^T (P_Z + (1 - \kappa) M_Z)  (y - X \hat\beta_\kclass(\kappa)) \\
        &= (y - X \hat\beta_\kclass(\kappa)) P_Z (y - X \hat\beta_\kclass(\kappa)) + (1 - \kappa) (y - X \hat\beta_\kclass(\kappa))^T M_Z (y - X \hat\beta_\kclass(\kappa)) \\
        &= \hat\sigma^2(\kappa) \cdot (\tilde\AR(\hat\beta_\kclass(\kappa)) + (n - k) (1 - \kappa)).
    \end{align*}
    This proves \eqref{eq:ci_closed_form:1}.
    If $\mW = 0$, then $\beta \in \CI_{\AR}(1 - \alpha)$ if and only if $\tilde\AR(\beta) \leq F^{-1}_{\chi^2(k)}(1 - \alpha)$.
    The expression of $\CI_{\AR}(1 - \alpha)$ follows from \eqref{eq:ci_closed_form:1} with 
    \begin{align*}(n - k) (\kappa_{\AR}(\alpha) - 1) = F^{-1}_{\chi^2(k)}(1 - \alpha) \Leftrightarrow \kappa_{\AR}(\alpha) = 1 + F^{-1}_{\chi^2(k)}(1 - \alpha) / (n-k).
    \end{align*}
    Also, $\beta \in \CI_{\LR}(1 - \alpha)$ if and only if $\tilde\AR(\beta) - \min_b \tilde\AR(b) \leq F^{-1}_{\chi^2(\mX)}(1 - \alpha) / (n - k)$.
    The expression of $\CI_{\LR}(1 - \alpha)$ thus follows from \eqref{eq:ci_closed_form:1} with $\kappa_{\LR}(\alpha) = 1 + (\min_b \tilde\AR(b) + F^{-1}_{\chi^2(\mX)}(1 - \alpha)) / (n-k) = \hat\kappa_\liml + F^{-1}_{\chi^2(\mX)}(1 - \alpha) / (n-k)$.

    Apply \cref{lem:kappa_pos_definite} with $X \leftarrow W$. This implies that $W^T (\kappa P_Z + (1 - \kappa) \Id) W$ is positive semi-definite for $\kappa \leq \kappa_\mathrm{max}$ and negative definite if $\kappa > \kappa_\mathrm{max}$.

    Calculate
    \begin{multline*}
        \{ \beta \in \BR^\mX \mid \min_\gamma \tilde{\AR} \left( \begin{pmatrix} \beta \\ \gamma \end{pmatrix} \right) \leq (n - k) \cdot (\kappa - 1) \}  \\
        \overset{\eqref{eq:ci_closed_form:1}}{=} \Big\{\beta \in \BR^\mX \mid \min_\gamma \left(\begin{pmatrix} \beta \\ \gamma \end{pmatrix} - \hat\beta_\kclass(\kappa) \right)^T (S^T (\kappa P_Z + (1 - \kappa) \Id) S) \left(\begin{pmatrix} \beta \\ \gamma \end{pmatrix} - \hat\beta_\kclass(\kappa) \right) \\
        \leq \hat\sigma^2(\kappa) \cdot \left((n - k) (\kappa - 1) - \tilde\AR(\hat\beta_\kclass(\kappa)) \right) \Big\} \\
        \overset{\cref{lem:projected_quadric}}{=}
        \begin{cases}
            \ \ \!\begin{aligned}
                \Big\{ \beta \in \BR^{\mX} \mid (\beta - &B \hat\beta_\kclass(\kappa))^T ( B (S^T (P_Z + (1 - \kappa) M_Z) S)^\dagger B^T)^{\dagger} \\
                &(\beta - B \hat\beta_\kclass(\kappa)) 
                 \leq \hat\sigma^2(\kappa) \cdot ((n - k) (\kappa - 1) - \tilde\AR(\hat\beta_\kclass(\kappa))) \Big\}
            \end{aligned} & \ \ \text{if } \kappa \leq \kappa_\mathrm{max} \\
            \ \ \BR^{\mX} & \ \ \text{if } \kappa > \kappa_\mathrm{max}
        \end{cases},
    \end{multline*}
    with $A = S^T (\kappa P_Z + (1 - \kappa) \Id) S$ and $A_{22} = W^T ( \kappa P_Z + (1 - \kappa) \Id) W$, where $\dagger$ denotes the Moore-Penrose pseudo-inverse.
    Again, the results for $\CI_{\AR}(1 - \alpha)$ and $\CI_{\LR}(1 - \alpha)$ follow from the cases via $\kappa = \kappa_{\AR}(\alpha)$ and $\kappa = \kappa_{\LR}(\alpha)$.
\end{proofEnd}%

A feature of the inverse Anderson-Rubin confidence sets is that they can be unbounded, particularly when instruments are weak.
This contrasts with the Wald test, whose confidence sets are always bounded ellipses.
Using \cref{prop:cis_closed_forms}, we can establish the following conditions for the inverse (subvector) Anderson-Rubin and likelihood-ratio test confidence sets to be nonempty and bounded.
\begin{theoremEnd}[malte, restate command=tctwo]{technical_condition}
    \label{tc:2}
    Let $S = \begin{pmatrix} X & W \end{pmatrix}$.
    \begin{enumerate}[(a)]
        \item \label{tc:2:a} If $\mW > 0$, then
            $$
                \lambda := \lambdamin{ [S^T M_Z S]^{-1} S^T P_Z S} < \lambdamin{ [X^T M_Z X]^{-1} X^T P_Z X},
            $$
            or, equivalently (using \cref{lem:kappa_pos_definite}),
            $$
                \lambdamin{ X^T (P_Z + (1 - \lambda) M_Z) X } > 0.
            $$
        \item Let $S^{(-i)}$ be $S$ with the $i$-th column removed. Then, for all $i = 1, \ldots, m$,
            $$
                \lambda < \lambdamin{ \left[S^{(-i)T} M_Z S^{(-i)}\right]^{-1} S^{(-i)T} P_Z S^{(-i)} }
            $$
            or, equivalently (using \cref{lem:kappa_pos_definite}),
            $$
                \lambdamin{ S^{(-i)T} (P_Z + (1 - \lambda) M_Z) S^{(-i)} } > 0.
            $$ \label{tc:2:b}
    \end{enumerate}
\end{theoremEnd}%

\begin{theoremEnd}[malte,category=confsets]{proposition}%
    \label{prop:cis_bounded}
    Assume the \cref{tc:2} holds.
    Let
    \begin{align*}
        J_\liml &= (k - \mW) \cdot \min_b \AR(b) \\
        &= (n - k) \ \lambdamin{ \left[\begin{pmatrix} X & W & y \end{pmatrix}^T M_Z \begin{pmatrix} X & W & y \end{pmatrix} \right]^{-1}\begin{pmatrix} X & W & y \end{pmatrix}^T P_Z \begin{pmatrix} X & W & y \end{pmatrix}}
    \end{align*}
    be the LIML variant of the J-statistic (see \cref{def:j_liml_statistic}) and let
    $$
        \mathrm{CD} = (n - k) \ \lambdamin{ \left[\begin{pmatrix} X & W \end{pmatrix}^T M_Z \begin{pmatrix} X & W \end{pmatrix} \right]^{-1}\begin{pmatrix} X & W \end{pmatrix}^T P_Z \begin{pmatrix} X & W \end{pmatrix}}.
    $$
    be the \citet{cragg1997inferring} test statistic of reduced rank (see \cref{def:cragg_donald_test_statistic}).

    \begin{itemize}
        \item The (subvector) inverse Anderson-Rubin test confidence set is nonempty and bounded if and only if
    \begin{align*}
        J_\liml \leq F^{-1}_{\chi^2(k - \mW)}(1 - \alpha) < \mathrm{CD} \ \Leftrightarrow \
        F_{\chi^2(k - \mW)} \left( J_\liml \right) \leq 1 - \alpha <  F_{\chi^2(k - \mW)} \left(\mathrm{CD} \right).
    \end{align*}
        \item The (subvector) inverse likelihood-ratio test confidence set is always nonempty. It is bounded if and only if
    \begin{align*}
        F^{-1}_{\chi^2(\mX)}(1 - \alpha) < \mathrm{CD} - J_\liml \ \Leftrightarrow \  1 - \alpha <  F_{\chi^2(\mX)} \left( \mathrm{CD} - J_\liml \right)
    \end{align*}

    \end{itemize}
\end{theoremEnd}%
\begin{proofEnd}%
    We prove the conditions for boundedness and non-emptiness separately.
    \paragraph*{Step 1: (non)emptiness of the confidence sets}
    The (subvector) inverse likelihood-ratio test confidence set always contains the LIML and hence is never empty.
    The (subvector) inverse Anderson-Rubin test confidence set is nonempty if and only if
    $$
    J_\liml = (k - \mW) \cdot \AR(\hat\beta_\liml) \leq F^{-1}_{\chi^2(k - \mW)}(1 - \alpha) \Leftrightarrow \alpha \leq 1 - F_{\chi^2(k - \mW)} \left( J_\liml  \right).
    $$

    \paragraph*{Step 2: Boundedness of the confidence sets}
    We show the boundedness equivalence simultaneously for the subvector inverse Anderson-Rubin and likelihood-ratio test confidence sets
    Setting $\kappa = \kappa_{\AR}(\alpha) = 1 + F^{-1}_{\chi^2(k - \mW)}(1 - \alpha)$ and $\kappa = \kappa_{\LR}(\alpha) = \hat\kappa_\liml + F^{-1}_{\chi^2(\mX)}(1 - \alpha) = 1 + (n - k )(J_\liml + F^{-1}_{\chi^2(\mX)}(1 - \alpha))$, we need to show that the confidence set is bounded if and only if $(n - k)(\kappa - 1) < \lambda$.
    By \cref{prop:cis_closed_forms}, the confidence sets are bounded if and only if $\kappa \leq \kappa_\mathrm{max}$ and $A(\kappa)$ is positive definite.

    \paragraph*{Step 2a: $(n - k)(\kappa - 1) < \mathrm{CD} \Leftrightarrow S^T (P_Z + (1 - \kappa) M_Z) S$ is positive definite}
    This follows from \cref{lem:kappa_pos_definite} with $X \leftarrow S = \begin{pmatrix} X & W \end{pmatrix}$.

    \paragraph*{Step 2b: $(n - k)(\kappa - 1) < \mathrm{CD}$ implies the confidence set is bounded.}

    By step 2a, $(n - k)(\kappa - 1) < \mathrm{CD} \Rightarrow S^T (P_Z + (1 - \kappa) M_Z) S$ is positive definite.
    By \citep[Proposition 2.1]{gallier2010schur}, the matrix $ S^T (P_Z + (1 - \kappa) M_Z) S$ is positive definite if and only if the supmatrix $W^T (P_Z + (1 - \kappa) M_Z) W$ and the Schur complement $A(\kappa)$ are positive definite.
    Applying \cref{lem:kappa_pos_definite} with $X \leftarrow W$ yields $\kappa < \kappa_\mathrm{max}$.

    \paragraph*{Step 2c: If $\kappa = \kappa_\mathrm{max}$ and \cref{tc:2} \ref{tc:2:a} holds, the confidence set is unbounded.}
    If $\kappa = \kappa_\mathrm{max}$, then $\lambdamin{S^T (P_Z + (1 - \kappa) M_Z) S} \leq \lambdamin{W^T (P_Z + (1 - \kappa) M_Z) W} = 0.$
    By \cref{tc:2} (\ref{tc:2:a}), this inequality is strict and thus $\lambdamin{S^T (P_Z + (1 - \kappa) M_Z) S} < 0$.
    Let $v = \begin{pmatrix} v_X \\ v_W \end{pmatrix}$ be an eigenvector corresponding to $\lambdamin{S^T (P_Z + (1 - \kappa) M_Z) S}$.
    Then $v_X \neq 0$, as else $0 > \lambdamin{S^T (P_Z + (1 - \kappa) M_Z) S} = v^T S^T (P_Z + (1 - \kappa) M_Z) S v / \|v\|^2 = v_W^T W^T (P_Z + (1 - \kappa) M_Z) W v_W / \|v_W \|^2 \leq \lambdamin{W^T (P_Z + (1 - \kappa) M_Z) W} = 0$.
    Write
    \begin{align*}
        \CI_{\AR}(1 - \alpha) &= \{ \beta\in\BR^\mX \mid \min_\gamma \begin{pmatrix} \beta \\ \gamma \end{pmatrix}^T S^T (P_Z + (1 - \kappa) M_Z) S \begin{pmatrix} \beta \\ \gamma \end{pmatrix} \leq \mathrm{const} \} + B \hat\beta_\kclass(\kappa) \\
        &\supseteq \{ t \cdot v_X \mid t^2 v^T S^T (P_Z + (1 - \kappa) M_Z) S v \leq \mathrm{const} \} + B \hat\beta_\kclass(\kappa).
    \end{align*}
    See also the proof of \cref{prop:cis_closed_forms}.
    The latter is equal to $\BR v_X$ if $\mathrm{const} \geq 0$ and $\left( (-\infty, t_\mathrm{min}] \cup [t_\mathrm{min}, \infty) \right) \cdot v_X + B \hat\beta_\kclass(\kappa)$ for $t_\mathrm{min} = \frac{1}{\|v\|} \sqrt{\frac{\mathrm{const}}{\lambdamin{S^T (P_Z + (1 - \kappa) M_Z) S}}}$ otherwise.
    In both cases, the confidence set is unbounded.

    \paragraph*{Step 2d: A bounded confidence set implies $(n - k)(\kappa - 1) < \mathrm{CD}$}
    If the confidence set is bounded, then $\kappa \leq \kappa_\mathrm{max}$ and $A(\kappa)$ is positive definite.
    By step 2c, the $\kappa < \kappa_\mathrm{max}$.
    Then by \cref{lem:kappa_pos_definite}, the supmatrix $W^T (P_Z + (1 - \kappa) M_Z) W$ is positive definite and by the if direction of \citep[Proposition 2.1]{gallier2010schur},  $S^T (P_Z + (1 - \kappa) M_Z) S$ is positive definite.
    By step 2a, $(n - k)(\kappa - 1) < \mathrm{CD}$.

\end{proofEnd}%

\begin{figure}[htbp]
    \centering
    \includegraphics[width=0.9\textwidth]{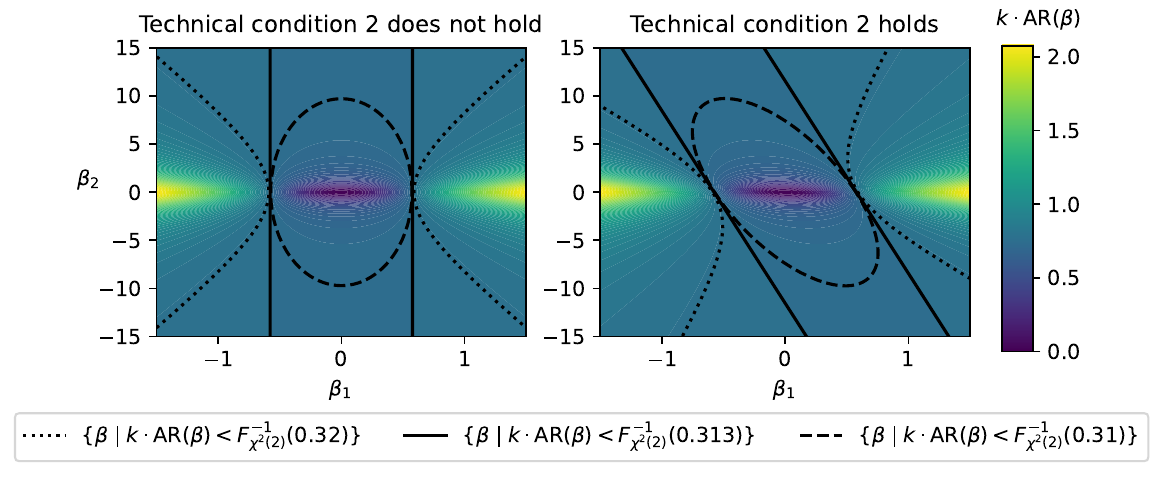}
    \caption{
        \label{fig:figure_tc3_counterexample}
        On the left $Z, X, y$ are as in the counterexample to \cref{tc:liml_theorem} below \cref{cor:liml_tsls_if_identified}, but with $y = (0, 0, 0, 0, 0, 1)^T$.
        On the right, $X \leftarrow X \begin{pmatrix} 1 & 0.05 \\ 0.05 & 1 \end{pmatrix}$.
        The confidence sets $\CI_{\AR}(1 - \alpha)$ for parameters $\beta_1$ and $\beta_2$ individually result from projecting the quadrics onto the respective axes.
        If $1 - \alpha = 0.32$, in both settings the confidence sets for both coefficients are unbounded.
        If $1 - \alpha = 0.31$, in both settings the confidence sets for both coefficients are bounded.
        On the left, if $1 - \alpha = 1 - F_{\chi^2(2)}^{-1}(0.25) \approx 0.313$,  the confidence set for $\beta_1$ is bounded, while the confidence set for $\beta_2$ is unbounded.
        This occurs as the principal axis of the quadric is aligned with the $\beta_2$ axis and \cref{tc:2} does not apply.
        On the right \cref{tc:2} applies and the confidence sets for both coefficients are unbounded.
    }
\end{figure}

A quick note about \cref{tc:2}.
This is only needed for the boundary point ${\mathrm{CD} = F^{-1}_{\chi^2(k - \mW)}(1 - \alpha)}$ for the inverse Anderson-Rubin test or ${\mathrm{CD} - J_\liml = F^{-1}_{\chi^2(\mX)}(1 - \alpha)}$ for the inverse likelihood-ratio test.
See \cref{fig:figure_tc3_counterexample} for a counterexample.
If the noise in \cref{model:1} is absolutely continuous with respect to the Lebesgue measure, \cref{tc:2} holds with probability one and hence, it will be satisfied in practice.
The condition is similar to \cref{tc:liml_theorem}, which is necessary for \cref{thm:liml_is_kclass} and implicitly assumed in the literature.
See also the counterexample below \cref{tc:liml_theorem}.

The condition for boundedness of the inverse Anderson-Rubin confidence set coincides with the condition for \cref{prop:k_class_well_defined}.
That is, the (subvector) inverse Anderson-Rubin test confidence set is an ellipsoid if and only if the k-class estimator $\hat\beta_\kclass(\kappa_{\AR}(\alpha))$ in the center minimizes
$$
    \beta \mapsto \kappa_{\AR}(\alpha) \| P_Z (y - X \beta) \|^2 + (1 - \kappa_{\AR}(\alpha)) \| y - X \beta \|^2.
$$

Notably, the conditions for (non)emptyness and (un)boundedness of the confidence sets depend on $X$ and $W$ only through $S = \begin{pmatrix} X & W \end{pmatrix}$.
That is, the conditions are invariant to the choice of how we split $S = \begin{pmatrix} X & W \end{pmatrix}$ and thus are the same if we look at different components of the causal parameter.

In practice, one typically starts with endogenous covariates $S$ and then, for each covariate $i = 1, \ldots, m$, separates $X \leftarrow S^{(i)}$ and $W \leftarrow S^{(-i)}$, yielding $p$-values and confidence sets (or confidence intervals) for each coefficient $\beta_0^{(i)}$ of $\beta_0$ separately.
The following corollary of \cref{prop:cis_bounded} gives a condition for these subvector inverse Anderson-Rubin test confidence sets to be (jointly) bounded and nonempty (and thus, confidence intervals).
\begin{theoremEnd}[malte,category=confsets]{corollary}[\citeauthor{londschien2024weak}, \citeyear{londschien2024weak}]
    \label{cor:ar_bounded_iff_rank_test_rejects}
    Consider structural equations $y_i = S_i^T \beta_0 + \varepsilon_i$ and $S_i = Z_i^T \Pi + V_i$.
    We are interested in inference for each component of the causal parameter $\beta$.
    Thus, for each covariate $i = 1, \ldots, m$, we separate $X \leftarrow S^{(i)}$ and $W \leftarrow S^{(-i)}$ and construct confidence intervals $\CI_{\AR}^{(i)}(1 - \alpha)$ and $\CI_{\LR}^{(i)}(1 - \alpha)$ for the $i$-th component of $\beta_0$.
    Then
    \begin{enumerate}[(a)]
        \item The subvector inverse Anderson-Rubin test confidence sets are jointly (un)bounded. That is, if $\CI_{\AR}^{(i)}(1 - \alpha)$ is (un)bounded for any $i$, then $\CI_{\AR}^{(i)}(1 - \alpha)$ is (un)bounded for all $i$.
        \item The subvector inverse Anderson-Rubin test confidence sets at level $\alpha$ are bounded (and thus confidence intervals) if and only if the Cragg-Donald test rejects the null hypothesis that $\Pi$ is of reduced rank at level $\alpha$. \label{cor:ar_bounded_iff_rank_test_rejects:b}
        \item The subvector inverse Anderson-Rubin test confidence sets are jointly (non)empty. That is, if $\CI_{\AR}^{(i)}(1 - \alpha)$ is (non)empty for any $i$, then $\CI_{\AR}^{(i)}(1 - \alpha)$ is (non)empty for all $i$.
        \item The subvector inverse Anderson-Rubin test confidence sets at level $\alpha$ are jointly empty if and only if the LIML variant of the J-statistic $J_\liml > F^{-1}_{\chi^2(k - m + 1)}(1 - \alpha)$. \label{cor:ar_bounded_iff_rank_test_rejects:d}
        \item The subvector inverse likelihood-ratio test confidence sets are jointly (un)bounded. That is, if $\CI_{\LR}^{(i)}(1 - \alpha)$ is (un)bounded for any $i$, then $\CI_{\LR}^{(i)}(1 - \alpha)$ is (un)bounded for all $i$.
    \end{enumerate}
\end{theoremEnd}%
\begin{proofEnd}%
    For each $i=1,\ldots,m$, the conditions for the subvector inverse Anderson-Rubin or likelihood-ratio test confidence sets to be bounded and nonempty follow from \cref{prop:cis_bounded}.
    That the confidence sets are jointly (un)bounded or (non)empty follows from the fact that these conditions do not depend on how $S = \begin{pmatrix} S^{(i)} & S^{(-i)} \end{pmatrix}$ is partitioned.
    \Cref{cor:ar_bounded_iff_rank_test_rejects:d} follows from \cref{cor:anderson_1951_rank_test} with $r=1$.
\end{proofEnd}%

\begin{figure}[!ht]
    \centering
    \includegraphics[width=\textwidth]{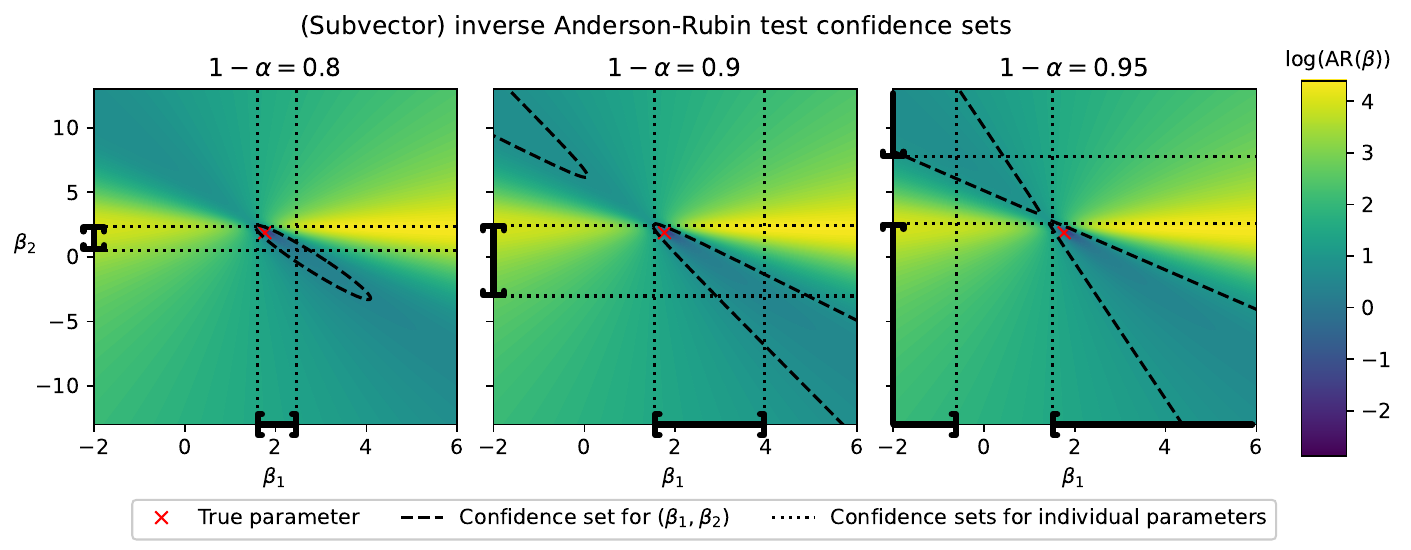}
    \caption{
        \label{fig:inverse_ar_different_alpha}
        The data was sampled from a Gaussian causal model with $n=100, k=3$, and $\mX = \mW = 1$.
        The $p$-value of the Cragg-Donald test of reduced rank is $0.07$.
        The confidence sets for $\beta = (\beta_1, \beta_2)$ are bounded for $1 - \alpha = 0.8$ and unbounded for $1 - \alpha = 0.9, 0.95$.
        The subvector confidence sets for the individual coefficients are jointly bounded for $1 - \alpha = 0.8, 0.9$ and unbounded for $1 - \alpha = 0.95$.
        Notably, for $1 - \alpha = 0.9$, the confidence set for $\beta = (\beta_1, \beta_2)$ is unbounded, while the subvector confidence sets for the individual coefficients are bounded.
    }
\end{figure}

\Cref{cor:ar_bounded_iff_rank_test_rejects} \ref{cor:ar_bounded_iff_rank_test_rejects:b} only holds if $\mX = 1$, that is, if we are testing a single component of the causal parameter.
Otherwise, the condition for the boundedness and the Cragg-Donald test use different critical values.
See also \cref{fig:inverse_ar_different_alpha} (middle panel), where the 2-dimensional 90\% confidence set for the causal parameter is unbounded, even though the $p$-value of the Cragg-Donald test is 0.07.
The 1-dimensional 90\% confidence sets are bounded.

On the other hand, the critical values $F^{-1}_{\chi^2(k-m + 1)}(1 - \alpha)$ in \cref{cor:ar_bounded_iff_rank_test_rejects} \ref{cor:ar_bounded_iff_rank_test_rejects:d} do not match those of the LIML variant of the J-statistic  $F^{-1}_{\chi^2(k-m)}(1 - \alpha)$ (see \cref{prop:j_liml_statistic_chi_squared} in \cref{sec:j_statistic}).
In particular, rejection of the LIML J-test does not imply that the inverse Anderson-Rubin confidence sets are empty.

Note that the Wald, likelihood-ratio, and Anderson-Rubin confidence sets are all centered around a k-class estimator (\cref{prop:cis_closed_forms}).
They differ only in the $\kappa$ parameter of $\hat\beta_\kclass(\kappa)$ and $A(\kappa)$ and the critical value.
This yields the following equality of confidence sets.
\begin{theoremEnd}[malte,category=confsets]{proposition}[\citeauthor{londschien2024weak}, \citeyear{londschien2024weak}]
    \label{prop:inverse_ar_equal_to_wald}
    Let $\alpha > 0$.
    Assume that $J_\liml \leq F^{-1}_{\chi^2(k-\mW)}(1 - \alpha) < \mathrm{CD}$.
    Let
    \begin{multline*}        
    s(\kappa) := 
    y^T (\Id_n - \kappa M_Z ) y %
    - y^T (\Id_n - \kappa M_Z ) X [X^T (\Id_n - \kappa M_Z ) X]^{-1} X^T (\Id_n - \kappa M_Z ) y.
    \end{multline*}
    Recall that $\kappa_{\AR}(\alpha) = 1 + F^{-1}_{\chi^2(k - \mW)}(1 - \alpha) / (n-k)$ and define
    \begin{align*}
    \alpha_{\Wald \mid \AR}(\alpha) &:= 1 - F_{\chi^2(\mX)}(-s(\kappa_{\AR}(\alpha)) / \hat\sigma^2_{\Wald}(\kappa_{\AR}(\alpha))) \text{ and }
    \\\alpha_{\LR \mid \AR}(\alpha) &:= 1 - F^{-1}_{\chi^2(\mX)}( F^{-1}_{\chi^2(k - \mW)}(1 - \alpha) - J_\liml),
    \end{align*}
    where $\hat\sigma^2_{\Wald}(\kappa) = \frac{1}{n-m}\| y - X \hat\beta_\kclass(\kappa) \|^2$. Then
    $$
    \CI_{\AR}(1 - \alpha) = \CI_{\LR}(1 - \alpha_{\LR \mid \AR}(\alpha)) = \CI_{\Wald_{\hat\beta_\kclass(\kappa_{\AR}(\alpha))}}(1 - \alpha_{\Wald \mid \AR}(\alpha)).
    $$
        
\end{theoremEnd}%
\begin{proofEnd}%
    We apply \cref{prop:cis_closed_forms}. Write $s := s(\kappa_{\AR}(\alpha))$.
    \paragraph*{Step 1: $J_\liml \leq F^{-1}_{\chi^2(k - \mW)}(1 - \alpha) < \lambda$ implies $s \leq 0$.}

    From 
    \begin{multline*}
    J_\liml \leq F^{-1}_{\chi^2(k - \mW)}(1 - \alpha) < \lambda \\
    \Leftrightarrow \lambdamin{ (S^T M_Z S)^{-1} S^T P_Z S } \leq \kappa_{\AR}(\alpha) - 1 < \lambdamin{ (\begin{pmatrix} S & y \end{pmatrix}^T M_Z \begin{pmatrix} S & y \end{pmatrix})^{-1} \begin{pmatrix} S & y \end{pmatrix}^T P_Z \begin{pmatrix} S & y \end{pmatrix} } \\
    \overset{\Cref{lem:kappa_pos_definite}}{\Leftrightarrow}
    \lambdamin{ S^T (P_Z - \kappa_{\AR}(\alpha) M_Z) S } \geq 0 > \lambdamin{ \begin{pmatrix} S & y \end{pmatrix}^T (P_Z - \kappa_{\AR}(\alpha) M_Z) \begin{pmatrix} S & y \end{pmatrix} }.
    \end{multline*}
    The eigenvalues of $\begin{pmatrix} S & y \end{pmatrix}^T (P_Z - \kappa_{\AR}(\alpha) M_Z) \begin{pmatrix} S & y \end{pmatrix}$ and $ S^T (P_Z - \kappa_{\AR}(\alpha) M_Z) S $ interleave and thus
    \begin{multline*}    
    \lambda_2 \left(\begin{pmatrix} S & y \end{pmatrix}^T (P_Z - \kappa_{\AR}(\alpha) M_Z) \begin{pmatrix} S & y \end{pmatrix} \right) \geq \lambdamin{ S^T (P_Z - \kappa_{\AR}(\alpha) M_Z) S } \geq 0 \\
    \Rightarrow \det( \begin{pmatrix} S & y \end{pmatrix}^T (P_Z - \kappa_{\AR}(\alpha) M_Z) \begin{pmatrix} S & y \end{pmatrix} ) \leq 0.
    \end{multline*}
    Applying the formula for the determinant of a block matrix, we get
    \begin{multline*}
        \det( S^T (P_Z - \kappa_{\AR}(\alpha) M_Z) S ) \cdot s = \det( \begin{pmatrix} S & y \end{pmatrix}^T (P_Z - \kappa_{\AR}(\alpha) M_Z) \begin{pmatrix} S & y \end{pmatrix} ) \Rightarrow s \leq 0.
    \end{multline*}

    \paragraph*{Step 2: $\CI_{\AR}(1 - \alpha) = \CI_{\Wald_{\hat\beta_\kclass(\kappa_{\AR}(\alpha))}}(1 - \alpha_{\Wald \mid \AR}(\alpha))$}

    We calculate
    \begin{multline*}
    \sigma_{\AR}^2(\kappa_{\AR}(\alpha)) \cdot ( F^{-1}_{\chi^2(k - \mW)}(1 - \alpha) - k \AR(\hat\beta_\kclass(\kappa_{\AR}(\alpha)))) \\
    = (\kappa_{\AR}(\alpha) - 1) \cdot \| M_Z (y - S \hat\beta_\kclass(\kappa_{\AR}(\alpha)) ) \|^2 - \| P_Z (y - S \hat\beta_\kclass(\kappa_{\AR}(\alpha)) ) \|^2 \\
    = \left( y^T - y^T \left(P_Z - \kappa_{\AR}(\alpha) M_Z \right) S \left( S^T \left( P_Z - \kappa_{\AR}(\alpha) M_Z \right) S \right)^{-1} S^T \right) \left(P_Z - \kappa_{\AR}(\alpha) M_Z \right) \\
    \left( y - S \left( S^T \left( P_Z - \kappa_{\AR}(\alpha) M_Z \right) S \right)^{-1} S^T \left( P_Z - \kappa_{\AR}(\alpha) M_Z \right) y \right)
    = - s.
    \end{multline*}
    
    Clearly $\hat\sigma^2_{\Wald}(\kappa_{\AR}(\alpha)) \geq 0$ and thus 
    \begin{multline*}        
    \hat\sigma^2_{\Wald}(\kappa_{\AR}(\alpha)) \cdot F^{-1}_{\chi^2(\mX)}(1 - \alpha_{\Wald \mid \AR}(\alpha)) = - s \\
    = \sigma_{\AR}^2(\kappa_{\AR}(\alpha)) \cdot ( F^{-1}_{\chi^2(k - \mW)}(1 - \alpha) - k \AR(\hat\beta_\kclass(\kappa_{\AR}(\alpha)))).
    \end{multline*}

    \paragraph*{Step 3: $\CI_{\AR}(1 - \alpha) = \CI_{\LR}(1 - \alpha_{\LR \mid \AR}(\alpha))$}

    From \cref{prop:cis_closed_forms}, the two confidence sets are equal if
    $$
    \kappa_{\AR}(\alpha) = \kappa_{\LR}(\alpha_{\LR \mid \AR}(\alpha)) \Leftrightarrow F^{-1}_{\chi^2(k - \mW)}(1 - \alpha) = J_\liml + F^{-1}_{\chi^2(\mX)}(1 - \alpha_{\LR \mid \AR}(\alpha)).
    $$
    By assumption $J_\liml \leq F^{-1}_{\chi^2(k - \mW)}(1 - \alpha)$, so we can solve for $\alpha_{\LR \mid \AR}(\alpha)$.
\end{proofEnd}%

\subsection{Application}
We compute 95\% confidence sets for the causal effect of education on log wages.
\begin{jupyternotebook}
\begin{tcolorbox}[breakable, size=fbox, boxrule=1pt, pad at break*=1mm,colback=cellbackground, colframe=cellborder]
\prompt{In}{incolor}{13}{\boxspacing}
\begin{Verbatim}[commandchars=\\\{\}]
\PY{k+kn}{from} \PY{n+nn}{ivmodels}\PY{n+nn}{.}\PY{n+nn}{tests} \PY{k+kn}{import} \PY{p}{(}
    \PY{n}{inverse\PYZus{}anderson\PYZus{}rubin\PYZus{}test}\PY{p}{,}
    \PY{n}{inverse\PYZus{}wald\PYZus{}test}\PY{p}{,}
    \PY{n}{inverse\PYZus{}likelihood\PYZus{}ratio\PYZus{}test}\PY{p}{,}
    \PY{n}{inverse\PYZus{}conditional\PYZus{}likelihood\PYZus{}ratio\PYZus{}test}\PY{p}{,}
    \PY{n}{inverse\PYZus{}lagrange\PYZus{}multiplier\PYZus{}test}
\PY{p}{)}

\PY{k}{for} \PY{n}{name}\PY{p}{,} \PY{n}{inverse\PYZus{}test} \PY{o+ow}{in} \PY{p}{[}
    \PY{p}{(}\PY{l+s+s2}{\PYZdq{}}\PY{l+s+s2}{Wald (TSLS)}\PY{l+s+s2}{\PYZdq{}}\PY{p}{,} \PY{n}{partial}\PY{p}{(}\PY{n}{inverse\PYZus{}wald\PYZus{}test}\PY{p}{,} \PY{n}{estimator}\PY{o}{=}\PY{l+s+s2}{\PYZdq{}}\PY{l+s+s2}{tsls}\PY{l+s+s2}{\PYZdq{}}\PY{p}{)}\PY{p}{)}\PY{p}{,}
    \PY{p}{(}\PY{l+s+s2}{\PYZdq{}}\PY{l+s+s2}{Wald (LIML)}\PY{l+s+s2}{\PYZdq{}}\PY{p}{,} \PY{n}{partial}\PY{p}{(}\PY{n}{inverse\PYZus{}wald\PYZus{}test}\PY{p}{,} \PY{n}{estimator}\PY{o}{=}\PY{l+s+s2}{\PYZdq{}}\PY{l+s+s2}{liml}\PY{l+s+s2}{\PYZdq{}}\PY{p}{)}\PY{p}{)}\PY{p}{,}
    \PY{p}{(}\PY{l+s+s2}{\PYZdq{}}\PY{l+s+s2}{AR}\PY{l+s+s2}{\PYZdq{}}\PY{p}{,} \PY{n}{inverse\PYZus{}anderson\PYZus{}rubin\PYZus{}test}\PY{p}{)}\PY{p}{,}
    \PY{p}{(}\PY{l+s+s2}{\PYZdq{}}\PY{l+s+s2}{LR}\PY{l+s+s2}{\PYZdq{}}\PY{p}{,} \PY{n}{inverse\PYZus{}likelihood\PYZus{}ratio\PYZus{}test}\PY{p}{)}\PY{p}{,}
    \PY{p}{(}\PY{l+s+s2}{\PYZdq{}}\PY{l+s+s2}{CLR}\PY{l+s+s2}{\PYZdq{}}\PY{p}{,} \PY{n}{inverse\PYZus{}conditional\PYZus{}likelihood\PYZus{}ratio\PYZus{}test}\PY{p}{)}\PY{p}{,}
    \PY{p}{(}\PY{l+s+s2}{\PYZdq{}}\PY{l+s+s2}{LM}\PY{l+s+s2}{\PYZdq{}}\PY{p}{,} \PY{n}{inverse\PYZus{}lagrange\PYZus{}multiplier\PYZus{}test}\PY{p}{)}\PY{p}{,}
\PY{p}{]}\PY{p}{:}
    \PY{n+nb}{print}\PY{p}{(}\PY{l+s+sa}{f}\PY{l+s+s2}{\PYZdq{}}\PY{l+s+si}{\PYZob{}}\PY{n}{name}\PY{l+s+si}{:}\PY{l+s+s2}{\PYZlt{}11}\PY{l+s+si}{\PYZcb{}}\PY{l+s+s2}{: }\PY{l+s+si}{\PYZob{}}\PY{n}{inverse\PYZus{}test}\PY{p}{(}\PY{n}{Z}\PY{o}{=}\PY{n}{Z}\PY{p}{,}\PY{+w}{ }\PY{n}{X}\PY{o}{=}\PY{n}{X}\PY{p}{,}\PY{+w}{ }\PY{n}{W}\PY{o}{=}\PY{n}{W}\PY{p}{,}\PY{+w}{ }\PY{n}{y}\PY{o}{=}\PY{n}{y}\PY{p}{,}\PY{+w}{ }\PY{n}{alpha}\PY{o}{=}\PY{l+m+mf}{0.05}\PY{p}{)}\PY{l+s+si}{:}\PY{l+s+s2}{.3f}\PY{l+s+si}{\PYZcb{}}\PY{l+s+s2}{\PYZdq{}}\PY{p}{)}
\end{Verbatim}
\end{tcolorbox}
\begin{Verbatim}[commandchars=\\\{\}]
Wald (TSLS): [0.058, 0.232]
Wald (LIML): [0.063, 0.282]
AR         : [0.083, 0.352]
LR         : [0.079, 0.368]
CLR        : [0.073, 0.396]
LM         : [-0.594, -0.059] U [0.061, 0.467]
\end{Verbatim}
\end{jupyternotebook}
The confidence sets are of varying size.
The inverse Anderson-Rubin test confidence set is the smallest among the weak-instrument-robust tests.
This occurs as \(J_\mathrm{LIML} = 4.284\) is substantial and \[
(k - m_w) \cdot \mathrm{AR}(\beta) = \mathrm{LR}(\beta) + J_\mathrm{LIML}
\] such that ``a little misspecification improves power'' as discussed
in \cref{sec:intro_to_iv_tests:tests}.

We observe a common phenomenon for the Lagrange multiplier test:
The confidence set obtained by inversion is disjoint.
This is due to the test being a function of the score, the derivative of the likelihood, such that the statistic is zero both at the minimum but also at the maximum of the likelihood.
The Lagrange multiplier test still has its uses, and might result in smaller confidence sets if one can discard one part of the confidence set a-priori, e.g., as prior knowledge dictates that \(\beta_0 > 0\).

We now numerically verify \cref{prop:cis_bounded}.
For \(\alpha_\mathrm{max} = 1 - F_{\chi^2(k - m_w)}(J_\mathrm{LIML})\), we expect the inverse Anderson-Rubin test confidence set to be just non-empty.
For any \(\alpha > \alpha_\mathrm{max}\), we expect the confidence set to be empty.
\begin{jupyternotebook}
    \begin{tcolorbox}[breakable, size=fbox, boxrule=1pt, pad at break*=1mm,colback=cellbackground, colframe=cellborder]
\prompt{In}{incolor}{14}{\boxspacing}
\begin{Verbatim}[commandchars=\\\{\}]
\PY{k+kn}{import} \PY{n+nn}{scipy}\PY{n+nn}{.}\PY{n+nn}{stats}

\PY{n}{alpha\PYZus{}max} \PY{o}{=} \PY{l+m+mi}{1} \PY{o}{\PYZhy{}} \PY{n}{scipy}\PY{o}{.}\PY{n}{stats}\PY{o}{.}\PY{n}{chi2}\PY{p}{(}\PY{l+m+mi}{5} \PY{o}{\PYZhy{}} \PY{l+m+mi}{2}\PY{p}{)}\PY{o}{.}\PY{n}{cdf}\PY{p}{(}\PY{n}{j\PYZus{}stat}\PY{p}{)}  \PY{c+c1}{\PYZsh{} not equal to j\PYZus{}pval}

\PY{n}{cs} \PY{o}{=} \PY{n}{inverse\PYZus{}anderson\PYZus{}rubin\PYZus{}test}\PY{p}{(}\PY{n}{Z}\PY{o}{=}\PY{n}{Z}\PY{p}{,} \PY{n}{X}\PY{o}{=}\PY{n}{X}\PY{p}{,} \PY{n}{y}\PY{o}{=}\PY{n}{y}\PY{p}{,} \PY{n}{W}\PY{o}{=}\PY{n}{W}\PY{p}{,} \PY{n}{alpha}\PY{o}{=}\PY{n}{alpha\PYZus{}max} \PY{o}{\PYZhy{}} \PY{l+m+mf}{1e\PYZhy{}6}\PY{p}{)}
\PY{n+nb}{print}\PY{p}{(}\PY{l+s+sa}{f}\PY{l+s+s2}{\PYZdq{}}\PY{l+s+s2}{Inverse AR (1\PYZhy{}alpha=}\PY{l+s+si}{\PYZob{}}\PY{l+m+mi}{1}\PY{+w}{ }\PY{o}{\PYZhy{}}\PY{+w}{ }\PY{n}{alpha\PYZus{}max}\PY{+w}{ }\PY{o}{\PYZhy{}}\PY{+w}{ }\PY{l+m+mf}{1e\PYZhy{}6}\PY{l+s+si}{:}\PY{l+s+s2}{.6f}\PY{l+s+si}{\PYZcb{}}\PY{l+s+s2}{): }\PY{l+s+si}{\PYZob{}}\PY{n}{cs}\PY{l+s+si}{:}\PY{l+s+s2}{.3f}\PY{l+s+si}{\PYZcb{}}\PY{l+s+s2}{\PYZdq{}}\PY{p}{)}
\PY{n}{cs} \PY{o}{=} \PY{n}{inverse\PYZus{}anderson\PYZus{}rubin\PYZus{}test}\PY{p}{(}\PY{n}{Z}\PY{o}{=}\PY{n}{Z}\PY{p}{,} \PY{n}{X}\PY{o}{=}\PY{n}{X}\PY{p}{,} \PY{n}{y}\PY{o}{=}\PY{n}{y}\PY{p}{,} \PY{n}{W}\PY{o}{=}\PY{n}{W}\PY{p}{,} \PY{n}{alpha}\PY{o}{=}\PY{n}{alpha\PYZus{}max} \PY{o}{+} \PY{l+m+mf}{1e\PYZhy{}6}\PY{p}{)}
\PY{n+nb}{print}\PY{p}{(}\PY{l+s+sa}{f}\PY{l+s+s2}{\PYZdq{}}\PY{l+s+s2}{Inverse AR (1\PYZhy{}alpha=}\PY{l+s+si}{\PYZob{}}\PY{l+m+mi}{1}\PY{+w}{ }\PY{o}{\PYZhy{}}\PY{+w}{ }\PY{n}{alpha\PYZus{}max}\PY{+w}{ }\PY{o}{+}\PY{+w}{ }\PY{l+m+mf}{1e\PYZhy{}6}\PY{l+s+si}{:}\PY{l+s+s2}{.6f}\PY{l+s+si}{\PYZcb{}}\PY{l+s+s2}{): }\PY{l+s+si}{\PYZob{}}\PY{n}{cs}\PY{l+s+si}{:}\PY{l+s+s2}{.3f}\PY{l+s+si}{\PYZcb{}}\PY{l+s+s2}{\PYZdq{}}\PY{p}{)}
\end{Verbatim}
\end{tcolorbox}
    \begin{Verbatim}[commandchars=\\\{\}]
Inverse AR (1-alpha=0.767640): [0.172, 0.173]
Inverse AR (1-alpha=0.767642): \ensuremath{\emptyset}
    \end{Verbatim}
\end{jupyternotebook}
For \(\alpha_\mathrm{min} = 1 - F_{\chi^2(k-{m_W})}^{-1}(\mathrm{CD})\), we
expect the inverse Anderson-Rubin test confidence set to be just
bounded. For \(\alpha > \alpha_\mathrm{min}\), the confidence set should
be unbounded.
\begin{jupyternotebook}
    \begin{tcolorbox}[breakable, size=fbox, boxrule=1pt, pad at break*=1mm,colback=cellbackground, colframe=cellborder]
\prompt{In}{incolor}{15}{\boxspacing}
\begin{Verbatim}[commandchars=\\\{\}]
\PY{n}{alpha\PYZus{}min} \PY{o}{=} \PY{n}{rank\PYZus{}pval}

\PY{c+c1}{\PYZsh{} The confidence set explodes at exactly alpha\PYZus{}min due to numerical instabilities}
\PY{n}{cs} \PY{o}{=} \PY{n}{inverse\PYZus{}anderson\PYZus{}rubin\PYZus{}test}\PY{p}{(}\PY{n}{Z}\PY{o}{=}\PY{n}{Z}\PY{p}{,} \PY{n}{X}\PY{o}{=}\PY{n}{X}\PY{p}{,} \PY{n}{y}\PY{o}{=}\PY{n}{y}\PY{p}{,} \PY{n}{W}\PY{o}{=}\PY{n}{W}\PY{p}{,} \PY{n}{alpha}\PY{o}{=}\PY{n}{alpha\PYZus{}min} \PY{o}{\PYZhy{}} \PY{l+m+mf}{1e\PYZhy{}6}\PY{p}{)}
\PY{n+nb}{print}\PY{p}{(}\PY{l+s+sa}{f}\PY{l+s+s2}{\PYZdq{}}\PY{l+s+s2}{Inverse AR (1\PYZhy{}alpha=}\PY{l+s+si}{\PYZob{}}\PY{l+m+mi}{1}\PY{+w}{ }\PY{o}{\PYZhy{}}\PY{+w}{ }\PY{n}{alpha\PYZus{}min}\PY{+w}{ }\PY{o}{\PYZhy{}}\PY{+w}{ }\PY{l+m+mf}{1e\PYZhy{}6}\PY{l+s+si}{:}\PY{l+s+s2}{.6f}\PY{l+s+si}{\PYZcb{}}\PY{l+s+s2}{): }\PY{l+s+si}{\PYZob{}}\PY{n}{cs}\PY{l+s+si}{:}\PY{l+s+s2}{.3f}\PY{l+s+si}{\PYZcb{}}\PY{l+s+s2}{\PYZdq{}}\PY{p}{)}
\PY{n}{cs} \PY{o}{=} \PY{n}{inverse\PYZus{}anderson\PYZus{}rubin\PYZus{}test}\PY{p}{(}\PY{n}{Z}\PY{o}{=}\PY{n}{Z}\PY{p}{,} \PY{n}{X}\PY{o}{=}\PY{n}{X}\PY{p}{,} \PY{n}{y}\PY{o}{=}\PY{n}{y}\PY{p}{,} \PY{n}{W}\PY{o}{=}\PY{n}{W}\PY{p}{,} \PY{n}{alpha}\PY{o}{=}\PY{n}{alpha\PYZus{}min} \PY{o}{+} \PY{l+m+mf}{1e\PYZhy{}6}\PY{p}{)}
\PY{n+nb}{print}\PY{p}{(}\PY{l+s+sa}{f}\PY{l+s+s2}{\PYZdq{}}\PY{l+s+s2}{Inverse AR (1\PYZhy{}alpha=}\PY{l+s+si}{\PYZob{}}\PY{l+m+mi}{1}\PY{+w}{ }\PY{o}{\PYZhy{}}\PY{+w}{ }\PY{n}{alpha\PYZus{}min}\PY{+w}{ }\PY{o}{+}\PY{+w}{ }\PY{l+m+mf}{1e\PYZhy{}6}\PY{l+s+si}{:}\PY{l+s+s2}{.6f}\PY{l+s+si}{\PYZcb{}}\PY{l+s+s2}{): }\PY{l+s+si}{\PYZob{}}\PY{n}{cs}\PY{l+s+si}{:}\PY{l+s+s2}{.3f}\PY{l+s+si}{\PYZcb{}}\PY{l+s+s2}{\PYZdq{}}\PY{p}{)}
\end{Verbatim}
\end{tcolorbox}

    \begin{Verbatim}[commandchars=\\\{\}]
Inverse AR (1-alpha=0.998638): [-inf, -1451.003] U [-0.005, inf]
Inverse AR (1-alpha=0.998640): [-0.005, 1452.363]
\end{Verbatim}
\end{jupyternotebook}
The confidence sets obtained using \(\chi^2(k - m_W)\) critical values
are more powerful than those proposed by \citet{dufour2005projection}. For
\(\alpha=0.005\), the former are bounded and do not include 0, whereas the
latter are unbounded.
\begin{jupyternotebook}
    \begin{tcolorbox}[breakable, size=fbox, boxrule=1pt, pad at break*=1mm,colback=cellbackground, colframe=cellborder]
\prompt{In}{incolor}{16}{\boxspacing}
\begin{Verbatim}[commandchars=\\\{\}]
\PY{n}{cs} \PY{o}{=} \PY{n}{inverse\PYZus{}anderson\PYZus{}rubin\PYZus{}test}\PY{p}{(}\PY{n}{Z}\PY{o}{=}\PY{n}{Z}\PY{p}{,} \PY{n}{X}\PY{o}{=}\PY{n}{X}\PY{p}{,} \PY{n}{y}\PY{o}{=}\PY{n}{y}\PY{p}{,} \PY{n}{W}\PY{o}{=}\PY{n}{W}\PY{p}{,} \PY{n}{alpha}\PY{o}{=}\PY{l+m+mf}{0.005}\PY{p}{)}
\PY{n+nb}{print}\PY{p}{(}\PY{l+s+sa}{f}\PY{l+s+s2}{\PYZdq{}}\PY{l+s+s2}{Inverse AR (1\PYZhy{}alpha=}\PY{l+s+si}{\PYZob{}}\PY{l+m+mi}{1}\PY{+w}{ }\PY{o}{\PYZhy{}}\PY{+w}{ }\PY{l+m+mf}{0.005}\PY{l+s+si}{\PYZcb{}}\PY{l+s+s2}{): }\PY{l+s+si}{\PYZob{}}\PY{n}{cs}\PY{l+s+si}{:}\PY{l+s+s2}{.3f}\PY{l+s+si}{\PYZcb{}}\PY{l+s+s2}{\PYZdq{}}\PY{p}{)}

\PY{n}{cs} \PY{o}{=} \PY{n}{inverse\PYZus{}anderson\PYZus{}rubin\PYZus{}test}\PY{p}{(}\PY{n}{Z}\PY{o}{=}\PY{n}{Z}\PY{p}{,} \PY{n}{X}\PY{o}{=}\PY{n}{XW}\PY{p}{,} \PY{n}{y}\PY{o}{=}\PY{n}{y}\PY{p}{,} \PY{n}{alpha}\PY{o}{=}\PY{l+m+mf}{0.005}\PY{p}{)}\PY{o}{.}\PY{n}{project}\PY{p}{(}\PY{p}{[}\PY{l+m+mi}{0}\PY{p}{]}\PY{p}{)}
\PY{n+nb}{print}\PY{p}{(}\PY{l+s+sa}{f}\PY{l+s+s2}{\PYZdq{}}\PY{l+s+s2}{Inverse AR (1\PYZhy{}alpha=}\PY{l+s+si}{\PYZob{}}\PY{l+m+mi}{1}\PY{+w}{ }\PY{o}{\PYZhy{}}\PY{+w}{ }\PY{l+m+mf}{0.005}\PY{l+s+si}{\PYZcb{}}\PY{l+s+s2}{) by projection: }\PY{l+s+si}{\PYZob{}}\PY{n}{cs}\PY{l+s+si}{:}\PY{l+s+s2}{.3f}\PY{l+s+si}{\PYZcb{}}\PY{l+s+s2}{\PYZdq{}}\PY{p}{)}
\end{Verbatim}
\end{tcolorbox}

    \begin{Verbatim}[commandchars=\\\{\}]
Inverse AR (1-alpha=0.995): [0.028, 0.932]
Inverse AR (1-alpha=0.995) by projection: [-inf, -1.822] U [-0.023, inf]
    \end{Verbatim}
\end{jupyternotebook}

\section{Included exogenous variables}
\label{sec:exogenous_variables}
In practice, one often has additional exogenous variables (controls) $C$ entering the model.
For ease of exposition, we reduced to the setting without such exogenous variables in \cref{model:0} and \ref{model:1} by considering the residuals $M_C Z$, $M_C X$, and $M_C y$ of $Z$, $X$, and $y$ after regressing out $C$.
In this section, we treat included exogenous regressors explicitly.

\subsection{Model and assumptions}
\label{sec:exogenous_variables:assumptions}
\begin{figure}[htbp]
    \centering
    \begin{subfigure}[b]{0.45\textwidth}
        \begin{tikzpicture}[
            node distance=2cm and 2cm,
            >=Stealth,
            every node/.style={draw, circle, minimum size=1cm, inner sep=0pt},
            dashednode/.style={draw, circle, minimum size=1cm, inner sep=0pt, dashed},
            dashedarrow/.style={->, dashed},
            dashedtwinarrow/.style={<->, dashed}
        ]

        \node (Z) at (0,2) {Z};
        \node (X) at (2,0) {X};
        \node (Y) at (5.5,0) {Y};
        \node (C) at (0, -0.5) {C};
        \node[dashednode] (U) at (3.5,2) {U};

        \draw[->] (Z) -- (X) node[pos=0.3, below, draw=none] {$\Pi_{ZX}$};
        \draw[->] (X) -- (Y) node[pos=0.4, below, draw=none, yshift=5] {$\beta_0$};
        \draw[dashedarrow] (U) -- (X);
        \draw[dashedarrow] (U) -- (Y);
        \draw[dashedtwinarrow] (Z) -- (C);
        \draw[->] (C) -- (X) node[pos=0.7, yshift=3, below, draw=none] {$\Pi_{CX}$};
        \draw[->, bend right=25] (C) edge node[below, pos=0.4, draw=none, yshift=5] {$\alpha_0$} (Y);
        \end{tikzpicture}
    \end{subfigure}
    \begin{subfigure}[b]{0.45\textwidth}
        \begin{tikzpicture}[
            node distance=2cm and 2cm,
            >=Stealth,
            every node/.style={draw, circle, minimum size=1cm, inner sep=0pt},
            dashednode/.style={draw, circle, minimum size=1cm, inner sep=0pt, dashed},
            dashedarrow/.style={->, dashed},
            dashedtwinarrow/.style={<->, dashed}
        ]

        \node (Z) at (0,2) {Z};
        \node (W) at (2,1.5) {W};
        \node (X) at (2,0) {X};
        \node (Y) at (5.5,0) {Y};
        \node[dashednode] (U) at (3.5,2) {U};
        \node (C) at (0, -0.5) {C};
        \node (D) at (1.25, -1.75) {D};

        \draw[->] (Z) -- (X); %
        \draw[->] (Z) -- (W); %
        \draw[->] (X) -- (Y) node[pos=0.5, below, draw=none, yshift=5] {$\beta_0$};
        \draw[->] (W) -- (Y) node[pos=0.4, below, draw=none, yshift=5] {$\gamma_0$};
        \draw[dashedtwinarrow] (W) -- (X);
        \draw[dashedarrow] (U) -- (W);
        \draw[dashedarrow] (U) -- (X);
        \draw[dashedarrow] (U) -- (Y);

        \draw[dashedtwinarrow] (Z) -- (C);
        \draw[dashedtwinarrow] (D) -- (C);
        \draw[dashedtwinarrow] (Z) -- (D);
    
        \draw[->] (C) -- (X);
        \draw[->] (C) -- (W);
        \draw[->] (D) -- (X);
        \draw[->, bend left=20] (D) to (W);

        \draw[->, bend right=25] (C) edge node[below, pos=0.4, draw=none, yshift=5] {$\alpha_0$} (Y);
        \draw[->, bend right=20] (D) edge node[below, midway, draw=none, yshift=5] {$\delta_0$} (Y);

        \end{tikzpicture}
    \end{subfigure}
    \caption{
        \label{fig:iv_graph_exogenous}
        Causal graphs visualizing \cref{model:1}.
        On the left, $\mW=\mD=0$.
        On the right, we split the endogenous variables $X$ into endogenous variables of interest $X$ and endogenous variables not of interest $W$, and split included exogenous variables $C$ into included exogenous variables of interest $D$ and included exogenous variables not of interest $C$.
        First-stage parameters are not labeled on the right to avoid clutter.
    }
\end{figure}
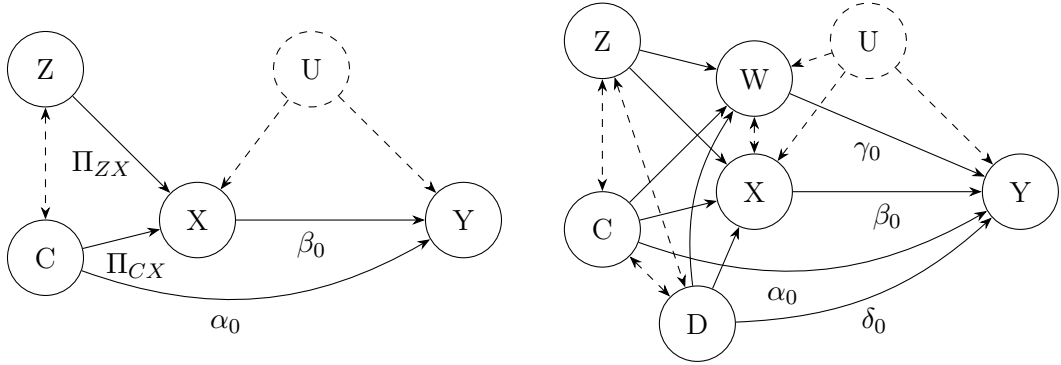

We extend \cref{model:1} and \cref{ass:1} to include exogenous regressors $C$ and $D$, where $C$ are included exogenous regressors whose causal parameter $\alpha_0$ we treat as a nuisance parameter (akin to $W$ and $\gamma_0$) and $D$ are included exogenous regressors whose causal parameter $\delta_0$ we wish to make inference for (akin to $X$ and $\beta_0$).
We illustrate this in \cref{fig:iv_graph_exogenous}.

\begin{model}
    \label{model:3}
    Let $y_i = X_i^T \beta_0 + W_i^T \gamma_0 + C_i^T \alpha_0 + D_i^T \delta + \varepsilon_i \in \BR$ with $X_i = Z_i^T \Pi_{ZX} + C_i^T \Pi_{CX} + D_i^T \Pi_{DX} + V_{X, i} \in \BR^\mX$ and $W_i = Z_i^T \Pi_{ZW} + C_i^T \Pi_{CW} + D_i^T \Pi_{DW} + V_{W, i} \in \BR^\mW$ for random vectors $Z_i \in \BR^k, C_i\in\BR^\mC, D_i \in \BR^\mD, V_{X, i}\in \BR^\mX,  V_{W, i}\in \BR^\mW$, and $\varepsilon_i \in \BR$ for $i=1\ldots, n$ and parameters $\Pi_{ZX} \in \BR^{k \times \mX}$, $\Pi_{CX} \in \BR^{\mC \times \mX}$, $\Pi_{DX} \in \BR^{\mD \times \mX}$, $\Pi_{ZW} \in \BR^{k \times \mW}$, $\Pi_{CW} \in \BR^{\mC \times \mW}$, $\Pi_{DW} \in \BR^{\mD \times \mW}$, $\beta_0 \in \BR^\mX$, $\gamma_0 \in \BR^\mW$, $\alpha_0 \in \BR^\mC$, and $\delta_0 \in \BR^\mD$.
    We call the $Z_i$ \emph{instruments}, the $X_i$ \emph{endogenous covariates of interest}, the $W_i$ \emph{endogenous covariates not of interest}, the $C_i$ \emph{exogenous (included) covariates not of interest}, the $D_i$ \emph{exogenous (included) covariates of interest}, and the $y_i$ \emph{outcomes}.
    The $V_{X, i}$, $V_{W, i}$n and $\varepsilon_i$ are \emph{errors}.
    These need not be independent across observations.
    Let $Z, X, W, C, D, y$ be the matrices of stacked observations

    In \emph{strong instrument asymptotics}, we assume that $\begin{pmatrix} \Pi_{ZX} & \Pi_{ZW} \end{pmatrix}$ is fixed and of full column rank $\mX + \mW$.
    In \emph{weak instrument asymptotics} \citep{staiger1997instrumental}, we assume $\sqrt{n} \, \begin{pmatrix} \Pi_{ZX} & \Pi_{ZW} \end{pmatrix}$ is fixed and of full column rank $\mX + \mW$.
    Thus, $\begin{pmatrix} \Pi_{ZX} & \Pi_{ZW} \end{pmatrix} = \CO(\frac{1}{\sqrt{n}})$.
    Both asymptotics imply that $k \geq \mX + \mW$.
\end{model}

\begin{theoremEnd}[malte,restate command=assumptiontwo]{assumption}
    \label{ass:2}
    Let
    \begin{align*}
    \Psi &:= \begin{pmatrix} \Psi_{\varepsilon} & \Psi_{V_X} & \Psi_{V_W} \end{pmatrix}\\
    &:= \left(\begin{pmatrix} Z & C & D \end{pmatrix}^T \begin{pmatrix} Z & C & D \end{pmatrix} \right)^{-1/2} \begin{pmatrix} Z & C & D \end{pmatrix}^T \begin{pmatrix} \varepsilon & V_X & V_W \end{pmatrix} \in \BR^{k \times (1 + \mX + \mW)}.
    \end{align*}
    Assume there exist $\Omega \in \BR^{1 + \mX + \mW}$ and $Q \in \BR^{(k + \mC + \mD) \times (k + \mC + \mD)}$ positive definite such that, as $n \to \infty$,
    \begin{align*}
        &\mathrm{(a)} \ \ \frac{1}{n} \begin{pmatrix}\varepsilon & V_X & V_W\end{pmatrix}^T \begin{pmatrix}\varepsilon & V_X & V_W \end{pmatrix} \toP \Omega, \\
        &\mathrm{(b)} \ \ \vecop(\Psi) \tod \CN(0, \Omega \otimes \Id_{k + \mC + \mD}), \text{ and }\\
        &\mathrm{(c)} \ \ \frac{1}{n} \begin{pmatrix} Z & C & D \end{pmatrix}^T \begin{pmatrix} Z & C & D \end{pmatrix}  \toP Q.
    \end{align*}
\end{theoremEnd}%

Note that \cref{ass:2} is equivalent to \cref{ass:1} after including the variables $C$ and $D$ into the instruments $Z \leftarrow \begin{pmatrix} Z & C & D \end{pmatrix}$.
However, if we furthermore include them into the endogenous variables $X \leftarrow \begin{pmatrix} X & D \end{pmatrix}, W \leftarrow \begin{pmatrix} W & C \end{pmatrix}$, then \cref{ass:1} no longer holds, as the last $\mD$ components of $V_X$ and the last $\mC$ components of $V_W$ are zero, and the corresponding columns of $\Psi$ are not Gaussian, but zero.

The following result shows that if \cref{ass:2} applies for $Z, C, D, X, W$, and $y$, then it also applies for $M_C Z, M_C D, M_C X, M_C W$, and $M_C y$ with $\mC = 0$.
This is a first step in allowing us to reduce to \cref{ass:1}.
\begin{theoremEnd}[malte,category=exogenous]{lemma}[\citeauthor{londschien2024weak}, \citeyear{londschien2024weak}]
    \label{lem:ass_2_reduction}
    If \Cref{ass:2} applies to $Z, X, W, C, D$, and $y$, then it also applies to $M_C Z, M_C X, M_C D,$ and $M_C y$, the residuals after regressing out $C$.
\end{theoremEnd}
\begin{proofEnd}
Let
$$
\tilde\Psi_{[Z, D]} := \begin{pmatrix} Z & D \end{pmatrix}^T \begin{pmatrix} \varepsilon & V_X & V_W \end{pmatrix},\ \tilde\Psi_C := C^T \begin{pmatrix} \varepsilon & V_X & V_W \end{pmatrix} \text{, and } Q = \begin{pmatrix} Q_{[Z, D]} & Q_{[Z, D], C} \\ Q_{C, [Z, D]} & Q_C \end{pmatrix}.
$$
Then,
$$
\frac{1}{\sqrt{n}} \tilde \Psi := \frac{1}{\sqrt{n}} \begin{pmatrix} \tilde\Psi_{[Z, D]} \\ \tilde \Psi_C \end{pmatrix} = \frac{1}{\sqrt{n}}\left(\begin{pmatrix} Z & D & C \end{pmatrix}^T \begin{pmatrix} Z & D & C \end{pmatrix} \right)^{1/2} \Psi  \tod \CN(0, \Omega \otimes Q)
$$
by \Cref{ass:2} (b) and (c).

\paragraph{(c)} Calculate
\begin{align*}
    \begin{pmatrix} M_C Z & M_C D \end{pmatrix}^T  \begin{pmatrix} M_C Z & M_C D \end{pmatrix} &= \begin{pmatrix} Z & D \end{pmatrix}^T  \begin{pmatrix} Z & D \end{pmatrix} - \begin{pmatrix} Z & D \end{pmatrix}^T C (C^T C)^{-1} C^T \begin{pmatrix} Z & D \end{pmatrix} \\
    &\overset{\Cref{ass:2} (c)} \toP Q_{[Z, D]} - Q_{[Z, D], C} Q_{C}^{-1} Q_{C, [Z, D]} =: \tilde Q.
\end{align*}
$\tilde Q$ is the Schur complement of $Q_{C}$ in $Q$ and thus positive definite.

\paragraph{(a)} Calculate
\begin{multline*}
    \frac{1}{n} \begin{pmatrix} M_C \varepsilon & M_C V_X & M_C V_W \end{pmatrix}^T \begin{pmatrix} M_C \varepsilon & M_C V_X & M_C V_W \end{pmatrix} \\
    = \frac{1}{n} \begin{pmatrix} \varepsilon & V_X & V_W \end{pmatrix}^T \begin{pmatrix} \varepsilon & V_X & V_W \end{pmatrix} -  \frac{1}{n} \begin{pmatrix} \varepsilon & V_X & V_W \end{pmatrix}^T P_C  \begin{pmatrix} \varepsilon & V_X & V_W \end{pmatrix},
\end{multline*}
where $ \begin{pmatrix} \varepsilon & V_X & V_W \end{pmatrix}^T P_C \begin{pmatrix} \varepsilon & V_X & V_W \end{pmatrix} = (\tilde \Psi_C)^T (Z^T Z)^{-1} (\tilde \Psi_C)\toP \frac{1}{n}(\tilde \Psi_C)^T Q_C^{-1} (\tilde \Psi_C) = O(1)$ by \Cref{ass:2} (b).

\paragraph*{(b)}
We have
\begin{align*}
\begin{pmatrix} M_C Z & M_C D \end{pmatrix}^T & \begin{pmatrix} M_C \varepsilon & M_C V_X & M_C V_W \end{pmatrix} \\
&= \begin{pmatrix} Z & D \end{pmatrix}^T \begin{pmatrix} \varepsilon & V_X & V_W \end{pmatrix}
 - \begin{pmatrix} Z & D \end{pmatrix}^T C (C^T C)^{-1} C^T \begin{pmatrix} \varepsilon & V_X & V_W \end{pmatrix} \\
 &= \tilde \Psi_{[Z, D]}
 - \begin{pmatrix} Z & D \end{pmatrix}^T C (C^T C)^{-1} \tilde \Psi_{C} \\
 &\toP  \tilde \Psi_{[Z, D]} - Q_{[Z, D], C} Q_{C}^{-1} {}_{C} \tilde \Psi_C
 = \begin{pmatrix} \Id & - Q_{[Z, D]} Q_C^{-1} \end{pmatrix} \tilde \Psi
\end{align*}
As $\frac{1}{\sqrt{n}} \Cov(\tilde \Psi) = \Omega \otimes Q$, we have 
$$
\frac{1}{\sqrt{n}} \Cov( \begin{pmatrix} \Id & - Q_{[Z, D]} Q_C^{-1} \end{pmatrix}  \tilde \Psi) = \Omega \otimes \begin{pmatrix} \Id & - Q_{[Z, D]} Q_C^{-1} \end{pmatrix} Q \begin{pmatrix} \Id & - Q_{[Z, D]} Q_C^{-1} \end{pmatrix} = \Omega \otimes \tilde Q.
$$
Together with (c) from above, this yields
$$
\left( \begin{pmatrix} M_C Z & M_C D \end{pmatrix}^T \begin{pmatrix} M_C Z & M_C D \end{pmatrix} \right)^{1/2} \begin{pmatrix} M_C Z & M_C D \end{pmatrix}^T \begin{pmatrix} M_C \varepsilon & M_C V_X & M_C V_W \end{pmatrix} \tod \CN(0, \Omega \otimes \Id).
$$
\end{proofEnd}

\subsection{Estimators}
For estimation purposes, there is no need to split $X$ into $(X, W)$ and $C$ into $(C, D)$.
For this subsection, let $\mW = \mD = 0$, as visualized in the left panel of \cref{fig:iv_graph_exogenous}.
\label{sec:exogenous_variables:estimators}
\begin{definition}
    Let $\kappa > 0$. The \emph{k-class estimator} for $(\beta_0, \alpha_0)$ is
    \begin{align*}
    \begin{pmatrix} \hat\beta_\kclass \\ \hat\alpha_\kclass \end{pmatrix} (\kappa) &:= \left( (X \ C)^T (\kappa P_{[Z, C]} + (1 - \kappa) \Id_n) (X \ C) \right)^{-1} (X \ C)^T ( \kappa P_{[Z, C]} + (1 - \kappa) \Id_n ) y \\
    &= \begin{pmatrix} X^T ( \kappa P_{[Z, C]} + (1 - \kappa) \Id_n ) X & X^T C \\ C^T X & C^T C \end{pmatrix}^{-1} \begin{pmatrix} X^T ( \kappa P_{[Z, C]} + (1 - \kappa) \Id_n ) y \\ C^T y \end{pmatrix}.
    \end{align*}
\end{definition}

\noindent That is, the k-class estimator is equal to the k-class estimator from \cref{def:kclass} after including the exogenous variables $C$ both as endogenous covariates and instruments.
The part $\hat\beta_\kclass(\kappa)$ of the k-class estimator corresponding to the endogenous covariates of interest $X$ is the same as the k-class estimator from \cref{def:kclass} after regressing out the exogenous variables $C$.

Assuming i.i.d.\ centered Gaussian errors $(\varepsilon_i, V_{X, i})$ with covariance $\Omega$ and independent of $(C_i, Z_i)$, the log-likelihood of observing $y$ and $X$ given $\beta, \alpha, \Pi_{ZX}, \Pi_{CX}, \Omega$, and $Z$ is
\begin{multline*}
\ell(y , X \mid \beta, \alpha, \Pi_{ZX}, \Pi_{CX}, \Omega, Z, C) 
= -\frac{n(m_x + 1)}{2} \log(2\pi) - \frac{n}{2} \log(\det(\Omega)) \\
- \frac{1}{2} \sum_{i=1}^n \begin{pmatrix} y_i - X_i^T \beta - C_i^T \alpha \\ X_i - Z_i^T \Pi_{ZX} - C_i^T \Pi_{CX} \end{pmatrix}^T \Omega^{-1} \begin{pmatrix} y_i - X_i^T \beta - C_i^T \alpha \\ X_i - Z_i^T \Pi_{ZX} - C_i^T \Pi_{CX} \end{pmatrix}.
\end{multline*}
\begin{definition}
    \label{def:liml_with_exogenous}
    The \emph{limited information maximum likelihood (LIML)} estimator of $(\beta_0, \alpha_0)$ is
    $$
    \begin{pmatrix} \hat\beta_\liml \\ \hat\alpha_\liml \end{pmatrix} := 
    \argmax_{\alpha \in \BR^\mC, \beta \in \BR^\mX} \max_{\substack{
        \Pi_{ZX} \in \BR^{k \times \mX}, \\ \Pi_{CX} \in \BR^{\mC \times \mX}\!,\, \Omega \succ 0}} \ell(y , X \mid \beta, \alpha, \Pi_{ZX}, \Pi_{CX}, \Omega, Z, C).
    $$
\end{definition}

\begin{theoremEnd}[malte,category=exogenous]{lemma}
    \label{lem:liml_likelihood_with_exogenous}
    Assume that $M_{[Z, C]} \begin{pmatrix} y & X \end{pmatrix}$ is of full column rank $\mX + 1$.
    Then,
    \begin{multline*}
        \max_{
            \substack{\Pi_{ZX} \in \BR^{k \times \mX} \\ \Pi_{CX} \in \BR^{\mC \times \mX}\!,\, \Omega \succ 0 }} \ell(y , X \mid \beta, \alpha, \Pi_{ZX}, \Pi_{CX}, \Omega, Z, C)  =
            -\frac{\mX + 1}{2}(n \log(2 \pi) + 1 - n \log(n)) \\
            - \frac{n}{2}\log(1 + \frac{(y - X \beta - C \alpha)^T P_{[Z, C]} (y - X \beta - C \alpha)}{(y - X \beta)^T M_{[Z, C]} (y - X \beta )})
            - \frac{n}{2} \log(\det(\begin{pmatrix} y \!\!& X \end{pmatrix}^T M_{[Z, C]} \begin{pmatrix} y \!\!& X \end{pmatrix})).
    \end{multline*}
    This is minimized with respect to $\alpha$ at $\hat\alpha_\liml(\beta) = (C^T C)^{-1} C^T (y - X \beta)$ with
    \begin{multline*}
        \max_{
            \substack{\alpha \in \BR^\mC, \, \Pi_{ZX} \in \BR^{k \times \mX}\\  \Pi_{CX} \in \BR^{\mC \times \mX} \!,\,\Omega \succ 0 }} \ell(y , X \mid \beta, \alpha, \Pi_{ZX}, \Pi_{CX}, \Omega, Z, C)  =
            -\frac{\mX + 1}{2}(n \log(2 \pi) + 1 - n \log(n)) \\
            - \frac{n}{2}\log(1 + \frac{(y - X \beta)^T P_{M_C Z} (y - X \beta)}{(y - X \beta)^T M_{[Z, C]} (y - X \beta)}) - \frac{n}{2} \log(\det(\begin{pmatrix} y \!\!& X \end{pmatrix}^T M_{[Z, C]} \begin{pmatrix} y \!\!& X \end{pmatrix})) \\
        = \max_{\Pi_{XZ}\!,\, \Omega \succ 0} \ell(M_C y, M_C X \mid \beta, \Pi_{ZX}, \Omega, M_C Z).
    \end{multline*}
    That is, the maximum likelihood of $\beta$ is equal to the likelihood in \cref{lem:liml_likelihood} with $Z \leftarrow M_C Z$, $X \leftarrow M_C X$, and $y \leftarrow M_C y$ replaced.
    Also,
    \begin{align*}
        \hat\Pi_\liml(\beta) &:= \argmax_{(\Pi_{ZX}\!,\, \Pi_{CX}) \in \BR^{(k + \mC) \times \mX}} \max_{\alpha \in \BR^{\mC}\!,\, \Omega \succ 0} \ell(y , X \mid \beta, \alpha, \Pi_{ZX}, \Pi_{CX}, \Omega, Z, C) \\
        &= \left[\begin{pmatrix} Z \!\!& C \end{pmatrix}^T \begin{pmatrix} Z \!\!& C \end{pmatrix} \right]^{-1} \begin{pmatrix} Z \!\!& C \end{pmatrix}^T \begin{pmatrix} X - M_C (y - X \beta) \frac{(y - X \beta)^T M_{[Z, C]} X}{(y - X \beta)^T M_{[Z, C]} (y - X \beta)} \end{pmatrix}
    \end{align*}
    and
    \begin{align*}
        \hat\Pi_{ZX, \liml}(\beta) &:= \argmax_{\Pi_{ZX} \in \BR^{k \times \mX}} \max_{\substack{\alpha \in \BR^{\mC},\, \Omega \succ 0\\ \Pi_{CX} \in \BR^{\mC \times \mX}}} \ell(y , X \mid \beta, \alpha, \Pi_{ZX}, \Pi_{CX}, \Omega, Z, C) \\
        &= [Z^T M_C Z]^{-1} Z^T M_C \begin{pmatrix} X - (y - X \beta) \frac{(y - X \beta)^T M_{[Z, C]} X}{(y - X \beta)^T M_{[Z, C]} (y - X \beta)} \end{pmatrix}.
    \end{align*}
\end{theoremEnd}%
\begin{proofEnd}%
    Follow the proof of \cref*{lem:liml_likelihood}, replacing $Z \leftarrow [Z, C]$, $\Pi \leftarrow \begin{pmatrix} \Pi_{ZX} & \Pi_{CX} \end{pmatrix}$, and $u(\beta) \leftarrow u(\alpha, \beta) = y - X \beta - C \alpha$.
    Plugging in the definition for $u$ into \cref{eq:detOmega} yields
    $$
    \det \left( \hat\Omega \right) = n^{-(\mX + 1)} \left(1 + \frac{
        (y - X \beta - C \alpha)^T P_{[Z, C]} (y - X \beta - C \alpha)
    }{
        (y - X \beta - C \alpha)^T M_{[Z, C]} (y - X \beta - C \alpha)
    } \right) \det \left( \begin{pmatrix} y^T \\ X^T \end{pmatrix} M_{[Z, C]} \begin{pmatrix} y & X \end{pmatrix} \right).
    $$
    The maximiser $\hat\alpha$ of this expression with respect to $\alpha$ satisfies
    \begin{align*}
    0 &= \frac{\dd}{\dd \alpha} \frac{
        (y - X \beta - C \alpha)^T P_{[Z, C]} (y - X \beta - C \alpha)
    }{
        (y - X \beta - C \alpha)^T M_{[Z, C]} (y - X \beta - C \alpha)
    } |_{{\alpha = \hat\alpha}} \\
    &= - 2 \frac{( y - X \beta - C \hat\alpha)^T P_{[Z, C]} C}{(y - X \beta - C  \hat\alpha)^T M_{[Z, C]} (y - X \beta - C  \hat\alpha)} \\
    &\Rightarrow \hat\alpha = (C^T C)^{-1} C^T (y - X \beta) \\
    &\Rightarrow y - X \beta - C \hat\alpha = M_C (y - X \beta).
    \end{align*}
    Again, plugging this into \cref{eq:detOmega}, using $M_{[Z, C]} (y - X \beta - C \alpha) = M_{[Z, C]} (y - X \beta)$ and $P_{[Z, C]} M_C = P_{M_C Z}$ yields the first result.

    The second result is the same as in \cref{lem:liml_likelihood} with the above replacements, using that $y - X \beta - C \hat\alpha = M_C (y - X \beta)$.
    The last equation follows from explicitly computing the inverse of the block matrix $\begin{pmatrix} Z & C \end{pmatrix}^T \begin{pmatrix} Z & C \end{pmatrix}$. 
    The last sentence follows from $M_{M_C Z} M_C = M_{[Z, C]}$.
\end{proofEnd}%

\begin{definition}
    \label{def:anderson_rubin_test_statistic_with_exogenous}
    The \emph{Anderson-Rubin (AR) test statistic} is
    $$
    \AR(\beta, \alpha) := \frac{n - k - \mC}{k + \mC} \frac{
        (y - X \beta - C \alpha )^T P_{[Z, C]} (y - X \beta - C \alpha)
    }{
        (y - X \beta)^T M_{[Z, C]} (y - X \beta)
    }.
    $$
\end{definition}%

\begin{theoremEnd}[malte,category=exogenous]{proposition}
    \label{thm:liml_is_kclass_with_exogenous}
    Assume that $M_{[Z, C]} \begin{pmatrix} y & X \end{pmatrix}$ has full column rank $\mX + 1$.
    Let
    $$
    \hat\kappa_\liml := \lambda_\mathrm{min} \left( \left[ \begin{pmatrix} y & X \end{pmatrix}^T M_{[Z, C]} \begin{pmatrix} y & X \end{pmatrix} \right]^{-1} \begin{pmatrix} y & X \end{pmatrix}^T P_{M_C Z} \begin{pmatrix} y & X \end{pmatrix} \right) + 1\geq 1
    $$
    Then,
    \begin{align*}
    \hat\beta_\liml &= \argmax_{\beta \in \BR^{\mX}}  \max_{
        \substack{\Pi_{ZX} \in \BR^{k \times \mX} , \Omega \succ 0\\ \alpha \in \BR^{\mC}\!,\, \Pi_{CX} \in \BR^{\mC \times \mX} }} \ell(y , X \mid \beta, \alpha, \Pi_{ZX}, \Pi_{CX}, \Omega) \\
        &= [X^T (M_C - \hat\kappa M_{[Z, C]}) X]^{-1} X^T (M_C - \hat\kappa M_{[Z, C]}) y
        = \hat\beta_\kclass(\hat\kappa_\liml)
    \end{align*}
    and $\AR(\hat\beta_\liml, \hat\alpha_\liml) = \frac{n - k - \mC}{k} (\hat\kappa_\liml - 1)$.
\end{theoremEnd}%
\begin{proofEnd}%
    By \cref{lem:liml_likelihood_with_exogenous}
    $$
        \max_{
            \substack{\Pi_{ZX} \in \BR^{k \times \mX} , \Omega \succ 0\\ \alpha \in \BR^{\mC}, \Pi_{CX} \in \BR^{\mC \times \mX} }} \ell(y , X \mid \beta, \alpha, \Pi_{ZX}, \Pi_{CX}, \Omega) = \max_{\Pi_{XZ}, \Omega \succ 0} \ell(M_C y, M_C X \mid \beta, \Pi_{ZX}, \Omega, M_C Z).
    $$
    We apply \cref{thm:liml_is_kclass} to the transformed model $M_C y = M_C X \beta + M_C Z \Pi_{XZ} + M_C \varepsilon$, which yields the desired result.
\end{proofEnd}%

That is, the k-class parameter $\hat\kappa_\liml$, which corresponds to the minimum of the Anderson-Rubin test statistic, is equal to the k-class parameter from \cref{def:kclass} after regressing out the exogenous variables $C$.

\subsection{Tests}
\label{sec:exogenous_variables:tests}
Now we discuss how to (jointly) make inference on the parameters of interest $\beta_0$, $\delta_0$.
For this, we include the exogenous variables $D$ as both endogenous covariates and instruments and consider the residuals after regressing out the exogenous variables $C$.
As mentioned in \cref{sec:exogenous_variables:assumptions}, \cref{ass:1} no longer applies, as the columns of $\Psi$ corresponding to $D$ are zero and thus non-Gaussian.
In the following, we discuss for each test whether the proofs in appendix \ref{app:proofs:tests} for size still apply.

\subsubsection*{The (conditional) likelihood ratio test}

The likelihood-ratio test has the correct size under strong-instrument asymptotics.
The proof for the limiting distribution of the conditional likelihood-ratio test requires inverting $\tilde X(\beta_0)^T M_Z \tilde X(\beta_0)$.
This is singular if included exogenous variables $D$ are included in both $X \leftarrow \begin{pmatrix}X & D \end{pmatrix}$ and $Z \leftarrow \begin{pmatrix} Z & D \end{pmatrix}$.
We prove the following.%
\begin{theoremEnd}[malte,category=exogenous]{proposition}%
    \label{prop:clr_test_statistic_with_exogenous}
    Assume $\mW = \mC = 0$.
    Let 
    \begin{multline*}
        \LR(\beta, \delta) = (n - k - m_d) \frac{(y - X \beta - D \delta)^T P_{[Z, D]}(y - X \beta - D \delta)}{(y - X \beta)^T M_{[Z, D]}(y - X \beta)} \\
        - (n - k - m_d)\min_{b \in \BR^\mX, d \in \BR^\mD} \frac{(y - X b - D d)^T P_{[Z, D]}(y - X b - D d)}{(y - X b)^T M_{[Z, D]}(y - X b)}.
    \end{multline*}
    Let $\tilde X(\beta) := X - (y - X \beta) \frac{(y - X \beta)^T M_{[Z, D]} X}{(y - X \beta )^T M_{[Z, D]} (y - X \beta)}$ and let
    $$
    s_\mathrm{min}(\beta) := (n - k - \mD) \lambdamin{ \left[\tilde X(\beta)^T M_{[Z, D]} \tilde X(\beta) \right]^{-1} \tilde X(\beta)^T P_{M_D Z} \tilde X(\beta)}.
    $$
    Then, under the null $(\beta, \delta) = (\beta_0, \delta_0)$ and both strong and weak instrument asymptotics, conditionally on $s_\mathrm{min}(\beta_0)$, the likelihood-ratio test statistic $\LR(\beta_0, \delta_0)$ is asymptotically bounded from above by a random variable with distribution
    \begin{align*}
    &Q_{\mD} + \frac{1}{2} \left(Q_{k - \mX} + Q_\mX - s_\mathrm{min}(\beta_0) + \sqrt{(Q_{k - \mX} + Q_\mX + s_\mathrm{min}(\beta_0))^2 - 4 Q_{k - \mX} s_\mathrm{min}(\beta_0)} \right) \\
    \ \ &= Q_{\mD} + \frac{1}{2} \left(Q_{k - \mX} + Q_\mX - s_\mathrm{min}(\beta_0) + \sqrt{(Q_{k - \mX} + Q_\mX - s_\mathrm{min}(\beta_0))^2 + 4 Q_{\mX} s_\mathrm{min}(\beta_0)} \right),
    \end{align*}
    where $Q_\mX \sim \chi^2(\mX)$, $Q_\mD \sim \chi^2(\mD)$, and $Q_{k - \mX} \sim \chi^2(k - \mX)$ and are independent.
\end{theoremEnd}%
\begin{proofEnd}%
    This proceeds in 5 steps.
    \paragraph*{Step 1:}
    Write $\tilde X = \tilde X(\beta_0)$.
    Then,
    \begin{equation} \label{eq:exog_clr1}
        \begin{pmatrix} y & X \end{pmatrix} \begin{pmatrix} 1 & 0 \\ - \beta_0 & \Id_\mX \end{pmatrix} 
        \begin{pmatrix}
             1 & -\frac{\varepsilon^T M_{[Z, D]} X}{\varepsilon^T M_{[Z, D]} \varepsilon} \\ 0 & \Id_\mX
         \end{pmatrix}
        =
        \begin{pmatrix} \varepsilon & X \end{pmatrix} 
        \begin{pmatrix}
            1 & -\frac{\varepsilon^T M_{[Z, D]} X}{\varepsilon^T M_{[Z, D]} \varepsilon} \\ 0 & \Id_\mX
        \end{pmatrix} =
        \begin{pmatrix} \varepsilon & \tilde X \end{pmatrix}.
    \end{equation}
    Note that $\varepsilon^T M_{[Z, D]} \tilde X = 0$ and thus
    \begin{equation}\label{eq:exog_clr2}
        \tilde \Omega := 
        \begin{pmatrix} \varepsilon & \tilde X \end{pmatrix}^T M_{[Z, D]} \begin{pmatrix} \varepsilon & \tilde X \end{pmatrix} = 
        \begin{pmatrix}
            \varepsilon^T M_{[Z, D]} \varepsilon & 0 \\
            0 & \tilde X^T M_{[Z, D]} \tilde X
        \end{pmatrix},
    \end{equation}
    We can calculate
    \begin{align*}
    &\min_{[b, d]} \frac{(y - X b - D d)^T P_{M_D Z}(y - X b - D d)}{(y - X b - D d)^T M_{[Z, D]}(y - X b - D d)}
    \\ &\overset{\text{Theorem \ref{thm:liml_is_kclass_with_exogenous}}}{=} \lambda_\text{min}
    \left(\left(\begin{pmatrix} X & y \end{pmatrix}^T M_{[Z, D]} \begin{pmatrix} X & y \end{pmatrix}\right)^{-1} \begin{pmatrix} X & y \end{pmatrix}^T P_{M_D Z} \begin{pmatrix} X & y \end{pmatrix}\right) \\
    &= \min \{\mu \in \BR \colon \det( 
        \mu \cdot \Id_{\mX+1} - \left(\begin{pmatrix} y & X \end{pmatrix}^T M_{[Z, D]} \begin{pmatrix} y & X \end{pmatrix}\right)^{-1} \begin{pmatrix} y & X \end{pmatrix}^T P_{M_D Z} \begin{pmatrix} y & X \end{pmatrix}
    ) = 0 \}\\
    &=
    \min \{ \mu \in \BR \colon \det( 
        \mu \cdot \begin{pmatrix} y & X \end{pmatrix}^T M_{[Z, D]} \begin{pmatrix} y & X \end{pmatrix} -  \begin{pmatrix} y & X \end{pmatrix}^T P_{M_D Z} \begin{pmatrix} y & X \end{pmatrix}
    ) = 0\}\\
    &\overset{\text{(\ref{eq:exog_clr1}, \ref{eq:exog_clr2})}}{=}
    \min \{ \mu \in \BR \colon \det( 
        \mu \cdot \Id_{\mX+1}
        - \tilde \Omega^{-1/2, T} \begin{pmatrix} \varepsilon & \tilde X \end{pmatrix}^T P_{M_D Z} \begin{pmatrix} \varepsilon & \tilde X \end{pmatrix}  \tilde \Omega^{-1/2}
    ) = 0\}.
    \end{align*}

    \paragraph*{Step 2:}
    Let $U D V = P_{M_D Z} \tilde X (\tilde X^T M_{[Z, D]} \tilde X)^{-1/2}$ be a singular value decomposition with $0 \leq d_1,\ldots, d_\mX$, the diagonal entries of $D^2 \in \BR^{\mX \times \mX}$, sorted ascending.
    Then $P_{P_{M_D Z} \tilde X} = U U^T$ and
    \begin{align*}
       \Sigma &:= \tilde \Omega^{-1/2, T} \begin{pmatrix} \varepsilon & \tilde X \end{pmatrix}^T P_{M_D Z} \begin{pmatrix} \varepsilon & \tilde X \end{pmatrix} \tilde \Omega^{-1/2} \numberthis\label{eq:exog_prop_clr_diagonal} \\
       &= 
        \begin{pmatrix} 1 & 0 \\ 0 & V^T \end{pmatrix}
        \begin{pmatrix} \varepsilon^T P_{M_D Z} \varepsilon / \varepsilon^T M_{[Z, D]} \varepsilon & \varepsilon^T P_{M_D Z} U D / \sqrt{\varepsilon^T M_{[Z, D]} \varepsilon} \\ D U^T P_{M_D Z} \varepsilon / \sqrt{\varepsilon^T M_{[Z, D]} \varepsilon} & D^2 \end{pmatrix}
        \begin{pmatrix} 1 & 0 \\ 0 & V \end{pmatrix}
    \end{align*}
    such that, for $\mu \notin \{d_1, \ldots, d_\mX\}$,
    \begin{align*}
        \phi_\Sigma(\mu) &:= \det(\mu \cdot \Id_{\mX+1} - \Sigma) 
        = \det \begin{pmatrix} \mu - \varepsilon^T P_{M_D Z} \varepsilon / \varepsilon^T M_{[Z, D]} \varepsilon & - \varepsilon^T P_{M_D Z} U D / \sqrt{\varepsilon^T M_{[Z, D]} \varepsilon} \\ - D U^T P_{M_D Z} \varepsilon /\sqrt{\varepsilon^T M_{[Z, D]} \varepsilon} & \mu \cdot \Id_\mX - D^2 \end{pmatrix} \\
        &= \det(\mu \cdot \Id_\mX - D^2) \cdot \left(\mu - \frac{\varepsilon^T P_{M_D Z} \varepsilon}{\varepsilon^T M_{[Z, D]} \varepsilon} - \frac{\varepsilon^T P_{M_D Z} U D (\mu \cdot \Id_\mX - D^2)^{-1} D U^T P_{M_D Z} \varepsilon}{\varepsilon^T M_{[Z, D]} \varepsilon} \right).
    \end{align*}
    Here we used the block matrix equality 
    $$
    \begin{pmatrix} A & B \\ C & D \end{pmatrix} \begin{pmatrix}\Id & 0 \\ -D^{-1} C & \Id \end{pmatrix} = \begin{pmatrix} A - B D^{-1} C & 0 \\ 0 & D \end{pmatrix}
    \Rightarrow \det \begin{pmatrix} A & B \\ C & D \end{pmatrix} = \det(D) \det( A - B D^{-1} C) 
    $$
    for invertible $D$.
    \paragraph*{Step 3:} Let $\tilde D := d_1 \cdot \Id_\mX$ and $Q := \begin{pmatrix} 1 & 0 \\ 0 & V^{-1} D^{-1} \cdot \tilde D V \end{pmatrix}$ and let $\tilde \Sigma := Q^T \Sigma Q$.
    As all nonzero entries of $D^{-1} \cdot \tilde D$ are bounded from above by 1 and $V$ is orthogonal,
    $$
    \lambda_\mathrm{min}(\tilde \Sigma) = \min_{x \neq 0} \frac{x^T \tilde\Sigma x}{x^T x} = \min_{x \neq 0} \frac{x^T Q^T \Sigma Q x}{x^T x} = \min_{x \neq 0} \frac{x^T \Sigma x}{x^T Q^T Q x} \geq \min_{x \neq 0} \frac{x^T \Sigma x}{x^T x} = \lambda_\mathrm{min}(\Sigma).
    $$
    \paragraph*{Step 4:}
    Compute
    $$
    \phi_{\tilde \Sigma}(\mu) = (\mu - d)^\mX \cdot (\mu - \frac{\varepsilon^T P_{M_D Z} \varepsilon}{\varepsilon^T M_{[Z, D]} \varepsilon} - \frac{d}{\mu - d} \frac{\varepsilon^T P_{P_{M_D Z} \tilde X} \varepsilon}{\varepsilon^T M_{[Z, D]} \varepsilon}),
    $$
    with 
    $$
    \phi_{\tilde \Sigma}(\mu) = 0 \Rightarrow (\mu = d) \  \vee \ (\mu \cdot (\mu - d) - (\mu - d) \cdot \frac{\varepsilon^T P_{M_D Z} \varepsilon}{\varepsilon^T M_{[Z, D]} \varepsilon} - d \cdot \frac{\varepsilon^T P_{P_{M_D Z} \tilde X} \varepsilon}{\varepsilon^T M_{[Z, D]} \varepsilon} = 0).
    $$
    The latter polynomial equation has solutions
    \begin{align*}
    \mu_\pm &= \frac{1}{2} \left(d + \frac{\varepsilon^T P_{M_D Z} \varepsilon}{\varepsilon^T M_{[Z, D]} \varepsilon} \pm \sqrt{ \left(d + \frac{\varepsilon^T P_{M_D Z} \varepsilon}{\varepsilon^T M_{[Z, D]} \varepsilon} \right)^2 - 4 d \frac{\varepsilon^T (P_{M_D Z} - P_{P_{M_D Z} \tilde X}) \varepsilon}{\varepsilon^T M_{[Z, D]} \varepsilon} }\right) \\
    &= \frac{1}{2} \left(d + \frac{\varepsilon^T P_{M_D Z} \varepsilon}{\varepsilon^T M_{[Z, D]} \varepsilon} \pm \sqrt{\left(d + \frac{\varepsilon^T P_{M_D Z} \varepsilon}{\varepsilon^T M_{[Z, D]} \varepsilon} - 2 \frac{\varepsilon^T P_{M_D Z} \varepsilon}{\varepsilon^T M_{[Z, D]} \varepsilon} \right)^2 + 4 d \frac{\varepsilon^T P_{P_{M_D Z} \tilde X} \varepsilon}{\varepsilon^T M_{[Z, D]} \varepsilon}} \right).
    \end{align*}
    If $d \leq \frac{\varepsilon^T P_{M_D Z} \varepsilon}{\varepsilon^T M_{[Z, D]} \varepsilon}$, then
    \begin{align*}
    \mu_{-} &\leq \frac{1}{2} \left(d + \frac{\varepsilon^T P_{M_D Z} \varepsilon}{\varepsilon^T M_{[Z, D]} \varepsilon} - \sqrt{ \left(d - \frac{\varepsilon^T P_{M_D Z} \varepsilon}{\varepsilon^T M_{[Z, D]} \varepsilon} \right)^2}\right) \\
    &= \frac{1}{2} \left(d + \frac{\varepsilon^T P_{M_D Z} \varepsilon}{\varepsilon^T M_{[Z, D]} \varepsilon} - (\frac{\varepsilon^T P_{M_D Z} \varepsilon}{\varepsilon^T M_{[Z, D]} \varepsilon} - d) \right) = d.
    \end{align*}
    If $d > \frac{\varepsilon^T P_{M_D Z} \varepsilon}{\varepsilon^T M_{[Z, D]} \varepsilon}$, then
    \begin{align*}
    \mu_{-} &\leq \frac{1}{2} \left(d + \frac{\varepsilon^T P_{M_D Z} \varepsilon}{\varepsilon^T M_{[Z, D]} \varepsilon} - \sqrt{(d - \frac{\varepsilon^T P_{M_D Z} \varepsilon}{\varepsilon^T M_{[Z, D]} \varepsilon})^2}\right) \\
    &= \frac{1}{2} \left( d + \frac{\varepsilon^T P_{M_D Z} \varepsilon}{\varepsilon^T M_{[Z, D]} \varepsilon} - (d - \frac{\varepsilon^T P_{M_D Z} \varepsilon}{\varepsilon^T M_{[Z, D]} \varepsilon}) \right) = \frac{\varepsilon^T P_{M_D Z} \varepsilon}{\varepsilon^T M_{[Z, D]} \varepsilon} < d.
    \end{align*}
    Thus, $\mu_{-} \leq d$ and thus $\lambda_\text{min}(\tilde \Sigma) = \mu_{-}$.

    \paragraph*{Step 5:}
    Putting everything together, we have
    \begin{align*}
        \CLR(\beta_0, \delta_0) &= (n - k - \mD) \left(\frac{\varepsilon^T P_D \varepsilon}{\varepsilon^T M_{[Z, D]} \varepsilon} + \frac{\varepsilon^T P_{M_D Z} \varepsilon}{\varepsilon^T M_{[Z, D]} \varepsilon} - \lambda_\text{min}( \Sigma) \right) \\
        &\leq (n - k - \mD) \left(\frac{\varepsilon^T P_D \varepsilon}{\varepsilon^T M_{[Z, D]} \varepsilon} + \frac{\varepsilon^T P_{M_D Z} \varepsilon}{\varepsilon^T M_{[Z, D]} \varepsilon} - \lambda_\text{min}( \tilde\Sigma) \right) \\
        &=  (n - k - \mD) \frac{\varepsilon^T P_D \varepsilon}{\varepsilon^T M_{[Z, D]} \varepsilon} + \frac{n - k - \mD}{2} \Biggl(\frac{\varepsilon^T P_{M_D Z} \varepsilon}{\varepsilon^T M_{[Z, D]} \varepsilon} - d  \ + \\
        &\hspace{2cm}  \sqrt{\left(d + \frac{\varepsilon^T P_{M_D Z} \varepsilon}{\varepsilon^T M_{[Z, D]} \varepsilon} \right)^2 - 4 d\frac{\varepsilon^T (P_{M_D Z} - P_{P_{M_D Z} \tilde X}) \varepsilon}{\varepsilon^T M_{[Z, D]} \varepsilon}}\Biggr) \\
        &= Q_\mD + \frac{1}{2}\left( Q_\mX + Q_{k-\mX} - d + \sqrt{ (Q_{k - \mX} + Q_\mX - d)^2 + 4 Q_\mX d} \right),
    \end{align*}
    where
    \begin{align*}
    &Q_\mD := (n - k - \mD) \frac{\varepsilon^T P_D \varepsilon}{\varepsilon^T M_{[Z, D]} \varepsilon} \tod \chi^2(\mD), \ 
    Q_{\mX} := (n - k - \mD) \frac{\varepsilon^T P_{P_{M_D Z} \tilde X} \varepsilon}{\varepsilon^T M_{[Z, D]} \varepsilon} \tod \chi^2(\mX), \text{ and } \\
    &Q_{k - \mX} := (n - k - \mD) \frac{\varepsilon^T (P_{M_D Z} - P_{P_{M_D Z} \tilde X}) \varepsilon}{\varepsilon^T M_{[Z, D]} \varepsilon} \tod \chi^2(k - \mX)
    \end{align*}
    are asymptotically independent and asymptotically independent of $s_\mathrm{min}(\beta_0)$.
\end{proofEnd}%

To the best of our knowledge, this has not been derived explicitly before.
The asymptotic distribution in \cref{prop:clr_test_statistic_with_exogenous} is equal to the distribution of ${Q_\mD} + {\Gamma(k - \mX, \mX, s_\mathrm{min}(\beta_0))}$, with $Q_{\mD} \sim \chi^2(\mD)$ independent of $\Gamma(k - \mX, \mX, s_\mathrm{min}(\beta_0))$ from \cref{prop:clr_test_statistic}.
If we applied \cref{prop:clr_test_statistic} by including $D$ in both $X$ and $Z$, we would obtain
\begin{multline*}
    \Gamma(k - \mX, \mD + \mX, \tilde s_\mathrm{min}(\beta_0)) = \frac{1}{2} \Big(Q_{k - \mX} + Q_{\mX} + Q_{\mD} - \tilde s_\mathrm{min}(\beta_0)\\
    + \sqrt{
        (Q_{k - \mX} + Q_{\mX} + Q_{\mD} - \tilde s_\mathrm{min}(\beta_0) )^2 + 4 (Q_{\mX} + Q_{\mD} ) \tilde s_\mathrm{min}(\beta_0) )
        } \Big)
\end{multline*}
where
$s_\mathrm{min}'(\beta_0)$
is the smallest generalized eigenvalue satisfying 
\begin{equation}
    \label{eq:tilde_s_min}
    s_\mathrm{min}'(\beta_0) \cdot \begin{pmatrix}\tilde X(\beta) \ D\end{pmatrix}^T M_{[Z \ D]} \begin{pmatrix}\tilde X(\beta) \ D\end{pmatrix} v = \begin{pmatrix}\tilde X(\beta) \ D\end{pmatrix}^T P_{[Z \ D]} \begin{pmatrix}\tilde X(\beta) \ D\end{pmatrix} v.
\end{equation}
The following proposition shows that $s_\mathrm{min}(\beta_0) = s_\mathrm{min}'(\beta_0)$, as long as $D$ is of full column rank.
\begin{theoremEnd}[malte,category=exogenous]{proposition}
    \label{prop:s_min_equivalence}
    Let $Z \in \BR^{n \times k}, \tilde X \in \BR^{n \times \mX}$, and $D \in \BR^{n \times \mD}$.
    Assume that $D$ is of full column rank.
    Then
    \begin{multline*}
        \left\{ \mu \in \BR \colon \det \left( \begin{pmatrix} \tilde X & D \end{pmatrix}^T P_{[Z, D]} \begin{pmatrix} \tilde X & D \end{pmatrix} - \mu \cdot \begin{pmatrix} \tilde X & D \end{pmatrix}^T M_{[Z, D]} \begin{pmatrix} \tilde X & D \end{pmatrix} \right) = 0 \right\} \\
          = \left\{ \mu \in \BR \colon \det \left( \tilde X P_{[M_D Z]} \tilde X - \mu \cdot \tilde X M_{[Z, D]} \tilde X  \right) = 0 \right\}.
    \end{multline*}
\end{theoremEnd}%
\begin{proofEnd}%
    Consider the block matrix equality
    \begin{equation}
        \label{eq:s_min_equivalence:1}
        \det \begin{pmatrix} A & B \\ C & D \end{pmatrix} = \det(D) \det( A - B D^{-1} C)
    \end{equation}
    for invertible $D$.
    Calculate
    \begin{align*}
    \det & \left( \begin{pmatrix} \tilde X & D \end{pmatrix}^T P_{[Z, D]} \begin{pmatrix} \tilde X & D \end{pmatrix} - \mu \cdot \begin{pmatrix} \tilde X & D \end{pmatrix}^T M_{[Z, D]} \begin{pmatrix} \tilde X & D \end{pmatrix} \right) \\
    &= \det \begin{pmatrix} \tilde X P_{[Z, D]} \tilde X - \mu \cdot \tilde X^T M_{[Z, D]} \tilde X & \tilde X D \\ D^T \tilde X & D^T D \end{pmatrix} \\
    &\overset{\eqref{eq:s_min_equivalence:1}}{=} \det(D^T D) \det\left( \tilde X P_{[Z, D]} \tilde X - \mu \cdot \tilde X^T M_{[Z, D]} \tilde X - \tilde X^T D (D^T D)^{-1} D^T \tilde X \right) \\
    &= \det(D^T D) \det\left( \tilde X P_{M_D Z} \tilde X - \mu \cdot \tilde X^T M_{[D, Z]}\tilde X \right) .
    \end{align*}
\end{proofEnd}%

\begin{theoremEnd}[malte,category=exogenous]{lemma}
    \label{lem:clr_dominance}
    Let $\mD > 0$ and $s > 0$. Then
    \begin{align*}
        &\Gamma(k - \mX, \mD + \mX, s) \\
        &= \frac{1}{2} \left(Q_{k - \mX} + Q_{\mX} + Q_{\mD} - s
        + \sqrt{(Q_{k - \mX} + Q_{\mX} + Q_{\mD} + s)^2 - 4 Q_{k - \mX} s}\right)\\
        &> Q_\mD + \frac{1}{2} \left(Q_{k - \mX} + Q_{\mX} - s + \sqrt{(Q_{k - \mX} + Q_{\mX} + s)^2 - 4 Q_{k - \mX} s}\right)\\
        &= Q_{\mD} + \Gamma(k - \mX, \mX, s).
    \end{align*}
\end{theoremEnd}%
\begin{proofEnd}%
    Let $a := Q_{k - \mX} + Q_{\mX} + s$, b := $Q_{\mD}$, and $c:= 4 Q_{k - \mX} s$.
    These are all positive almost surely.
    Also, $a^2 - c = (Q_\mX + s)^2 + 2 (Q_\mX + s) \cdot Q_{k - \mX} + Q_{k - \mX}^2 - 4 Q_{k - \mX} s = (Q_\mX + s - Q_{k - \mX})^2 + 4 (Q_\mX) \cdot Q_{k - \mX} > 0$.
    We show that $b + \sqrt{a^2 - c} < \sqrt{(a + b)^2 - c}$ for all $a, b, c > 0$ such that $a^2 - c > 0$.
    \begin{align*}
        &\ a > \sqrt{a^2 - c} \\
        &\Rightarrow 2ab > 2b \sqrt{a^2 - c} &\mid \cdot 2 b > 0 \\
        &\Rightarrow (a + b)^2 - c = a^2 + 2ab + b^2 - c \\
        &\hspace{1cm} > a^2 - c + 2b \sqrt{a^2 - c} + b^2 = (b + \sqrt{a^2 - c})^2 &\mid + a^2 + b^2 - c\\
        &\Rightarrow \sqrt{(a + b)^2 - c} > b + \sqrt{a^2 - c}. &\mid \sqrt{\ } \text{ and } (a^2 + b^2) - c > a^2 - c > 0
    \end{align*}
    Thus
    \begin{align*}
        &\Gamma(k - \mX, \mD + \mX, s) \\
        &= \frac{1}{2} \Big(Q_{k - \mX} + Q_{\mX} + Q_{\mD} - s
        + \sqrt{(Q_{k - \mX} + Q_{\mX} + Q_{\mD} + s)^2 - 4 Q_{k - \mX} s}\Big)\\
        &> \frac{1}{2} \Big(Q_{k - \mX} + Q_{\mX} + Q_{\mD} - s + Q_{\mD} + \sqrt{(Q_{k - \mX} + Q_{\mX} + s)^2 - 4 Q_{k - \mX} s}\Big)\\
        &= Q_{\mD} + \Gamma(k - \mX, \mX, s).
    \end{align*}
\end{proofEnd}%

Thus, the distribution $\Gamma(k - \mX, \mX + \mD, s_\mathrm{min}(\beta_0))$ that would be obtained by applying \cref{prop:clr_test_statistic} after including $D$ in the instruments $Z$ and endogenous variables $X$ is stricly larger than $Q_\mD + \Gamma(k - \mX, \mX, s_\mathrm{min}(\beta_0))$, the bounding distribution of \cref{prop:clr_test_statistic_with_exogenous}.
Thus, applying \cref{prop:clr_test_statistic} leads to a conservative, but size correct test.
The test resulting from directly applying \cref{prop:clr_test_statistic_with_exogenous} is strictly more powerful.

Recall \citeauthor{kleibergen2021efficient}'s \citeyearpar{kleibergen2021efficient} subvector variant of the conditional likelihood-ratio test.
\subvectorclr*
\noindent Furthermore, \citet[][page 80]{kleibergen2021efficient} writes
\begin{quote}
    When we want to test a hypothesis on the parameters of the included exogenous variables, we just include them as elements of $X$.
\end{quote}
This would lead to a limiting distribution of $\LR(\beta_0, \delta_0) \leq \Gamma(k - m, \mD + \mX, \tilde s_\mathrm{min}(\beta_0))$, where $\lambda_1, \lambda_2$ are the smallest and second smallest elements of
\begin{multline*}
\left\{ \mu \colon \det\left[  \mu \cdot \begin{pmatrix} X \!\!& W \!\!&D \!\!& y \end{pmatrix}^T P_{[D, Z]} \begin{pmatrix} X \!\! & W \!\! & D \!\! & y \end{pmatrix} - \begin{pmatrix} X \!\! & W \!\! & D \!\! & y \end{pmatrix}^T M_{[Z, D]} \begin{pmatrix} X \!\! & W \!\! & D \!\! & y \end{pmatrix} \right] = 0 \right\} \\
=
\left\{ \mu \colon \det\left[  \mu \cdot \begin{pmatrix} X & W & y \end{pmatrix}^T P_{[M_D Z]} \begin{pmatrix} X & W & y \end{pmatrix} - \begin{pmatrix} X & W & y \end{pmatrix}^T M_{[Z, D]} \begin{pmatrix} X & W & y \end{pmatrix} \right] = 0 \right\}
\end{multline*}
(using \cref{prop:s_min_equivalence}),
\begin{align*}
\mu(\beta, \delta) &= \lambda_\mathrm{min} \Big(
    \left[ \begin{pmatrix} W & y - X \beta  \end{pmatrix}^T M_{[Z, D]} \begin{pmatrix} W & y - X \beta \end{pmatrix} \right]^{-1} \\&\hspace{3cm}  \begin{pmatrix} W & y - X \beta - D \delta \end{pmatrix}^T P_{[Z, D]} \begin{pmatrix} W & y - X \beta - D \delta \end{pmatrix}
\Big),
\end{align*}
and $\tilde s_\mathrm{min}(\beta, \delta) = \lambda_1 + \lambda_2 - \mu(\beta, \delta) \overset{\cref{thm:liml_is_kclass_with_exogenous}}{=} \lambda_2 - \LR(\beta_0, \delta_0)$.

Considering \cref{prop:clr_test_statistic_with_exogenous}, one could conjecture that the more powerful $\LR(\beta_0, \delta_0) \leq Q_\mD + \Gamma(k - \mX - \mW, \mX, \tilde s_\mathrm{min}(\beta_0))$ would also hold.
With $\mX = 0$, this would imply
\begin{align*}
\LR(\delta_0) &\leq  Q_\mD + \Gamma(k - \mW, 0, \tilde s_\mathrm{min}(\beta_0)) = Q_{\mD} + \max\left\{ Q_{k - \mW} - \tilde s_\mathrm{min}(\delta_0), 0 \right\}.
\end{align*}

\subsection{Application}

We repeat the analysis on the \citep{card1993using} dataset, this time treating the included exogenous regressors explicitly.
We are particularly interested in the causal effect of race (\texttt{black}) on log-hourly wages.
\begin{jupyternotebook}
\begin{tcolorbox}[breakable, size=fbox, boxrule=1pt, pad at break*=1mm,colback=cellbackground, colframe=cellborder]
\prompt{In}{incolor}{1}{\boxspacing}
\begin{Verbatim}[commandchars=\\\{\}]
\PY{p}{[}\PY{o}{.}\PY{o}{.}\PY{o}{.}\PY{p}{]}  \PY{c+c1}{\PYZsh{} Load data. See appendix A for details.}

\PY{c+c1}{\PYZsh{} construct potential experience and its square}
\PY{n}{df}\PY{p}{[}\PY{l+s+s2}{\PYZdq{}}\PY{l+s+s2}{exp76}\PY{l+s+s2}{\PYZdq{}}\PY{p}{]} \PY{o}{=} \PY{n}{df}\PY{p}{[}\PY{l+s+s2}{\PYZdq{}}\PY{l+s+s2}{age76}\PY{l+s+s2}{\PYZdq{}}\PY{p}{]} \PY{o}{\PYZhy{}} \PY{n}{df}\PY{p}{[}\PY{l+s+s2}{\PYZdq{}}\PY{l+s+s2}{ed76}\PY{l+s+s2}{\PYZdq{}}\PY{p}{]} \PY{o}{\PYZhy{}} \PY{l+m+mi}{6}
\PY{n}{df}\PY{p}{[}\PY{l+s+s2}{\PYZdq{}}\PY{l+s+s2}{exp762}\PY{l+s+s2}{\PYZdq{}}\PY{p}{]} \PY{o}{=} \PY{n}{df}\PY{p}{[}\PY{l+s+s2}{\PYZdq{}}\PY{l+s+s2}{exp76}\PY{l+s+s2}{\PYZdq{}}\PY{p}{]} \PY{o}{*}\PY{o}{*} \PY{l+m+mi}{2}
\PY{n}{df}\PY{p}{[}\PY{l+s+s2}{\PYZdq{}}\PY{l+s+s2}{age762}\PY{l+s+s2}{\PYZdq{}}\PY{p}{]} \PY{o}{=} \PY{n}{df}\PY{p}{[}\PY{l+s+s2}{\PYZdq{}}\PY{l+s+s2}{age76}\PY{l+s+s2}{\PYZdq{}}\PY{p}{]} \PY{o}{*}\PY{o}{*} \PY{l+m+mi}{2}

\PY{c+c1}{\PYZsh{} endogenous variables: years of education, experience, experience squared}
\PY{n}{X} \PY{o}{=} \PY{n}{df}\PY{p}{[}\PY{p}{[}\PY{l+s+s2}{\PYZdq{}}\PY{l+s+s2}{ed76}\PY{l+s+s2}{\PYZdq{}}\PY{p}{,} \PY{l+s+s2}{\PYZdq{}}\PY{l+s+s2}{exp76}\PY{l+s+s2}{\PYZdq{}}\PY{p}{,} \PY{l+s+s2}{\PYZdq{}}\PY{l+s+s2}{exp762}\PY{l+s+s2}{\PYZdq{}}\PY{p}{]}\PY{p}{]}
\PY{n}{y} \PY{o}{=} \PY{n}{df}\PY{p}{[}\PY{l+s+s2}{\PYZdq{}}\PY{l+s+s2}{lwage76}\PY{l+s+s2}{\PYZdq{}}\PY{p}{]}  \PY{c+c1}{\PYZsh{} outcome: log wage}
\PY{c+c1}{\PYZsh{} included exogenous variables: indicators for family background,}
\PY{c+c1}{\PYZsh{} region, and race.}
\PY{n}{C} \PY{o}{=} \PY{n}{df}\PY{p}{[}\PY{n}{family} \PY{o}{+} \PY{n}{indicators}\PY{p}{]}
\PY{c+c1}{\PYZsh{} instruments: proximity to colleges, age, and age squared}
\PY{n}{Z} \PY{o}{=} \PY{n}{df}\PY{p}{[}\PY{p}{[}\PY{l+s+s2}{\PYZdq{}}\PY{l+s+s2}{nearc4a}\PY{l+s+s2}{\PYZdq{}}\PY{p}{,} \PY{l+s+s2}{\PYZdq{}}\PY{l+s+s2}{nearc4b}\PY{l+s+s2}{\PYZdq{}}\PY{p}{,} \PY{l+s+s2}{\PYZdq{}}\PY{l+s+s2}{nearc2}\PY{l+s+s2}{\PYZdq{}}\PY{p}{,} \PY{l+s+s2}{\PYZdq{}}\PY{l+s+s2}{age76}\PY{l+s+s2}{\PYZdq{}}\PY{p}{,} \PY{l+s+s2}{\PYZdq{}}\PY{l+s+s2}{age762}\PY{l+s+s2}{\PYZdq{}}\PY{p}{]}\PY{p}{]}
\end{Verbatim}
\end{tcolorbox}
\end{jupyternotebook}

We compare the ordinary least-squares (ols), two-stage least-squares (TSLS), and limited information maximum likelihood (LIML) estimators.

\begin{jupyternotebook}
    \begin{tcolorbox}[breakable, size=fbox, boxrule=1pt, pad at break*=1mm,colback=cellbackground, colframe=cellborder]
\prompt{In}{incolor}{2}{\boxspacing}
\begin{Verbatim}[commandchars=\\\{\}]
\PY{k+kn}{from} \PY{n+nn}{ivmodels} \PY{k+kn}{import} \PY{n}{KClass}

\PY{n}{ols} \PY{o}{=} \PY{n}{KClass}\PY{p}{(}\PY{n}{kappa}\PY{o}{=}\PY{l+s+s2}{\PYZdq{}}\PY{l+s+s2}{ols}\PY{l+s+s2}{\PYZdq{}}\PY{p}{)}\PY{o}{.}\PY{n}{fit}\PY{p}{(}\PY{n}{Z}\PY{o}{=}\PY{k+kc}{None}\PY{p}{,} \PY{n}{X}\PY{o}{=}\PY{n}{X}\PY{p}{,} \PY{n}{C}\PY{o}{=}\PY{n}{C}\PY{p}{,} \PY{n}{y}\PY{o}{=}\PY{n}{y}\PY{p}{)}
\PY{n}{ols}\PY{o}{.}\PY{n}{named\PYZus{}coef\PYZus{}}\PY{p}{[}\PY{p}{:}\PY{l+m+mi}{5}\PY{p}{]}
\end{Verbatim}
\end{tcolorbox}

            \begin{tcolorbox}[breakable, size=fbox, boxrule=.5pt, pad at break*=1mm, opacityfill=0]
\prompt{Out}{outcolor}{2}{\boxspacing}
\begin{Verbatim}[commandchars=\\\{\}]
intercept    4.040851
ed76         0.072634
exp76        0.084529
exp762      -0.002290
black       -0.189408
Name: coefficients, dtype: float64
\end{Verbatim}
\end{tcolorbox}

    \begin{tcolorbox}[breakable, size=fbox, boxrule=1pt, pad at break*=1mm,colback=cellbackground, colframe=cellborder]
\prompt{In}{incolor}{3}{\boxspacing}
\begin{Verbatim}[commandchars=\\\{\}]
\PY{n}{tsls} \PY{o}{=} \PY{n}{KClass}\PY{p}{(}\PY{n}{kappa}\PY{o}{=}\PY{l+s+s2}{\PYZdq{}}\PY{l+s+s2}{tsls}\PY{l+s+s2}{\PYZdq{}}\PY{p}{)}\PY{o}{.}\PY{n}{fit}\PY{p}{(}\PY{n}{Z}\PY{o}{=}\PY{n}{Z}\PY{p}{,} \PY{n}{X}\PY{o}{=}\PY{n}{X}\PY{p}{,} \PY{n}{C}\PY{o}{=}\PY{n}{C}\PY{p}{,} \PY{n}{y}\PY{o}{=}\PY{n}{y}\PY{p}{)}
\PY{n}{tsls}\PY{o}{.}\PY{n}{named\PYZus{}coef\PYZus{}}\PY{p}{[}\PY{p}{:}\PY{l+m+mi}{5}\PY{p}{]}
\end{Verbatim}
\end{tcolorbox}

            \begin{tcolorbox}[breakable, size=fbox, boxrule=.5pt, pad at break*=1mm, opacityfill=0]
\prompt{Out}{outcolor}{3}{\boxspacing}
\begin{Verbatim}[commandchars=\\\{\}]
intercept    3.011786
ed76         0.144954
exp76        0.061604
exp762      -0.001196
black       -0.159219
Name: coefficients, dtype: float64
\end{Verbatim}
\end{tcolorbox}
        
    \begin{tcolorbox}[breakable, size=fbox, boxrule=1pt, pad at break*=1mm,colback=cellbackground, colframe=cellborder]
\prompt{In}{incolor}{4}{\boxspacing}
\begin{Verbatim}[commandchars=\\\{\}]
\PY{n}{liml} \PY{o}{=} \PY{n}{KClass}\PY{p}{(}\PY{n}{kappa}\PY{o}{=}\PY{l+s+s2}{\PYZdq{}}\PY{l+s+s2}{liml}\PY{l+s+s2}{\PYZdq{}}\PY{p}{)}\PY{o}{.}\PY{n}{fit}\PY{p}{(}\PY{n}{Z}\PY{o}{=}\PY{n}{Z}\PY{p}{,} \PY{n}{X}\PY{o}{=}\PY{n}{X}\PY{p}{,} \PY{n}{C}\PY{o}{=}\PY{n}{C}\PY{p}{,} \PY{n}{y}\PY{o}{=}\PY{n}{y}\PY{p}{)}
\PY{n}{liml}\PY{o}{.}\PY{n}{named\PYZus{}coef\PYZus{}}\PY{p}{[}\PY{p}{:}\PY{l+m+mi}{5}\PY{p}{]}
\end{Verbatim}
\end{tcolorbox}

            \begin{tcolorbox}[breakable, size=fbox, boxrule=.5pt, pad at break*=1mm, opacityfill=0]
\prompt{Out}{outcolor}{4}{\boxspacing}
\begin{Verbatim}[commandchars=\\\{\}]
intercept    2.627637
ed76         0.172352
exp76        0.051571
exp762      -0.000713
black       -0.147746
Name: coefficients, dtype: float64
\end{Verbatim}
\end{tcolorbox}
\end{jupyternotebook}

\noindent All coefficients, except for the intercept which differs as \(C\) is not
centered, are exactly the same as in section 3.

We test the null hypothesis
\[H_0 \colon \text{ The causal effect of race on log wages } \delta_0 = 0\]
using the subvector Wald, Anderson-Rubin, likelihood-ratio, and lagrange
multiplier tests.

\begin{jupyternotebook}
    \begin{tcolorbox}[breakable, size=fbox, boxrule=1pt, pad at break*=1mm,colback=cellbackground, colframe=cellborder]
\prompt{In}{incolor}{5}{\boxspacing}
\begin{Verbatim}[commandchars=\\\{\}]
\PY{k+kn}{from} \PY{n+nn}{functools} \PY{k+kn}{import} \PY{n}{partial}
\PY{k+kn}{from} \PY{n+nn}{ivmodels}\PY{n+nn}{.}\PY{n+nn}{tests} \PY{k+kn}{import} \PY{p}{(}
    \PY{n}{anderson\PYZus{}rubin\PYZus{}test}\PY{p}{,}
    \PY{n}{conditional\PYZus{}likelihood\PYZus{}ratio\PYZus{}test}\PY{p}{,}
    \PY{n}{lagrange\PYZus{}multiplier\PYZus{}test}\PY{p}{,}
    \PY{n}{likelihood\PYZus{}ratio\PYZus{}test}\PY{p}{,}
    \PY{n}{wald\PYZus{}test}\PY{p}{,}
\PY{p}{)}

\PY{n}{C}\PY{p}{,} \PY{n}{D} \PY{o}{=} \PY{n}{C}\PY{o}{.}\PY{n}{drop}\PY{p}{(}\PY{n}{columns}\PY{o}{=}\PY{l+s+s2}{\PYZdq{}}\PY{l+s+s2}{black}\PY{l+s+s2}{\PYZdq{}}\PY{p}{)}\PY{p}{,} \PY{n}{C}\PY{p}{[}\PY{p}{[}\PY{l+s+s2}{\PYZdq{}}\PY{l+s+s2}{black}\PY{l+s+s2}{\PYZdq{}}\PY{p}{]}\PY{p}{]}

\PY{k}{for} \PY{n}{test}\PY{p}{,} \PY{n}{name} \PY{o+ow}{in} \PY{p}{[}
    \PY{p}{(}\PY{n}{partial}\PY{p}{(}\PY{n}{wald\PYZus{}test}\PY{p}{,} \PY{n}{estimator}\PY{o}{=}\PY{l+s+s2}{\PYZdq{}}\PY{l+s+s2}{tsls}\PY{l+s+s2}{\PYZdq{}}\PY{p}{)}\PY{p}{,} \PY{l+s+s2}{\PYZdq{}}\PY{l+s+s2}{Wald (TSLS)}\PY{l+s+s2}{\PYZdq{}}\PY{p}{)}\PY{p}{,}
    \PY{p}{(}\PY{n}{partial}\PY{p}{(}\PY{n}{wald\PYZus{}test}\PY{p}{,} \PY{n}{estimator}\PY{o}{=}\PY{l+s+s2}{\PYZdq{}}\PY{l+s+s2}{liml}\PY{l+s+s2}{\PYZdq{}}\PY{p}{)}\PY{p}{,} \PY{l+s+s2}{\PYZdq{}}\PY{l+s+s2}{Wald (LIML)}\PY{l+s+s2}{\PYZdq{}}\PY{p}{)}\PY{p}{,}
    \PY{p}{(}\PY{n}{anderson\PYZus{}rubin\PYZus{}test}\PY{p}{,} \PY{l+s+s2}{\PYZdq{}}\PY{l+s+s2}{AR}\PY{l+s+s2}{\PYZdq{}}\PY{p}{)}\PY{p}{,}
    \PY{p}{(}\PY{n}{likelihood\PYZus{}ratio\PYZus{}test}\PY{p}{,} \PY{l+s+s2}{\PYZdq{}}\PY{l+s+s2}{LR}\PY{l+s+s2}{\PYZdq{}}\PY{p}{)}\PY{p}{,}
    \PY{p}{(}\PY{n}{conditional\PYZus{}likelihood\PYZus{}ratio\PYZus{}test}\PY{p}{,} \PY{l+s+s2}{\PYZdq{}}\PY{l+s+s2}{CLR}\PY{l+s+s2}{\PYZdq{}}\PY{p}{)}\PY{p}{,}
    \PY{p}{(}\PY{n}{lagrange\PYZus{}multiplier\PYZus{}test}\PY{p}{,} \PY{l+s+s2}{\PYZdq{}}\PY{l+s+s2}{LM}\PY{l+s+s2}{\PYZdq{}}\PY{p}{)}\PY{p}{,}
\PY{p}{]}\PY{p}{:}
    \PY{n}{stat}\PY{p}{,} \PY{n}{pval} \PY{o}{=} \PY{n}{test}\PY{p}{(}\PY{n}{Z}\PY{o}{=}\PY{n}{Z}\PY{p}{,} \PY{n}{X}\PY{o}{=}\PY{k+kc}{None}\PY{p}{,} \PY{n}{W}\PY{o}{=}\PY{n}{X}\PY{p}{,} \PY{n}{C}\PY{o}{=}\PY{n}{C}\PY{p}{,} \PY{n}{D}\PY{o}{=}\PY{n}{D}\PY{p}{,} \PY{n}{y}\PY{o}{=}\PY{n}{y}\PY{p}{,} \PY{n}{beta}\PY{o}{=}\PY{n}{np}\PY{o}{.}\PY{n}{array}\PY{p}{(}\PY{p}{[}\PY{l+m+mf}{0.}\PY{p}{]}\PY{p}{)}\PY{p}{)}
    \PY{n+nb}{print}\PY{p}{(}\PY{l+s+sa}{f}\PY{l+s+s2}{\PYZdq{}}\PY{l+s+si}{\PYZob{}}\PY{n}{name}\PY{l+s+si}{:}\PY{l+s+s2}{\PYZlt{}11}\PY{l+s+si}{\PYZcb{}}\PY{l+s+s2}{: statistic=}\PY{l+s+si}{\PYZob{}}\PY{n}{stat}\PY{l+s+si}{:}\PY{l+s+s2}{5.2f}\PY{l+s+si}{\PYZcb{}}\PY{l+s+s2}{, p\PYZhy{}value=}\PY{l+s+si}{\PYZob{}}\PY{n}{pval}\PY{l+s+si}{:}\PY{l+s+s2}{.3g}\PY{l+s+si}{\PYZcb{}}\PY{l+s+s2}{\PYZdq{}}\PY{p}{)}
\end{Verbatim}
\end{tcolorbox}
    \begin{Verbatim}[commandchars=\\\{\}]
Wald (TSLS): statistic=31.60, p-value=1.89e-08
Wald (LIML): statistic=20.26, p-value=6.75e-06
AR         : statistic= 3.27, p-value=0.0204
LR         : statistic= 5.55, p-value=0.0185
CLR        : statistic= 5.55, p-value=0.0275
LM         : statistic= 4.03, p-value=0.0448
    \end{Verbatim}
\end{jupyternotebook}

    The causal effect of race on log-hourly wages is significant at the 5\%
level for all tests. The Wald-based \(p\)-values are much smaller than
those resulting from the likelihood-ratio and robust tests.

\begin{jupyternotebook}
    \begin{tcolorbox}[breakable, size=fbox, boxrule=1pt, pad at break*=1mm,colback=cellbackground, colframe=cellborder]
\prompt{In}{incolor}{6}{\boxspacing}
\begin{Verbatim}[commandchars=\\\{\}]
\PY{k+kn}{from} \PY{n+nn}{ivmodels}\PY{n+nn}{.}\PY{n+nn}{tests} \PY{k+kn}{import} \PY{n}{j\PYZus{}test}\PY{p}{,} \PY{n}{rank\PYZus{}test}

\PY{n}{DC} \PY{o}{=} \PY{n}{pd}\PY{o}{.}\PY{n}{concat}\PY{p}{(}\PY{p}{[}\PY{n}{D}\PY{p}{,} \PY{n}{C}\PY{p}{]}\PY{p}{,} \PY{n}{axis}\PY{o}{=}\PY{l+m+mi}{1}\PY{p}{)}

\PY{n}{j\PYZus{}stat}\PY{p}{,} \PY{n}{j\PYZus{}pval} \PY{o}{=} \PY{n}{j\PYZus{}test}\PY{p}{(}\PY{n}{Z}\PY{o}{=}\PY{n}{Z}\PY{p}{,} \PY{n}{X}\PY{o}{=}\PY{n}{X}\PY{p}{,} \PY{n}{C}\PY{o}{=}\PY{n}{DC}\PY{p}{,} \PY{n}{y}\PY{o}{=}\PY{n}{y}\PY{p}{,} \PY{n}{estimator}\PY{o}{=}\PY{l+s+s2}{\PYZdq{}}\PY{l+s+s2}{liml}\PY{l+s+s2}{\PYZdq{}}\PY{p}{)}
\PY{n+nb}{print}\PY{p}{(}\PY{l+s+sa}{f}\PY{l+s+s2}{\PYZdq{}}\PY{l+s+s2}{J\PYZhy{}statistic   : }\PY{l+s+si}{\PYZob{}}\PY{n}{j\PYZus{}stat}\PY{l+s+si}{:}\PY{l+s+s2}{6.3f}\PY{l+s+si}{\PYZcb{}}\PY{l+s+s2}{, p\PYZhy{}value: }\PY{l+s+si}{\PYZob{}}\PY{n}{j\PYZus{}pval}\PY{l+s+si}{:}\PY{l+s+s2}{.4f}\PY{l+s+si}{\PYZcb{}}\PY{l+s+s2}{\PYZdq{}}\PY{p}{)}
\PY{n}{rank\PYZus{}stat}\PY{p}{,} \PY{n}{rank\PYZus{}pval} \PY{o}{=} \PY{n}{rank\PYZus{}test}\PY{p}{(}\PY{n}{Z}\PY{o}{=}\PY{n}{Z}\PY{p}{,} \PY{n}{X}\PY{o}{=}\PY{n}{X}\PY{p}{,} \PY{n}{C}\PY{o}{=}\PY{n}{DC}\PY{p}{)}
\PY{n+nb}{print}\PY{p}{(}\PY{l+s+sa}{f}\PY{l+s+s2}{\PYZdq{}}\PY{l+s+s2}{Rank statistic: }\PY{l+s+si}{\PYZob{}}\PY{n}{rank\PYZus{}stat}\PY{l+s+si}{:}\PY{l+s+s2}{6.3f}\PY{l+s+si}{\PYZcb{}}\PY{l+s+s2}{, p\PYZhy{}value: }\PY{l+s+si}{\PYZob{}}\PY{n}{rank\PYZus{}pval}\PY{l+s+si}{:}\PY{l+s+s2}{.4f}\PY{l+s+si}{\PYZcb{}}\PY{l+s+s2}{\PYZdq{}}\PY{p}{)}
\end{Verbatim}
\end{tcolorbox}

    \begin{Verbatim}[commandchars=\\\{\}]
J-statistic   :  4.247, p-value: 0.1196
Rank statistic: 15.478, p-value: 0.0015
    \end{Verbatim}
\end{jupyternotebook}
    These are almost the same as in section 5, except that we now use the
more accurate \(n - k - m_c - m_d\) degrees of freedom in the numerator,
leading to slightly smaller test statistics.

\begin{jupyternotebook}
\begin{tcolorbox}[breakable, size=fbox, boxrule=1pt, pad at break*=1mm,colback=cellbackground, colframe=cellborder]
\prompt{In}{incolor}{7}{\boxspacing}
\begin{Verbatim}[commandchars=\\\{\}]
\PY{k+kn}{from} \PY{n+nn}{ivmodels}\PY{n+nn}{.}\PY{n+nn}{tests} \PY{k+kn}{import} \PY{p}{(}
    \PY{n}{inverse\PYZus{}anderson\PYZus{}rubin\PYZus{}test}\PY{p}{,}
    \PY{n}{inverse\PYZus{}conditional\PYZus{}likelihood\PYZus{}ratio\PYZus{}test}\PY{p}{,}
    \PY{n}{inverse\PYZus{}lagrange\PYZus{}multiplier\PYZus{}test}\PY{p}{,}
    \PY{n}{inverse\PYZus{}likelihood\PYZus{}ratio\PYZus{}test}\PY{p}{,}
    \PY{n}{inverse\PYZus{}wald\PYZus{}test}\PY{p}{,}
\PY{p}{)}

\PY{k}{for} \PY{n}{name}\PY{p}{,} \PY{n}{inverse\PYZus{}test} \PY{o+ow}{in} \PY{p}{[}
    \PY{p}{(}\PY{l+s+s2}{\PYZdq{}}\PY{l+s+s2}{Wald (TSLS)}\PY{l+s+s2}{\PYZdq{}}\PY{p}{,} \PY{n}{partial}\PY{p}{(}\PY{n}{inverse\PYZus{}wald\PYZus{}test}\PY{p}{,} \PY{n}{estimator}\PY{o}{=}\PY{l+s+s2}{\PYZdq{}}\PY{l+s+s2}{tsls}\PY{l+s+s2}{\PYZdq{}}\PY{p}{)}\PY{p}{)}\PY{p}{,}
    \PY{p}{(}\PY{l+s+s2}{\PYZdq{}}\PY{l+s+s2}{Wald (LIML)}\PY{l+s+s2}{\PYZdq{}}\PY{p}{,} \PY{n}{partial}\PY{p}{(}\PY{n}{inverse\PYZus{}wald\PYZus{}test}\PY{p}{,} \PY{n}{estimator}\PY{o}{=}\PY{l+s+s2}{\PYZdq{}}\PY{l+s+s2}{liml}\PY{l+s+s2}{\PYZdq{}}\PY{p}{)}\PY{p}{)}\PY{p}{,}
    \PY{p}{(}\PY{l+s+s2}{\PYZdq{}}\PY{l+s+s2}{AR}\PY{l+s+s2}{\PYZdq{}}\PY{p}{,} \PY{n}{inverse\PYZus{}anderson\PYZus{}rubin\PYZus{}test}\PY{p}{)}\PY{p}{,}
    \PY{p}{(}\PY{l+s+s2}{\PYZdq{}}\PY{l+s+s2}{CLR}\PY{l+s+s2}{\PYZdq{}}\PY{p}{,} \PY{n}{inverse\PYZus{}conditional\PYZus{}likelihood\PYZus{}ratio\PYZus{}test}\PY{p}{)}\PY{p}{,}
    \PY{p}{(}\PY{l+s+s2}{\PYZdq{}}\PY{l+s+s2}{LR}\PY{l+s+s2}{\PYZdq{}}\PY{p}{,} \PY{n}{inverse\PYZus{}likelihood\PYZus{}ratio\PYZus{}test}\PY{p}{)}\PY{p}{,}
    \PY{p}{(}\PY{l+s+s2}{\PYZdq{}}\PY{l+s+s2}{LM}\PY{l+s+s2}{\PYZdq{}}\PY{p}{,} \PY{n}{inverse\PYZus{}lagrange\PYZus{}multiplier\PYZus{}test}\PY{p}{)}\PY{p}{,}
\PY{p}{]}\PY{p}{:}
    \PY{n+nb}{print}\PY{p}{(}\PY{l+s+sa}{f}\PY{l+s+s2}{\PYZdq{}}\PY{l+s+si}{\PYZob{}}\PY{n}{name}\PY{l+s+si}{:}\PY{l+s+s2}{\PYZlt{}11}\PY{l+s+si}{\PYZcb{}}\PY{l+s+s2}{: }\PY{l+s+si}{\PYZob{}}\PY{n}{inverse\PYZus{}test}\PY{p}{(}\PY{n}{Z}\PY{p}{,}\PY{+w}{ }\PY{n}{X}\PY{o}{=}\PY{k+kc}{None}\PY{p}{,}\PY{+w}{ }\PY{n}{W}\PY{o}{=}\PY{n}{X}\PY{p}{,}\PY{+w}{ }\PY{n}{D}\PY{o}{=}\PY{n}{D}\PY{p}{,}\PY{+w}{ }\PY{n}{C}\PY{o}{=}\PY{n}{C}\PY{p}{,}\PY{+w}{ }\PY{n}{y}\PY{o}{=}\PY{n}{y}\PY{p}{)}\PY{l+s+si}{:}\PY{l+s+s2}{.3f}\PY{l+s+si}{\PYZcb{}}\PY{l+s+s2}{\PYZdq{}}\PY{p}{)}
\end{Verbatim}
\end{tcolorbox}
    \begin{Verbatim}[commandchars=\\\{\}]
Wald (TSLS): [-0.215, -0.104]
Wald (LIML): [-0.212, -0.083]
AR         : [-0.202, -0.055]
CLR        : [-0.207, -0.036]
LR         : [-0.204, -0.049]
LM         : [-0.490, -0.011]
    \end{Verbatim}
\end{jupyternotebook}
\section*{Acknowledgements}
Malte Londschien is supported by the ETH Foundations of Data Science and the ETH AI Center.
Malte Londschien would like to thank Cyrill Scheidegger, Fabio Sigrist, Felix Kuchelmeister, Frank Kleibergen, Gianna Wolfisberg, Jonas Peters, Juan Gamella, Leonard Henckel, Markus Ulmer, Maybritt Schillinger, Michael Law, Peter Bühlmann, and Zijian Guo for helpful discussions and comments.

\bibliography{bib}

\appendix
\section{Code to load the \citet{card1993using} data set}
\label{sec:omitted_code}
Below is the code to load the \citet{card1993using} dataset used in the application.%

\begin{jupyternotebook}
    \begin{tcolorbox}[breakable, size=fbox, boxrule=1pt, pad at break*=1mm,colback=cellbackground, colframe=cellborder]
\prompt{In}{incolor}{1}{\boxspacing}
\begin{Verbatim}[commandchars=\\\{\}]
\PY{k+kn}{from} \PY{n+nn}{io} \PY{k+kn}{import} \PY{n}{BytesIO}
\PY{k+kn}{from} \PY{n+nn}{zipfile} \PY{k+kn}{import} \PY{n}{ZipFile}

\PY{k+kn}{import} \PY{n+nn}{numpy} \PY{k}{as} \PY{n+nn}{np}
\PY{k+kn}{import} \PY{n+nn}{pandas} \PY{k}{as} \PY{n+nn}{pd}
\PY{k+kn}{import} \PY{n+nn}{requests}

\PY{n}{url} \PY{o}{=} \PY{l+s+s2}{\PYZdq{}}\PY{l+s+s2}{https://davidcard.berkeley.edu/data\PYZus{}sets/proximity.zip}\PY{l+s+s2}{\PYZdq{}}
\PY{n}{content} \PY{o}{=} \PY{n}{requests}\PY{o}{.}\PY{n}{get}\PY{p}{(}\PY{n}{url}\PY{p}{)}\PY{o}{.}\PY{n}{content}

\PY{c+c1}{\PYZsh{} From code\PYZus{}bk.txt in the zip file}
\PY{n}{colspec} \PY{o}{=} \PY{p}{\PYZob{}}
    \PY{l+s+s2}{\PYZdq{}}\PY{l+s+s2}{id}\PY{l+s+s2}{\PYZdq{}}\PY{p}{:} \PY{p}{(}\PY{l+m+mi}{1}\PY{p}{,} \PY{l+m+mi}{5}\PY{p}{)}\PY{p}{,}  \PY{c+c1}{\PYZsh{} sequential id runs from 1 to 5225}
    \PY{l+s+s2}{\PYZdq{}}\PY{l+s+s2}{nearc2}\PY{l+s+s2}{\PYZdq{}}\PY{p}{:} \PY{p}{(}\PY{l+m+mi}{7}\PY{p}{,} \PY{l+m+mi}{7}\PY{p}{)}\PY{p}{,}  \PY{c+c1}{\PYZsh{} grew up near 2\PYZhy{}yr college}
    \PY{l+s+s2}{\PYZdq{}}\PY{l+s+s2}{nearc4}\PY{l+s+s2}{\PYZdq{}}\PY{p}{:} \PY{p}{(}\PY{l+m+mi}{10}\PY{p}{,} \PY{l+m+mi}{10}\PY{p}{)}\PY{p}{,}  \PY{c+c1}{\PYZsh{} grew up near 4\PYZhy{}yr college}
    \PY{l+s+s2}{\PYZdq{}}\PY{l+s+s2}{nearc4a}\PY{l+s+s2}{\PYZdq{}}\PY{p}{:} \PY{p}{(}\PY{l+m+mi}{12}\PY{p}{,} \PY{l+m+mi}{13}\PY{p}{)}\PY{p}{,}  \PY{c+c1}{\PYZsh{} grew up near 4\PYZhy{}yr public college}
    \PY{l+s+s2}{\PYZdq{}}\PY{l+s+s2}{nearc4b}\PY{l+s+s2}{\PYZdq{}}\PY{p}{:} \PY{p}{(}\PY{l+m+mi}{15}\PY{p}{,} \PY{l+m+mi}{16}\PY{p}{)}\PY{p}{,}  \PY{c+c1}{\PYZsh{} grew up near 4\PYZhy{}yr priv college}
    \PY{l+s+s2}{\PYZdq{}}\PY{l+s+s2}{ed76}\PY{l+s+s2}{\PYZdq{}}\PY{p}{:} \PY{p}{(}\PY{l+m+mi}{18}\PY{p}{,} \PY{l+m+mi}{19}\PY{p}{)}\PY{p}{,}  \PY{c+c1}{\PYZsh{} educ in 1976}
    \PY{l+s+s2}{\PYZdq{}}\PY{l+s+s2}{ed66}\PY{l+s+s2}{\PYZdq{}}\PY{p}{:} \PY{p}{(}\PY{l+m+mi}{21}\PY{p}{,} \PY{l+m+mi}{22}\PY{p}{)}\PY{p}{,}  \PY{c+c1}{\PYZsh{} educ in 1966}
    \PY{l+s+s2}{\PYZdq{}}\PY{l+s+s2}{age76}\PY{l+s+s2}{\PYZdq{}}\PY{p}{:} \PY{p}{(}\PY{l+m+mi}{24}\PY{p}{,} \PY{l+m+mi}{25}\PY{p}{)}\PY{p}{,}  \PY{c+c1}{\PYZsh{} age in 1976}
    \PY{l+s+s2}{\PYZdq{}}\PY{l+s+s2}{daded}\PY{l+s+s2}{\PYZdq{}}\PY{p}{:} \PY{p}{(}\PY{l+m+mi}{27}\PY{p}{,} \PY{l+m+mi}{31}\PY{p}{)}\PY{p}{,}  \PY{c+c1}{\PYZsh{} dads education missing=avg}
    \PY{l+s+s2}{\PYZdq{}}\PY{l+s+s2}{nodaded}\PY{l+s+s2}{\PYZdq{}}\PY{p}{:} \PY{p}{(}\PY{l+m+mi}{33}\PY{p}{,} \PY{l+m+mi}{33}\PY{p}{)}\PY{p}{,}  \PY{c+c1}{\PYZsh{} 1 if dad ed imputed}
    \PY{l+s+s2}{\PYZdq{}}\PY{l+s+s2}{momed}\PY{l+s+s2}{\PYZdq{}}\PY{p}{:} \PY{p}{(}\PY{l+m+mi}{35}\PY{p}{,} \PY{l+m+mi}{39}\PY{p}{)}\PY{p}{,}  \PY{c+c1}{\PYZsh{} moms education}
    \PY{l+s+s2}{\PYZdq{}}\PY{l+s+s2}{nomomed}\PY{l+s+s2}{\PYZdq{}}\PY{p}{:} \PY{p}{(}\PY{l+m+mi}{41}\PY{p}{,} \PY{l+m+mi}{41}\PY{p}{)}\PY{p}{,}  \PY{c+c1}{\PYZsh{} 1 if mom ed imputed}
    \PY{l+s+s2}{\PYZdq{}}\PY{l+s+s2}{weight}\PY{l+s+s2}{\PYZdq{}}\PY{p}{:} \PY{p}{(}\PY{l+m+mi}{43}\PY{p}{,} \PY{l+m+mi}{54}\PY{p}{)}\PY{p}{,}  \PY{c+c1}{\PYZsh{} nls weight for 1976 cross\PYZhy{}section}
    \PY{l+s+s2}{\PYZdq{}}\PY{l+s+s2}{momdad14}\PY{l+s+s2}{\PYZdq{}}\PY{p}{:} \PY{p}{(}\PY{l+m+mi}{56}\PY{p}{,} \PY{l+m+mi}{56}\PY{p}{)}\PY{p}{,}  \PY{c+c1}{\PYZsh{} 1 if live with mom and dad age 14}
    \PY{l+s+s2}{\PYZdq{}}\PY{l+s+s2}{sinmom14}\PY{l+s+s2}{\PYZdq{}}\PY{p}{:} \PY{p}{(}\PY{l+m+mi}{58}\PY{p}{,} \PY{l+m+mi}{58}\PY{p}{)}\PY{p}{,}  \PY{c+c1}{\PYZsh{} lived with single mom age 14}
    \PY{l+s+s2}{\PYZdq{}}\PY{l+s+s2}{step14}\PY{l+s+s2}{\PYZdq{}}\PY{p}{:} \PY{p}{(}\PY{l+m+mi}{60}\PY{p}{,} \PY{l+m+mi}{60}\PY{p}{)}\PY{p}{,}  \PY{c+c1}{\PYZsh{} lived step parent age 14}
    \PY{l+s+s2}{\PYZdq{}}\PY{l+s+s2}{reg661}\PY{l+s+s2}{\PYZdq{}}\PY{p}{:} \PY{p}{(}\PY{l+m+mi}{62}\PY{p}{,} \PY{l+m+mi}{62}\PY{p}{)}\PY{p}{,}  \PY{c+c1}{\PYZsh{} dummy for region=1 in 1966}
    \PY{l+s+s2}{\PYZdq{}}\PY{l+s+s2}{reg662}\PY{l+s+s2}{\PYZdq{}}\PY{p}{:} \PY{p}{(}\PY{l+m+mi}{64}\PY{p}{,} \PY{l+m+mi}{64}\PY{p}{)}\PY{p}{,}  \PY{c+c1}{\PYZsh{} dummy for region=2 in 1966}
    \PY{l+s+s2}{\PYZdq{}}\PY{l+s+s2}{reg663}\PY{l+s+s2}{\PYZdq{}}\PY{p}{:} \PY{p}{(}\PY{l+m+mi}{66}\PY{p}{,} \PY{l+m+mi}{66}\PY{p}{)}\PY{p}{,}  \PY{c+c1}{\PYZsh{} dummy for region=3 in 1966}
    \PY{l+s+s2}{\PYZdq{}}\PY{l+s+s2}{reg664}\PY{l+s+s2}{\PYZdq{}}\PY{p}{:} \PY{p}{(}\PY{l+m+mi}{68}\PY{p}{,} \PY{l+m+mi}{68}\PY{p}{)}\PY{p}{,}
    \PY{l+s+s2}{\PYZdq{}}\PY{l+s+s2}{reg665}\PY{l+s+s2}{\PYZdq{}}\PY{p}{:} \PY{p}{(}\PY{l+m+mi}{70}\PY{p}{,} \PY{l+m+mi}{70}\PY{p}{)}\PY{p}{,}
    \PY{l+s+s2}{\PYZdq{}}\PY{l+s+s2}{reg666}\PY{l+s+s2}{\PYZdq{}}\PY{p}{:} \PY{p}{(}\PY{l+m+mi}{72}\PY{p}{,} \PY{l+m+mi}{72}\PY{p}{)}\PY{p}{,}
    \PY{l+s+s2}{\PYZdq{}}\PY{l+s+s2}{reg667}\PY{l+s+s2}{\PYZdq{}}\PY{p}{:} \PY{p}{(}\PY{l+m+mi}{74}\PY{p}{,} \PY{l+m+mi}{74}\PY{p}{)}\PY{p}{,}
    \PY{l+s+s2}{\PYZdq{}}\PY{l+s+s2}{reg668}\PY{l+s+s2}{\PYZdq{}}\PY{p}{:} \PY{p}{(}\PY{l+m+mi}{76}\PY{p}{,} \PY{l+m+mi}{76}\PY{p}{)}\PY{p}{,}
    \PY{l+s+s2}{\PYZdq{}}\PY{l+s+s2}{reg669}\PY{l+s+s2}{\PYZdq{}}\PY{p}{:} \PY{p}{(}\PY{l+m+mi}{78}\PY{p}{,} \PY{l+m+mi}{78}\PY{p}{)}\PY{p}{,}  \PY{c+c1}{\PYZsh{} dummy for region=9 in 1966}
    \PY{l+s+s2}{\PYZdq{}}\PY{l+s+s2}{south66}\PY{l+s+s2}{\PYZdq{}}\PY{p}{:} \PY{p}{(}\PY{l+m+mi}{80}\PY{p}{,} \PY{l+m+mi}{80}\PY{p}{)}\PY{p}{,}  \PY{c+c1}{\PYZsh{} lived in south in 1966}
    \PY{l+s+s2}{\PYZdq{}}\PY{l+s+s2}{work76}\PY{l+s+s2}{\PYZdq{}}\PY{p}{:} \PY{p}{(}\PY{l+m+mi}{82}\PY{p}{,} \PY{l+m+mi}{82}\PY{p}{)}\PY{p}{,}  \PY{c+c1}{\PYZsh{} worked in 1976}
    \PY{l+s+s2}{\PYZdq{}}\PY{l+s+s2}{work78}\PY{l+s+s2}{\PYZdq{}}\PY{p}{:} \PY{p}{(}\PY{l+m+mi}{84}\PY{p}{,} \PY{l+m+mi}{84}\PY{p}{)}\PY{p}{,}  \PY{c+c1}{\PYZsh{} worked in 1978}
    \PY{l+s+s2}{\PYZdq{}}\PY{l+s+s2}{lwage76}\PY{l+s+s2}{\PYZdq{}}\PY{p}{:} \PY{p}{(}\PY{l+m+mi}{86}\PY{p}{,} \PY{l+m+mi}{97}\PY{p}{)}\PY{p}{,}  \PY{c+c1}{\PYZsh{} log wage (outliers trimmed) 1976}
    \PY{l+s+s2}{\PYZdq{}}\PY{l+s+s2}{lwage78}\PY{l+s+s2}{\PYZdq{}}\PY{p}{:} \PY{p}{(}\PY{l+m+mi}{99}\PY{p}{,} \PY{l+m+mi}{110}\PY{p}{)}\PY{p}{,}  \PY{c+c1}{\PYZsh{} log wage in 1978 outliers trimmed}
    \PY{l+s+s2}{\PYZdq{}}\PY{l+s+s2}{famed}\PY{l+s+s2}{\PYZdq{}}\PY{p}{:} \PY{p}{(}\PY{l+m+mi}{112}\PY{p}{,} \PY{l+m+mi}{112}\PY{p}{)}\PY{p}{,}  \PY{c+c1}{\PYZsh{} mom\PYZhy{}dad education class 1\PYZhy{}9}
    \PY{l+s+s2}{\PYZdq{}}\PY{l+s+s2}{black}\PY{l+s+s2}{\PYZdq{}}\PY{p}{:} \PY{p}{(}\PY{l+m+mi}{114}\PY{p}{,} \PY{l+m+mi}{114}\PY{p}{)}\PY{p}{,}  \PY{c+c1}{\PYZsh{} 1 if black}
    \PY{l+s+s2}{\PYZdq{}}\PY{l+s+s2}{smsa76r}\PY{l+s+s2}{\PYZdq{}}\PY{p}{:} \PY{p}{(}\PY{l+m+mi}{116}\PY{p}{,} \PY{l+m+mi}{116}\PY{p}{)}\PY{p}{,}  \PY{c+c1}{\PYZsh{} in smsa in 1976}
    \PY{l+s+s2}{\PYZdq{}}\PY{l+s+s2}{smsa78r}\PY{l+s+s2}{\PYZdq{}}\PY{p}{:} \PY{p}{(}\PY{l+m+mi}{118}\PY{p}{,} \PY{l+m+mi}{118}\PY{p}{)}\PY{p}{,}  \PY{c+c1}{\PYZsh{} in smsa in 1978}
    \PY{l+s+s2}{\PYZdq{}}\PY{l+s+s2}{reg76r}\PY{l+s+s2}{\PYZdq{}}\PY{p}{:} \PY{p}{(}\PY{l+m+mi}{120}\PY{p}{,} \PY{l+m+mi}{120}\PY{p}{)}\PY{p}{,}  \PY{c+c1}{\PYZsh{} in south in 1976}
    \PY{l+s+s2}{\PYZdq{}}\PY{l+s+s2}{reg78r}\PY{l+s+s2}{\PYZdq{}}\PY{p}{:} \PY{p}{(}\PY{l+m+mi}{122}\PY{p}{,} \PY{l+m+mi}{122}\PY{p}{)}\PY{p}{,}  \PY{c+c1}{\PYZsh{} in south in 1978}
    \PY{l+s+s2}{\PYZdq{}}\PY{l+s+s2}{reg80r}\PY{l+s+s2}{\PYZdq{}}\PY{p}{:} \PY{p}{(}\PY{l+m+mi}{124}\PY{p}{,} \PY{l+m+mi}{124}\PY{p}{)}\PY{p}{,}  \PY{c+c1}{\PYZsh{} in south in 1980}
    \PY{l+s+s2}{\PYZdq{}}\PY{l+s+s2}{smsa66r}\PY{l+s+s2}{\PYZdq{}}\PY{p}{:} \PY{p}{(}\PY{l+m+mi}{126}\PY{p}{,} \PY{l+m+mi}{126}\PY{p}{)}\PY{p}{,}  \PY{c+c1}{\PYZsh{} in smsa in 1966}
    \PY{l+s+s2}{\PYZdq{}}\PY{l+s+s2}{wage76}\PY{l+s+s2}{\PYZdq{}}\PY{p}{:} \PY{p}{(}\PY{l+m+mi}{128}\PY{p}{,} \PY{l+m+mi}{132}\PY{p}{)}\PY{p}{,}  \PY{c+c1}{\PYZsh{} raw wage cents per hour 1976}
    \PY{l+s+s2}{\PYZdq{}}\PY{l+s+s2}{wage78}\PY{l+s+s2}{\PYZdq{}}\PY{p}{:} \PY{p}{(}\PY{l+m+mi}{134}\PY{p}{,} \PY{l+m+mi}{138}\PY{p}{)}\PY{p}{,}
    \PY{l+s+s2}{\PYZdq{}}\PY{l+s+s2}{wage80}\PY{l+s+s2}{\PYZdq{}}\PY{p}{:} \PY{p}{(}\PY{l+m+mi}{140}\PY{p}{,} \PY{l+m+mi}{144}\PY{p}{)}\PY{p}{,}
    \PY{l+s+s2}{\PYZdq{}}\PY{l+s+s2}{noint78}\PY{l+s+s2}{\PYZdq{}}\PY{p}{:} \PY{p}{(}\PY{l+m+mi}{146}\PY{p}{,} \PY{l+m+mi}{146}\PY{p}{)}\PY{p}{,}  \PY{c+c1}{\PYZsh{} 1 if noninterview in 78}
    \PY{l+s+s2}{\PYZdq{}}\PY{l+s+s2}{noint80}\PY{l+s+s2}{\PYZdq{}}\PY{p}{:} \PY{p}{(}\PY{l+m+mi}{148}\PY{p}{,} \PY{l+m+mi}{148}\PY{p}{)}\PY{p}{,}
    \PY{l+s+s2}{\PYZdq{}}\PY{l+s+s2}{enroll76}\PY{l+s+s2}{\PYZdq{}}\PY{p}{:} \PY{p}{(}\PY{l+m+mi}{150}\PY{p}{,} \PY{l+m+mi}{150}\PY{p}{)}\PY{p}{,}  \PY{c+c1}{\PYZsh{} 1 if enrolled in 76}
    \PY{l+s+s2}{\PYZdq{}}\PY{l+s+s2}{enroll78}\PY{l+s+s2}{\PYZdq{}}\PY{p}{:} \PY{p}{(}\PY{l+m+mi}{152}\PY{p}{,} \PY{l+m+mi}{152}\PY{p}{)}\PY{p}{,}
    \PY{l+s+s2}{\PYZdq{}}\PY{l+s+s2}{enroll80}\PY{l+s+s2}{\PYZdq{}}\PY{p}{:} \PY{p}{(}\PY{l+m+mi}{154}\PY{p}{,} \PY{l+m+mi}{154}\PY{p}{)}\PY{p}{,}
    \PY{l+s+s2}{\PYZdq{}}\PY{l+s+s2}{kww}\PY{l+s+s2}{\PYZdq{}}\PY{p}{:} \PY{p}{(}\PY{l+m+mi}{156}\PY{p}{,} \PY{l+m+mi}{157}\PY{p}{)}\PY{p}{,}  \PY{c+c1}{\PYZsh{} the kww score}
    \PY{l+s+s2}{\PYZdq{}}\PY{l+s+s2}{iq}\PY{l+s+s2}{\PYZdq{}}\PY{p}{:} \PY{p}{(}\PY{l+m+mi}{159}\PY{p}{,} \PY{l+m+mi}{161}\PY{p}{)}\PY{p}{,}  \PY{c+c1}{\PYZsh{} a normed iq score}
    \PY{l+s+s2}{\PYZdq{}}\PY{l+s+s2}{marsta76}\PY{l+s+s2}{\PYZdq{}}\PY{p}{:} \PY{p}{(}\PY{l+m+mi}{163}\PY{p}{,} \PY{l+m+mi}{163}\PY{p}{)}\PY{p}{,}  \PY{c+c1}{\PYZsh{} mar status in 1976 1=married, sp. present}
    \PY{l+s+s2}{\PYZdq{}}\PY{l+s+s2}{marsta78}\PY{l+s+s2}{\PYZdq{}}\PY{p}{:} \PY{p}{(}\PY{l+m+mi}{165}\PY{p}{,} \PY{l+m+mi}{165}\PY{p}{)}\PY{p}{,}
    \PY{l+s+s2}{\PYZdq{}}\PY{l+s+s2}{marsta80}\PY{l+s+s2}{\PYZdq{}}\PY{p}{:} \PY{p}{(}\PY{l+m+mi}{167}\PY{p}{,} \PY{l+m+mi}{167}\PY{p}{)}\PY{p}{,}
    \PY{l+s+s2}{\PYZdq{}}\PY{l+s+s2}{libcrd14}\PY{l+s+s2}{\PYZdq{}}\PY{p}{:} \PY{p}{(}\PY{l+m+mi}{169}\PY{p}{,} \PY{l+m+mi}{169}\PY{p}{)}\PY{p}{,}  \PY{c+c1}{\PYZsh{} 1 if lib card in home age 14}
\PY{p}{\PYZcb{}}

\PY{k}{with} \PY{n}{ZipFile}\PY{p}{(}\PY{n}{BytesIO}\PY{p}{(}\PY{n}{content}\PY{p}{)}\PY{p}{)}\PY{o}{.}\PY{n}{open}\PY{p}{(}\PY{l+s+s2}{\PYZdq{}}\PY{l+s+s2}{nls.dat}\PY{l+s+s2}{\PYZdq{}}\PY{p}{)} \PY{k}{as} \PY{n}{file}\PY{p}{:}
    \PY{n}{df} \PY{o}{=} \PY{n}{pd}\PY{o}{.}\PY{n}{read\PYZus{}fwf}\PY{p}{(}
        \PY{n}{file}\PY{p}{,}
        \PY{n}{names}\PY{o}{=}\PY{n}{colspec}\PY{o}{.}\PY{n}{keys}\PY{p}{(}\PY{p}{)}\PY{p}{,}
        \PY{c+c1}{\PYZsh{} pandas expects [from, to[ values, starting at 0}
        \PY{n}{colspecs}\PY{o}{=}\PY{p}{[}\PY{p}{(}\PY{n}{f} \PY{o}{\PYZhy{}} \PY{l+m+mi}{1}\PY{p}{,} \PY{n}{t}\PY{p}{)} \PY{k}{for} \PY{p}{(}\PY{n}{f}\PY{p}{,} \PY{n}{t}\PY{p}{)} \PY{o+ow}{in} \PY{n}{colspec}\PY{o}{.}\PY{n}{values}\PY{p}{(}\PY{p}{)}\PY{p}{]}\PY{p}{,}
        \PY{n}{na\PYZus{}values}\PY{o}{=}\PY{l+s+s2}{\PYZdq{}}\PY{l+s+s2}{.}\PY{l+s+s2}{\PYZdq{}}\PY{p}{,}
    \PY{p}{)}

\PY{n}{df} \PY{o}{=} \PY{n}{df}\PY{p}{[}\PY{k}{lambda} \PY{n}{x}\PY{p}{:} \PY{n}{x}\PY{p}{[}\PY{l+s+s2}{\PYZdq{}}\PY{l+s+s2}{lwage76}\PY{l+s+s2}{\PYZdq{}}\PY{p}{]}\PY{o}{.}\PY{n}{notna}\PY{p}{(}\PY{p}{)}\PY{p}{]}\PY{o}{.}\PY{n}{set\PYZus{}index}\PY{p}{(}\PY{l+s+s2}{\PYZdq{}}\PY{l+s+s2}{id}\PY{l+s+s2}{\PYZdq{}}\PY{p}{)}

\PY{c+c1}{\PYZsh{} construct potential experience and its square}
\PY{n}{df}\PY{p}{[}\PY{l+s+s2}{\PYZdq{}}\PY{l+s+s2}{exp76}\PY{l+s+s2}{\PYZdq{}}\PY{p}{]} \PY{o}{=} \PY{n}{df}\PY{p}{[}\PY{l+s+s2}{\PYZdq{}}\PY{l+s+s2}{age76}\PY{l+s+s2}{\PYZdq{}}\PY{p}{]} \PY{o}{\PYZhy{}} \PY{n}{df}\PY{p}{[}\PY{l+s+s2}{\PYZdq{}}\PY{l+s+s2}{ed76}\PY{l+s+s2}{\PYZdq{}}\PY{p}{]} \PY{o}{\PYZhy{}} \PY{l+m+mi}{6}
\PY{n}{df}\PY{p}{[}\PY{l+s+s2}{\PYZdq{}}\PY{l+s+s2}{exp762}\PY{l+s+s2}{\PYZdq{}}\PY{p}{]} \PY{o}{=} \PY{n}{df}\PY{p}{[}\PY{l+s+s2}{\PYZdq{}}\PY{l+s+s2}{exp76}\PY{l+s+s2}{\PYZdq{}}\PY{p}{]} \PY{o}{*}\PY{o}{*} \PY{l+m+mi}{2}
\PY{n}{df}\PY{p}{[}\PY{l+s+s2}{\PYZdq{}}\PY{l+s+s2}{age762}\PY{l+s+s2}{\PYZdq{}}\PY{p}{]} \PY{o}{=} \PY{n}{df}\PY{p}{[}\PY{l+s+s2}{\PYZdq{}}\PY{l+s+s2}{age76}\PY{l+s+s2}{\PYZdq{}}\PY{p}{]} \PY{o}{*}\PY{o}{*} \PY{l+m+mi}{2}

\PY{n}{df}\PY{p}{[}\PY{l+s+s2}{\PYZdq{}}\PY{l+s+s2}{f1}\PY{l+s+s2}{\PYZdq{}}\PY{p}{]} \PY{o}{=} \PY{n}{df}\PY{p}{[}\PY{l+s+s2}{\PYZdq{}}\PY{l+s+s2}{famed}\PY{l+s+s2}{\PYZdq{}}\PY{p}{]}\PY{o}{.}\PY{n}{eq}\PY{p}{(}\PY{l+m+mi}{1}\PY{p}{)}\PY{o}{.}\PY{n}{astype}\PY{p}{(}\PY{l+s+s2}{\PYZdq{}}\PY{l+s+s2}{float}\PY{l+s+s2}{\PYZdq{}}\PY{p}{)}  \PY{c+c1}{\PYZsh{} mom and dad both \PYZgt{} 12 yrs ed}
\PY{n}{df}\PY{p}{[}\PY{l+s+s2}{\PYZdq{}}\PY{l+s+s2}{f2}\PY{l+s+s2}{\PYZdq{}}\PY{p}{]} \PY{o}{=} \PY{n}{df}\PY{p}{[}\PY{l+s+s2}{\PYZdq{}}\PY{l+s+s2}{famed}\PY{l+s+s2}{\PYZdq{}}\PY{p}{]}\PY{o}{.}\PY{n}{eq}\PY{p}{(}\PY{l+m+mi}{2}\PY{p}{)}\PY{o}{.}\PY{n}{astype}\PY{p}{(}\PY{l+s+s2}{\PYZdq{}}\PY{l+s+s2}{float}\PY{l+s+s2}{\PYZdq{}}\PY{p}{)}  \PY{c+c1}{\PYZsh{} mom\PYZam{}dad \PYZgt{}=12 and not both exactly 12}
\PY{n}{df}\PY{p}{[}\PY{l+s+s2}{\PYZdq{}}\PY{l+s+s2}{f3}\PY{l+s+s2}{\PYZdq{}}\PY{p}{]} \PY{o}{=} \PY{n}{df}\PY{p}{[}\PY{l+s+s2}{\PYZdq{}}\PY{l+s+s2}{famed}\PY{l+s+s2}{\PYZdq{}}\PY{p}{]}\PY{o}{.}\PY{n}{eq}\PY{p}{(}\PY{l+m+mi}{3}\PY{p}{)}\PY{o}{.}\PY{n}{astype}\PY{p}{(}\PY{l+s+s2}{\PYZdq{}}\PY{l+s+s2}{float}\PY{l+s+s2}{\PYZdq{}}\PY{p}{)}  \PY{c+c1}{\PYZsh{} mom=dad=12}
\PY{n}{df}\PY{p}{[}\PY{l+s+s2}{\PYZdq{}}\PY{l+s+s2}{f4}\PY{l+s+s2}{\PYZdq{}}\PY{p}{]} \PY{o}{=} \PY{n}{df}\PY{p}{[}\PY{l+s+s2}{\PYZdq{}}\PY{l+s+s2}{famed}\PY{l+s+s2}{\PYZdq{}}\PY{p}{]}\PY{o}{.}\PY{n}{eq}\PY{p}{(}\PY{l+m+mi}{4}\PY{p}{)}\PY{o}{.}\PY{n}{astype}\PY{p}{(}\PY{l+s+s2}{\PYZdq{}}\PY{l+s+s2}{float}\PY{l+s+s2}{\PYZdq{}}\PY{p}{)}  \PY{c+c1}{\PYZsh{} mom \PYZgt{}=12 and dad missing}
\PY{n}{df}\PY{p}{[}\PY{l+s+s2}{\PYZdq{}}\PY{l+s+s2}{f5}\PY{l+s+s2}{\PYZdq{}}\PY{p}{]} \PY{o}{=} \PY{n}{df}\PY{p}{[}\PY{l+s+s2}{\PYZdq{}}\PY{l+s+s2}{famed}\PY{l+s+s2}{\PYZdq{}}\PY{p}{]}\PY{o}{.}\PY{n}{eq}\PY{p}{(}\PY{l+m+mi}{5}\PY{p}{)}\PY{o}{.}\PY{n}{astype}\PY{p}{(}\PY{l+s+s2}{\PYZdq{}}\PY{l+s+s2}{float}\PY{l+s+s2}{\PYZdq{}}\PY{p}{)}  \PY{c+c1}{\PYZsh{} father \PYZgt{}=12 and mom not in f1\PYZhy{}f4}
\PY{n}{df}\PY{p}{[}\PY{l+s+s2}{\PYZdq{}}\PY{l+s+s2}{f6}\PY{l+s+s2}{\PYZdq{}}\PY{p}{]} \PY{o}{=} \PY{n}{df}\PY{p}{[}\PY{l+s+s2}{\PYZdq{}}\PY{l+s+s2}{famed}\PY{l+s+s2}{\PYZdq{}}\PY{p}{]}\PY{o}{.}\PY{n}{eq}\PY{p}{(}\PY{l+m+mi}{6}\PY{p}{)}\PY{o}{.}\PY{n}{astype}\PY{p}{(}\PY{l+s+s2}{\PYZdq{}}\PY{l+s+s2}{float}\PY{l+s+s2}{\PYZdq{}}\PY{p}{)}  \PY{c+c1}{\PYZsh{} mom\PYZgt{}=12 and dad nonmissing}
\PY{n}{df}\PY{p}{[}\PY{l+s+s2}{\PYZdq{}}\PY{l+s+s2}{f7}\PY{l+s+s2}{\PYZdq{}}\PY{p}{]} \PY{o}{=} \PY{n}{df}\PY{p}{[}\PY{l+s+s2}{\PYZdq{}}\PY{l+s+s2}{famed}\PY{l+s+s2}{\PYZdq{}}\PY{p}{]}\PY{o}{.}\PY{n}{eq}\PY{p}{(}\PY{l+m+mi}{7}\PY{p}{)}\PY{o}{.}\PY{n}{astype}\PY{p}{(}\PY{l+s+s2}{\PYZdq{}}\PY{l+s+s2}{float}\PY{l+s+s2}{\PYZdq{}}\PY{p}{)}  \PY{c+c1}{\PYZsh{} mom and dad both \PYZgt{}=9}
\PY{n}{df}\PY{p}{[}\PY{l+s+s2}{\PYZdq{}}\PY{l+s+s2}{f8}\PY{l+s+s2}{\PYZdq{}}\PY{p}{]} \PY{o}{=} \PY{n}{df}\PY{p}{[}\PY{l+s+s2}{\PYZdq{}}\PY{l+s+s2}{famed}\PY{l+s+s2}{\PYZdq{}}\PY{p}{]}\PY{o}{.}\PY{n}{eq}\PY{p}{(}\PY{l+m+mi}{8}\PY{p}{)}\PY{o}{.}\PY{n}{astype}\PY{p}{(}\PY{l+s+s2}{\PYZdq{}}\PY{l+s+s2}{float}\PY{l+s+s2}{\PYZdq{}}\PY{p}{)}  \PY{c+c1}{\PYZsh{} mom and dad both nonmissing}

\PY{n}{indicators} \PY{o}{=} \PY{p}{[}\PY{l+s+s2}{\PYZdq{}}\PY{l+s+s2}{black}\PY{l+s+s2}{\PYZdq{}}\PY{p}{,} \PY{l+s+s2}{\PYZdq{}}\PY{l+s+s2}{smsa66r}\PY{l+s+s2}{\PYZdq{}}\PY{p}{,} \PY{l+s+s2}{\PYZdq{}}\PY{l+s+s2}{smsa76r}\PY{l+s+s2}{\PYZdq{}}\PY{p}{,} \PY{l+s+s2}{\PYZdq{}}\PY{l+s+s2}{reg76r}\PY{l+s+s2}{\PYZdq{}}\PY{p}{]}
\PY{c+c1}{\PYZsh{} exclude reg669, as sum(reg661, ..., reg669) = 1}
\PY{n}{indicators} \PY{o}{+}\PY{o}{=} \PY{p}{[}\PY{l+s+sa}{f}\PY{l+s+s2}{\PYZdq{}}\PY{l+s+s2}{reg66}\PY{l+s+si}{\PYZob{}}\PY{n}{i}\PY{l+s+si}{\PYZcb{}}\PY{l+s+s2}{\PYZdq{}} \PY{k}{for} \PY{n}{i} \PY{o+ow}{in} \PY{n+nb}{range}\PY{p}{(}\PY{l+m+mi}{1}\PY{p}{,} \PY{l+m+mi}{9}\PY{p}{)}\PY{p}{]}

\PY{n}{family} \PY{o}{=} \PY{p}{[}\PY{l+s+s2}{\PYZdq{}}\PY{l+s+s2}{daded}\PY{l+s+s2}{\PYZdq{}}\PY{p}{,} \PY{l+s+s2}{\PYZdq{}}\PY{l+s+s2}{momed}\PY{l+s+s2}{\PYZdq{}}\PY{p}{,} \PY{l+s+s2}{\PYZdq{}}\PY{l+s+s2}{nodaded}\PY{l+s+s2}{\PYZdq{}}\PY{p}{,} \PY{l+s+s2}{\PYZdq{}}\PY{l+s+s2}{nomomed}\PY{l+s+s2}{\PYZdq{}}\PY{p}{,} \PY{l+s+s2}{\PYZdq{}}\PY{l+s+s2}{famed}\PY{l+s+s2}{\PYZdq{}}\PY{p}{,} \PY{l+s+s2}{\PYZdq{}}\PY{l+s+s2}{momdad14}\PY{l+s+s2}{\PYZdq{}}\PY{p}{,} \PY{l+s+s2}{\PYZdq{}}\PY{l+s+s2}{sinmom14}\PY{l+s+s2}{\PYZdq{}}\PY{p}{]}
\PY{n}{fs} \PY{o}{=} \PY{p}{[}\PY{l+s+sa}{f}\PY{l+s+s2}{\PYZdq{}}\PY{l+s+s2}{f}\PY{l+s+si}{\PYZob{}}\PY{n}{i}\PY{l+s+si}{\PYZcb{}}\PY{l+s+s2}{\PYZdq{}} \PY{k}{for} \PY{n}{i} \PY{o+ow}{in} \PY{n+nb}{range}\PY{p}{(}\PY{l+m+mi}{1}\PY{p}{,} \PY{l+m+mi}{8}\PY{p}{)}\PY{p}{]}  \PY{c+c1}{\PYZsh{} exclude f8 as sum(f1, ..., f8) = 1}
\PY{n}{family} \PY{o}{+}\PY{o}{=} \PY{n}{fs}

\PY{c+c1}{\PYZsh{} endogenous variables: years of education, experience, experience squared}
\PY{n}{X} \PY{o}{=} \PY{n}{df}\PY{p}{[}\PY{p}{[}\PY{l+s+s2}{\PYZdq{}}\PY{l+s+s2}{ed76}\PY{l+s+s2}{\PYZdq{}}\PY{p}{,} \PY{l+s+s2}{\PYZdq{}}\PY{l+s+s2}{exp76}\PY{l+s+s2}{\PYZdq{}}\PY{p}{,} \PY{l+s+s2}{\PYZdq{}}\PY{l+s+s2}{exp762}\PY{l+s+s2}{\PYZdq{}}\PY{p}{]}\PY{p}{]}
\PY{n}{y} \PY{o}{=} \PY{n}{df}\PY{p}{[}\PY{l+s+s2}{\PYZdq{}}\PY{l+s+s2}{lwage76}\PY{l+s+s2}{\PYZdq{}}\PY{p}{]}  \PY{c+c1}{\PYZsh{} outcome: log wage}
\PY{c+c1}{\PYZsh{} included exog. variables: indicators for family background, region, race}
\PY{n}{C} \PY{o}{=} \PY{n}{df}\PY{p}{[}\PY{n}{indicators} \PY{o}{+} \PY{n}{family}\PY{p}{]}
\PY{c+c1}{\PYZsh{} instruments: proximity to colleges, age, and age squared}
\PY{n}{Z} \PY{o}{=} \PY{n}{df}\PY{p}{[}\PY{p}{[}\PY{l+s+s2}{\PYZdq{}}\PY{l+s+s2}{nearc4a}\PY{l+s+s2}{\PYZdq{}}\PY{p}{,} \PY{l+s+s2}{\PYZdq{}}\PY{l+s+s2}{nearc4b}\PY{l+s+s2}{\PYZdq{}}\PY{p}{,} \PY{l+s+s2}{\PYZdq{}}\PY{l+s+s2}{nearc2}\PY{l+s+s2}{\PYZdq{}}\PY{p}{,} \PY{l+s+s2}{\PYZdq{}}\PY{l+s+s2}{age76}\PY{l+s+s2}{\PYZdq{}}\PY{p}{,} \PY{l+s+s2}{\PYZdq{}}\PY{l+s+s2}{age762}\PY{l+s+s2}{\PYZdq{}}\PY{p}{]}\PY{p}{]}
\end{Verbatim}
\end{tcolorbox}
\end{jupyternotebook}

\section{Approximating the critical value function of the conditional likelihood ratio test}
\label{sec:approximate_clr_critical_value_function}
To implement the conditional likelihood-ratio tests of \citet{moreira2003conditional} and \citet{kleibergen2021efficient}, one needs to approximate the cumulative distribution function of the limiting distribution.
For this, \citeauthor*{moreira2003conditional} proposes to simulate i.i.d.\ draws from the limiting distribution and to use the empirical distribution function as an approximation.
For $m = \mX = 1$, they include a table for the critical values for various values of $k, s_\mathrm{min}$, and $\alpha$.

\citet{hillier2009conditional} proposes the following more efficient approximation for general $\mX =: p \geq 1$.
\begin{theoremEnd}[malte,category=clr_critical_value]{proposition}[\citeauthor{hillier2009conditional}, \citeyear{hillier2009conditional}]%
    \label{prop:clr_critical_value_function_approximation}
    Let $q \geq p > 0$ be whole numbers, let $\lambda \geq 0$ and $z > 0$.
    Write $a := \lambda / (z + \lambda)$.
    Then
    $$
    \BP[ \Gamma(q - p, p, \lambda) \leq z] = (1 - a)^{p/2} \sum_{j=0}^\infty \frac{a^j (p/2)_j}{j!}F_{\chi^2(p + 2 j)}(z + \lambda),
    $$
    where $(x)_j := \Gamma(x+j) / \Gamma(x) = \prod_{i=0}^{j-1} (x + i)$ is the Pochhammer symbol and $F_{\chi^2(p + 2 j)}$ is the cumulative distribution function of a chi-squared random variable with $p + 2 j$ degrees of freedom.
\end{theoremEnd}%
\begin{proofEnd}%
    This is Equation 28 of \citep{hillier2009conditional}.
\end{proofEnd}%

\noindent In their appendix, \citet{hillier2009conditional} also derives a bound for the approximation error when truncating the above series.
However, their bound is not tight enough for practical purposes and incorrect for $p=1$.
We prove a tighter bound in the following.
\begin{theoremEnd}[malte,category=clr_critical_value]{proposition}%
    \label{prop:clr_critical_value_function_tolerance}
    The error by approximating the power series in \cref{prop:clr_critical_value_function_approximation}
    with
    $$
        (1 - a)^{p/2} \sum_{j=0}^J \frac{(a)^j (p/2)_j}{j!} F_{\chi^2(q + 2 j)}(z + \lambda)
    $$
    is
    \begin{align*}
    (1 - a)^{p/2} \sum_{j=J+1}^\infty &\frac{(a)^j (p/2)_j}{j!} F_{\chi^2(q + 2 j)}(z + \lambda) \\
    &\leq F_{\chi^2(q + 2 J + 2)}(z + \lambda) \frac{a^{J+1} \ (p/2)_{J+1}}{(J+1)!} (1 + 2 / \sqrt{ -\log(a)} ).
    \end{align*}
\end{theoremEnd}%
\begin{proofEnd}%
    A similar statement is made in Section 5 of \citep{hillier2009conditional}.
    However, their bound is not tight enough for practical purposes and is incorrect for $p=1$.

    First, note that  $F_{\chi^2(q + 2 j)}(z + \lambda) \leq F_{\chi^2(q + 2 J + 2)}(z + \lambda)$ for $j \geq J + 1$ and thus
    \begin{align*}
        \sum_{j=J+1}^\infty \frac{a^j (p/2)_j}{j!} F_{\chi^2(q + 2 j)}(z + \lambda) \leq F_{\chi^2(q + 2 J + 2)}(z + \lambda) \sum_{j=J+1}^\infty \frac{a^j (p/2)_j}{j!}.
    \end{align*}

    Then, write
    \begin{equation*}
        (n)_{j + l} = \frac{\Gamma(n + j + l)}{\Gamma(n)} = \frac{\Gamma(n + j + l)}{\Gamma(n + j)} \frac{\Gamma(n + j)}{\Gamma(n)} = (n)_j (n + j)_l
    \end{equation*}
    and
    \begin{equation*}
        j! = (1)_j \ \Rightarrow \ (j + l)! = (1)_{j + l} = (1)_j (1 + j)_l.
    \end{equation*}
    Thus,
    \begin{align*}
        \sum_{j=J+1}^\infty \frac{a^j (p/2)_j}{j!}
        = \sum_{l = 0}^\infty \frac{a^{l + J +1} (p/2)_{l + J + 1}}{(l + J + 1)!}
        = \frac{a^{J+1} \ (p/2)_{J+1}}{(J+1)!} \sum_{l = 0}^\infty \frac{a^l \ (p/2 + J + 1)_l }{(J + 2)_l}.
    \end{align*}
    Euler's transformation formula for the hypergeometric function ${}_2F_1(a, b, c, x)$ is
    \begin{align*}
        \sum_{j = 0}^\infty \frac{(a)_j (b)_j}{(c)_j} \frac{x^j}{j!} = (1 - x)^{c - a - b} \sum_{j = 0}^\infty \frac{(c - a)_j (c - b)_j}{(c)_j} \frac{x^j}{j!}.
    \end{align*}
    We apply this with $a = p / 2 + J + 1, b = 1, c = J + 2$, and $x = a$. Then,
    \begin{align*}
        \sum_{l = 0}^\infty \frac{a^l \ (p/2 + J + 1)_l }{(J + 2)_l} = ( 1 - a )^{-p/2} \sum_{l = 0}^\infty \frac{(1 - p/2)_l (J + 1)_l}{(J + 2)_l} \frac{a^l}{l!}
        \leq ( 1 - a )^{-p/2} \sum_{l = 0}^\infty (1/2)_l \frac{a^l}{l!}.
    \end{align*}
    We rewrite $\frac{(1 / 2)_l}{l!} = \frac{\Gamma(l + 1/2)}{\Gamma(1/2) \Gamma(l + 1)}$ and use the fact that $\Gamma(1/2) = \sqrt{\pi}$ and $\frac{\Gamma(n + 1/2)}{\Gamma(n + 1)} \leq n^{-1/2}$ for $n \in \BZ^{\geq 1}$.
    Then,
    \begin{align*}
        \sum_{l = 0}^\infty (1/2)_l \frac{a^l}{l!} 
        &\leq 1 + \sum_{l = 1}^\infty \frac{a^l}{\sqrt{l} \sqrt{\pi}} \leq
        1 + \frac{1}{\sqrt{\pi}} \int_{l=1}^\infty \frac{a^l}{\sqrt{l}} \mathrm{d}l \\
        &\leq 1 + \frac{1}{\sqrt{\pi}} \frac{ \sqrt{\pi} (1 - \mathrm{erf}(\sqrt{-\log(a)}))}{\sqrt{-\log(a)}} \leq 1 + \frac{2}{\sqrt{-\log(a)}},
    \end{align*}
    where $-1 < \erf(x) < 1$ is the Gauss error function.

    Combining the above yields
    \begin{align*}
        (1 - a)^{p/2} \sum_{j=J+1}^\infty &\frac{a^j (p/2)_j}{j!} F_{\chi^2(q + 2 j)}(z + \lambda) \\
        &\leq F_{\chi^2(q + 2 J + 2)}(z + \lambda) \frac{a^{J+1} \ (p/2)_{J+1}}{(J+1)!} (1 + \frac{2}{ \sqrt{ -\log(a)} })
    \end{align*}
\end{proofEnd}%

The convergence of the power series in \cref{prop:clr_critical_value_function_approximation} is slow when identification is reasonably large, say $\lambda = s_\mathrm{min}(\beta) = 1000$.
For example if additionally $q=k = 20, p=m_x = 5$ and $z = 5$, the bound from \cref{prop:clr_critical_value_function_tolerance} is smaller than $10^{-6}$ only for $J \geq 5495$.
The approximation error itself is smaller than $10^{-6}$ for $J \geq 5227$.
On the other hand, the identification $\lambda = s_\mathrm{min}(\beta) = 1000$ is not large enough for the $\chi^2(\mX)$ approximation to be close.
Here, the approximation error is around $0.01$.

In the \texttt{ivmodels} software package, we thus approximate the cumulative distribution function of $\Gamma(q - p, p, \lambda)$ by numerically integrating a variable transformed variant of Equation (27)
$$
\BP[ \Gamma(q - p, p, \lambda) \leq z] = \BE_{x \sim \mathrm{Beta}((q - p)/2, p/2)}[ F_{\chi^2(q)}(z / (1 - a x)) ]
$$
of \citep{hillier2009conditional}, where $a = \lambda / (z + \lambda)$.
For this, we substitute $y := 1 - a x$, let $\alpha := (q - p)/2, \beta = p/2$ and use $f_{B(\alpha, \beta)}(x) = \frac{1}{B(\alpha, \beta)} x^{\alpha - 1} (1 - x)^{\beta - 1}$ and $F_{\chi^2(q)}(z) = \frac{1}{\Gamma(q/2)} \gamma(q/2, z/2)$, where $\Gamma$ denotes the gamma function and $\gamma$ is the lower incomplete gamma function.
The integral becomes
\begin{multline*}    
\BE_{x \sim \mathrm{Beta}(\alpha, \beta)}[ F_{\chi^2(q)}(z / (1 - a x)) ] = \frac{a^{-q / 2 + 1}}{B(\alpha, \beta) \Gamma(q / 2)} \int_{1 - a}^1 (y - (1 - a))^{\alpha - 1} (1 - y)^{\beta - 1} \gamma(q / 2, z / (2 y)) \, \dd y.
\end{multline*}
Integrals of this form $\int_a^b (x - a)^\alpha (b - x)^\beta f(x) \dd x$ can be efficiently approximated by the \texttt{qawse} routine of \texttt{QUADPACK}.
Even for difficult settings as above with $q=k=20, p=\mX = 5$,$z=5$, and $\lambda = s_\mathrm{min}(\beta) = 1000$, the integral can be approximated up to an error of $10^{-6}$ with around 50 evaluations of $\gamma$.

\section{Proofs}
\label{app:proofs}
\subsection{Auxiliary results}
\label{app:proofs:auxiliary}
\begin{theoremEnd}[malte_all_end]{lemma}[\citeauthor{londschien2024weak}, \citeyear{londschien2024weak}]
    \label{lem:kappa_pos_definite}
    Assume that $M_Z X$ is of full column rank.
    Let $\lambda_1 := \lambdamin{ (X^T M_Z X)^{-1} X^T X} = \lambdamin{ (X^T M_Z X)^{-1} X^T P_Z X} + 1$.
    \begin{itemize}
    \item Then $M(\kappa) := X^T (\kappa P_Z + (1 - \kappa) \Id) X$ is positive definite if and only if $\kappa < \lambda_1$.
    \item The matrix $M(\kappa)$ is singular positive semi-definite if and only if $\kappa = \lambda_1$.
    \item The matrix $M(\kappa)$ has at least one negative eigenvalue if and only if $\kappa > \lambda_1$.
    \end{itemize}
\end{theoremEnd}%
\begin{proofEnd}%
    Let $v$ be an eigenvector corresponding to an eigenvalue $\lambda$ of $(X^T M_Z X)^{-1} X^T X$.
    Then
    $$
        (X^T M_Z X)^{-1} X^T X v = \lambda v \Leftrightarrow \lambda X^T M_Z X v = X^T X v \Leftrightarrow X^T (\Id - \lambda M_Z) X v = 0.
    $$
    That is, the eigenvalues of $(X^T M_Z X)^{-1} X^T X v$ correspond to $\kappa$ such that $X^T (\Id - \kappa M_Z) X v = M(\kappa) v =  0$.
    Also, as $X^T M_Z X \preceq X^T X$, we have $\lambda \geq 1$.

    Note that, as $M_Z X$ is of full column rank, $X^T M_Z X$ is positive definite.
    Write $M(\kappa) = X^T (P_Z + (1 - \lambda) M_Z) X + (\lambda - \kappa) X^T M_Z X$ such that for any eigenvector, eigenvalue pair $(v, \lambda)$ of $(X^T M_Z X)^{-1} X^T X$, we have $v^T M(\kappa) v = (\lambda - \kappa) v^T X^T M_Z Xv$.
    \begin{itemize}
    \item If $\kappa < \lambda_1 \leq \lambda$, then $v^T M(\kappa) v > 0$ by positive definiteness of $X^T M_Z X$ and thus $M(\kappa)$ is positive definite.
    \item If $\kappa \leq \lambda_1 \leq \lambda$, then $v^T M(\kappa) v_1 \geq 0$ and thus $M(\kappa)$ is positive semi-definite.
    Let $v_1$ be an eigenvector corresponding to $\lambda_1$. Then $v_1^T M(\lambda_1) v_1 = 0$ and thus $M(\lambda_1)$ is not positive definite.
    \item Let $v_1$ be the eigenvector corresponding to $\lambda_1$.
    If $\kappa > \lambda_1$, then $v^T M(\kappa) v = (\lambda_1 - \kappa) v^T X^T M_Z X v < 0$ and thus $M(\kappa)$ has at least one negative eigenvalue.
    \end{itemize}
    The other direction follows as we exhaust all possible cases.
\end{proofEnd}%

\begin{theoremEnd}[malte_all_end]{lemma}[\citeauthor{londschien2024weak}, \citeyear{londschien2024weak}]
    \label{lemma:ar_statistic_derivative}
    Let $\tX(\beta) := X - (y - X \beta) \frac{(y - X \beta)^T M_Z X}{(y - X \beta)^T M_Z (y - X \beta)}$.
    Then,
    $$
    \frac{k}{n-k} \frac{\dd}{\dd \beta} \AR(\beta) = -2 \frac{(y - X \beta)^T P_Z \tilde X(\beta)_j}{(y - X \beta)^T M_Z (y - X \beta)}
    $$
\end{theoremEnd}%
\begin{proofEnd}%
    Write $u = y - X \beta$. Calculate
    \begin{align*}
    \frac{\dd}{\dd \beta_j} \frac{u^T P_Z u}{u^T M_Z u} &=
    -2 \frac{u^T P_Z X_j}{u^T M_Z u} + 2 \frac{u^T P_Z u}{u^T M_Z u} \frac{u^T M_Z X_j}{u^T M_Z u} \\
    &= -2 \frac{u^T P_Z \tilde X(\beta)_j}{u^T M_Z u}.
    \end{align*}
\end{proofEnd}%

\begin{theoremEnd}[malte_all_end]{lemma}[\citeauthor{londschien2024weak}, \citeyear{londschien2024weak}]
    \label{lem:projected_quadric}
    Let $n\geq m$, $A = \begin{pmatrix} A_{11} & A_{12} \\ A_{21} & A_{22} \end{pmatrix} \in \BR^{n \times n}$ with $A_{11} \in \BR^m$, let $z_0 \in \BR^n$ and $c \in \BR$.
    Let $B \in \BR^{m \times n}$ have ones on the diagonal and zeros elsewhere.
    Then
    \begin{align*}
        &\ \  \{ x \in \BR^{m} \mid \min_{y \in \BR^{n - m}} \left(\begin{pmatrix} x \\ y \end{pmatrix} - z_0 \right)^T A \left(\begin{pmatrix} x \\ y \end{pmatrix} - z_0 \right) \leq c \} \\
        &= \begin{cases}
            \ \ \{ x \in \BR^m \mid x^T (A_{11} - A_{12} A_{22}^{\dagger} A_{21}) x \leq c \} + B z_0 & \ \ \text{if } \lambdamin{A_{22}} \geq 0 \\
            \ \ \BR^{m} & \ \ \text{if } \lambdamin{A_{22}} < 0
            \end{cases} \numberthis \label{eq:lem:projected_quadric:1} \\
        &= \begin{cases}
            \ \ \{ x \in \BR^m \mid (x - B z_0)^T (B A^{\dagger} B^T)^{\dagger} (x - B z_0) \leq c \} &  \ \ \text{if } \lambdamin{A_{22}} \geq 0 \\
            \ \ \BR^{m} & \ \  \text{if } \lambdamin{A_{22}} < 0,
            \end{cases} \numberthis \label{eq:lem:projected_quadric:2}
    \end{align*}
    where $\dagger$ denotes the Moore-Penrose pseudo-inverse.
\end{theoremEnd}%
\begin{proofEnd}%
        First,
        $$
            \{ x \in \BR^{m} \mid \min_{y \in \BR^{n - m}} \left(\begin{pmatrix} x \\ y \end{pmatrix} - z_0 \right)^T A \left(\begin{pmatrix} x \\ y \end{pmatrix} - z_0 \right) \leq c \} = \{ \tilde x \in \BR^{m} \mid \min_{y \in \BR^{n - m}} \begin{pmatrix} \tilde x \\ y \end{pmatrix}^T A \begin{pmatrix} \tilde x \\ y \end{pmatrix} \leq c \} + B z_0
        $$
        via a change of variables $\tilde x = x - B z_0$.
        \paragraph*{Case 1: $A_{22}$ is negative definite.}
        Let $v \in \BR^{n - m}$ be an eigenvector of $A_{22}$ with eigenvalue $\lambda < 0$.
        Then,
        $$
            \inf_{y \in \BR^{n-m}} \begin{pmatrix} x \\ y \end{pmatrix}^T A \begin{pmatrix} x \\ y \end{pmatrix} \leq \inf_{t \in \BR} \begin{pmatrix} x \\ tv \end{pmatrix}^T A \begin{pmatrix} x \\ tv \end{pmatrix} = \inf_t x^T A_{11} x+ x^T A_{12} v t + \lambda \| v \| t^2 = -\infty
        $$
        Thus, the set is $\BR^m$.

        \paragraph*{Case 2: $A_{22}$ is positive semi-definite.}
        Then, any minimizer $y^\star(x)$ of $y \mapsto \begin{pmatrix} x \\ y \end{pmatrix}^T A \begin{pmatrix} x \\ y \end{pmatrix}$ satisfies $0 = \frac{\dd}{\dd y} \left[ \begin{pmatrix} x \\ y \end{pmatrix}^T A \begin{pmatrix} x \\ y \end{pmatrix}\right] = 2 A_{21} x + 2 A_{22} y^\star(x)$ and $\min_y \begin{pmatrix} x \\ y \end{pmatrix}^T A \begin{pmatrix} x \\ y \end{pmatrix}= x(A_{11} - A_{12} A_{22}^\dagger A_{21}) x$.
        This proves \eqref{eq:lem:projected_quadric:1}.
        \eqref{eq:lem:projected_quadric:2} follows from the formula for the pseudo-inverse of a block matrix in \citep{rohde1965generalized}.
\end{proofEnd}%

\subsection{Proofs of \cref{sec:intro_to_iv_tests:model}}
\label{app:proofs:model}
\printProofs[model]
\subsection{Proofs of \cref{sec:intro_to_iv_tests:estimators}}
\label{app:proofs:estimators}
\printProofs[estimators]
\subsection{Proofs of \cref{sec:intro_to_iv_tests:tests}}
\label{app:proofs:tests}
\printProofs[tests]
\subsection{Proofs of \cref{sec:intro_to_iv_tests:auxiliary_tests}}
\label{app:proofs:auxiliary_tests}
\printProofs[tests2]
\printProofs[cyrill]
\subsection{Proofs of \cref{sec:intro_to_iv_tests:confidence_sets}}
\label{app:proofs:confsets}
\printProofs[confsets]
\subsection{Proofs of \cref{sec:exogenous_variables}}
\label{app:proofs:exogenous}
\printProofs[exogenous]
\subsection{Proofs of appendix \ref{sec:approximate_clr_critical_value_function}}
\printProofs[clr_critical_value]
\end{document}